\documentclass[a4paper,11pt]{article}
\pdfoutput=1

\usepackage[utf8]{inputenc}
\usepackage[english, italian]{babel}
\usepackage[T1]{fontenc}

\usepackage[titletoc,toc,title]{appendix}

\usepackage{graphicx}
\usepackage{amsmath}
\usepackage{amsfonts}
\usepackage{float}

\usepackage{booktabs}
\usepackage{array}
\usepackage{paralist}
\usepackage{verbatim}
\usepackage{subfig}
\usepackage{multirow}
\usepackage{empheq}

\usepackage{feynmf}
\usepackage{rotating}

\include{caption}
\include{subfig}

\usepackage{adjustbox}

\usepackage{color} 
\usepackage{cite}

\usepackage{paracol}

\usepackage{hyperref}
\newcommand{\mail}[1]{\href{mailto:#1}{\texttt{#1}}}

\renewcommand{\epsilon}{\varepsilon}
\newcommand{\ep}{\varepsilon}
\renewcommand{\theta}{\vartheta}
\renewcommand{\rho}{\varrho}
\renewcommand{\phi}{\varphi}

	{\left\lbrace\begin{array}{@{}l@{}}}%
	{\end{array}\right.}

\newcommand{\imineq}[2]{\vcenter{\hbox{\includegraphics[height=#2ex]{#1}}}}

\newsavebox{\overlongequation}

\begin{document}
\selectlanguage{english}

\setlength{\unitlength}{1.3cm} 
\begin{titlepage}
\vspace*{-1cm}
\begin{flushright}
CERN-TH-2017-247\\TTP17-047
\end{flushright}                                
\vskip 3.5cm
\begin{center}
\boldmath
 
{\Large\bf Three-loop mixed QCD-electroweak  corrections to Higgs boson gluon fusion\\[3mm] }
\unboldmath
\vskip 1.cm
{\large Marco Bonetti}$^{a,}$
\footnote{\mail{ marco.bonetti@kit.edu}},
{\large Kirill Melnikov}$^{a,}$
\footnote{\mail{kirill.melnikov@kit.edu}} and
{\large Lorenzo Tancredi}$^{b,}$
\footnote{\mail{lorenzo.tancredi@cern.ch}} 
\vskip .7cm
{\it $^a$ Institute for Theoretical Particle Physics, KIT, 76128 Karlsruhe, 
Germany } \\
{\it $^b$ CERN Theory Division, CH-1211, Geneva 23, Switzerland } 
\end{center}
\vskip 2.6cm

\begin{abstract}
We compute the contribution of  three-loop mixed QCD-electroweak 
corrections ($\alpha_S^2\alpha^2$) to the $gg \to H$ scattering amplitude.
We employ  the method of differential equations 
to compute the relevant integrals and  express them in terms of Goncharov polylogarithms.

\vskip .7cm 
{\it Key words}: Higgs gluon fusion, electroweak corrections, 3-loop computations.
\end{abstract}
\vfill
\end{titlepage}                                                                
\newpage


\section{Introduction}\setcounter{equation}{0} 
\numberwithin{equation}{section}
\label{intro}
It is an open question if 
the scalar boson discovered by ATLAS and CMS collaborations 
in 2012  is indeed the Higgs boson  of the Standard Model.  To answer this 
question, Higgs  boson production cross sections and decay rates are measured 
experimentally   and  compared to theoretical predictions.  
Thanks to the renormalizability of the Standard Model,
properties of the Higgs 
boson, including its quantum numbers and its couplings to gauge bosons and matter 
particles, are completely fixed.  As a result,   it is   
possible, at least in principle,   to describe 
 production and decay  of the Higgs boson in  the Standard Model 
with unlimited  precision.
 
On the other hand, if the  Standard Model is not a complete theory,  New Physics 
at scale $\Lambda$ that couples to the Higgs boson is expected to modify its 
couplings to matter and gauge particles 
by an amount $\delta g/g \sim {\cal O}(v^2/\Lambda^2)$  where 
$v = 246~{\rm GeV}$ is the  scale of electroweak symmetry breaking.  Taking  
$\Lambda = 1~{\rm TeV}$, we find a typical   modification in Higgs couplings of 
about $5\%$, which sets a precision goal for both experimental measurements and 
theoretical computations. 

The major Higgs-boson production channel at the LHC is 
gluon fusion $gg \to H$;  it contributes  more than  $90\%$ to the total Higgs production 
cross section at  $13\,\mathrm{TeV}$. As discussed in 
Refs.~\cite{Anastasiou:2016cez,Patrignani:2016xqp}, almost 
$5\%$ of the gluon fusion cross section  comes from mixed QCD-electroweak (QCD-EW) 
contributions
where the Higgs boson  couples to electroweak gauge bosons that, later,  convert to gluons 
through a quark loop.    The remaining $95\%$ of the $gg \to H$ cross section
is generated by pure QCD interactions.  

The importance of the $gg \to H$ channel for the investigation 
of the Higgs boson properties motivated its extensive studies in the past.
Since the Higgs boson 
is light and the top quarks are relatively heavy, it is possible to  integrate 
them out and work with an effective Lagrangian where the $gg H$ coupling is point-like. 
The $gg \to H$ production cross section in  the $m_t \to \infty$ approximation 
is known through N$^3$LO in QCD while other, smaller contributions, are known 
less precisely.   

At present,  the theoretical 
uncertainty in $gg \to H$ cross section 
originates from (see \cite{Anastasiou:2016cez,Patrignani:2016xqp})
the ${\cal O}(2\%)$ residual scale 
uncertainty  in pure QCD contributions, the ${\cal O}(1\%)$ uncertainty caused 
by unknown  mass effects of $b$ and $c$ quarks in higher orders of QCD perturbation theory, 
and  the ${\cal O}(1\%)$ uncertainty in QCD-EW contributions.
The latter uncertainty is particularly peculiar since computation of QCD corrections 
to mixed QCD-EW contributions exists \cite{Anastasiou:2008tj}.
However, although  the leading order  QCD-EW 
$\mathcal{O}(\alpha_S^{~}\alpha^2)$
 contribution   is known for 
arbitrary values of the Higgs and electroweak gauge boson 
masses~\cite{Aglietti:2006yd,Aglietti:2004nj,Aglietti:2004ki,Bonetti:2016brm}, 
the next-to-leading order (NLO)  QCD corrections to it  have only been 
computed  in the unphysical  limit 
$m_H \ll m_{W,Z}$ \cite{Anastasiou:2008tj}. 
It turns out \cite{Anastasiou:2008tj} that in this 
approximation  the QCD  corrections to mixed  QCD-EW contributions  
are large and, in fact,  similar  to  NLO corrections  to leading QCD contributions. 
Nevertheless, the magnitude of  QCD 
corrections and the approximate nature of the calculation of 
Ref.~\cite{Anastasiou:2008tj} are major reasons for assigning a large  uncertainty 
to the QCD-EW contribution to $gg \to H$ in Ref.~\cite{Anastasiou:2016cez}. 
To remove this  theoretical 
uncertainty, the NLO QCD corrections to QCD-EW contribution to $gg \to H$ 
should be re-computed 
for physical values of $m_H,m_W$ and $m_Z$.

The goal of this paper is to present one of the ingredients of such a computation -- 
the virtual QCD corrections to mixed QCD-EW contribution 
 to $gg \to H$ amplitude. The NLO QCD-EW corrections are given by  
three-loop  $\mathcal{O}(\alpha_S^2\alpha^2)$ diagrams where gluons couple to 
electroweak gauge bosons through quark 
loops,  and  electroweak gauge bosons fuse into Higgs boson.  
In principle, all quark flavors   
contribute to mixed QCD-EW corrections 
but, since the Higgs boson is 
light, the contribution of top quark loops is small~\cite{Anastasiou:2008tj}. For this 
reason we focus on the contribution of massless  quarks in what follows. 

To compute the three-loop mixed QCD-EW contribution to  the $gg \to H$ amplitude, 
we use the method of differential equations~\cite{Kotikov:1990kg,Remiddi:1997ny,Gehrmann:1999as} to 
calculate the relevant master integrals (MIs). We  employ the concept of canonical 
basis introduced in~Ref.\cite{Henn:2013pwa}, and express
the MIs  in terms of Goncharov's multiple
 polylogarithms (GPLs)~\cite{Remiddi:1999ew, Goncharov, Vollinga:2004sn}, up to weight 6. 
We use the mathematical limit of small Higgs mass $m_H \ll m_{W,Z}$ to determine the boundary 
values for the solutions of the differential equations.  Our final three-loop 
result for QCD-EW contributions to  $gg \to H$  amplitude is expressed in terms of 
GPLs up to weight five. 

The paper is organized as follows. 
We introduce the gluon fusion amplitude $gg \to H$ and discuss 
 the QCD-EW contributions  in  Section~\ref{glufu}.
In Section~\ref{topo} we describe the classes of integrals 
that contribute to the  QCD-EW amplitude. 
In  Section~\ref{deq} we  discuss differential equations for MIs and their simplification to a canonical form.  We introduce the large-mass expansion  
in Section~\ref{bclme} and explain how we use it to provide boundary conditions for the solutions 
of the differential equations. 
The final result for the $gg \to H$ amplitude is given in 
Section~\ref{ampli}. We conclude in Section~\ref{sec:con}.
Additional information about integral topologies and 
MIs can be found, respectively, in Appendices \ref{Topologies} and \ref{Master_Integrals}.
We provide the analytic results for leading and next-to-leading order QCD-EW contributions to $gg \to H$ amplitude in the ancillary file.

Additional material that includes the expressions for MIs, the basis of canonical functions, the matrix of coefficients of the differential equations and the integration constants is available at \url{https://www.ttp.kit.edu/_media/progdata/2017/ttp17-047/anc.tar.gz}.

\section{QCD-EW contribution to \boldmath{$gg \to H$} amplitude}
\label{glufu}

\begin{figure}[t]
      \centering
      \includegraphics[width=0.50\textwidth]{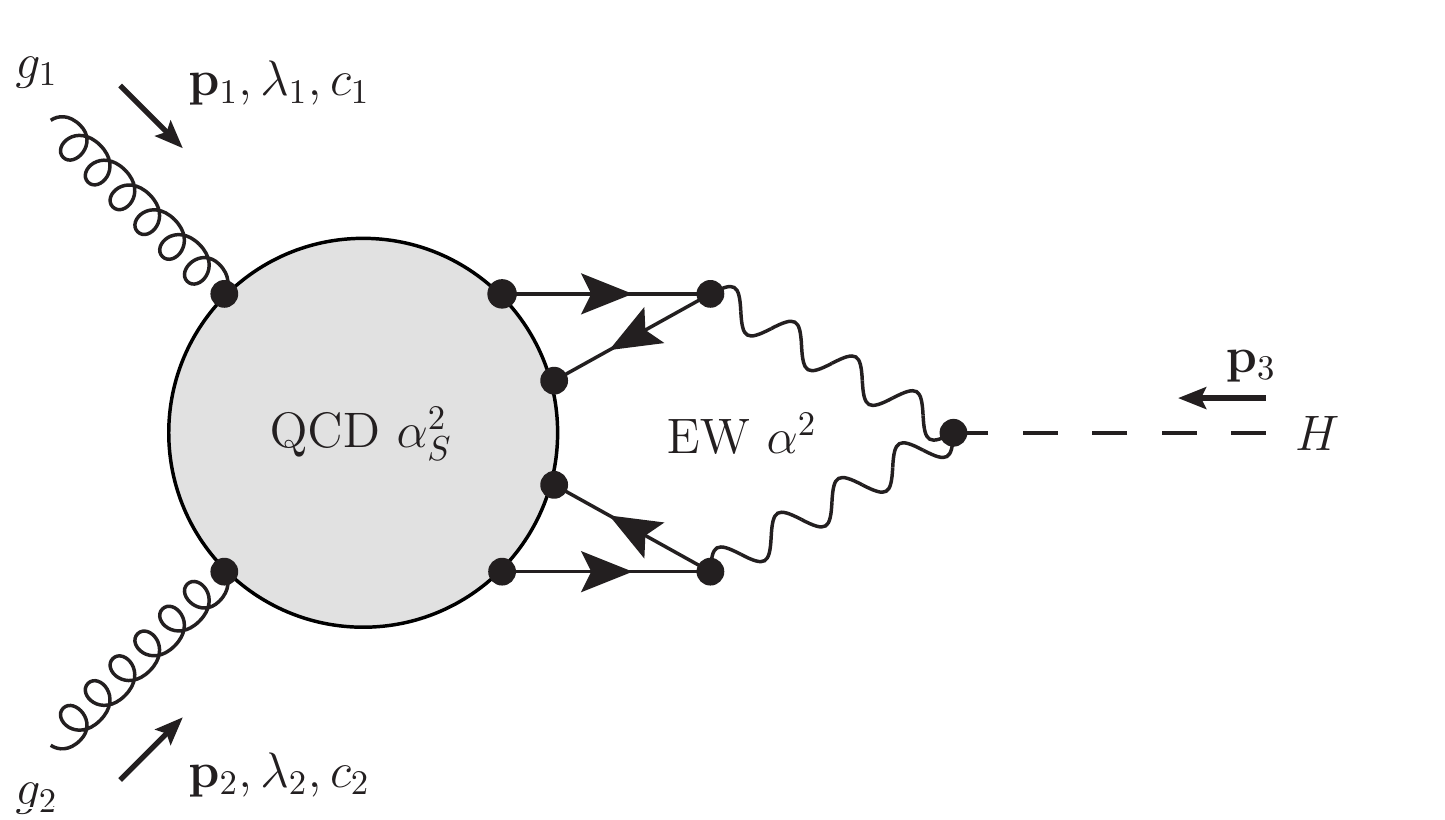}
      \caption{
Mixed QCD-EW corrections to Higgs boson production 
in gluon fusion. The powers of the couplings refer to  NLO contributions.}
      \label{Gluon_f}
\end{figure}

We show the mixed QCD-electroweak corrections to the $gg \to H$ amplitude 
in Fig.~\ref{Gluon_f}. 
They appear for the first time at two loops. The two  gluons annihilate 
to electroweak vector bosons through a quark loop. 
The vector bosons fuse into a Higgs boson in a second step.  
These leading order  QCD-electroweak contributions are well-known~\cite{Aglietti:2006yd,Aglietti:2004nj,Aglietti:2004ki,Bonetti:2016brm}. 
Our goal in this paper is to 
compute the QCD  corrections to the leading order QCD-EW contribution 
to $gg \to H$ amplitude. 

Both $W$ and $Z$ bosons  appear in diagrams that contribute to the QCD-EW amplitude. 
In the case of $W$-bosons, we  account for 
quarks of the first two generations; 
when considering $Z$-bosons, we  include, in addition,   bottom quarks. 
We treat all quarks as  massless.  To write an  expression 
for  the $gg \to H $ amplitude we assume that  
the two incoming  gluons 
 $g_1$ and $g_2$  carry  on-shell momenta $p_{1,2}$, color indices   $c_{1,2}$ and 
polarization labels $\lambda_{1,2}$.
The momentum of the outgoing Higgs boson is taken to 
be $p_3=-p_1-p_2$, with $p_3^2 = m_H^2 = s$.

QCD gauge invariance ensures that the 
$gg \to H$ amplitude depends on a  single form factor
\begin{equation}
 \mathcal{M}^{c_1c_2}_{\lambda_1\lambda_2}=\mathcal{F}\left(s,m_W^2,m_Z^2\right) \;
\delta^{c_1 c_2} \; \epsilon_{\lambda_1}(\mathbf{p}_1) \cdot \epsilon_{\lambda_2}(\mathbf{p}_2),
\end{equation}
where $\epsilon_{\lambda_{1,2}}$ are
gluon polarization vectors that satisfy the transversality 
conditions  $\epsilon_{\lambda_i} \cdot p_1 = \epsilon_{\lambda_i} \cdot p_2 = 0$, 
for $i = 1,2$. 
It is convenient to construct a  projection operator that can be applied 
to individual Feynman diagrams to extract their contributions to the  form factor
$\mathcal{F}$. 
The projection operator reads
\begin{gather}
 \mathbb{P}_{c_1c_2}^{\lambda_1\lambda_2}=
\frac{\epsilon^{*\lambda_1}(\mathbf{p}_1)
 \cdot \epsilon^{*\lambda_2}(\mathbf{p}_2)
 }{d-2} \frac{\delta_{c_1c_2}}{N_c^2-1},
\end{gather}
where $d = 4-2\epsilon$ is the space-time dimensionality. 
Using the standard expression for the 
 sum over polarizations  for each of the two gluons,  consistent with 
the transversality conditions  
\begin{equation}
\sum \limits_{\lambda_i = \pm}^{} 
\epsilon^\mu_{\lambda_i}(\mathbf{p}_i) \epsilon^{*\lambda_i,\nu}(\mathbf{p}_i)
= -g^{\mu \nu} + \frac{p_1^\mu p_2^\nu+p_1^\nu p_2^\mu}{p_1\cdot p_2},	
\end{equation}
it is easy to show that the following equation holds
\begin{equation}
 \mathcal{F}\left(s,m_W^2,m_Z^2\right)=\sum_{\lambda_1,\lambda_2,c_1,c_2}\mathbb{P}_{c_1,c_2}^{\lambda_1\lambda_2}\mathcal{M}_{\lambda_1\lambda_2}^
{c_1,c_2}.
\end{equation}

We can further simplify this equation if we  write  the amplitude as  
\begin{equation}
 \mathcal{M}^{c_1c_2}_{\lambda_1\lambda_2}= {\cal A}^{c_1 c_2}_{\mu \nu} 
\epsilon^\mu_{\lambda_1}(\mathbf{p}_1) \epsilon^\nu_{\lambda_2}(\mathbf{p}_2),
\end{equation}
apply the projection operator and sum over helicity labels of the 
two gluons, to obtain
\begin{equation} 
 \mathcal{F}\left(s,m_W^2,m_Z^2\right)
= \frac{{\cal A}^{c_1 c_2}_{\mu \nu} \delta_{c_1 c_2}}{(N_c^2-1)(d-2)} 
\left ( - g^{\mu \nu} + \frac{p_1^\mu p_2^\nu+p_1^\nu p_2^\mu}{p_1\cdot p_2} \right ).
\label{eq2.6}
\end{equation}
This formula can be used to extract the contribution of individual diagrams to 
the gauge-invariant form factor $ \mathcal{F}$. 

\begin{figure}[h]
      \centering
      \subfloat
      {\includegraphics[width=0.28\textwidth]{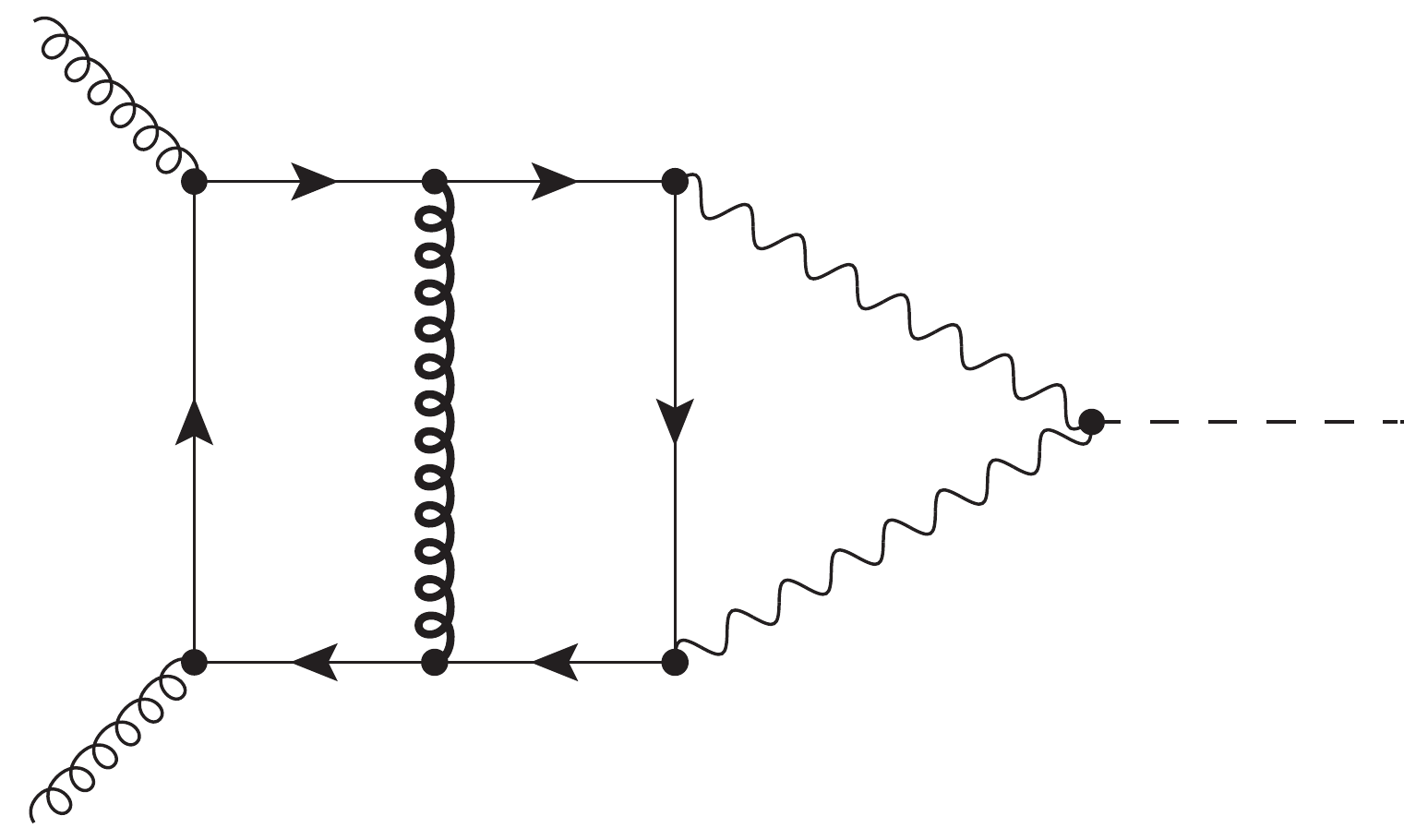}}
      \qquad
      \subfloat
      {\includegraphics[width=0.28\textwidth]{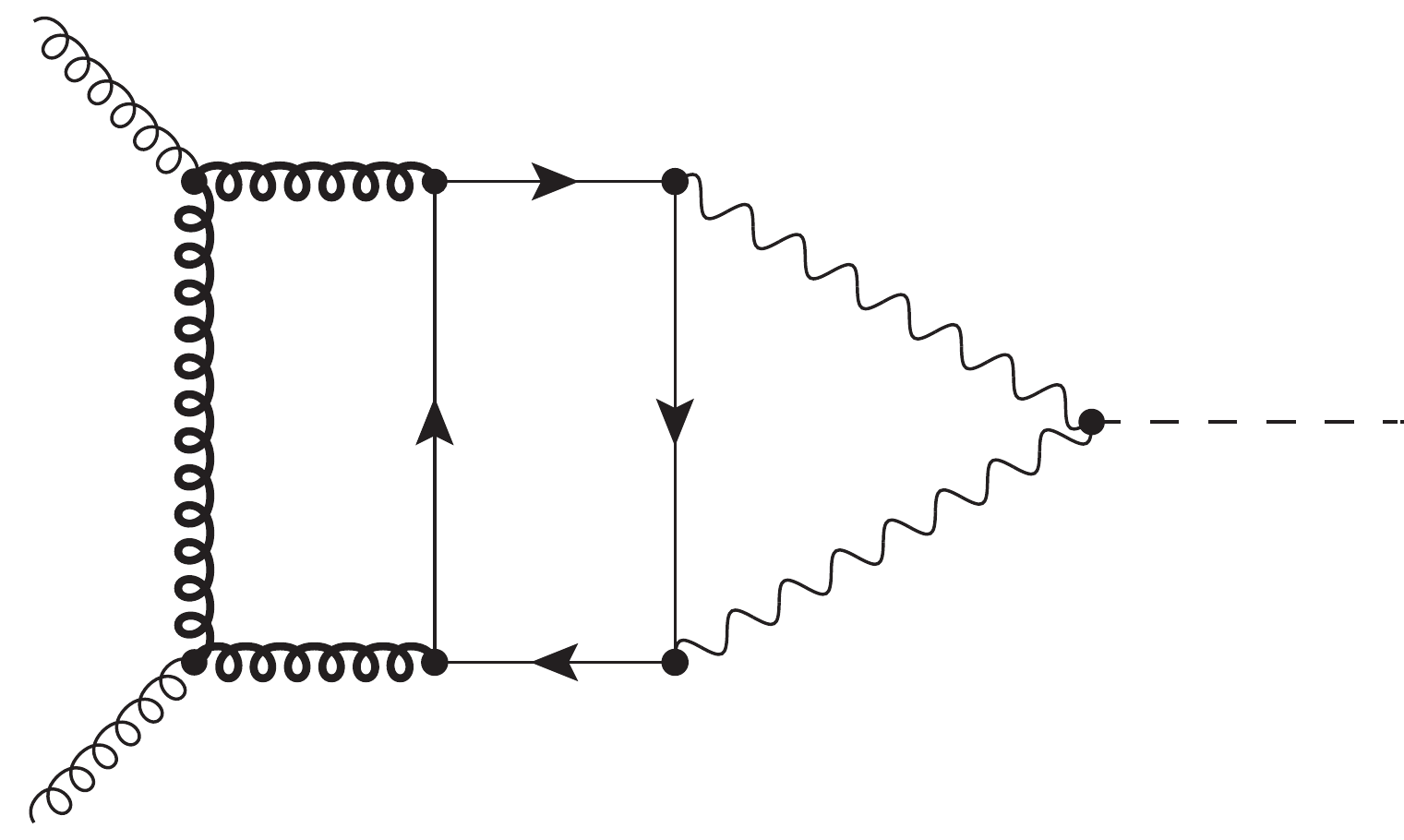}}
      \qquad
      \subfloat
      {\includegraphics[width=0.28\textwidth]{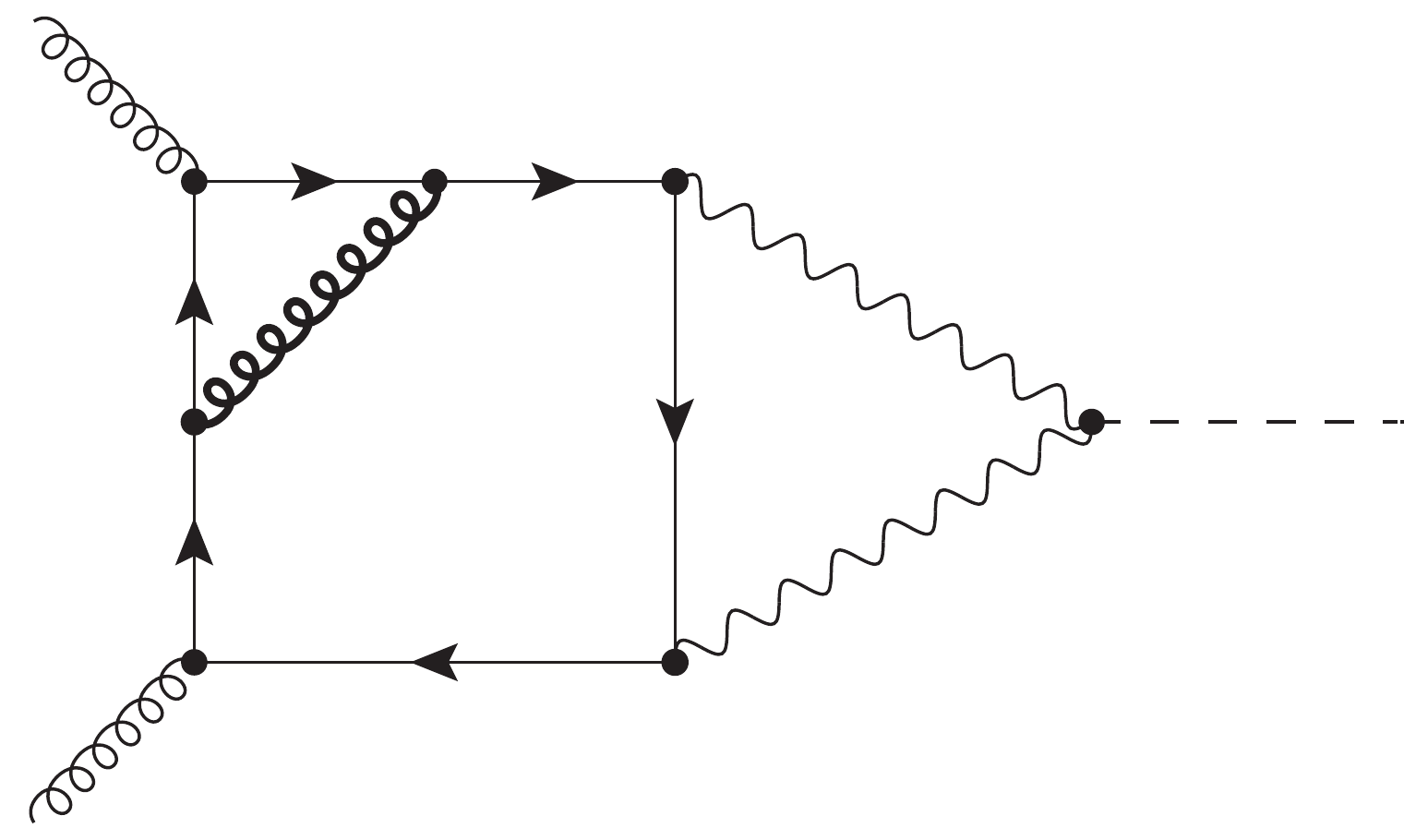}}
      \caption{Examples of QCD corrections to the LO QCD-EW Higgs gluon fusion.}
      \label{NLO}
\end{figure}

\begin{figure}[t]
      \centering
      \subfloat[][Quark loop connection.]
      {\label{NU2}\includegraphics[height=0.14\textheight]{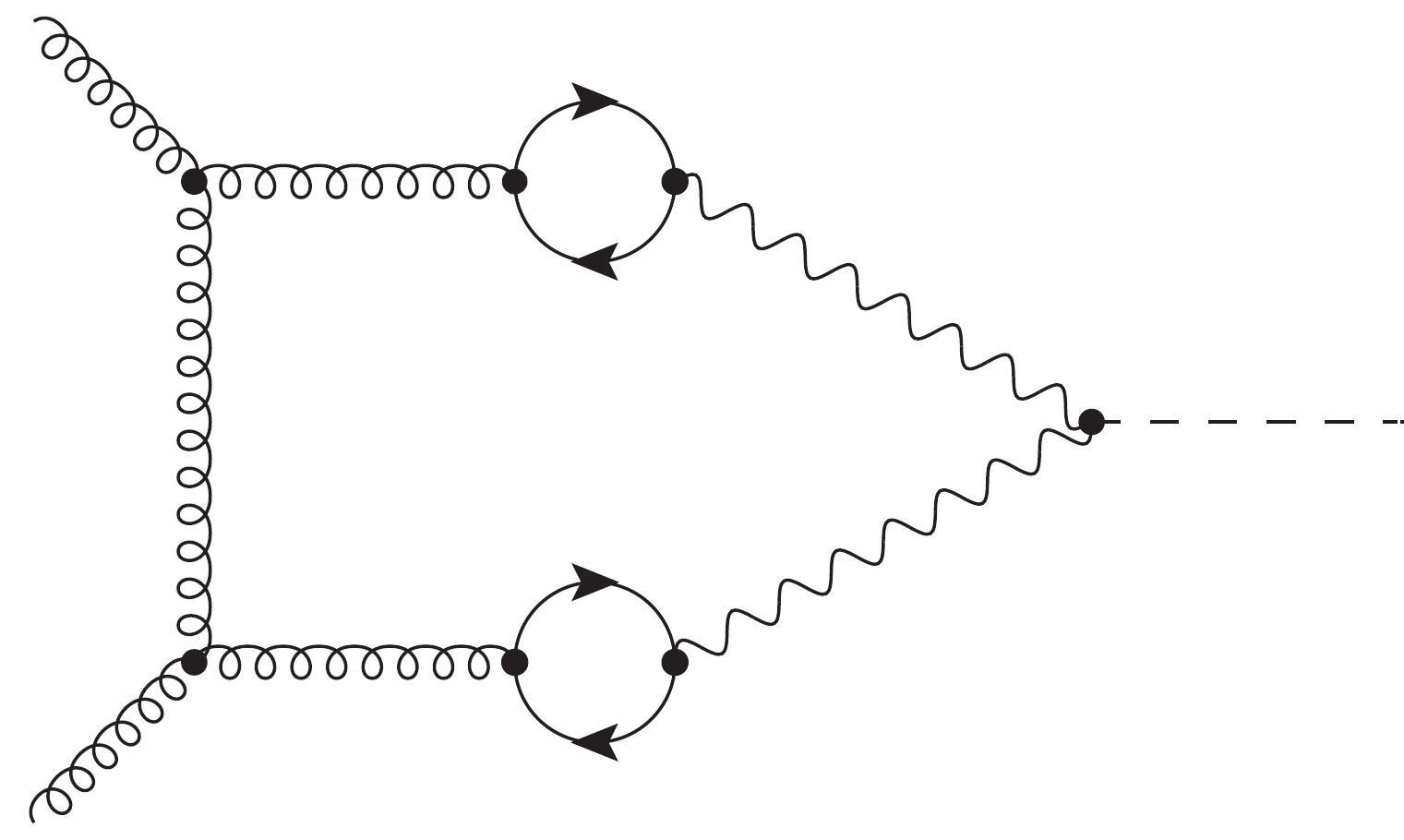}}
      \qquad
      \subfloat[][Two separate quark loops.]
      {\label{NUB}\includegraphics[height=0.14\textheight]{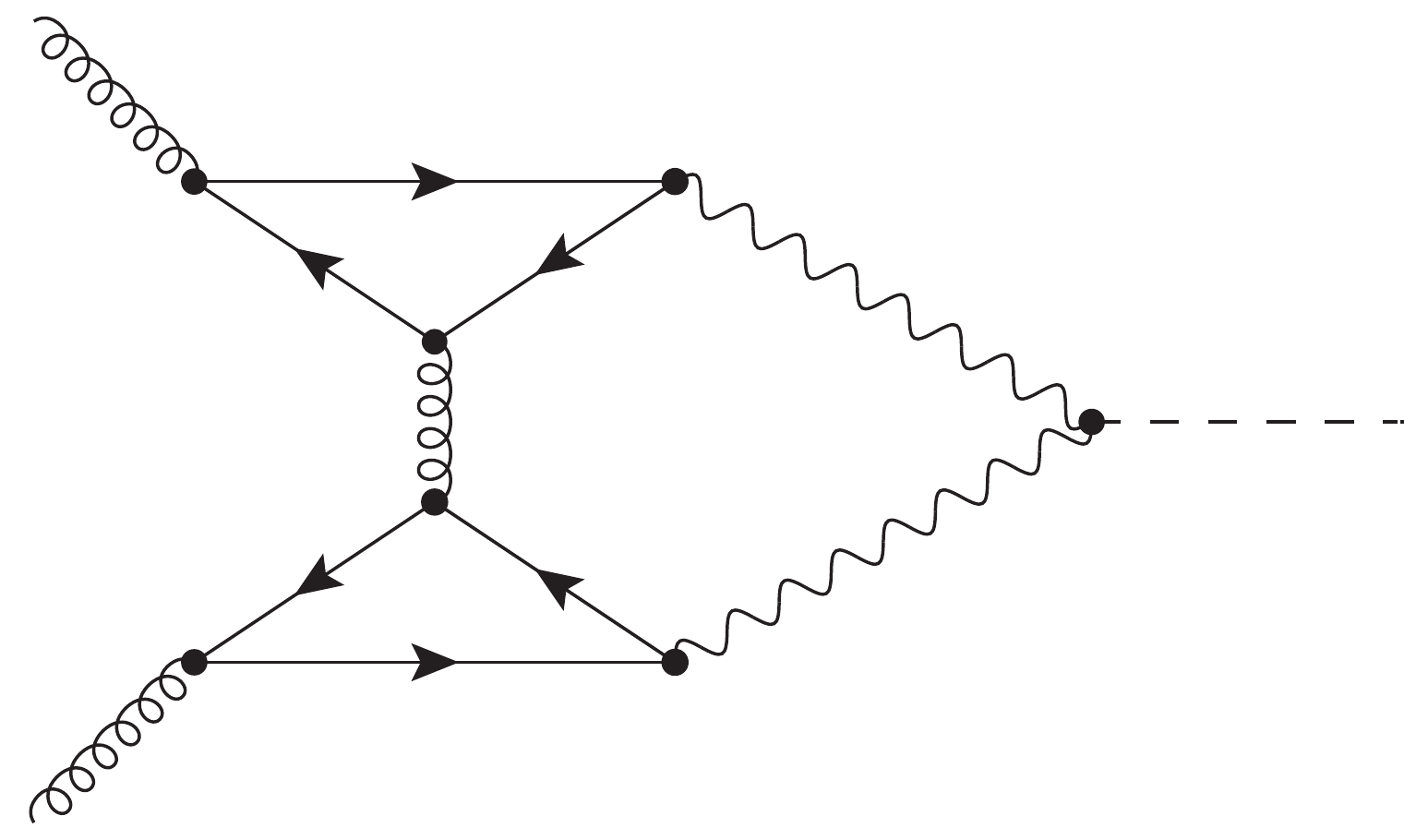}}
      \caption{Typical diagrams that may provide  $Z$- but not 
$W$-boson contributions to mixed QCD-EW corrections.  }
      \label{NULL}
\end{figure}

Examples of three loop diagrams that contribute to the NLO QCD corrections 
to the mixed QCD-electroweak $gg \to H$ amplitude are shown in Fig.~\ref{NLO}. 
The majority of the diagrams contributes both to 
$Z$ and $W$ channels with a few exceptions 
that contribute only in the $Z$ channel; some examples 
are shown in Fig.~\ref{NULL}.  However, all 
diagrams that distinguish $Z$ and $W$ contributions 
vanish either because of color algebra (Fig.~\ref{NU2}), or 
Furry's theorem (vector current contribution in case of Fig.~\ref{NUB}) 
or the fact that we consider complete generations of massless quarks (Fig.~\ref{NUB}, 
axial current). 
We note that in the latter case, because of  the axial 
anomaly, the contribution of the $b$-quark is not well defined without 
the top quark; in this paper we ignore this issue and completely 
discard all contributions that lead to diagrams of the 
type (Fig.~\ref{NUB}) but we include the  $b$-quark 
in all other three-loop QCD-EW diagrams  that we consider in this paper. 
After these simplifications, there remain   47 three-loop 
non-vanishing  diagrams for both  $W$-bosons and  the $Z$-bosons.  
All the relevant contributions can be obtained by considering diagrams 
where a massive vector boson interacts with the quark loop through 
a vector current.

\section{Topologies and master integrals}
\label{topo}

It is straightforward to  compute contributions of individual diagrams 
to the invariant form factor $\mathcal{F}$ using 
Eq.~(\ref{eq2.6}). Upon doing so, we 
find that  the form factor $\mathcal{F}$ is given by a linear combination 
of Feynman integrals that we write as follows
\begin{equation}
  \int \frac{\mathrm{d}^d k_1 \,\mathrm{d}^d k_2 \,\mathrm{d}^d k_3}{\left[{\mathrm{i}\mathrm{\pi}^{d/2}\Gamma(1+\epsilon)}\right]^3}\,\prod_{j=1}^{J}\frac{1}{[j]^{a_j}}.
\end{equation}
The $[j]$s denote inverse Feynman propagators that we 
specify in Appendix~\ref{Topologies}. Since 
$W$ and $Z$ bosons never appear in the same Feynman diagram, 
we  use a generic notation $M$ for the mass of the vector bosons 
when we discuss  Feynman diagrams and integrals.  
All Feynman integrals are analytic functions in the variables 
$s = (p_1 + p_2)^2$  and $M^2$ . 
Discontinuities common to all parent topologies occur  at $s \ge 0$ (on-shell massless intermediate states), $s \ge M^2$ 
(production of an on-shell vector boson plus two massless quarks), 
and $s \ge 4M^2$ (production of an on-shell pair of vector bosons).

To compute the different Feynman integrals contributing to the 
form factor, we identify eleven different topologies; these topologies can 
be written using three different sets of the inverse Feynman propagators. 
The topologies are described in Appendix~\ref{Topologies}.
We use the program \texttt{Reduze2} \cite{vonManteuffel:2012np} to express all Feynman 
integrals that contribute to the invariant form factor $\mathcal{F}$ 
through 95 master integrals (MIs). The full list of MIs can be found in 
Appendix~\ref{Master_Integrals}.

\section{Differential equations}
\label{deq}

\subsection{General considerations}
To calculate the MIs, we employ the differential equations method. The MIs 
depend on two dimensionfull variables $s$ and $M^2$ and it is convenient to trade 
them for one dimensionfull and one dimensionless variable.  We take $s$ to be the dimensionfull variable, 
extract the mass dimensions of the MIs and write 
\begin{equation}
 I_n(\epsilon,s,y) = (-s)^{-a_n-3\epsilon} J_n(\epsilon,y).
\end{equation}
The dimensionless variable $y$ reads 
\begin{equation}
  y=\frac{\sqrt{1-4{M^2}/{s}}-1}{\sqrt{1-4{M^2}/{s}}+1}.
\end{equation}
Values of the variable $y$ for different relations between $M^2$ an $s$ are shown in Table~\ref{tabvar}, together with prescriptions for analytic continuation.

\begin{table}[h]
\centering
\begin{tabular}{ccccc}
\toprule
\multirow{2}*{Variable}						&\multirow{2}*{Prescription}	&\multirow{2}*{Euclidean region}	&\multicolumn{2}{c}{Minkowski region}							\\
\cmidrule(lr){4-5}
								&				&		&Below threshold				&Above threshold		\\
\midrule
  $s$								&$s+\mathrm{i}\,0$	&$-\infty<s<0$		&$0<s<4M^2$							&$4M^2<s<+\infty$		\\
  $y=\frac{\sqrt{1-4{M^2}/{s}}-1}{\sqrt{1-4{M^2}/{s}}+1}$	&$y+\mathrm{i}\,0$	&$0<y<1$		&$\mathrm{e}^{\mathrm{i}\theta}$, $0<\theta<\mathrm{\pi}$	&$-1<y<0$			\\
\bottomrule
\end{tabular}
\caption{Prescriptions for analytic continuation and different kinematic regions for  $s$ and $y$.}\label{tabvar}
\end{table}

To study the non-trivial dependence of the MIs on $y$,  we differentiate the vector of 
rescaled MIs $\mathbf{J}(\epsilon,y)$ with respect to $y$,  
express the resulting integrals again through MIs 
and  obtain a closed system of  differential equations 
\begin{equation}
 \frac{\mathrm{d}\mathbf{J}(\epsilon,y)}{\mathrm{d}y}=A(\epsilon,y)\mathbf{J}(\epsilon,y).
\end{equation}

To compute the MIs we need to solve this system of differential equations. To do so, 
we first determine a \emph{canonical basis} of MIs.  With the integrals 
in the canonical basis, the system of differential equations assumes the 
$\epsilon$-homogeneous form \cite{Henn:2013pwa}
\begin{equation}
 \frac{\mathrm{d}\mathbf{F}(\epsilon,y)}{\mathrm{d}y}=\epsilon \mathcal{A}(y)\mathbf{F}(\epsilon,y).
 \label{eq:can}
\end{equation}

The necessary and sufficient conditions for the existence of a canonical basis are not known, 
but it is expected that for all cases where  solutions can be written in terms of Chen iterated 
integrals,  a canonical basis exists. Procedures to determine canonical bases were 
discussed in Ref.~\cite{Henn:2013pwa,Henn:2014qga,Argeri:2014qva,Gehrmann:2014bfa,Lee:2014ioa,
Meyer:2016slj,Adams:2017tga}.

Once the canonical basis is found and  integrals are normalized in such a way that 
the finite $\epsilon$-limit exists 
$\lim \limits_{\epsilon \to 0} \mathbf{F}(\epsilon,y) = F_0\,,$ 
it becomes 
straightforward to obtain  the solution to the system of differential equations 
as series expansion in the dimensional regularization parameter $\epsilon$.
An important property of the canonical form of 
differential equations in Eq.~\eqref{eq:can} is that 
the matrix $\mathcal{A}(y)$ contains only a finite number 
of simple poles in the variable  $y$~\cite{Henn:2014qga}
\begin{equation}
\frac{ \mathrm{d}\mathbf{F}(\epsilon,y)}{
\mathrm{d} y}
=\epsilon \sum_{k=1}^K \frac{B_k}{y-y_k} \mathbf{F}(\epsilon,y),
\label{eq4.5}
\end{equation}
where $B_k$ are $y$-independent matrices. Equations  
of the type shown in Eq.(\ref{eq4.5}) are called \emph{fuchsian} equations.

The solution of Eq.(\ref{eq4.5})  is  obtained upon iterative integration 
\begin{equation}
  \label{4_6}
  \begin{split}
 \mathbf{F}(\epsilon,y)	= \quad &\mathbf{F}_0^{(0)}+ \\
                        +\,\, & \epsilon \left[\int_0^y\mathcal{A}(\tau_1)\mathbf{F}_0^{(0)}\,\mathrm{d}\tau_1+\mathbf{F}_0^{(1)}\right]+	\\
                        +\,\, & \epsilon^2 \left[\int_0^y\mathcal{A}(\tau_1)\int_0^{\tau_1}\mathcal{A}(\tau_2)\mathbf{F}_0^{(0)}\,\mathrm{d}\tau_2\mathrm{d}\tau_1+\int_0^y\mathcal{A}(\tau_1)\mathbf{F}_0^{(1)}\,\mathrm{d}\tau_1+\mathbf{F}_0^{(2)}\right]+ \\
                        +\,\, & O\left(\epsilon^3\right)\,.
  \end{split}
\end{equation}

Thanks to the fuchsian form of the matrix $\mathcal{A}$, the nested integrations in Eq.(\ref{4_6}) 
can be performed in terms of multiple polylogarithms~\cite{Nielsen, Kummer}, best known in the 
physics literature as Goncharov's 
polylogarithms (GPLs) 
\cite{Goncharov, Remiddi:1999ew, Gehrmann:2000zt, Vollinga:2004sn}. 
GPLs are defined recursively using the following equations
\begin{equation}
 G(\mathbf{m}_{w};y):=
 \begin{cases}
  \frac{1}{w!}\log^w y	&\text{if $\mathbf{m}_w=(0,\dots,0)$}	\\
  \int \limits_{0}^{y} \frac{ \mathrm{d}\tau	 }{\tau - m_w} \; G(\mathbf{m}_{w-1};\tau) 
&\text{if $\mathbf{m}_w\neq(0,\dots,0)$}
 \end{cases},
\end{equation}
where $\mathbf{m}_w=(m_w,\mathbf{m}_{w-1})$. 
The weight $w$ of a GPL is the length of the vector $\mathbf{m}_w$.
GPLs evaluated at 
rational points are expressed by constants of 
the same weight. In Section~\ref{bclme}, we use   
this property 
to efficiently fix 
the integration constants $\mathbf{F}_0^{(n)}$, c.f. Eq.(\ref{4_6}).

\subsection{The system of differential equations}

We are now in position to discuss the system of differential equations 
and the steps that are required to  transform it  into a canonical form.  
To achieve that, we used a combination of  different methods that we now explain. 

The matrix $A(\epsilon,y)$ that appears 
on the right hand side of the differential equation has a block-diagonal 
form. This form implies that differential equations for simpler MIs close and that simpler integrals appear as inhomogeneous terms 
in differential equations for more complex ones. 
This form of the differential equations suggests that one should start 
with bringing simplest  topologies to a canonical form and then moving 
on to more complex topologies, step by step. 

For MIs with four, five or  six propagators  
coupled together in at most $3\times3$ blocks, we  
first adjusted powers of propagators  and introduced $\epsilon$-dependent  prefactors, 
as described in Ref.~\cite{Argeri:2014qva,Gehrmann:2014bfa}.  It is well understood 
how to do this to reach  canonical form for  simple integrals, such 
as  bubbles  and triangles where the guiding principle is to ensure that  
integrals  are free from ultraviolet divergences. 
After that, the  candidate integrals  are multiplied 
by a polynomial in $\epsilon$ of the form 
$\epsilon^{a_1} ( c_1 + c_2 \epsilon)^{a_2}$, to bring the system into a 
linear-$\epsilon$ 
form 
\begin{equation}
 \frac{\mathrm{d}\tilde{\mathbf{F}}(\epsilon,y)}{\mathrm{d}y} = \left[ A_0(y) + \epsilon A_1(y) \right]\tilde{\mathbf{F}}(\epsilon,y).
\end{equation}
We then integrate out  the $\ep$-independent part of the differential equation 
and obtain the canonical form 
\begin{gather}
 \frac{\mathrm{d}\mathbf{F}(\epsilon,y)}{\mathrm{d}y} = \epsilon \hat{S}_{A_0}^{-1}(y)A_1(y)\hat{S}_{A_0}(y)\mathbf{F}(\epsilon,y),	\\
 \hat{S}_{A_0}(y)=P_y\mathrm{e}^{\int A_0(\tau)\,\mathrm{d}\tau}=\sum_{k=0}^{+\infty}\int_{y_0}^y A_0(\tau_1) \int_{y_0}^{\tau_1} A_0(\tau_2) \dots \int_{y_0}^{\tau_{k-1}} A_0(\tau_k)\,\mathrm{d}\tau_k \dots \mathrm{d}\tau_2\mathrm{d}\tau_1,	\\
 \mathbf{F}(\epsilon,y) = \hat{S}_{A_0}^{-1}(y) \tilde{\mathbf{F}}(\epsilon,y).
\end{gather}
We stress that it is not always possible to integrate out 
$A_0(y)$ in the presence of off-diagonal terms and 
that, even if  successful, this procedure is not guaranteed to keep 
the Fuchsian form of the system of differential equations. Nevertheless, 
it provides a practical way to achieve canonical form for some differential 
equations relevant for our  problem. 

For larger sets of coupled MIs 
 or for MIs with a
larger  number of denominators, 
the differential equations become  too complex  to reach the 
linear-$\epsilon$ form  by a clever  choice of propagator powers  and prefactors. 
For these differential equations, 
we employ the  algebraic algorithm presented in 
\cite{Lee:2014ioa} and, in particular, its computer-algebraic 
implementation \texttt{Fuchsia} \cite{Gituliar:2017vzm}. We also use 
techniques described in Ref.~\cite{Meyer:2016slj} and their implementation  
in the  \texttt{Mathematica} package \texttt{CANONICA} \cite{Meyer:2017joq}.
Despite the power and versatility of these algorithms, 
their blind application to our system of differential equations 
results either in a failure of the procedure to reach a canonical fuchsian form, 
or in an incorrect redefinition of the MIs that have 
already been fixed. In the second case, 
the new system shows the correct differential structure, but the constants 
$\mathbf{F}_0^{(n)}$ are, in general, not of an uniform weight $n$. Again, by carefully selecting the MIs 
and choosing their prefactors,  it is possible to overcome these problems. 

A useful  procedure to select candidate integrals  for the canonical basis
is to inspect their generalized unitarity cuts, as explained in \cite{Henn:2014qga}.
The idea is that  if one replaces a propagator in an integral by a delta-function
$1/(p^2-m^2) \to \delta(p^2 - m^2)$, 
the differential equation that this integral satisfies does not change 
except that all integrals on the right hand side of the differential 
equation where this propagator is not present have to be set to zero. 
Thanks to this observation, by cutting a MI  in different ways 
we  can inspect different subsets of the differential equation that it satisfies. 
By replacing all propagators of a given integral 
 with the $\delta$-functions (the maximal 
cut), 
we obtain the differential equation which involves only  MIs of the 
highest topology since all other integrals drop out~\cite{Primo:2016ebd}. 
The analysis of the cuts can be simplified by use of the so-called
Baikov representation~\cite{Baikov:1996rk,Frellesvig:2017aai}.
The upshot of these discussions is that 
if all the cut integrals are of the $\mathrm{d}\log$-iterated form
\begin{equation}
 I = C(\mathbf{x}) \int \dots \int\int \,\mathrm{d}\log R_n(\mathbf{x},a_1, \dots, a_n) \,\mathrm{d}\log R_{n-1}(\mathbf{x},a_1, \dots, a_{n-1}) \dots \,\mathrm{d}\log R_{1}(\mathbf{x},a_1),
\end{equation}
where $C(\mathbf{x})$ is a function of the kinematic invariants 
$\mathbf{x}$ and $R(\mathbf{x},a_1, \dots, a_k)$ 
are rational  functions of $\mathbf{x}$ and of the integration variables $a_i$, 
the integral $I$ is a valid candidate for the canonical basis. 

For our purposes, we require  integrals with large number of propagators; 
they satisfy complex 
differential equations with  up to four MIs coupled. 
For this reasons,  the study of all the different cuts is prohibitive
and in most of the cases we limit ourselves to the maximal cut.
As explained above, the latter only allows us to inspect the part of the differential equation 
that corresponds to the highest  topology. Once the $\mathrm{d}\log$-form is achieved 
for the maximal cut of all the coupled highest-level  MIs, 
the homogeneous part of the equations is ensured to be canonical.
Of course, integrals of lower topologies 
that appear in the differential equations  still do not have a a canonical  form, 
but the study of the maximal cut has proven to be sufficient to provide a 
starting points  for a successful application of either \texttt{Fuchsia} 
or \texttt{CANONICA}.

Performing  manipulations that we just described, we were able to 
transform the system of differential equations into a canonical Fuchsian 
form. We write it as
\begin{equation}
 \label{4_15}
 \mathrm{d}\mathbf{F}(\epsilon,y) = \epsilon \left[ B_0\, \mathrm{d}\log y + B_1\, \mathrm{d}\log (1-y) + B_{-1}\, \mathrm{d}\log (1+y) + B_r\, \mathrm{d}\log (1-y+y^2) \right]\mathbf{F}(\epsilon,y).
\end{equation}
We have chosen  arguments of the logarithms to ensure that the logarithms are 
real-valued  for $0<y<1$ ($s<0$). It follows from this form that  integration 
kernels are identical to  what was found in the calculation 
of MIs for leading order  QCD-electroweak  amplitude  \cite{Bonetti:2016brm}.
They read
\begin{equation}
 f(0;y)=\frac{1}{y}, \quad
 f(1;y)=\frac{1}{y-1}, \quad
 f(-1;y)=\frac{1}{y+1}, \quad
 f(r;y)=\frac{2y-1}{y^2-y+1},
\end{equation}
where the function $f(r;y)$ is a short-hand notation for  a particular 
linear combination of two integration kernels
\begin{equation}
 f(r;y)=\frac{2y-1}{y^2-y+1}=\frac{1}{y - r_+} + \frac{1}{y - r_-}=f(r_+;y)+f(r_-;y), 
\;\;\;\; r_{\pm} = \mathrm{e}^{\pm \mathrm{i}\mathrm{\pi}/3},
\end{equation}
that  make the  alphabet complex-valued. Note that this expansion is only needed for 
numerical evaluations, while the integration of the differential equation  can be performed 
using the quadratic  form that we  associate 
with the alphabet  symbol $r$~\cite{Ablinger:2011te, vonManteuffel:2013vja}. Indeed,  thanks 
to the linearity of both the 
differential equations and the definition of the GPLs, it is immediately  possible to  
express the obtained generalized GPL in terms of the canonical ones
\begin{equation}
 G(\dots,r,\dots;x)=G(\dots,r_-,\dots;x)+G(\dots,r_+,\dots;x)\,.
\end{equation}

\section{Boundary conditions and the large-mass expansion}
\label{bclme}

To fully solve  the system of linear differential equations Eq.~(\ref{4_15}),  
we need to determine  the integration constants $\mathbf{F}_0^{(n)}$.  The easiest 
way to fix them is to compute the MIs at some kinematic point 
$y_0$ using alternative methods and then compare the result of the computation 
with the solution of the differential equations. 
As we already explained in Ref.~\cite{Bonetti:2016brm}, it is convenient to 
compute the required integrals at $y=1$, corresponding to $s/M^2=0$. 
This kinematic condition can be studied 
using the large-mass expansion procedure~\cite{Smirnov:2002pj} 
to independently compute the MIs in this limit. In this Section 
we explain how to apply  the large-mass expansion procedure to obtain 
boundary values for the MIs. 

A typical MI that we need to compute depends on two different scales: the energy of 
the
external gluons  $p_1 \sim p_2 \sim \sqrt{s}$, and the mass of the electroweak vector bosons $M$.  We consider the hypothetical limit 
$M^2 \gg s$.  To provide a non-vanishing 
contribution in dimensional regularization, the 
 loop momentum $k$ that  flows  through any sub-set  of internal lines has 
to scale either as $k \sim \sqrt{s}$ or as $k \sim M$. All possible combinations 
of scalings for the three loop momenta 
are allowed and have to  be considered,  provided that momentum conservation is satisfied at each  vertex.\footnote{Indeed, since  large momentum 
cannot be created, destroyed or provided by external legs,  
it must form at least  one closed flow  along the internal 
lines of an integral.} Once a valid momentum scaling is 
identified for a particular Feynman 
integral,  the {\it integrand}   is Taylor-expanded in all 
small momenta (both external and loop ones) and then integrated. 
This procedure is performed for all possible  momenta 
assignments and the results are summed over to obtain the large-mass 
expansion of the original integral. 

As an example of the large-mass expansion procedure, we consider MI $I_7$
\begin{equation}
 \label{MI7}
 I_7=\imineq{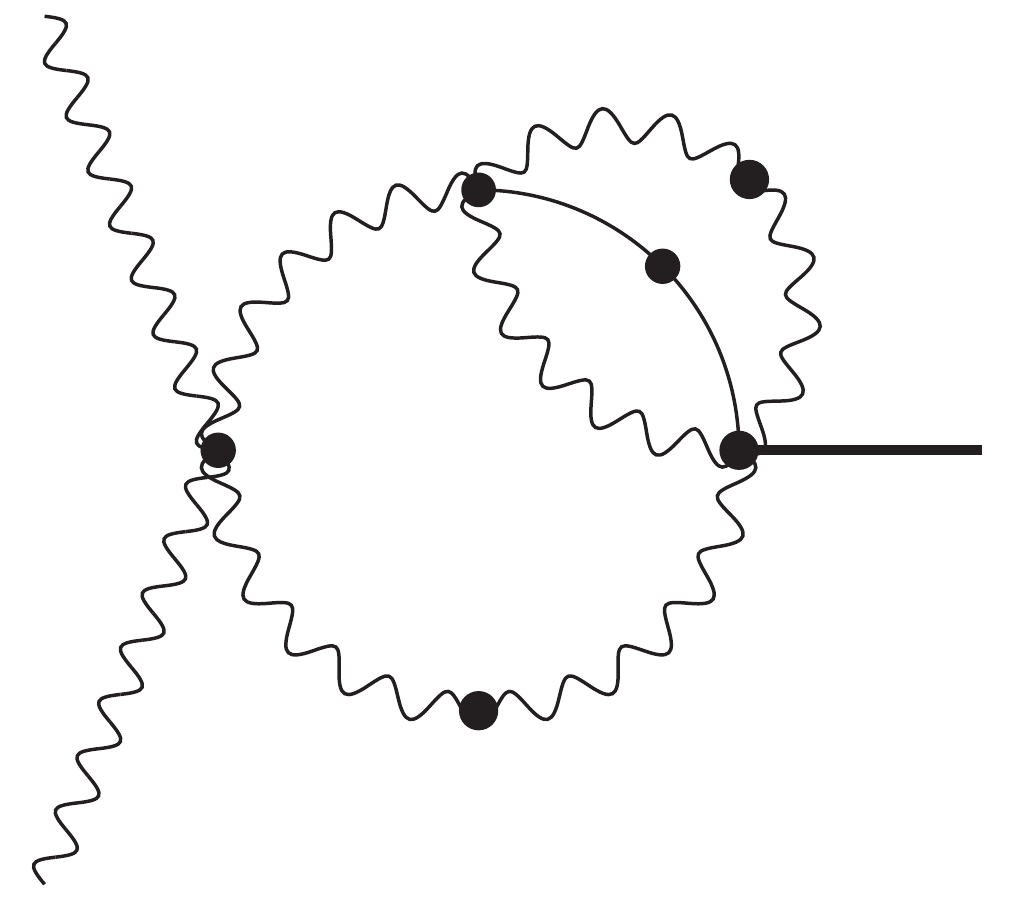}{12},
\end{equation}
where a dot  on a  line implies that the corresponding propagator  is  raised to second 
power. 
Among all allowed choices of large and small momenta only two give non-zero 
contributions. Indeed, it is possible to have large momenta  flowing  in 
the two-loop self-energy sub-graph and small momentum flowing 
in the ``outer'' loop, or to have all the three loop momenta to be large. 
It is clear that lines with large momenta ``decouple'' from adjacent propagators: 
they become tadpoles that multiply  a diagram obtained by 
shrinking the corresponding ``large momentum'' 
lines to a point.  For the two possible kinematic regions 
described above, the large mass expansion reads 
(bold internal lines in the original integral 
indicate the flow of large momenta)
\begin{align}
 \imineq{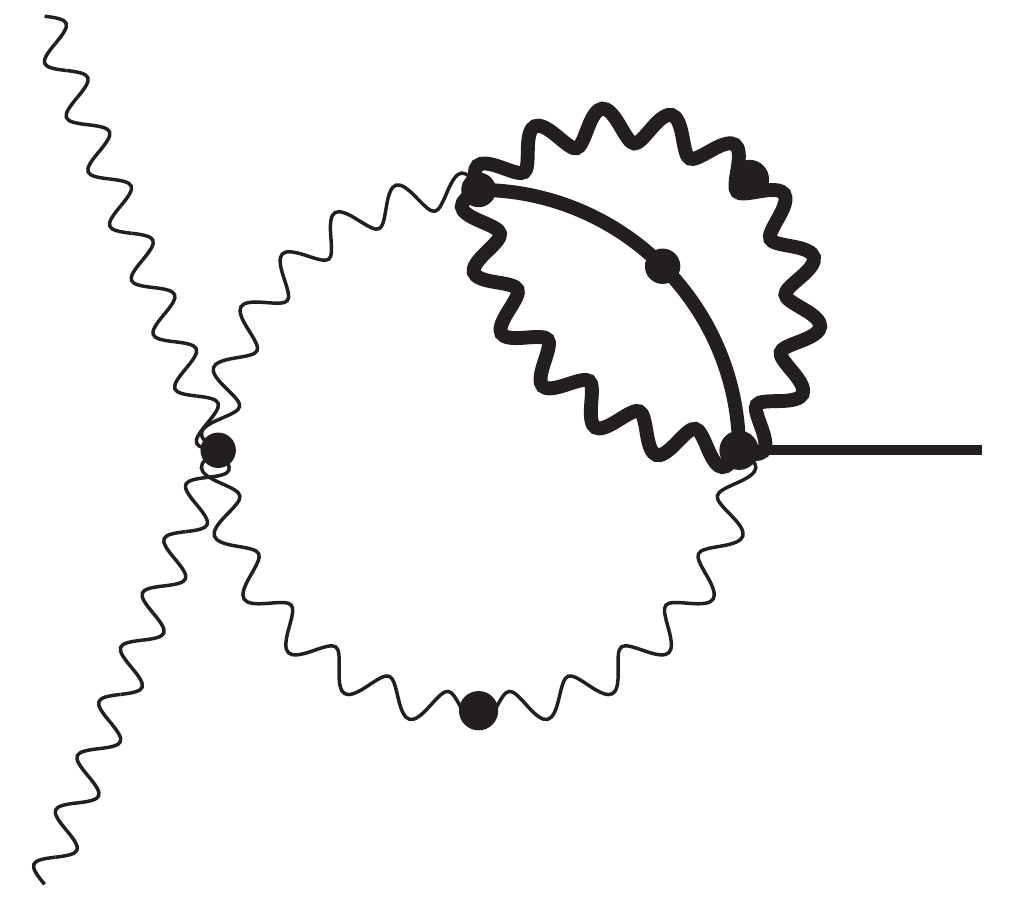}{10} \quad\to\quad& \imineq{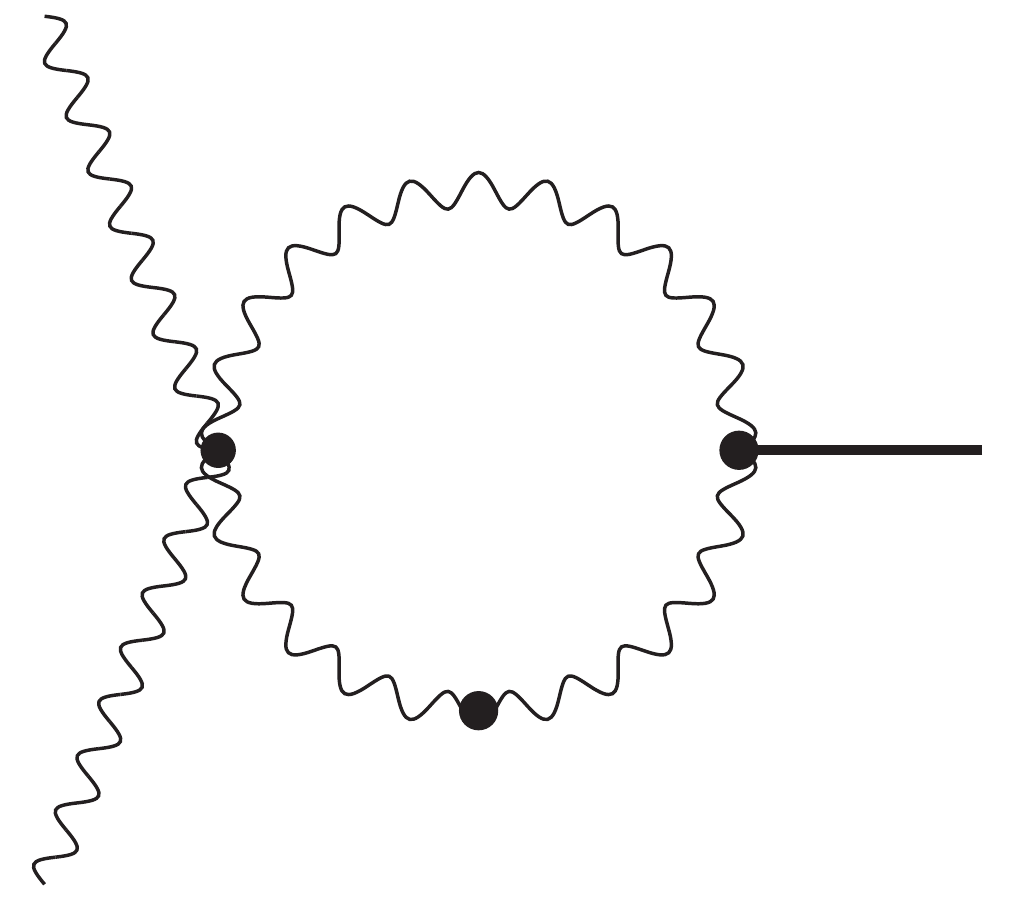}{10} \quad\times\quad \imineq{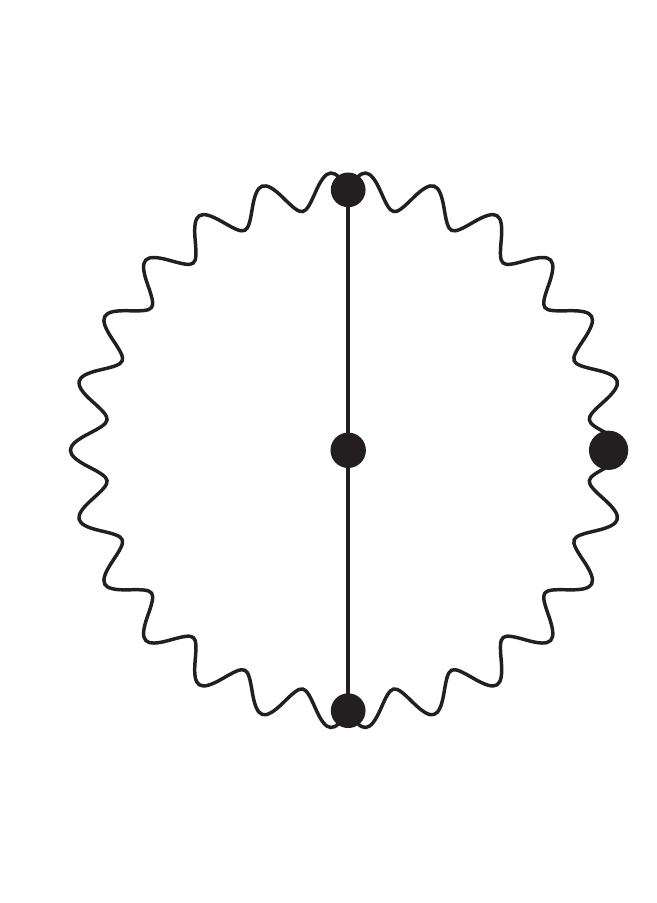}{10},	\\
 \imineq{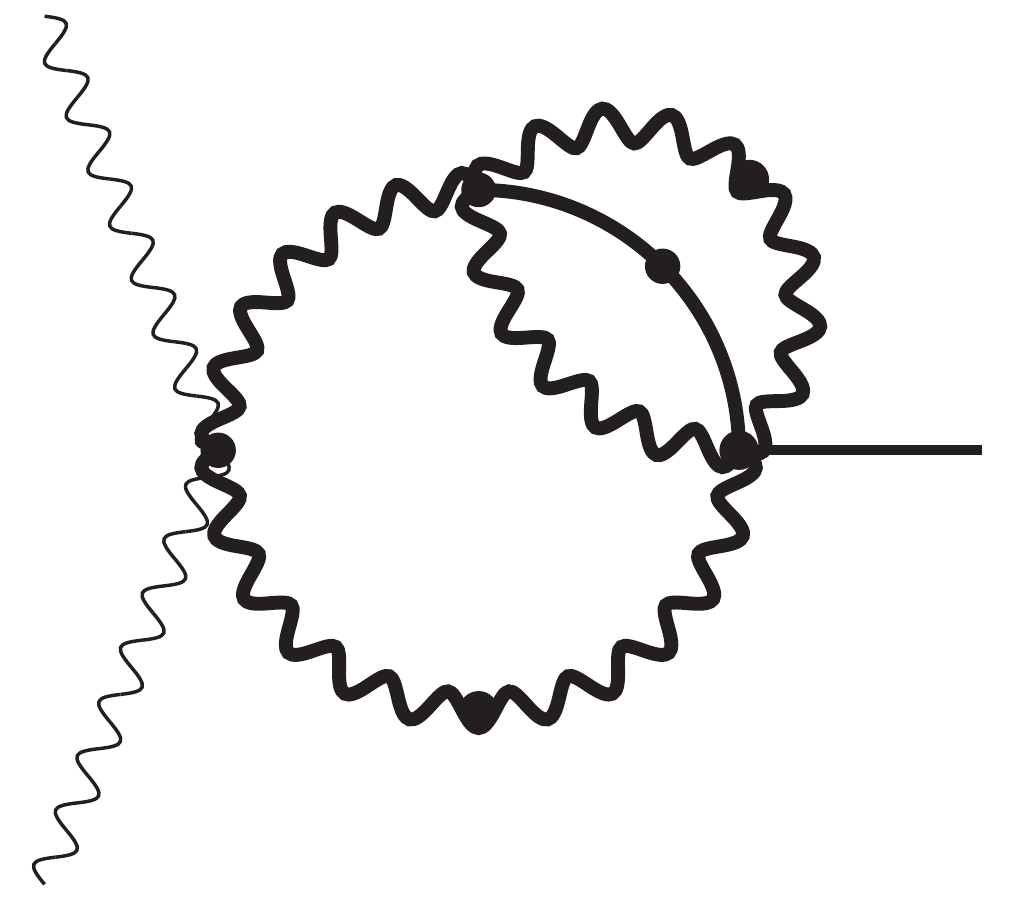}{10} \quad\to\quad& \imineq{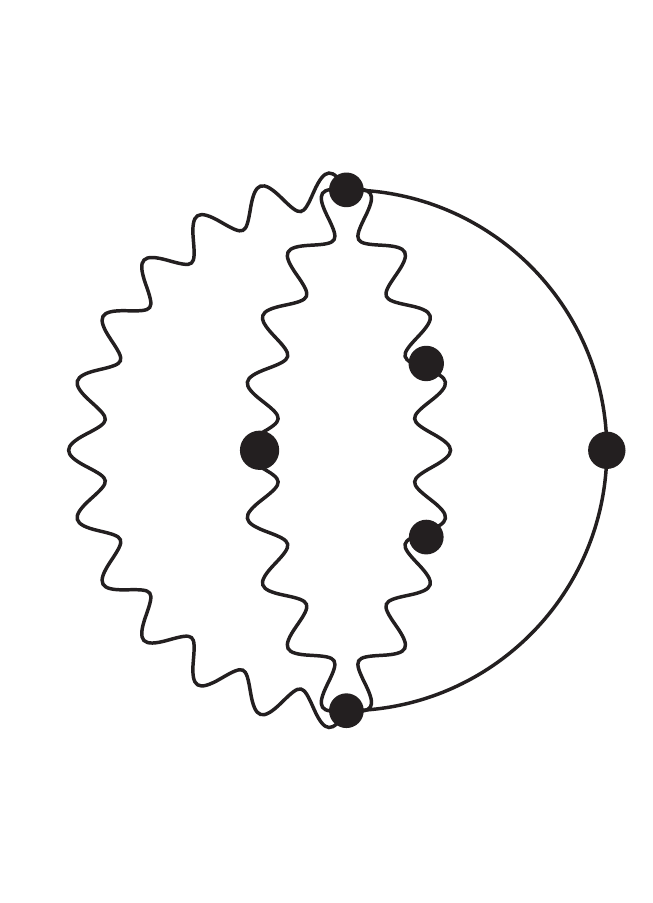}{10} \quad+\quad s\frac{2(1+\epsilon)}{2-\epsilon}\,\imineq{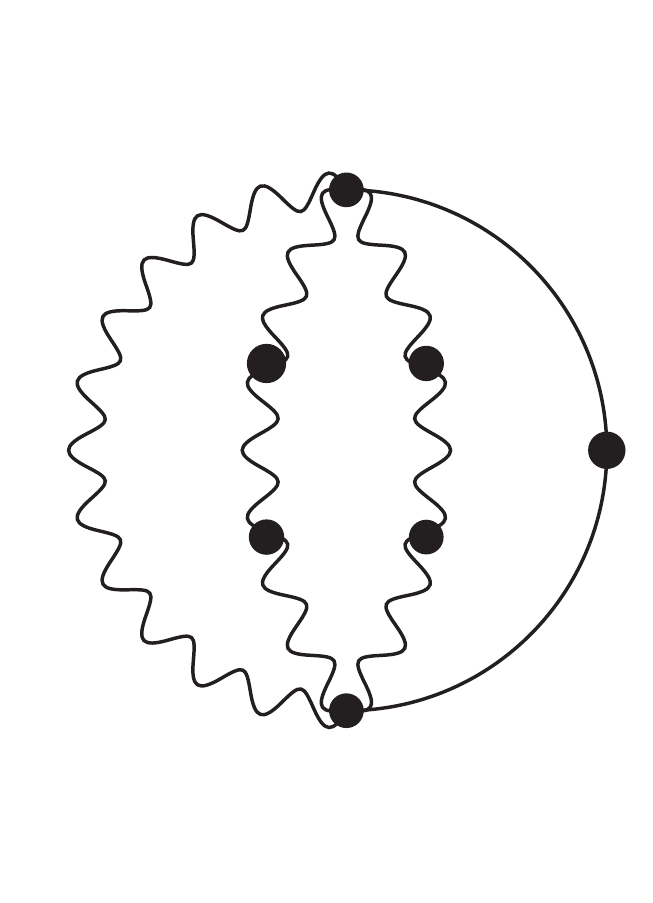}{10} \quad+\quad \mathcal{O}\left(\frac{(-s)^2}{(M^2)^{4}}\right).
\end{align}
We note that in  the first case the result is exact since, upon expansion, 
 the tadpole produces a term that 
cancels one of the propagators of the massless bubble.
We obtain the large-mass expansion of the integral~(\ref{MI7}) 
by summing over the two possible momenta flows described above
\begin{equation}
 \lim_{M^2 \gg s}\imineq{MI7}{10} = \left[\, \imineq{BMI7}{10} \,\times\, \imineq{T2MI7}{10} \,\right] + \left[\, \imineq{T3MI7}{10} \,+\, s\frac{2(1+\epsilon)}{2-\epsilon}\,\imineq{T3MI7_2}{10} \, \right]
\,+\, \mathcal{O}\left(\frac{s^2}{M^8}\right). 
\end{equation}
All  integrals that appear on the right hand side of this equation can easily be evaluated.

Clearly, the number of different momenta flows that need to be considered 
increases as one moves from relatively simple to more complex  integrals. 
Nevertheless, the complexity remains manageable.   Indeed, once the large mass 
expansion is performed, the most complex integrals are 
 three-loop tadpoles and two-loop massless triangles that are well-known, 
see e.g. Refs.~\cite{Gehrmann:2005pd,Steinhauser:2000ry}.

The large-mass expansions of the MIs computed 
to relevant   order in $s/M^2$ expansion, is 
compared with  the $y \to 1$ limit  of the solutions 
of the differential equations; requiring that the two agree, 
we obtain all the integration 
constants  $\mathbf{F}_0^{(n)}$, $n \le 6$, c.f. Eq.(\ref{4_6}).
Since analytic expressions for GPLs with letters 
$\mathrm{e}^{\pm \mathrm{i}\mathrm{\pi}/3}$, 
evaluated at  $y=1$, are unknown at present, we follow a numerical approach to determine the constants. 
The fact that the integrals of the canonical basis are of uniform weight
is crucial, since in this case 
only a small variety of  constants can appear at each order 
of their $\epsilon$-expansion. In particular, 
$\mathbf{F}_0^{(1)} = 0$, $\mathbf{F}_0^{(2,3,4)}$ must be proportional 
to  $\mathrm{\pi}^2$,  $\zeta(3)$ and $\mathrm{\pi}^4$, respectively,  $\mathbf{F}_0^{(5)}$ should be given by a  linear combination of 
 $\mathrm{\pi}^2 \zeta(3)$ and $ \zeta(5)$, and $\mathbf{F}_0^{(6)}$ by a  
linear combination of 
 $\mathrm{\pi}^6$ and $ \zeta^2(3)$.
We evaluate with high precision the canonical integrals  written 
in terms of GPLs at $y=1$ and 
match their numerical values to  rational linear combinations 
of analytic constant using  PSLQ algorithm, to at least 750 digits.
As a final check,  all MIs 
 have been compared for at least two different values of 
$s$ and $M^2$ to the numerical 
results obtained using the programs 
\texttt{SecDec} \cite{Borowka:2015mxa} and 
\texttt{pySecDec} \cite{Borowka:2017idc}. In all cases agreement was found.

\section{The final result for the \boldmath{$gg \to H$} amplitude}
\label{ampli}

As explained in Section~\ref{glufu}, the mixed QCD-electroweak contributions to 
$gg \to H$ amplitude are parametrized 
in terms of a single form factor.
The form factor receives contributions from diagrams with $W$ and $Z$ bosons; 
such diagrams appear for the first time at two loops.  
We account for quarks of the first two generations and include $b$-quark  contributions 
in diagrams with $Z$-boson.\footnote{As we already mentioned, we discard certain 
anomalous-type diagrams also for the third generation, see 
Section~\ref{glufu}.}  All quarks are taken to be massless. 
The  CKM matrix is approximated by the identity matrix.

Computation of three-loop contribution to the form factor yields divergent results. The divergences 
are of both ultraviolet and infrared origin. The ultraviolet divergences are removed 
by ultraviolet renormalization; the infrared divergences remain but will, eventually, get 
canceled by the real emission contributions when infrared safe observables are computed.

To remove the ultraviolet divergences, 
 we only need to  renormalize the strong coupling constant;  this is so because for massless 
quarks both vector and axial currents are conserved and, for this reason, weak 
couplings do not need an ultraviolet renormalization. 
To renormalize the strong coupling constant, we use the relation between 
bare and ${\overline {\rm MS}}$ coupling constants  that reads 
\begin{equation}
  \alpha_S^{(0)} \mu_0^{2\ep} =
\frac{\mathrm{e}^{\epsilon\gamma_E}}{(4\mathrm{\pi} )^\epsilon}\alpha_S^{~}(\mu) \mu^{2 \ep} 
\left(1
-\frac{\beta_0}{\epsilon} \frac{\alpha_S^{~}}{2\mathrm{\pi}}\right)
+\mathcal{O}\left(\alpha_S^3\right),
\end{equation}
where $\gamma_E$ is the Euler--Mascheroni constant and $\beta_0$ is the 
first coefficient of the QCD $\beta$-function. It reads 
\begin{equation}
  \beta_0=\frac{11}{6}C_A-\frac{2}{3}T_F N_f,
\end{equation}
where  $C_A=N_c=3$, $T_F=1/2$ and $N_f=5$ is the number of massless fermions that 
contribute to the renormalization of the QCD coupling constant.  

We write the UV-renormalized form factor as 
\begin{equation}
\label{eq6.3}
{\cal F} = -\mathrm{i} \frac{\alpha^2 \alpha_S^{~}(\mu) v }{64 \mathrm{\pi} \sin^4 \theta_W }
\sum \limits_{i=W,Z}^{} C_i A(m_i^2/s,\mu^2/s),
\end{equation}
where  $\alpha$ is the QED fine structure constant, $\theta_W$ is the Weinberg angle 
and $v = m_W \sin \theta_W/\sqrt{\mathrm{\pi} \alpha}$ is the Higgs field vacuum expectation 
value. In addition, 
\begin{equation}
C_W = 4,\;\;\;\;
\;\;\; C_Z = \frac{2}{ \cos^4 \theta_W} \left ( \frac{5}{4}
- \frac{7}{3} \sin^2 \theta_W + \frac{22}{9} \sin^4 \theta_W \right ), 
\end{equation}

The two terms in Eq.(\ref{eq6.3}) denote contributions of diagrams with $W$ and $Z$ bosons. 
The function $A$ can be written as an expansion in the strong coupling constant 
\begin{equation}
A(m^2/s,\mu^2/s) = A_{\rm LO} + \frac{\alpha_S^{~}(\mu)}{2\mathrm{\pi}} A_{\rm NLO} +{\cal O}(\alpha_S^2).
\end{equation}

It is interesting to remark that individual integrals that contribute to $gg \to H$ 
amplitude are strongly divergent; the strongest divergence is ${\cal O}(\ep^{-6})$. 
When the integrals are combined to produce a form factor, all divergences 
stronger  than ${\cal O}(\ep^{-2})$ cancel out. The ultraviolet renormalization 
described above removes some of the $1/\ep$ poles but infra-red singularities 
that start at $1/\ep^2$ still remain. 
The general structure of these singularities in QCD is described by Catani's formula \cite{Catani:1998bh}.
This formula, applied to the form factor that describes mixed QCD-electroweak corrections 
to $gg \to H$ amplitude reads 
\begin{equation}
\label{eq6.6}
A_{\rm NLO} = \mathbf{I}^{(1)} A_{\rm LO} + A_{\rm NLO}^{\rm fin},
\end{equation}
where 
\begin{equation}
\mathbf{I}^{(1)}=\left(\frac{-s-i0}{\mu^2}\right)^{-\epsilon}
\frac{\mathrm{e}^{\epsilon\gamma_E}}{\Gamma(1-\epsilon)}
\left[-\frac{C_A}{\epsilon^2}-\frac{\beta_0}{\epsilon}\right].
\end{equation}
Note that the leading order amplitude in Eq.(\ref{eq6.6}) 
is needed through order 
${\cal O}(\ep^2)$; it was computed in Ref.~\cite{Bonetti:2016brm}.
An  analytic expression for the 
finite part of the NLO amplitude $A_{\rm NLO}^{\rm fin}$ 
is given in the ancillary file.   

For future reference, we give numerical values of  the two functions $A_{\rm LO}$ and $A_{\rm NLO}^{\rm fin}$, 
for physical values of the Higgs boson and vector boson masses. Taking 
$m_H = 125.09~{\rm GeV}$,  $m_W=80.385\,\textrm{GeV}$, $m_Z=91.1876\,\textrm{GeV}$ (see~\cite{Patrignani:2016xqp}), $N_c=3$, 
$N_f=5$ and $\mu=m_H$, we find

\begin{equation}
  \label{eq:numres}
  \arraycolsep=1.4pt\def\arraystretch{2.2}
  \begin{array}{lcll}
A_\text{LO}(m_Z^2/m_H^2,1)		&=	&-6.880846	&- \mathrm{i}\, 0.5784119\,,
\\
A_\text{LO}(m_W^2/m_H^2,1)		&=	&-10.71693	&- \mathrm{i}\, 2.302953\,,
\\
A^\text{fin}_\text{NLO}(m_Z^2/m_H^2,1)	&=	&-2.975801	&- \mathrm{i}\, 41.19509\,,
\\
A^\text{fin}_\text{NLO}(m_W^2/m_H^2,1)	&=	&-11.31557	&- \mathrm{i}\, 54.02989\,.
  \end{array}
\end{equation} 

\begin{figure}[t]
      \centering
      \subfloat[][Cut for $s > 0$ at LO.]
      {\label{CUTA}\includegraphics[width=0.28\textwidth]{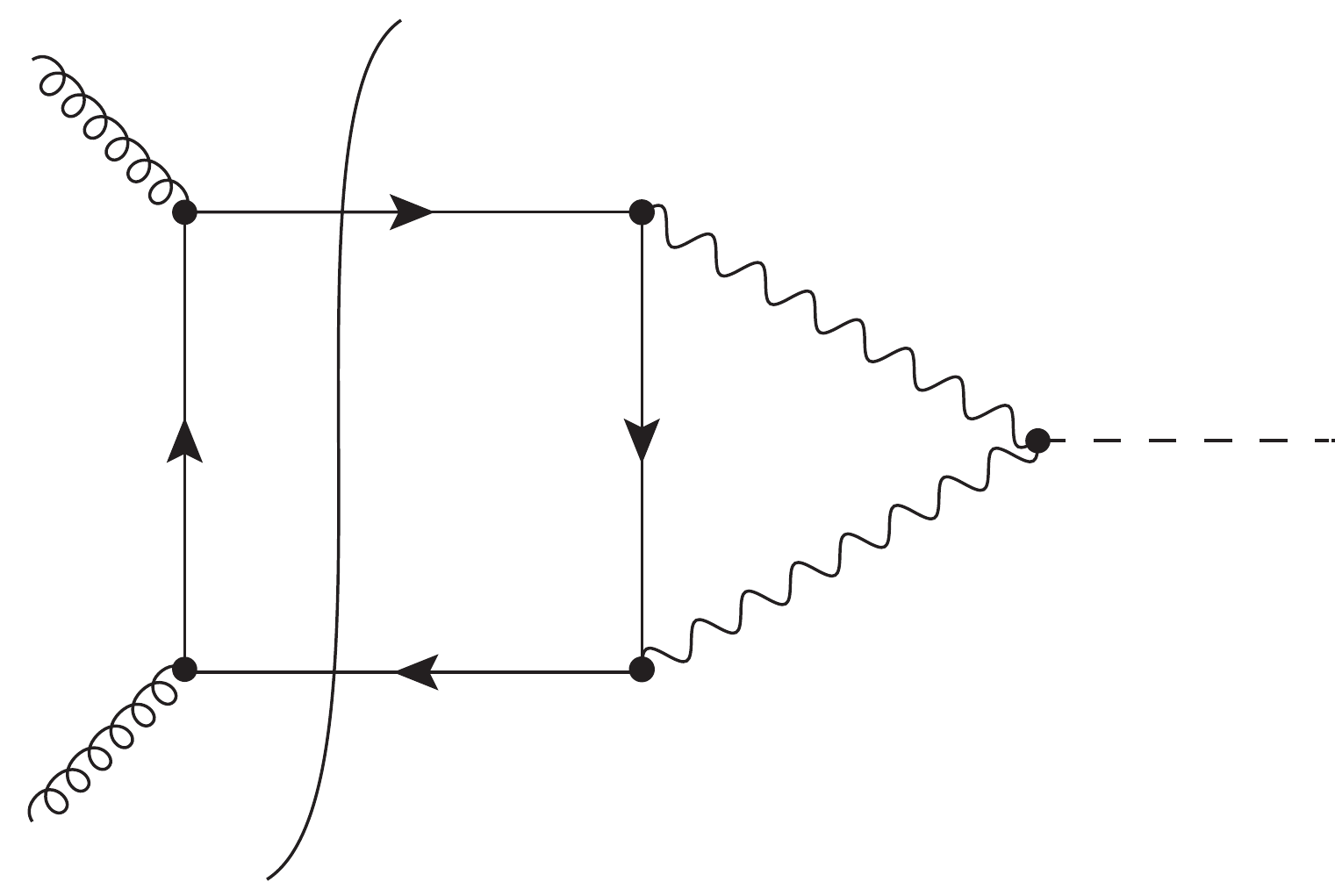}}
      \qquad \qquad 
      \subfloat[][Non-zero contribution to the cut for $s>0$ at NLO.]
      {\label{CUTB}\includegraphics[width=0.28\textwidth]{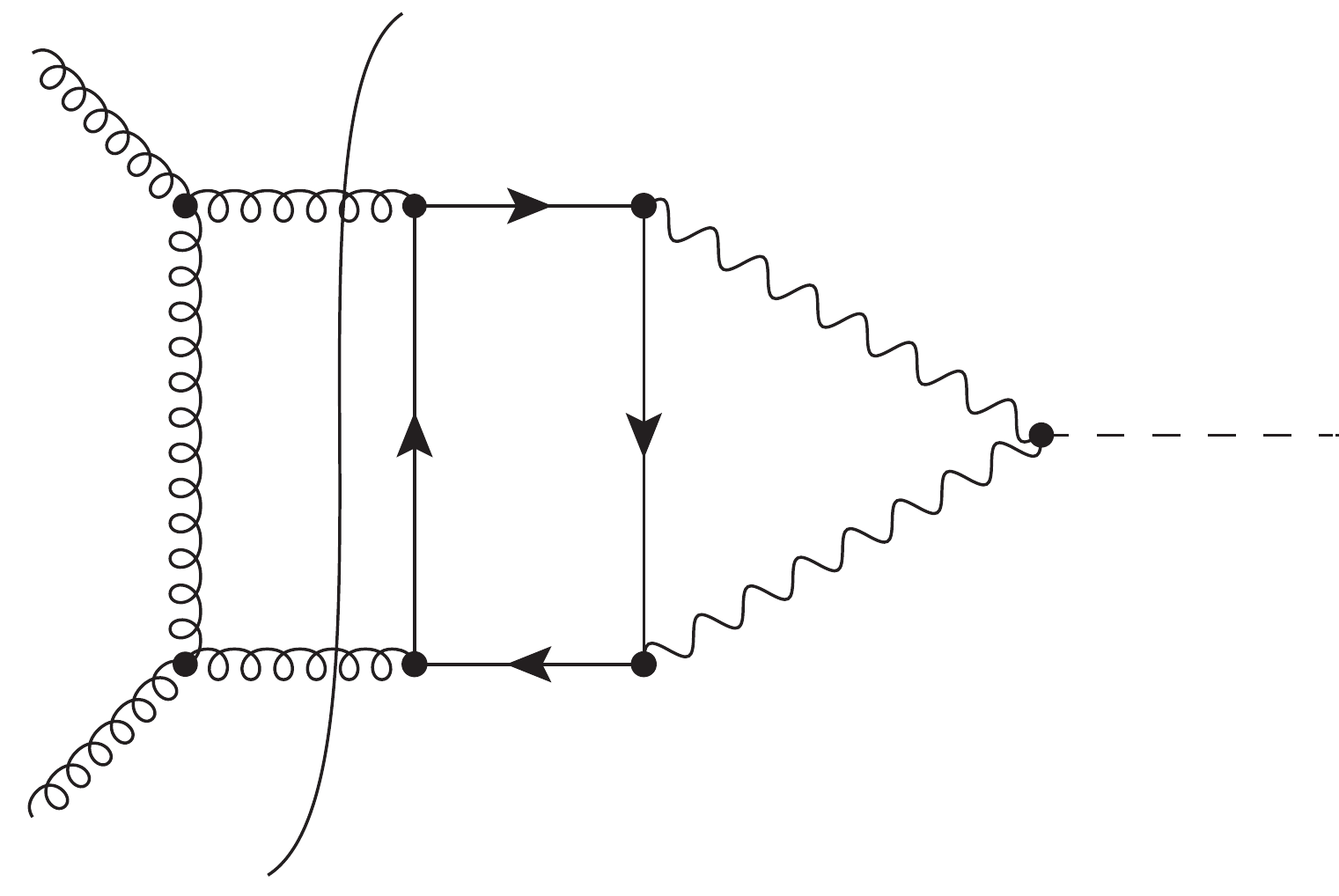}}
      \caption{Examples of cut-diagrams contributing to the discontinuity for $0<s< m_{W,Z}^2$.}
      \label{Cut}
\end{figure}

Before concluding, it is interesting to point 
out a feature of the $gg \to H$ scattering
amplitude, seen as a function of  $s = m_H^2\,.$
Naively, one would expect that both the LO and the NLO scattering amplitudes 
have discontinuities at 
$s=0$, $s=m_V^2$ and $s=4 m_V^2$, where $V=W,Z$ respectively. 
Physically, $s = m_H^2 > m_Z^2 > m_W^2$, so 
that the first two cuts $s=0$ and $s=m_V^2$ should both contribute to the
imaginary parts in Eq.~\eqref{eq:numres}.
However,   comparing absolute values of  
real and imaginary parts in Eq.~\eqref{eq:numres},
the difference in the relative importance of 
imaginary parts at NLO compared to LO is striking.
The reason for this can be traced back to the fact 
that, contrary to expectations, 
the LO amplitude is actually {\it real} in the region $0< s< m_V^2$, 
and that the discontinuity
at  $s=0$ 
contributes only to the imaginary part of NLO amplitude. 
This is easy to understand.
Indeed, the imaginary part of the LO amplitude in this region 
is obtained by cutting through the fermion loop 
and schematically corresponds to the process 
$gg \to q \bar{q} \quad | \quad q \bar{q} \to H \,$, see Fig.~\ref{CUTA}.
Clearly, since the Higgs boson cannot couple 
to massless fermions, this contribution must vanish. We emphasize that 
this is a feature of the {\it amplitude} that does not hold 
at the level of  individual MIs. 
The same is not true at NLO, 
where more cuts contribute to the discontinuity that starts at  $s=0$. 
In particular, a cut through a gluon loop provides 
a non-vanishing  contribution to the imaginary part 
of the amplitude for  $0 < s < m_V^2$ from the re-scattering process
$gg \to gg \quad | \quad gg \to H ,$ see Fig.~\ref{CUTB}.

\section{Conclusions} 
\label{sec:con}

We computed  the three-loop virtual contributions to 
next-to-leading order mixed QCD-electroweak corrections to Higgs boson 
production amplitude in the annihilation of two gluons $gg \to H$.  The analytic 
result for the amplitude is obtained for arbitrary relation between the Higgs 
boson  and the electroweak gauge boson masses.  We computed the required integrals 
using the method  of differential equations and obtained the boundary constants 
using the large-mass expansion valid in the limit $m_H \ll m_{W,Z}$. The finite part of the 
three-loop amplitude  is written in terms of  Goncharov polylogarithms and is 
easy to evaluate numerically. 

To understand the impact of the mixed QCD-electroweak corrections on the Higgs 
boson production cross section, the virtual corrections computed in this paper 
will have to be combined with the real emission contributions of the type $gg \to Hg$, 
$qg  \to Hq$ where, again, the Higgs boson couples to electroweak vector bosons.  
The computation of the relevant real emission amplitude is non-trivial as it involves 
two-loop box diagrams with both internal and external masses. Nevertheless, 
it is conceivable that, with the current computational technology, 
analytic results for these contributions can be obtained. 

\section*{Acknowledgements}
We are grateful to O.~Gituliar for his help with 
the program \texttt{Fuchsia}. 
We wish to thank Robbert Rietkerk and Christopher Wever 
for their help and support during all steps of this project.
Finally, we thank Erich Weihs for providing us with his 
Mathematica routines to numerically evaluate GPLs with 
arbitrary precision using Ginac~\cite{Vollinga:2004sn}.

The work of M.B. was supported by a graduate fellowship 
from Karlsruhe Graduate School “Collider Physics 
at the highest energies and at the highest precision”.

\begin{appendix}
\section{Topologies}
\label{Topologies}
All the Feynman integrals appearing in the amplitude can be regrouped into three families, according to what propagators they feature. Table~\ref{tabprop} shows the sets of inverse propagators for each family.

\begin{figure}[H]
      \centering
      \subfloat
      {\includegraphics[height=0.15\textheight]{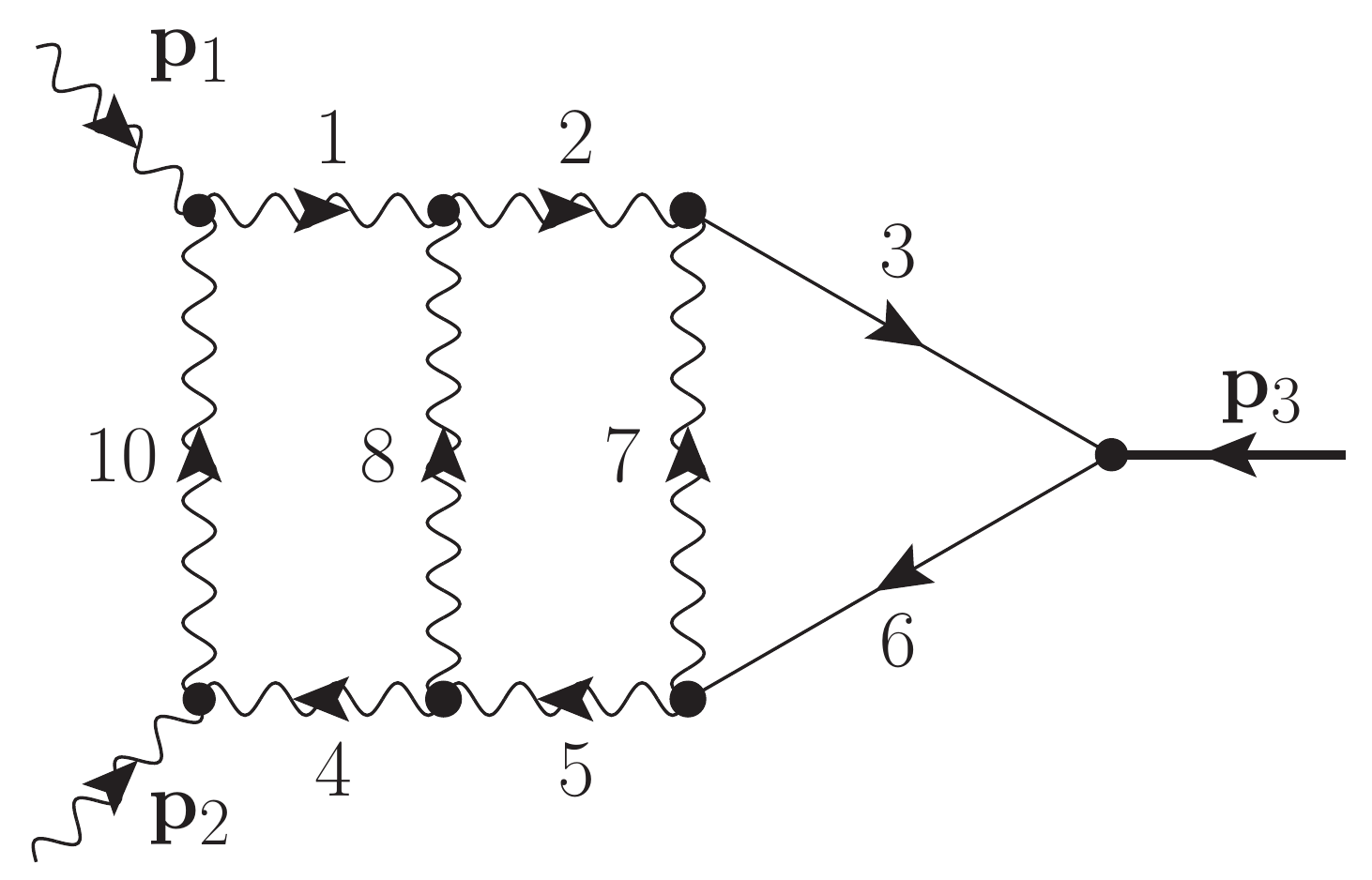}}
      \qquad
      \subfloat
      {\includegraphics[height=0.15\textheight]{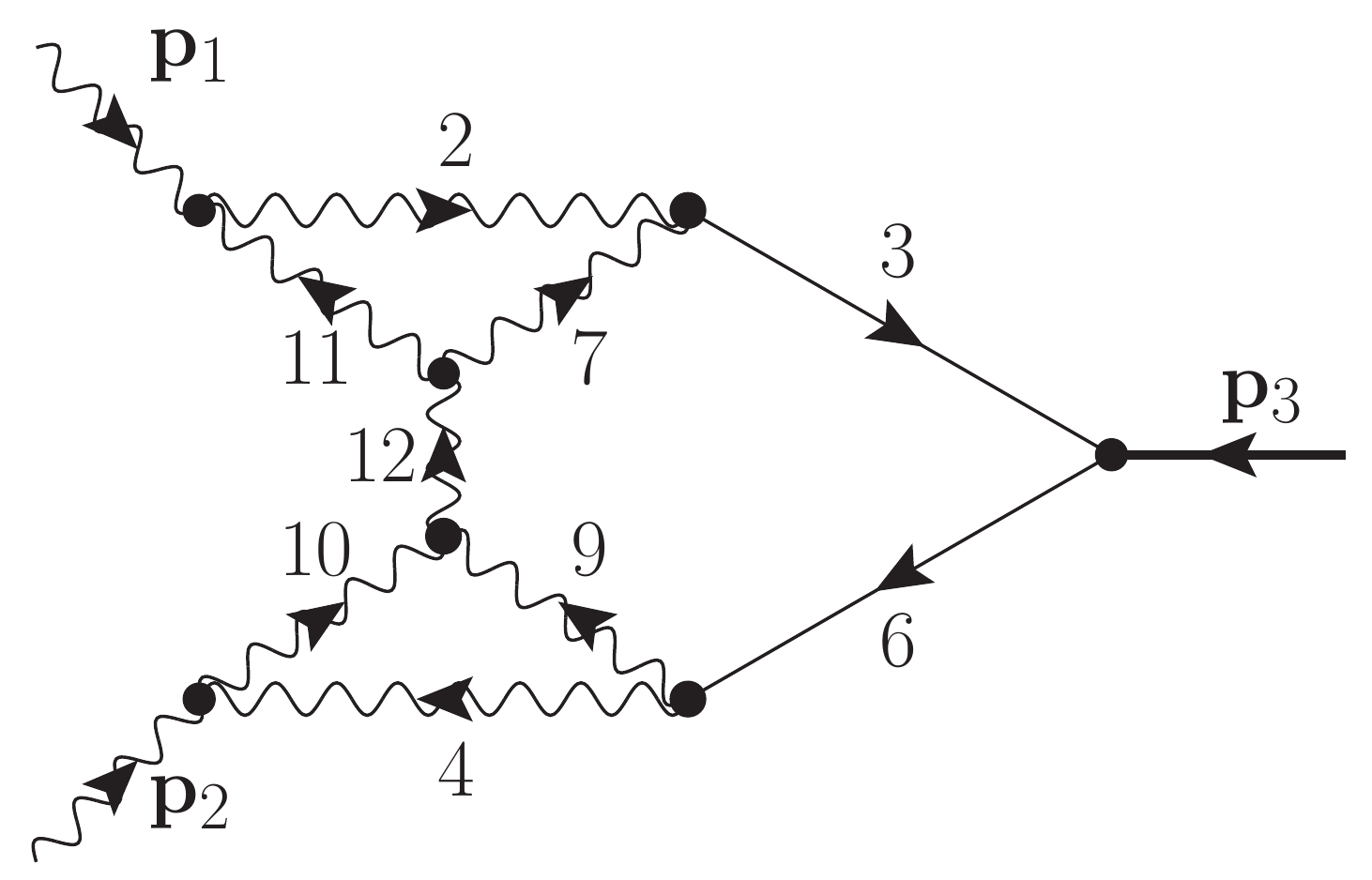}}
      \qquad
      \subfloat
      {\includegraphics[height=0.15\textheight]{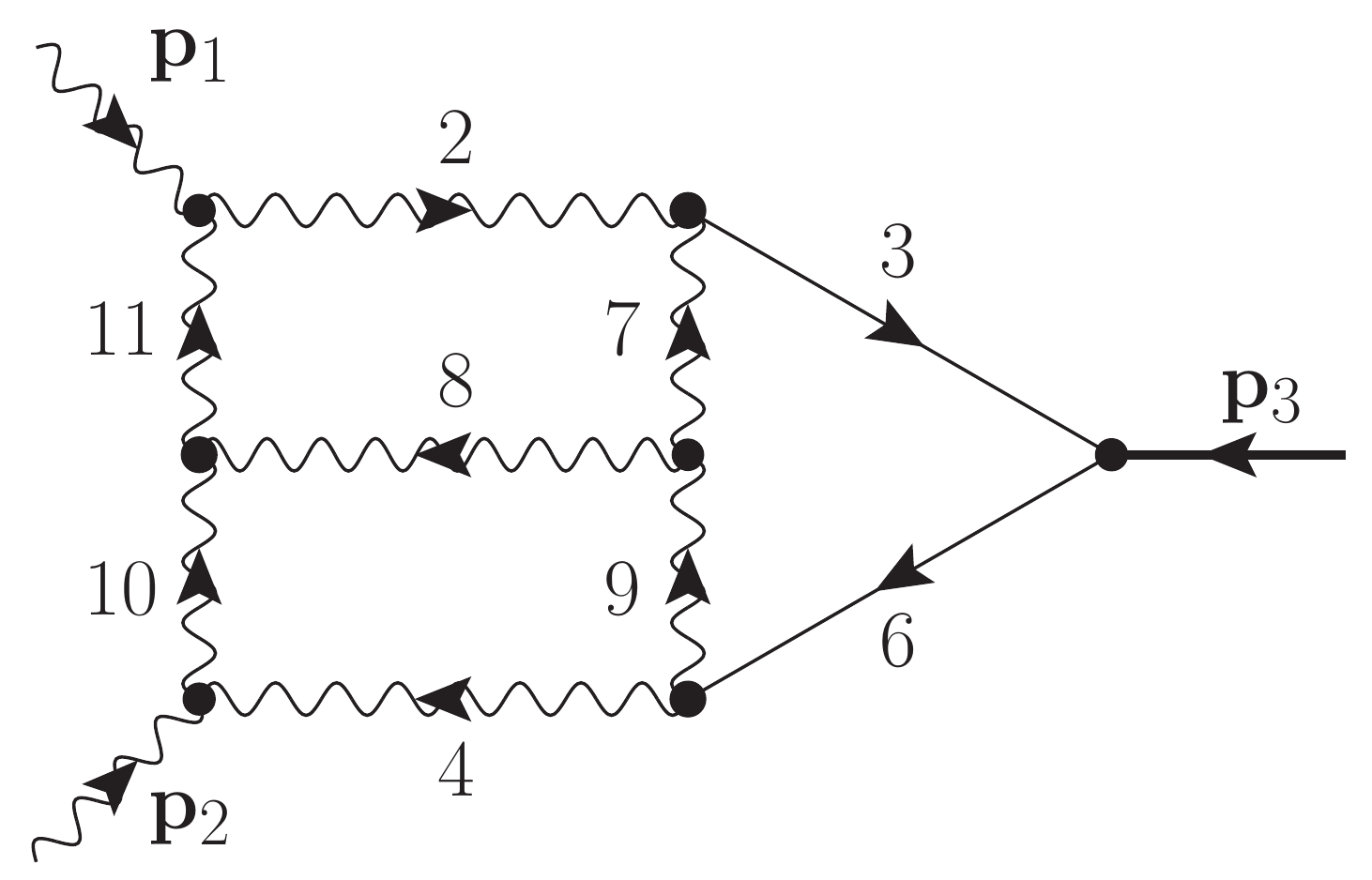}}
      \qquad
      \subfloat
      {\includegraphics[height=0.15\textheight]{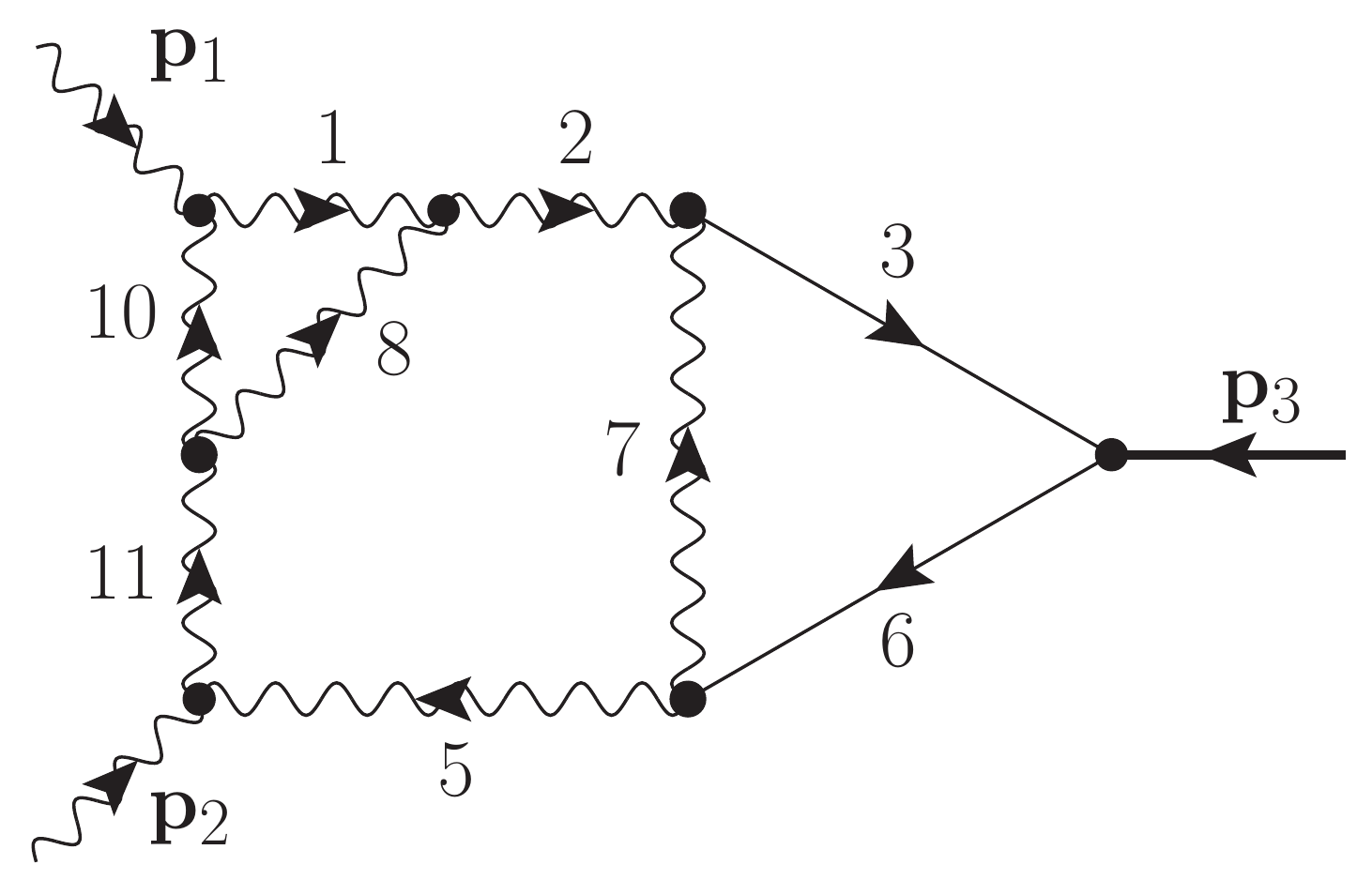}}
      \qquad
      \subfloat
      {\includegraphics[height=0.15\textheight]{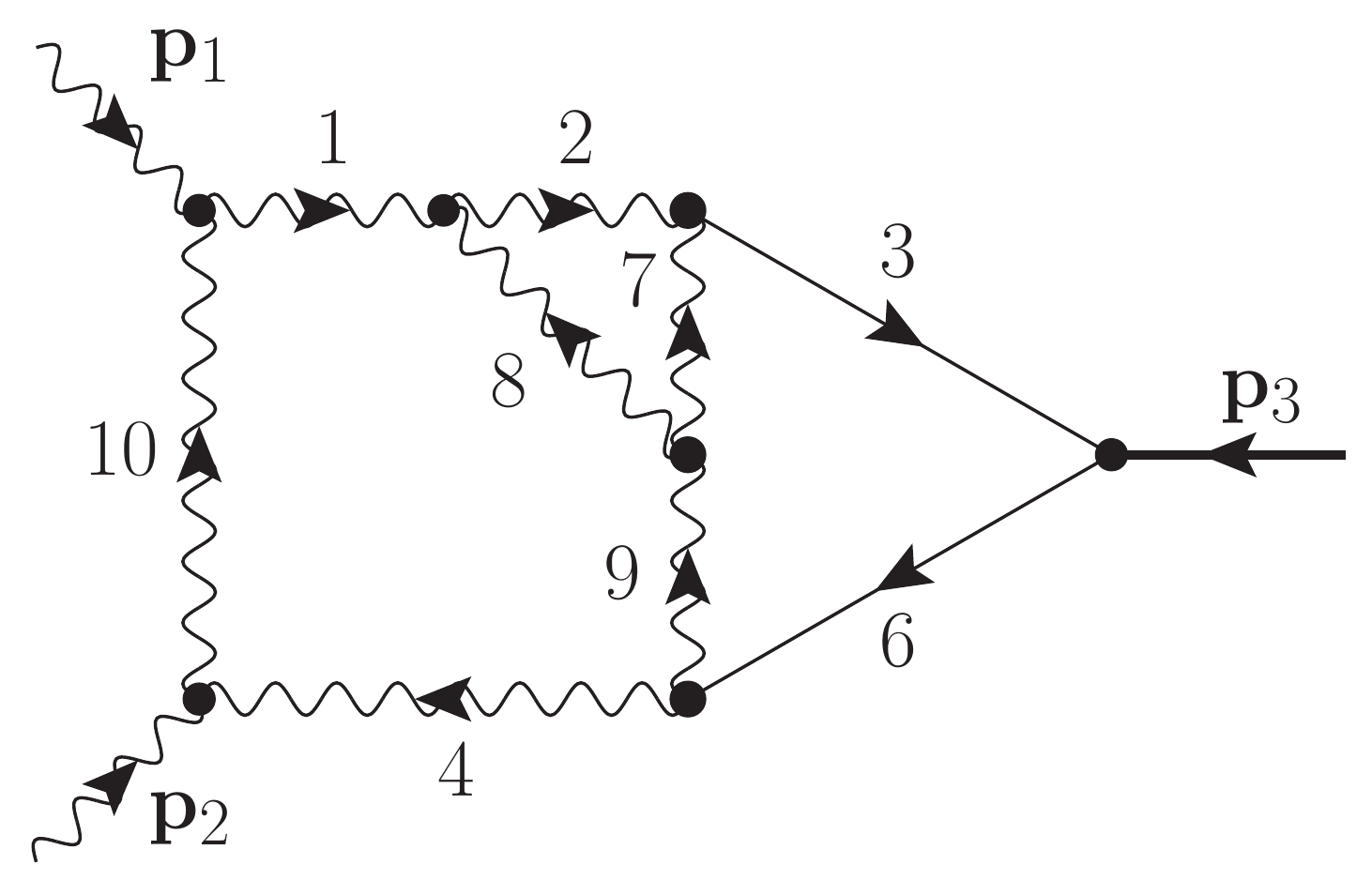}}
      \caption{Planar parent topologies (PP).}
      \label{PP}
\end{figure}

\begin{figure}[H]
      \centering
      \subfloat
      {\includegraphics[height=0.15\textheight]{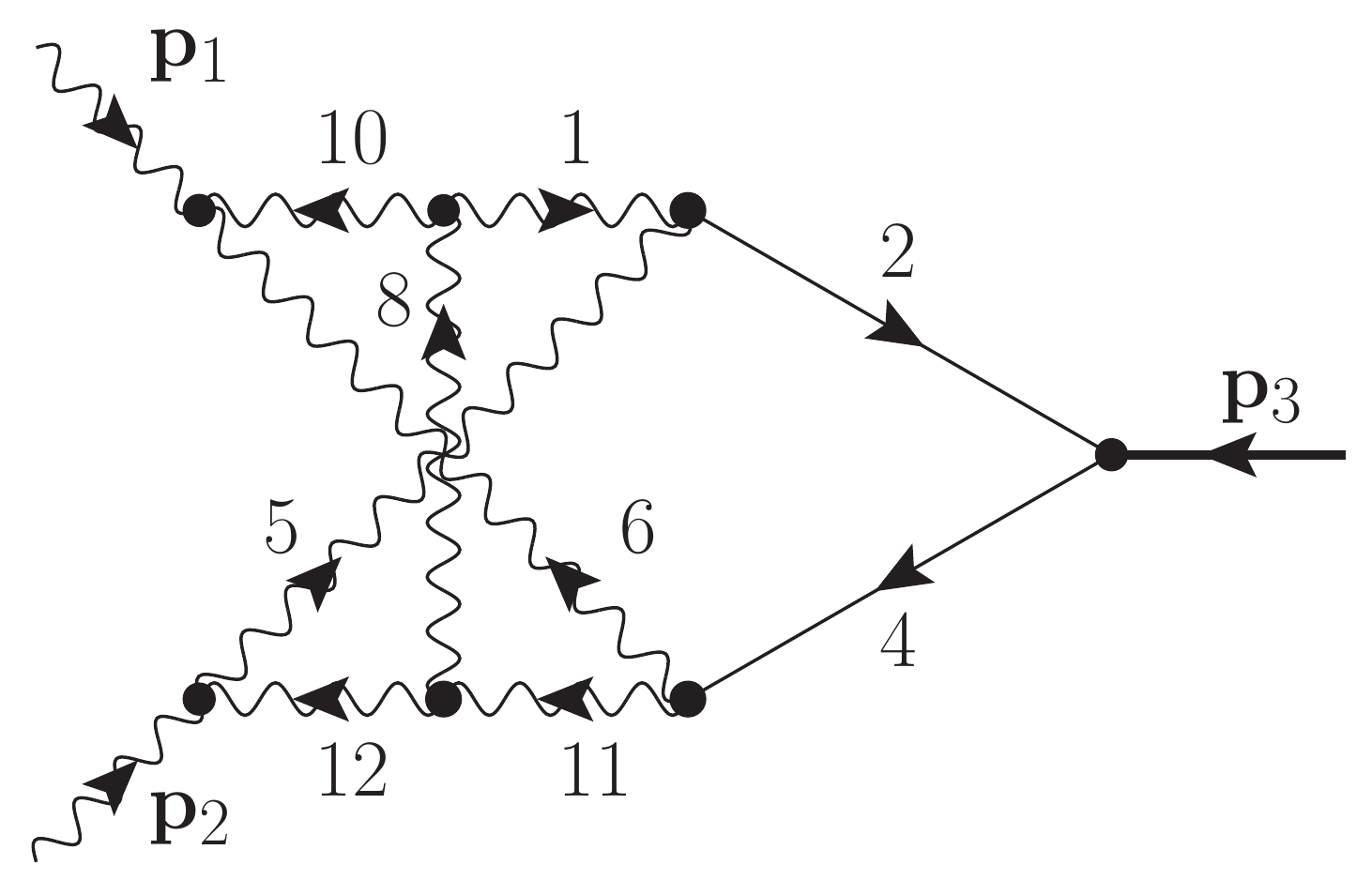}}
      \qquad
      \subfloat
      {\includegraphics[height=0.15\textheight]{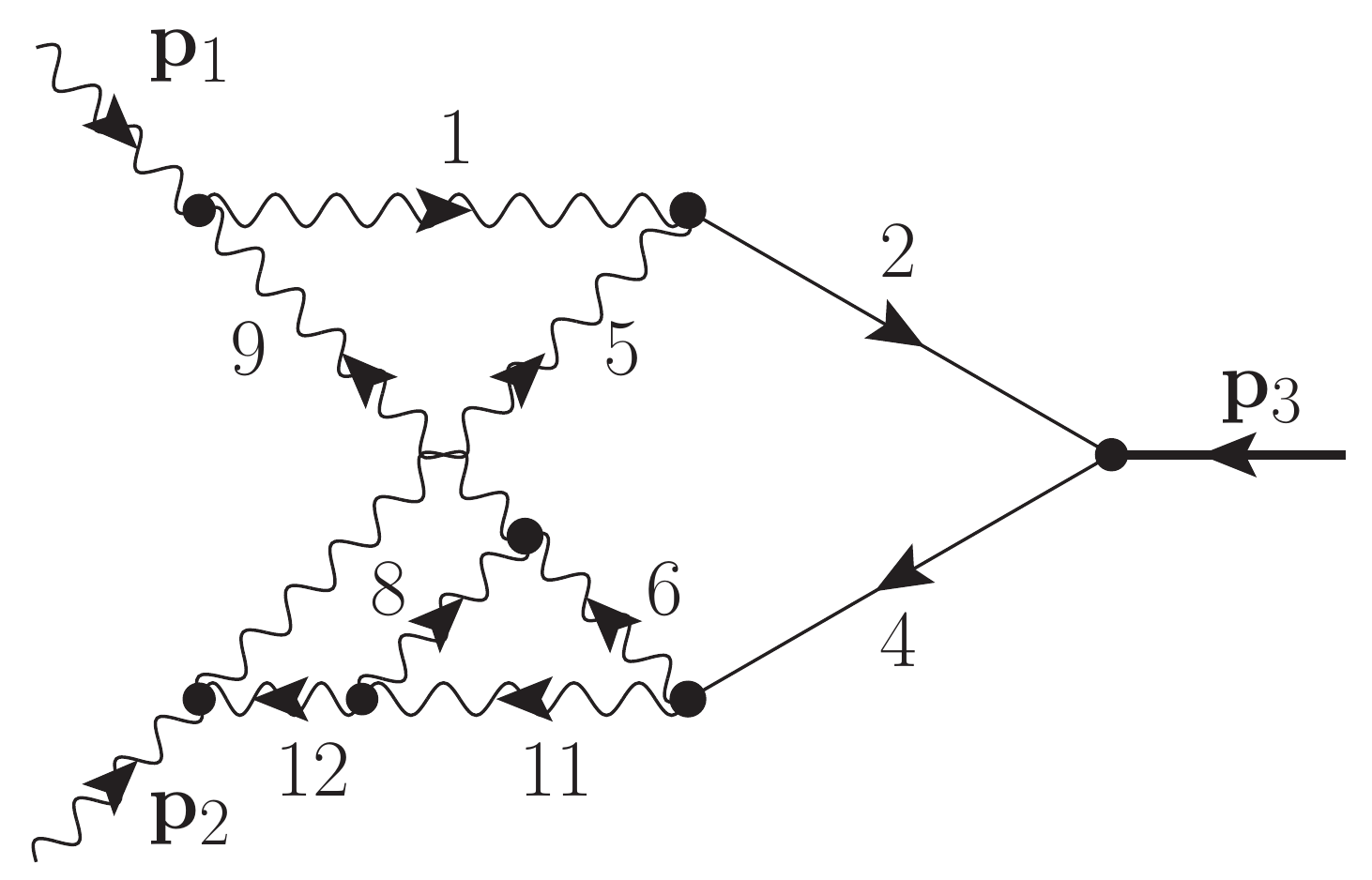}}
      \\
      \subfloat
      {\includegraphics[height=0.15\textheight]{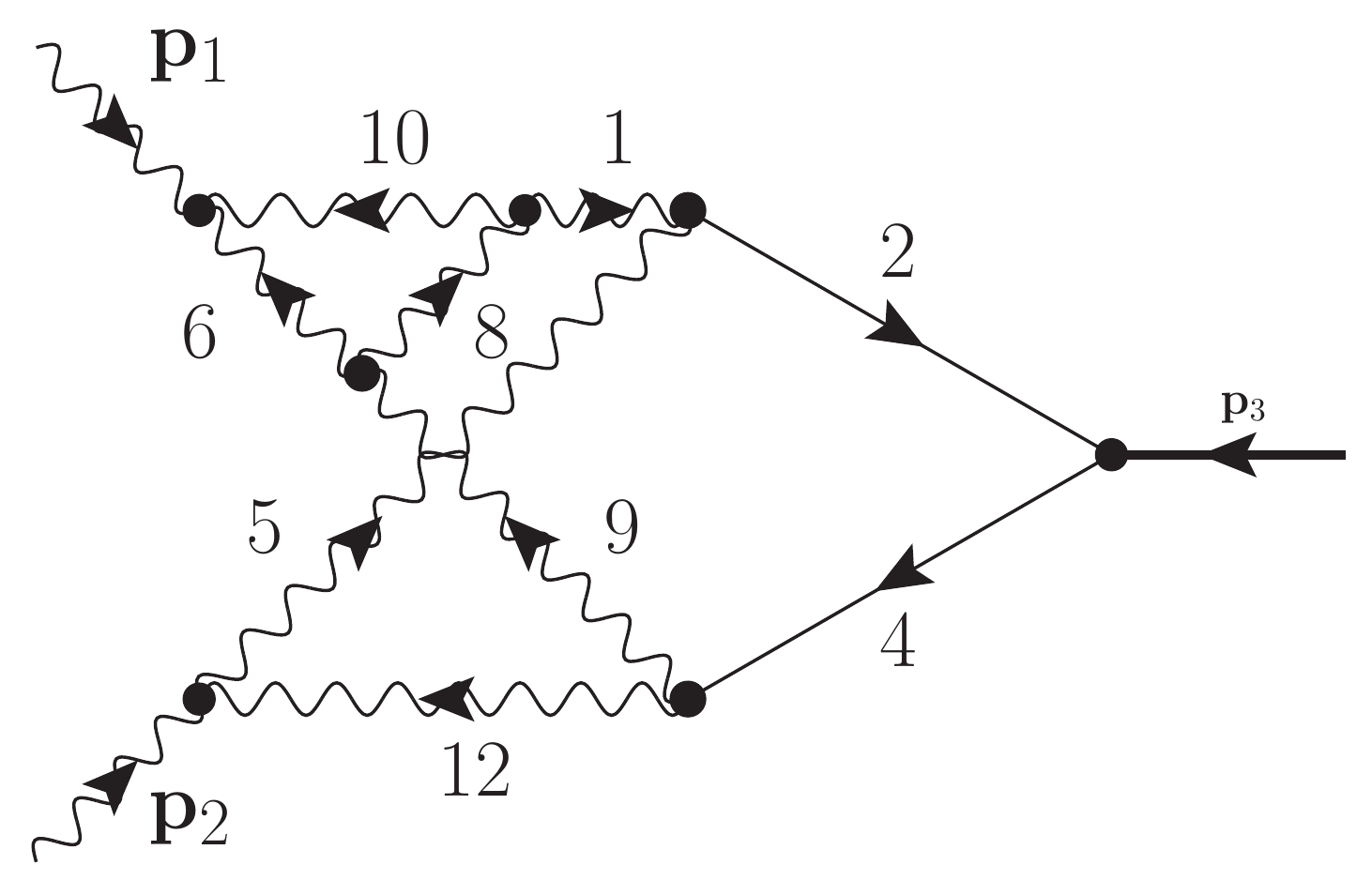}}
      \qquad
      \subfloat
      {\includegraphics[height=0.15\textheight]{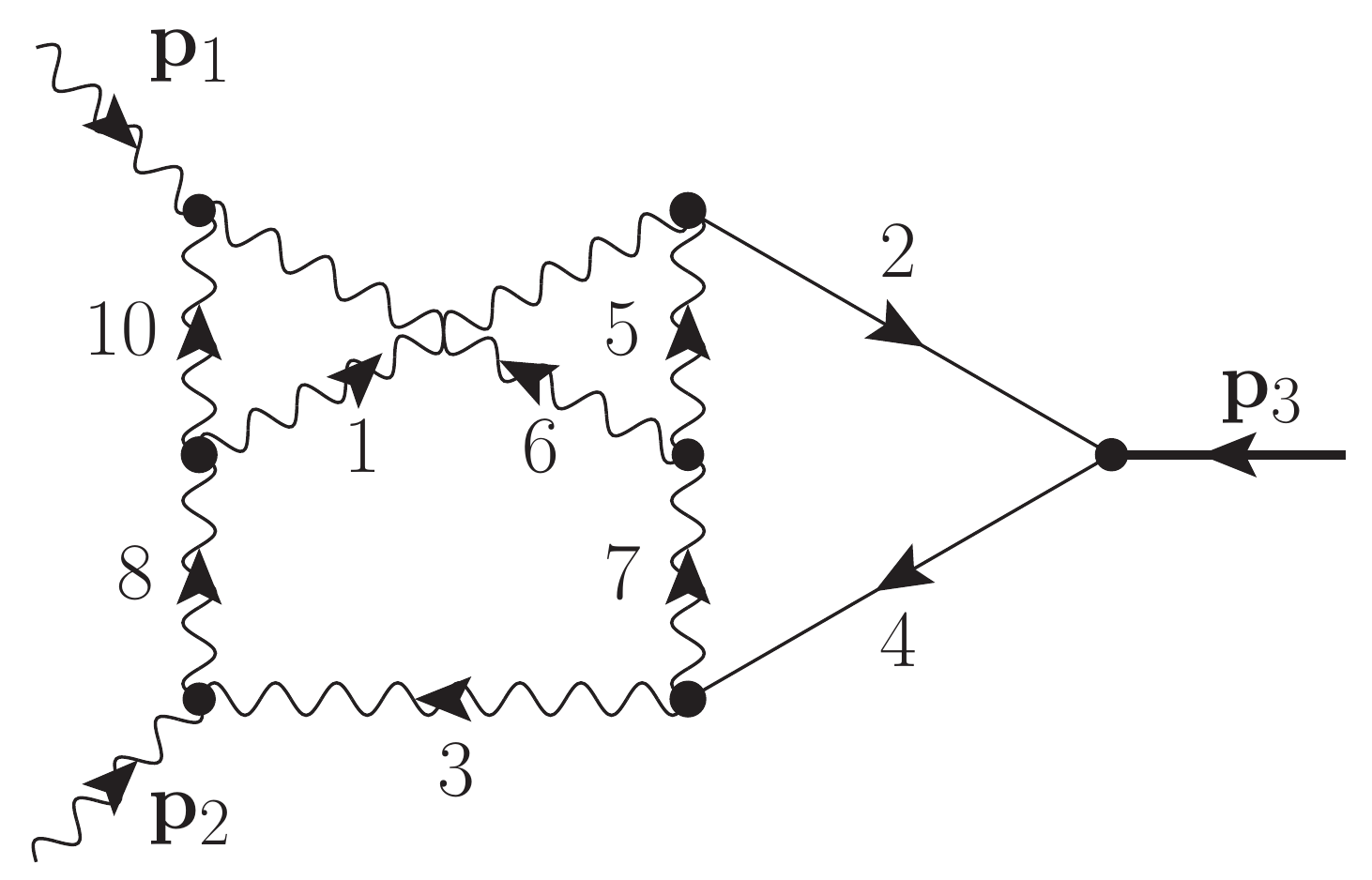}}
      \caption{Non-planar A-type parent topologies (NA).}
      \label{NA}
\end{figure}

\begin{figure}[H]
      \centering
      \subfloat
      {\includegraphics[height=0.15\textheight]{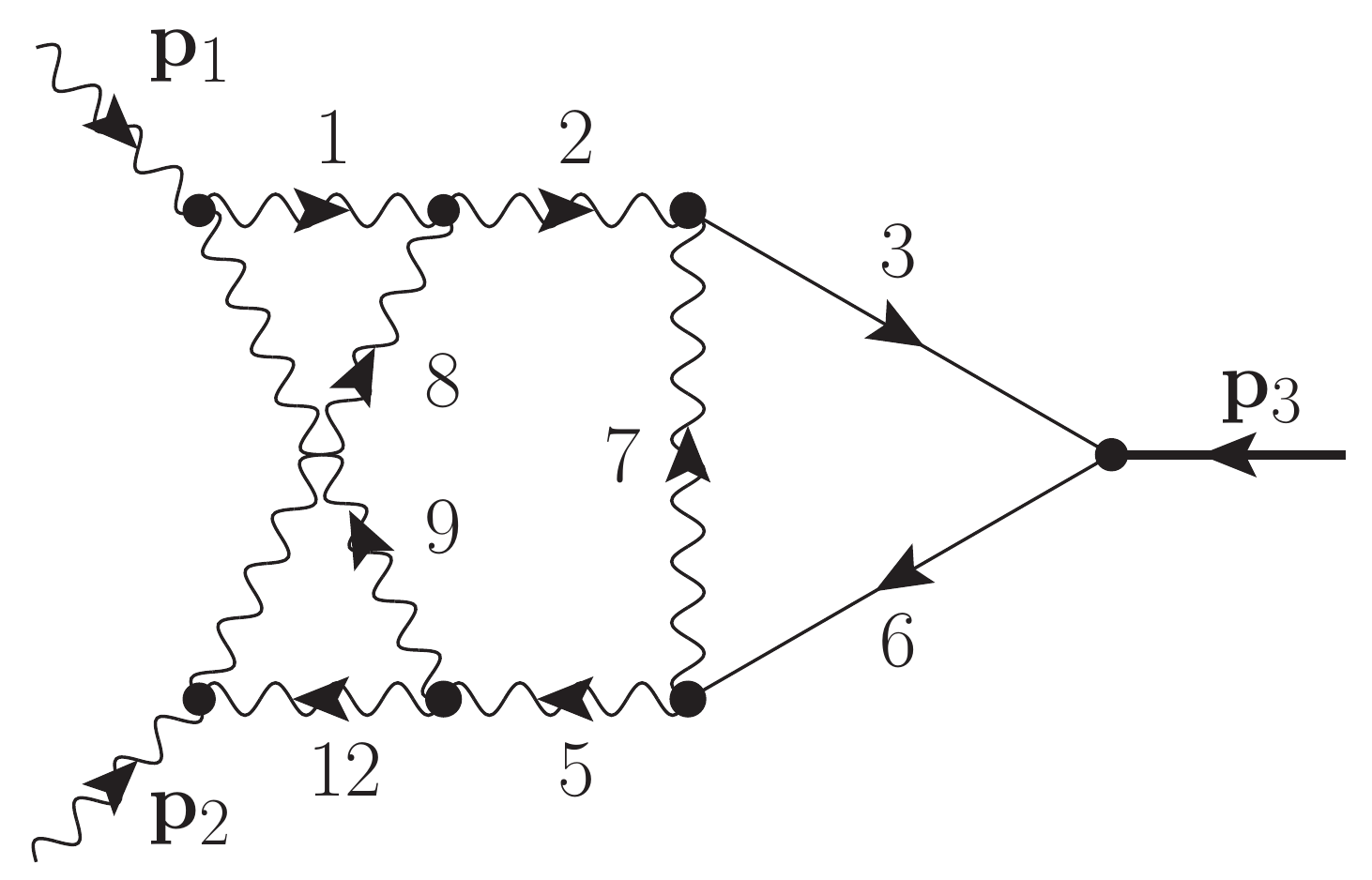}}
      \qquad
      \subfloat
      {\includegraphics[height=0.15\textheight]{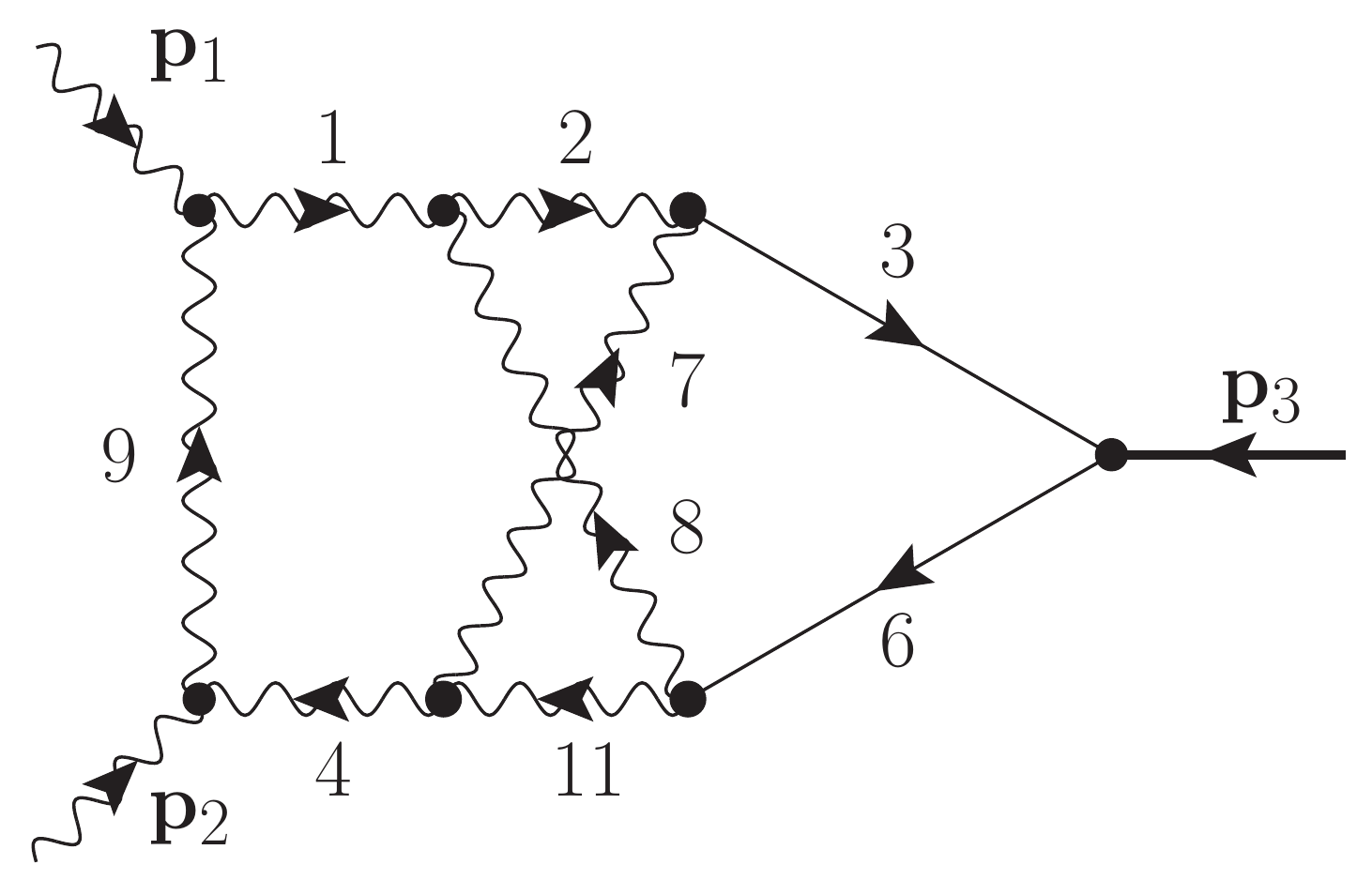}}
      \caption{Non-planar B-type parent topologies (NB).}
      \label{NB}
\end{figure}

The parent topologies of each family are depicted in Figs.~\ref{PP}, \ref{NA} and~\ref{NB}. Solid lines are massive, wavy lines are massless.

\begin{table}[h]
\centering
\begin{tabular}{clll}
\toprule
  \multirow{2}*{Label}	&\multicolumn{3}{c}{Families of denominators}	\\
				\cmidrule(lr){2-4}
				&Planar (PP, Fig.~\ref{PP})		&Non-planar, A (NA, Fig.~\ref{NA})	&Non-planar, B (NB, Fig.~\ref{NB})		\\
\midrule
  $[1]$				&$(k_1)^2$			&$(k_2)^2$			&$(k_1)^2$			\\[2ex]
  $[2]$				&$(k_2)^2$			&$(k_3)^2 \,\,-\,\, M^2$	&$(k_2)^2$			\\[2ex]
  $[3]$				&$(k_3)^2 \,\,-\,\, M^2$	&$(k_1+p_3)^2$			&$(k_3)^2 \,\,-\,\, M^2$	\\[2ex]
  $[4]$				&$(k_1+p_3)^2$			&$(k_3+p_3)^2 \,\,-\,\, M^2$	&$(k_1+p_3)^2$			\\[2ex]
  $[5]$				&$(k_2+p_3)^2$			&$(k_3-k_2)^2$			&$(k_2+p_3)^2$			\\[2ex]
  $[6]$				&$(k_3+p_3)^2 \,\,-\,\, M^2$	&$(k_2-k_1)^2$			&$(k_3+p_3)^2 \,\,-\,\, M^2$	\\[2ex]
  $[7]$				&$(k_3-k_2)^2$			&$(k_3-k_1)^2$			&$(k_3-k_2)^2$			\\[2ex]
  $[8]$				&$(k_2-k_1)^2$			&$(k_1-p_1)^2$			&$(k_2-k_1)^2$			\\[2ex]
  $[9]$				&$(k_3-k_1)^2$			&$(k_2-p_1)^2$			&$(k_1-p_1)^2$			\\[2ex]
  $[10]$			&$(k_1-p_1)^2$			&$(k_1-k_2-p_1)^2$		&$(k_3-p_1)^2$			\\[2ex]
  $[11]$			&$(k_2-p_1)^2$			&$(k_1-k_2+k_3+p_3)^2$		&$(k_1-k_2+k_3+p_3)^2$		\\[2ex]
  $[12]$			&$(k_3-p_1)^2$			&$(k_3-k_2-p_2)^2$		&$(k_2-k_1-p_2)^2$		\\
\bottomrule
\end{tabular}
\caption{The three families of denominators appearing in the amplitude.}
\label{tabprop}
\end{table}


\section{Master integrals}
\label{Master_Integrals}
\begin{figure}[H]
      \captionsetup[subfigure]{labelformat=empty}
      \centering
      \subfloat[][$I_{1}$ (PP)]
      {\includegraphics[width=0.16\textwidth]{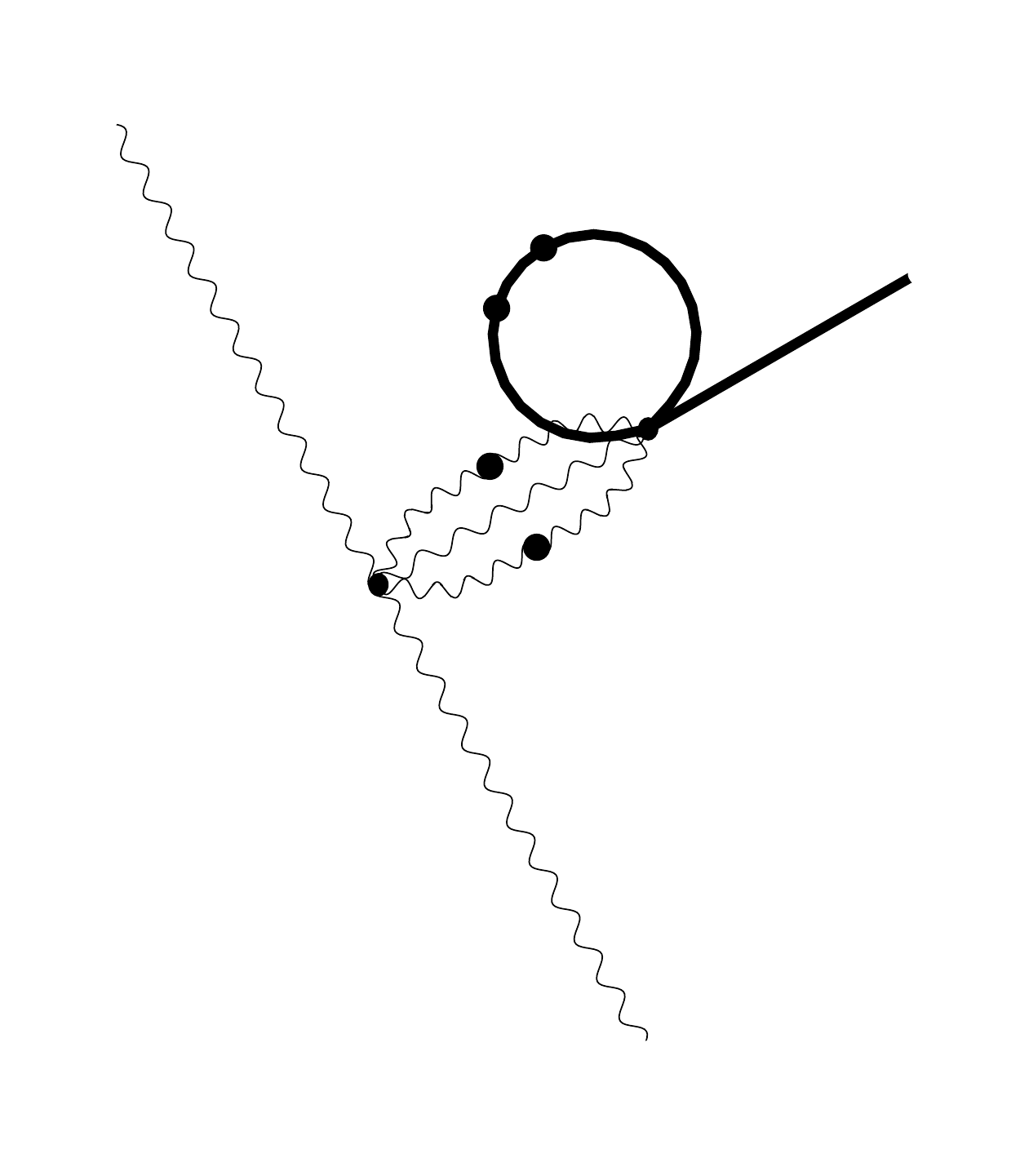}}
      \subfloat[][$I_{2}$ (PP)]
      {\includegraphics[width=0.16\textwidth]{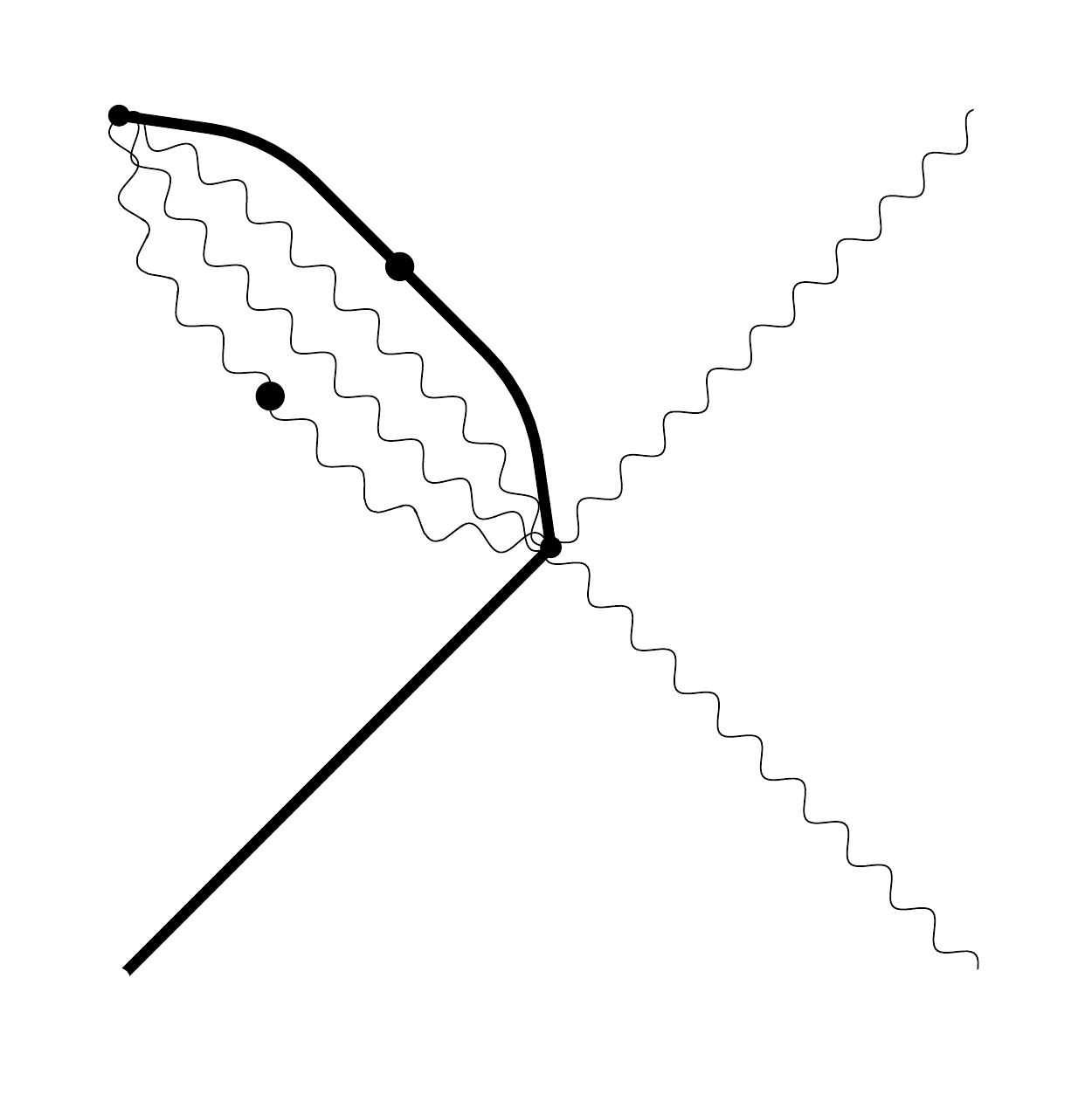}}
      \subfloat[][$I_{3}$ (PP)]
      {\includegraphics[width=0.16\textwidth]{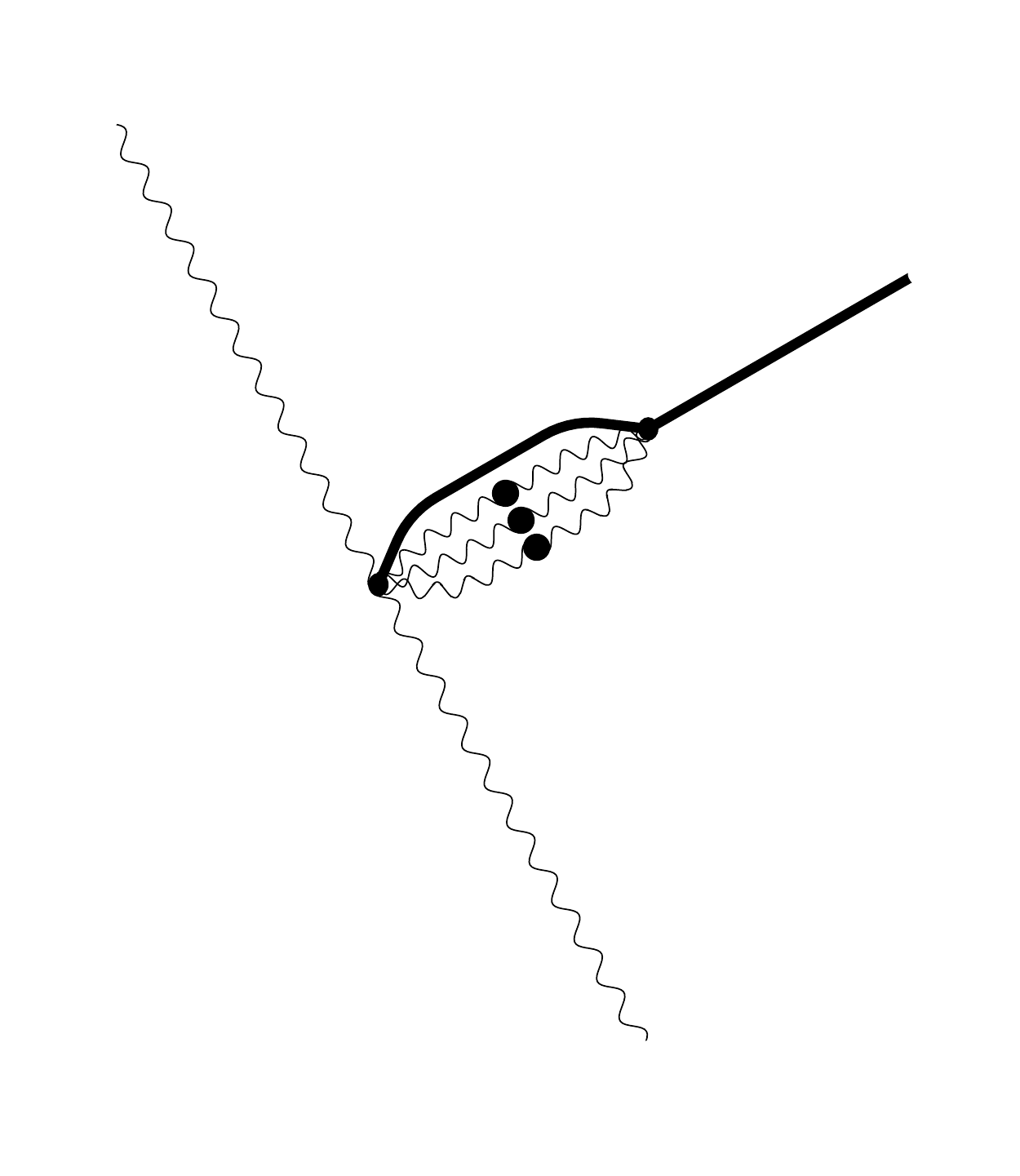}}
      \subfloat[][$I_{4}$ (PP)]
      {\includegraphics[width=0.16\textwidth]{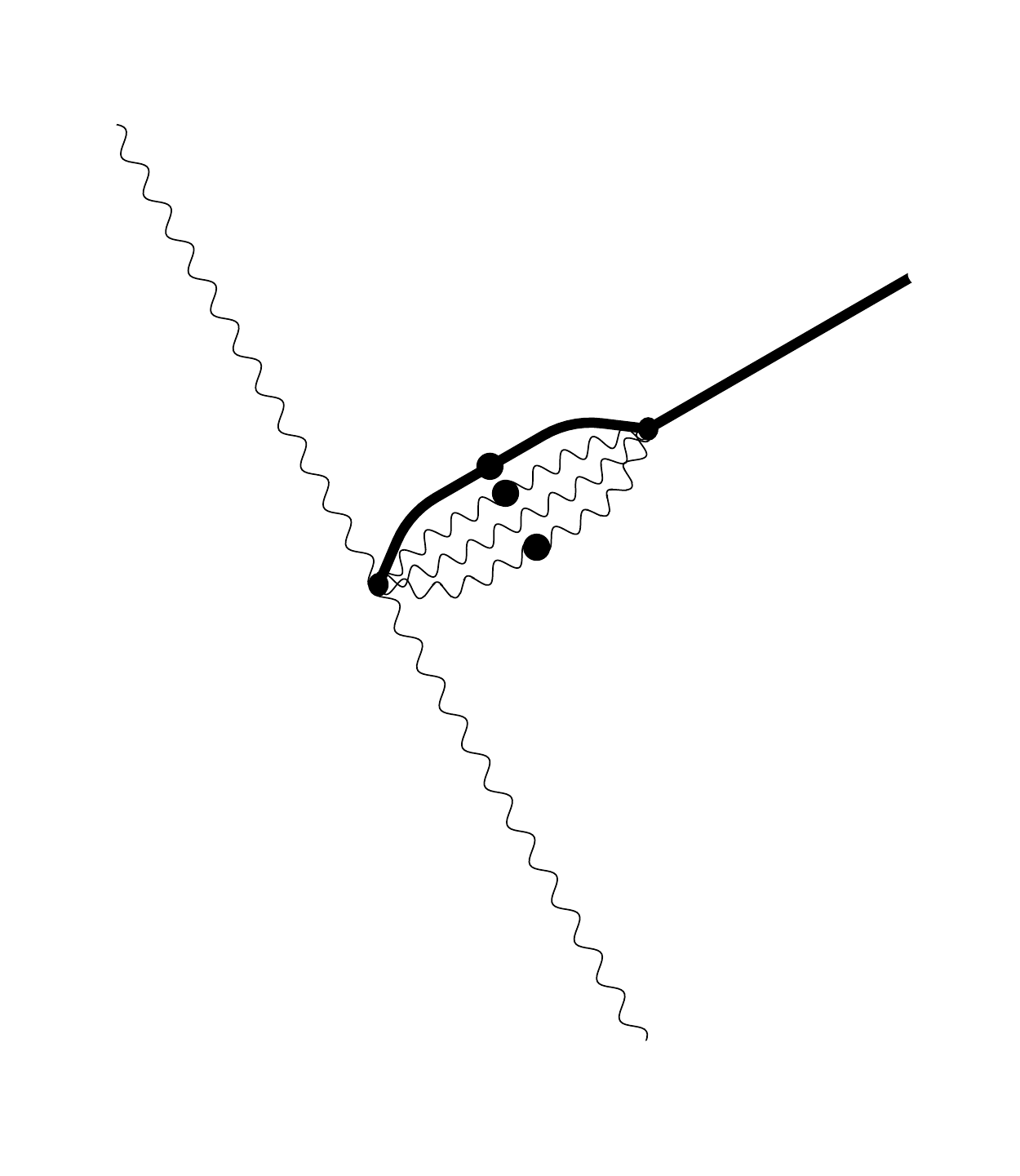}}
      \subfloat[][$I_{5}$ (PP)]
      {\includegraphics[width=0.16\textwidth]{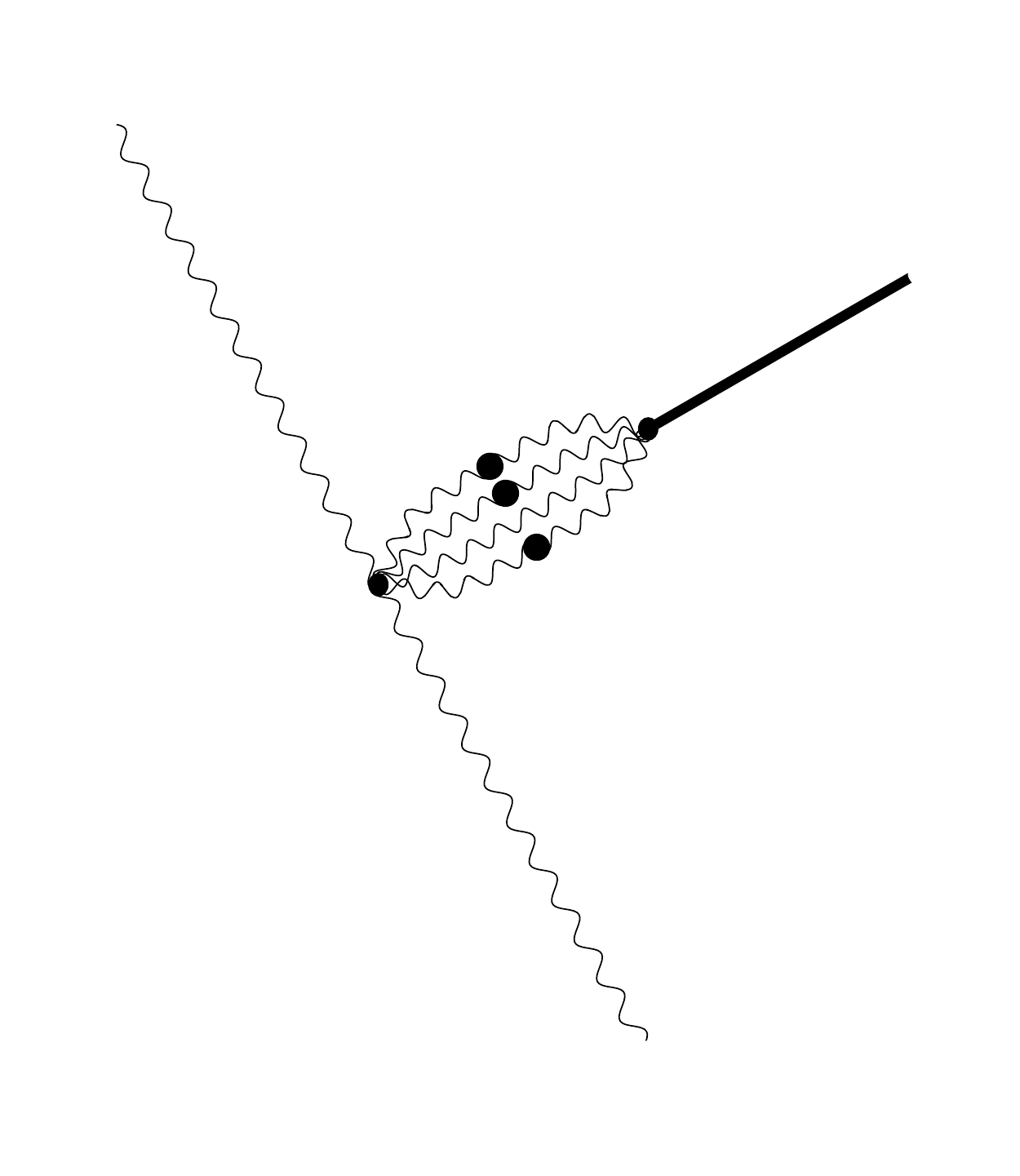}}
      \subfloat[][$I_{6}$ (PP)]
      {\includegraphics[width=0.16\textwidth]{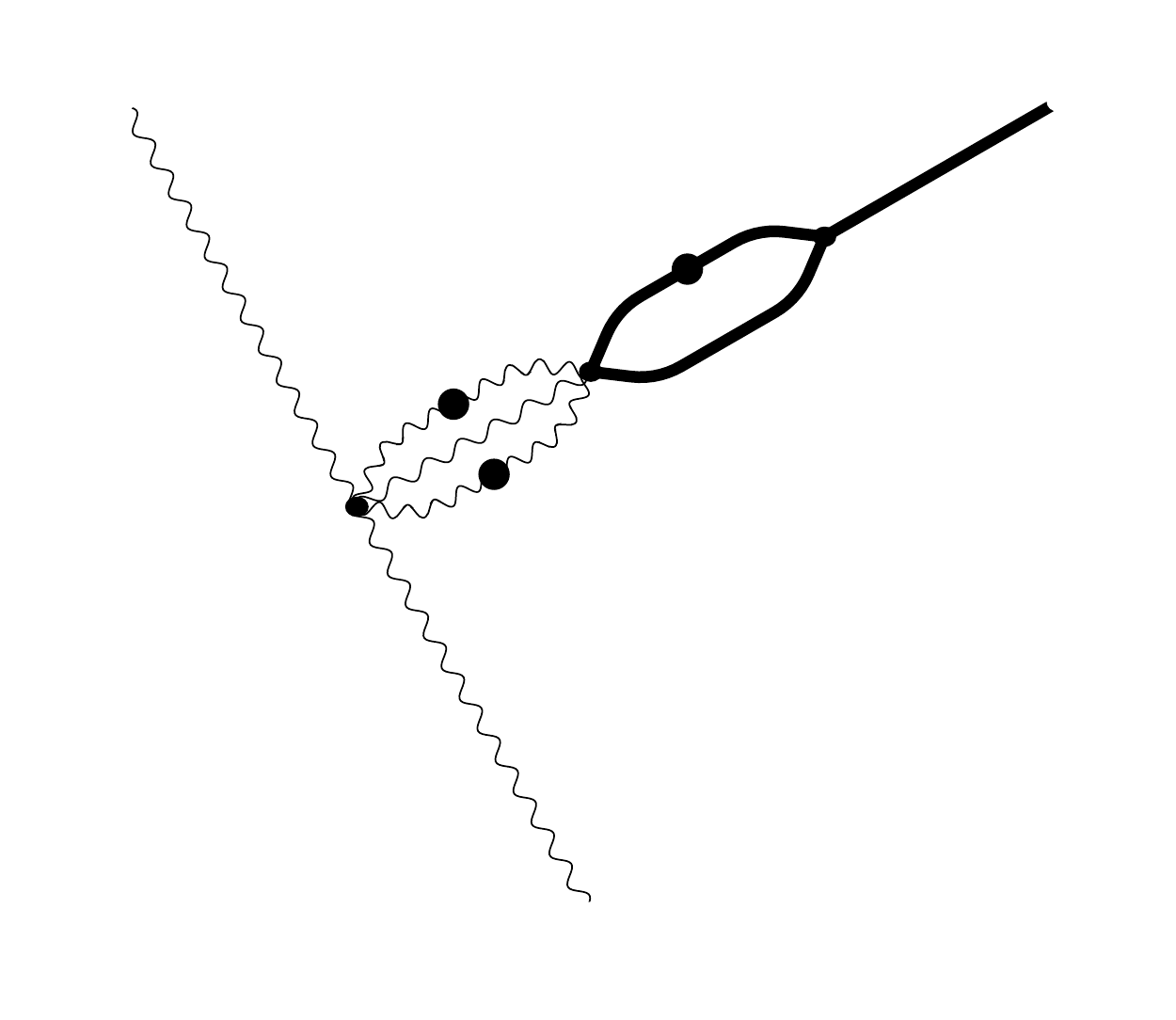}}
      \\

      \subfloat[][$I_{7}$ (PP)]
      {\includegraphics[width=0.16\textwidth]{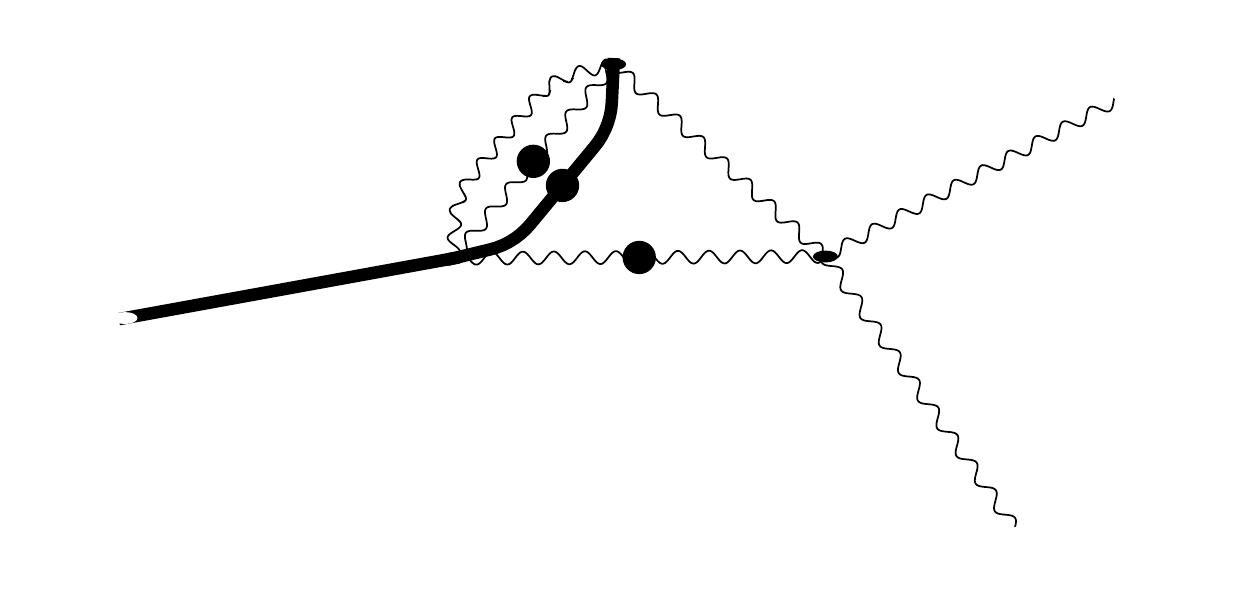}}
      \subfloat[][$I_{8}$ (PP)]
      {\includegraphics[width=0.16\textwidth]{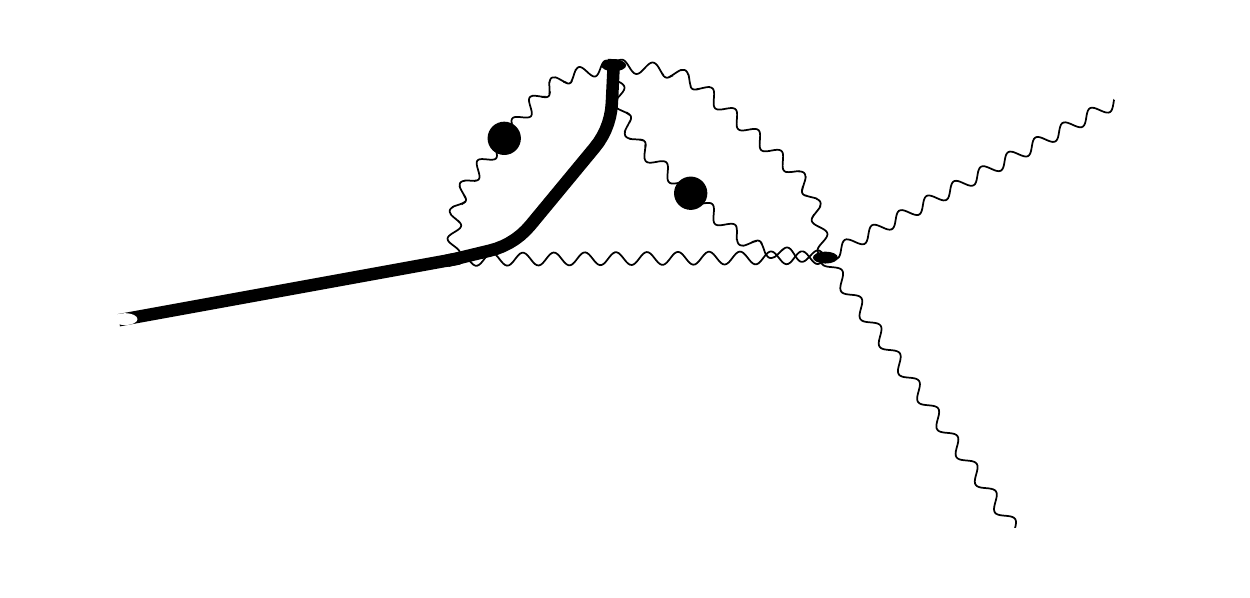}}
      \subfloat[][$I_{9}$ (PP)]
      {\includegraphics[width=0.16\textwidth]{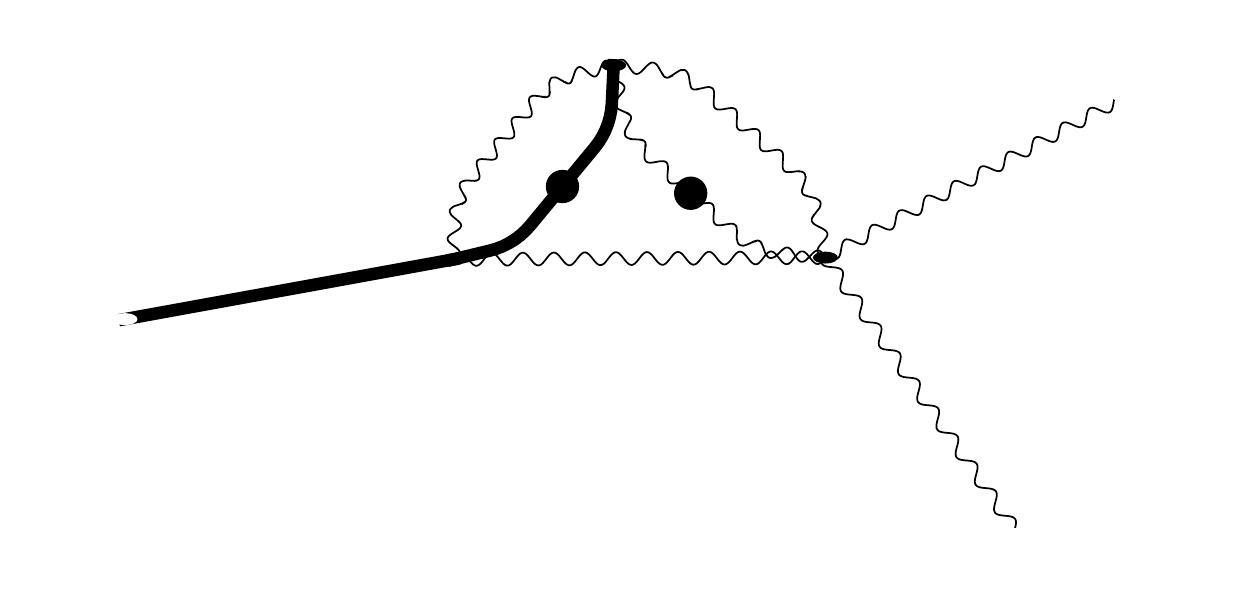}}
      \subfloat[][$I_{10}$ (PP)]
      {\includegraphics[width=0.16\textwidth]{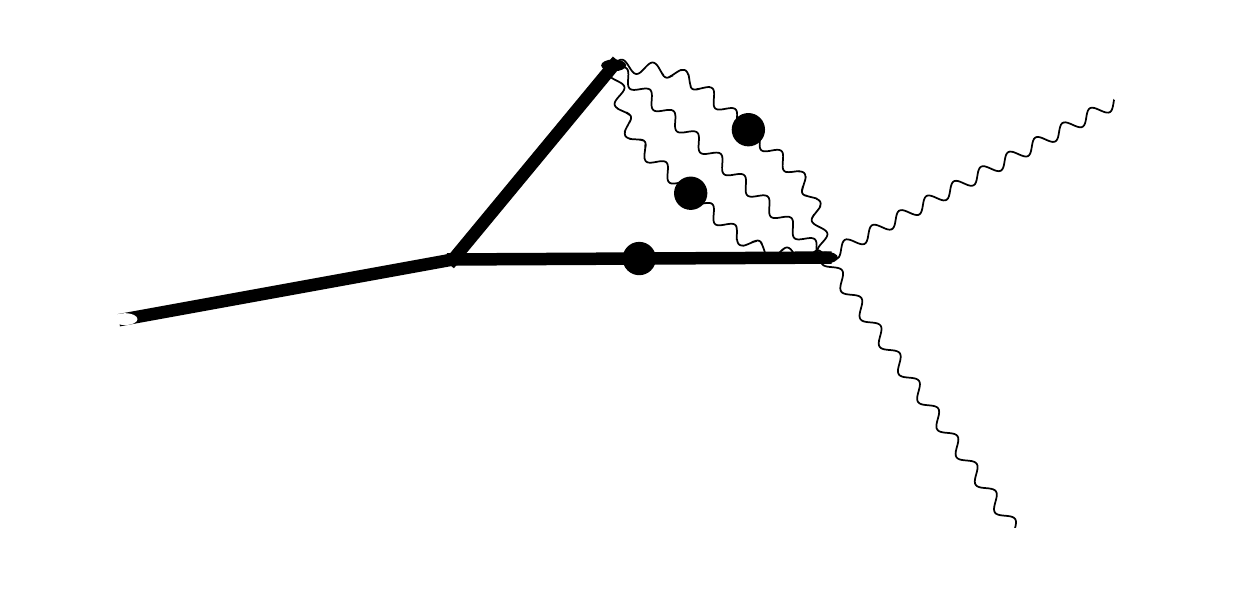}}
      \subfloat[][$I_{11}^*$ (PP)]
      {\includegraphics[width=0.16\textwidth]{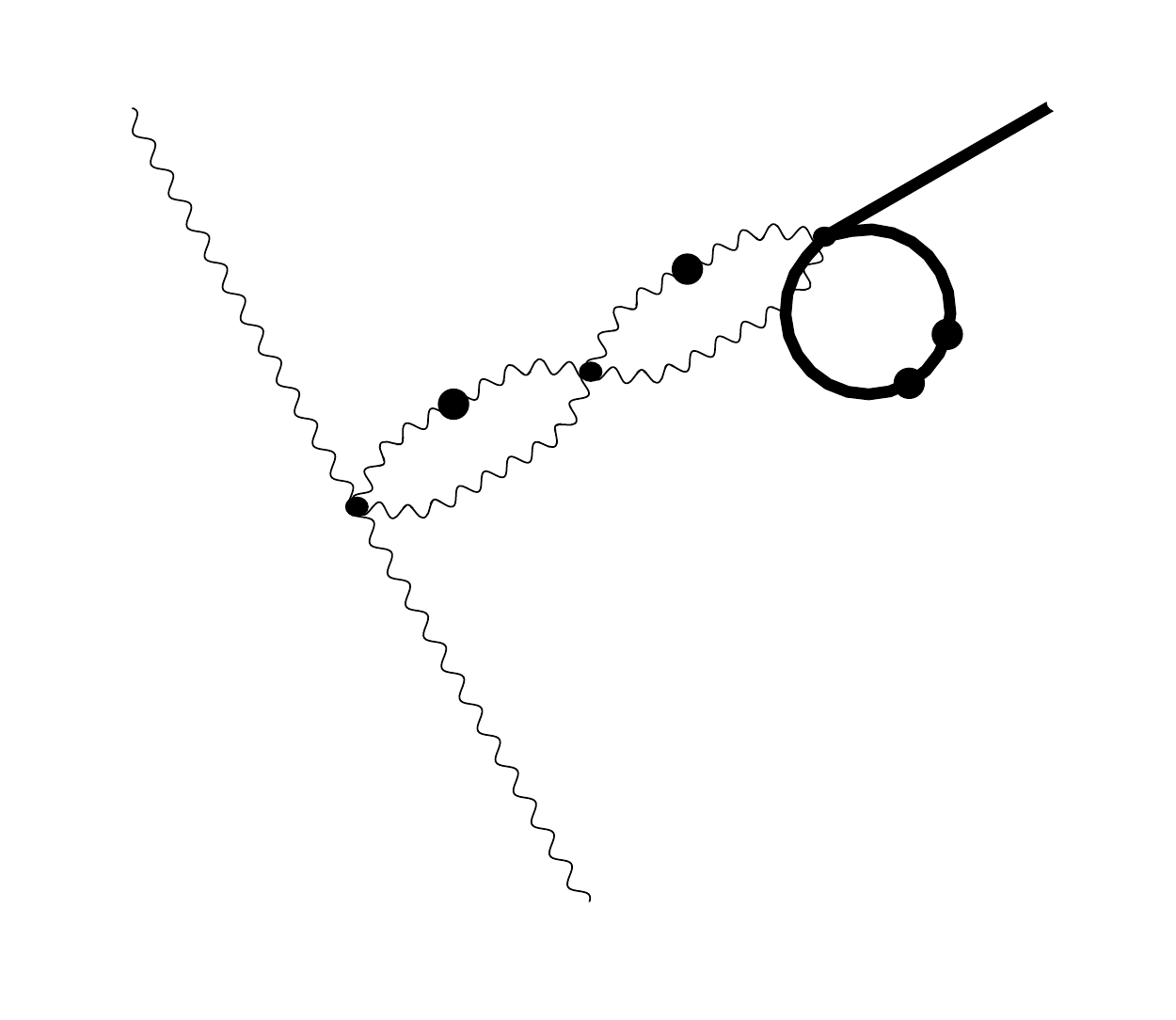}}
      \subfloat[][$I_{12}$ (PP)]
      {\includegraphics[width=0.16\textwidth]{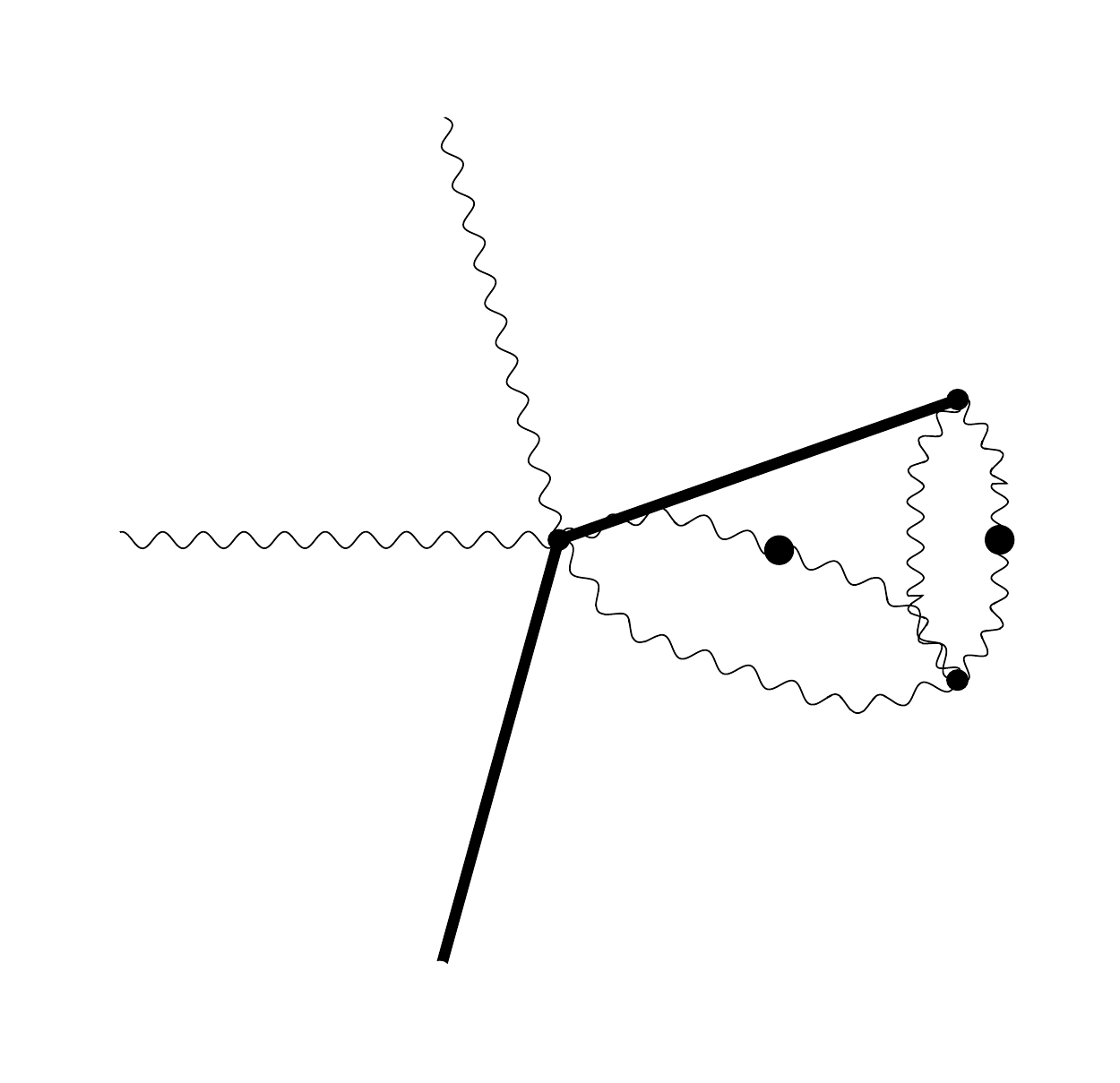}}
      \\

      \subfloat[][$I_{13}$ (PP)]
      {\includegraphics[width=0.16\textwidth]{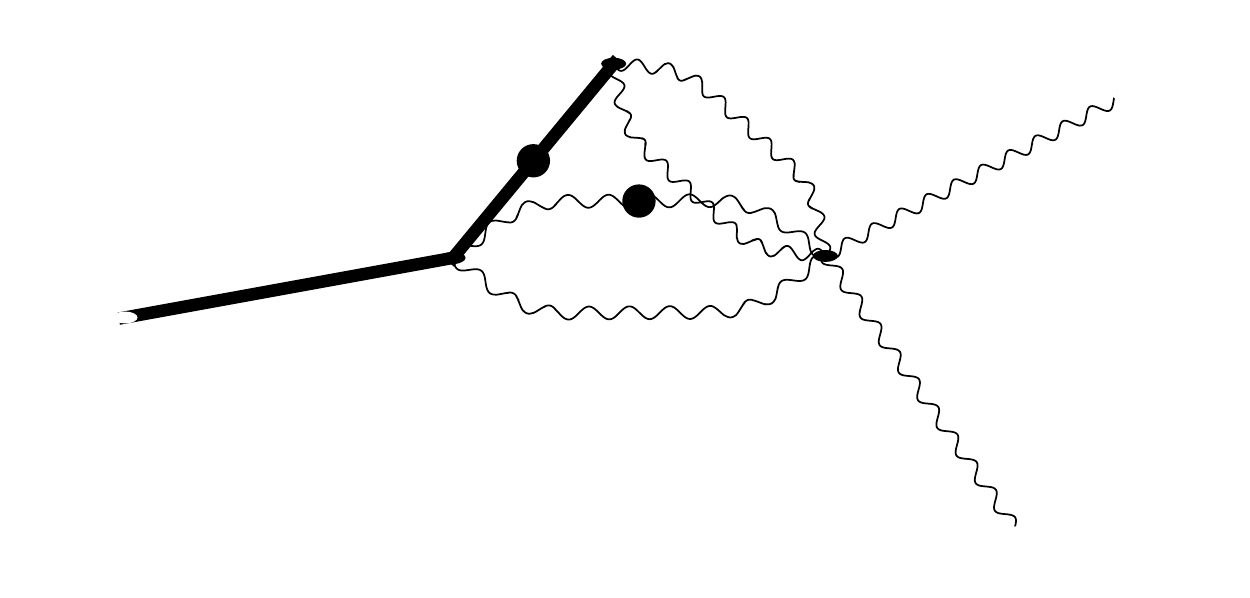}}
      \subfloat[][$I_{14}$ (PP)]
      {\includegraphics[width=0.16\textwidth]{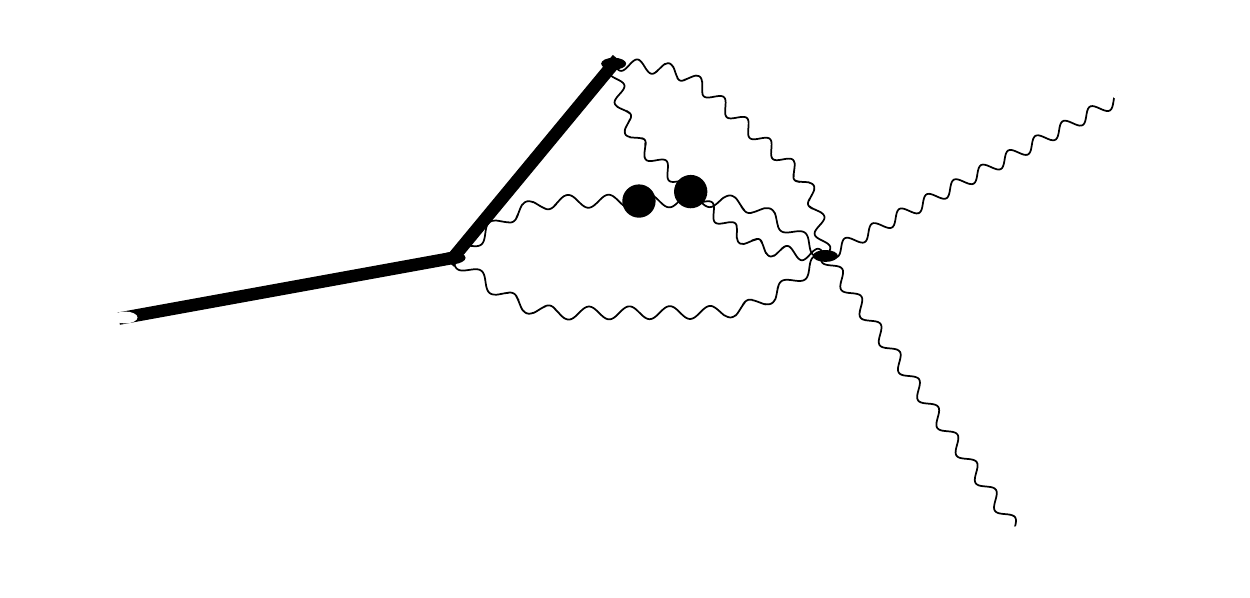}}
      \subfloat[][$I_{15}$ (PP)]
      {\includegraphics[width=0.16\textwidth]{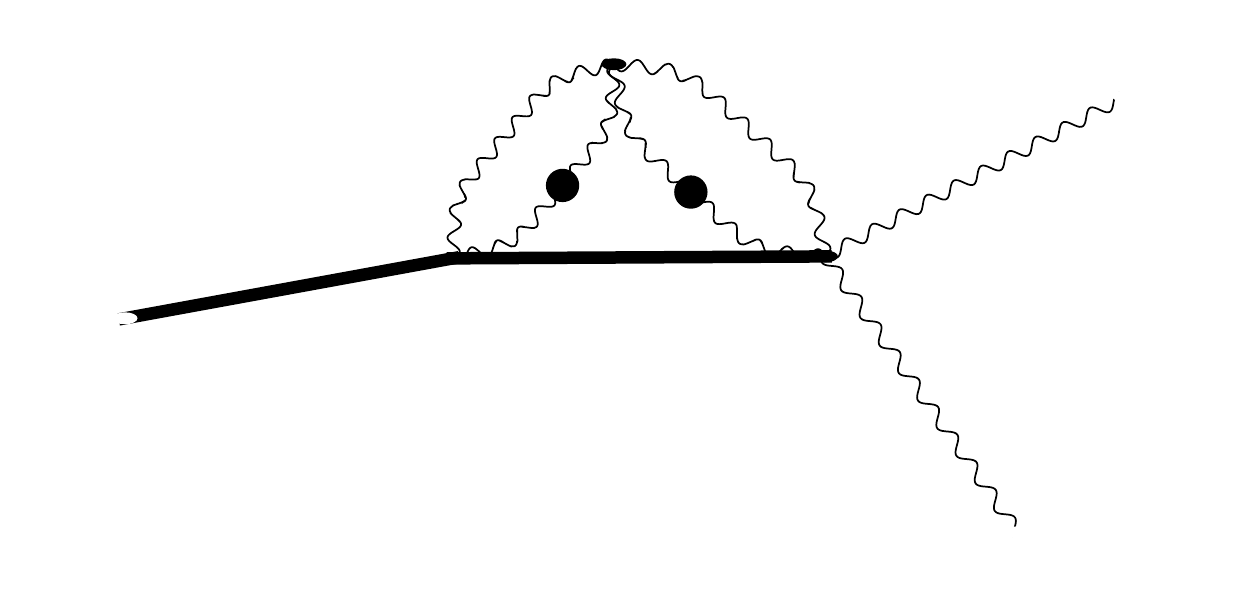}}
      \subfloat[][$I_{16}$ (PP)]
      {\includegraphics[width=0.16\textwidth]{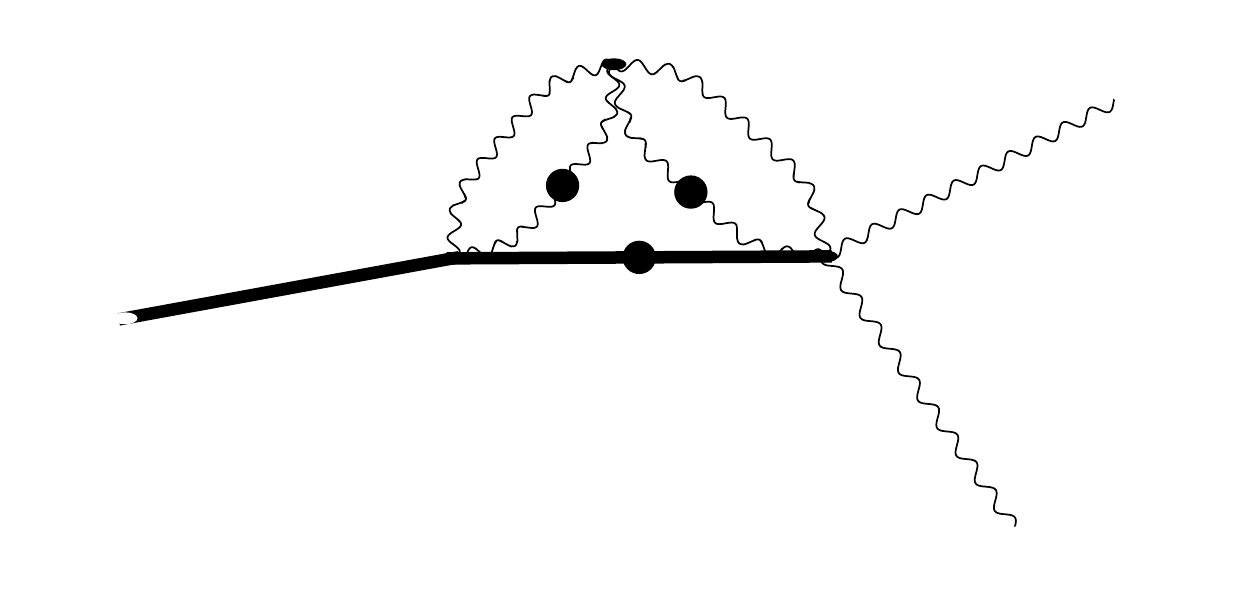}}
      \subfloat[][$I_{17}^*$ (PP)]
      {\includegraphics[width=0.16\textwidth]{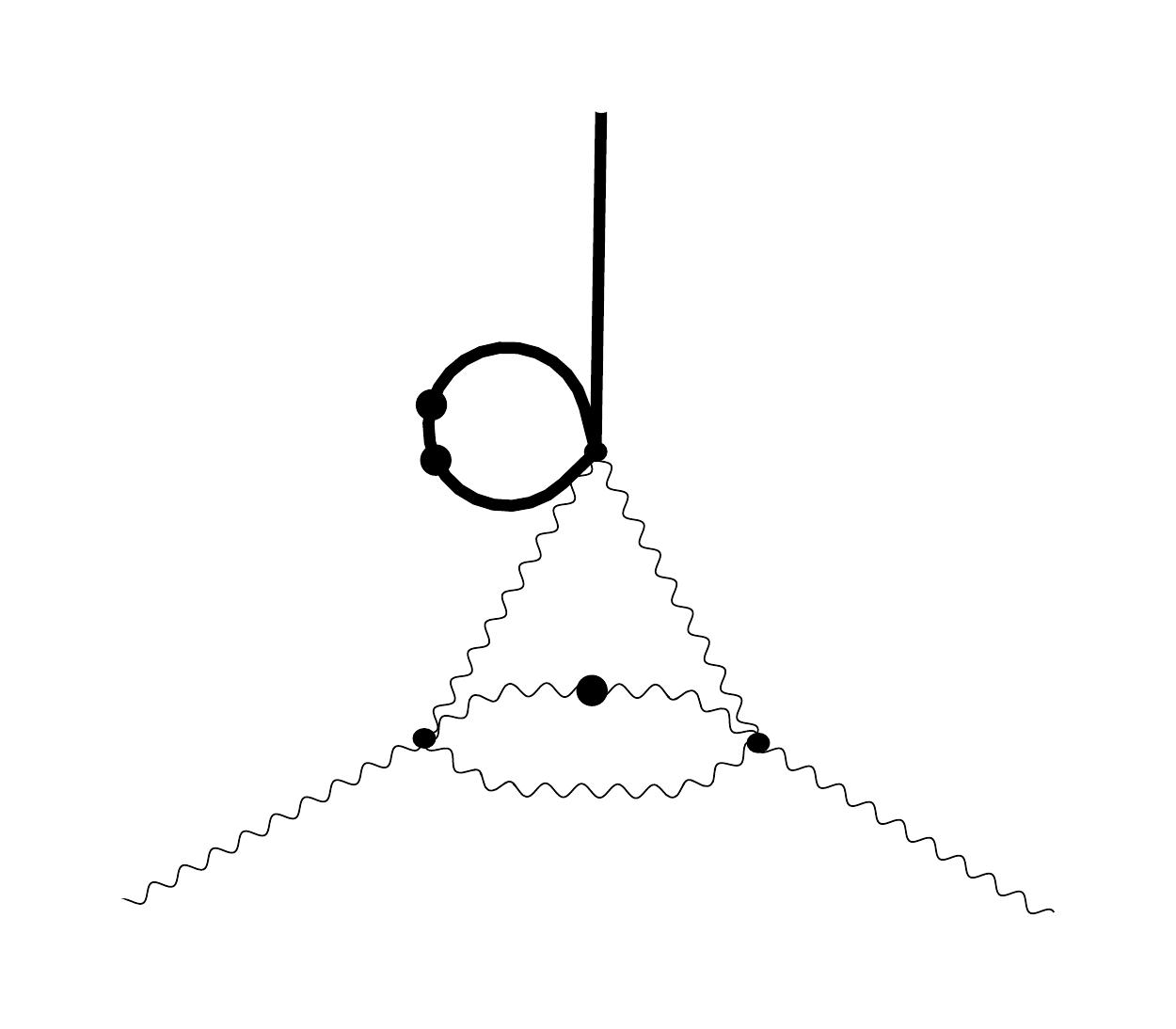}}
      \subfloat[][$I_{18}$ (PP)]
      {\includegraphics[width=0.16\textwidth]{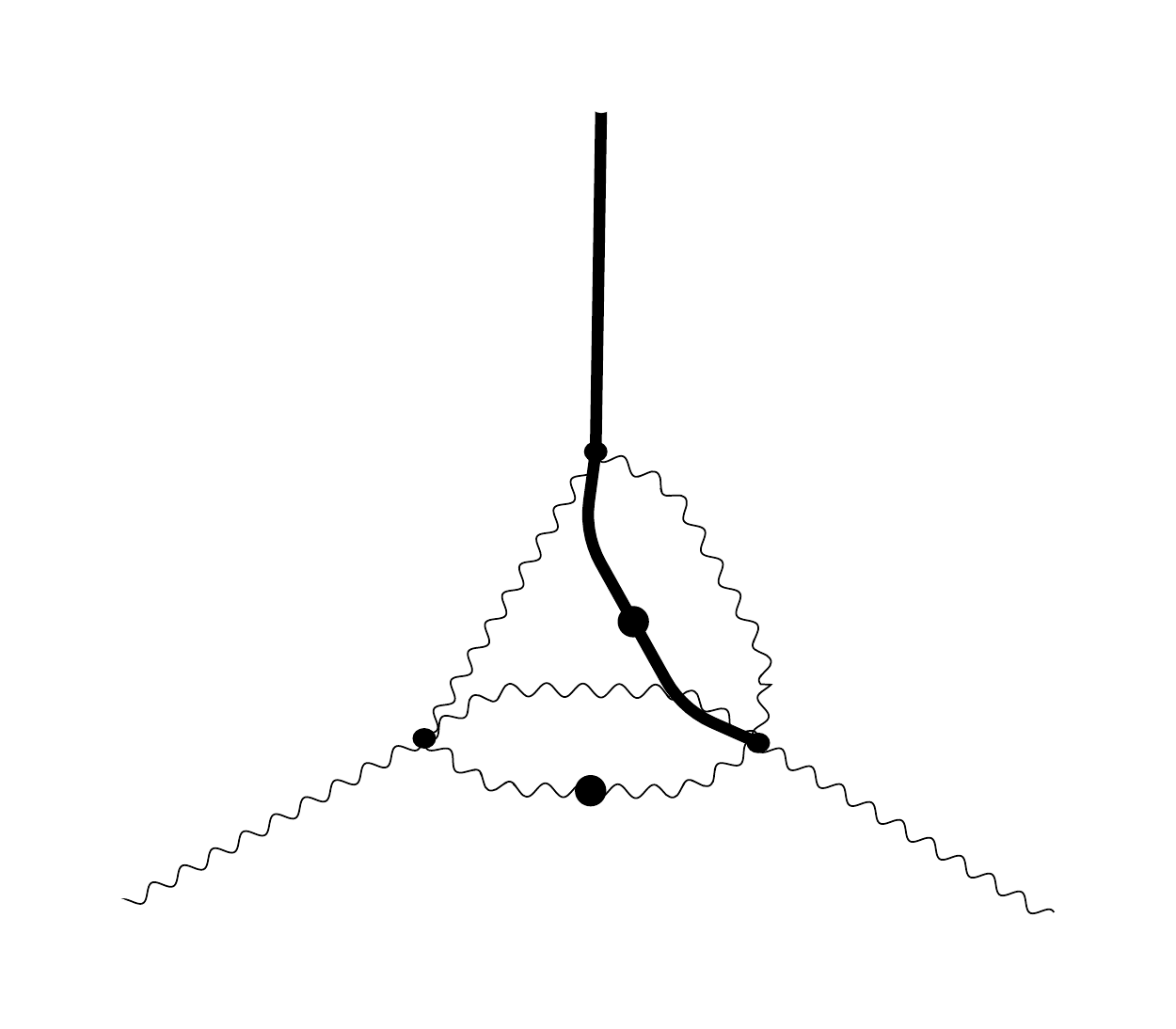}}
      \\

      \subfloat[][$I_{19}$ (PP)]
      {\includegraphics[width=0.16\textwidth]{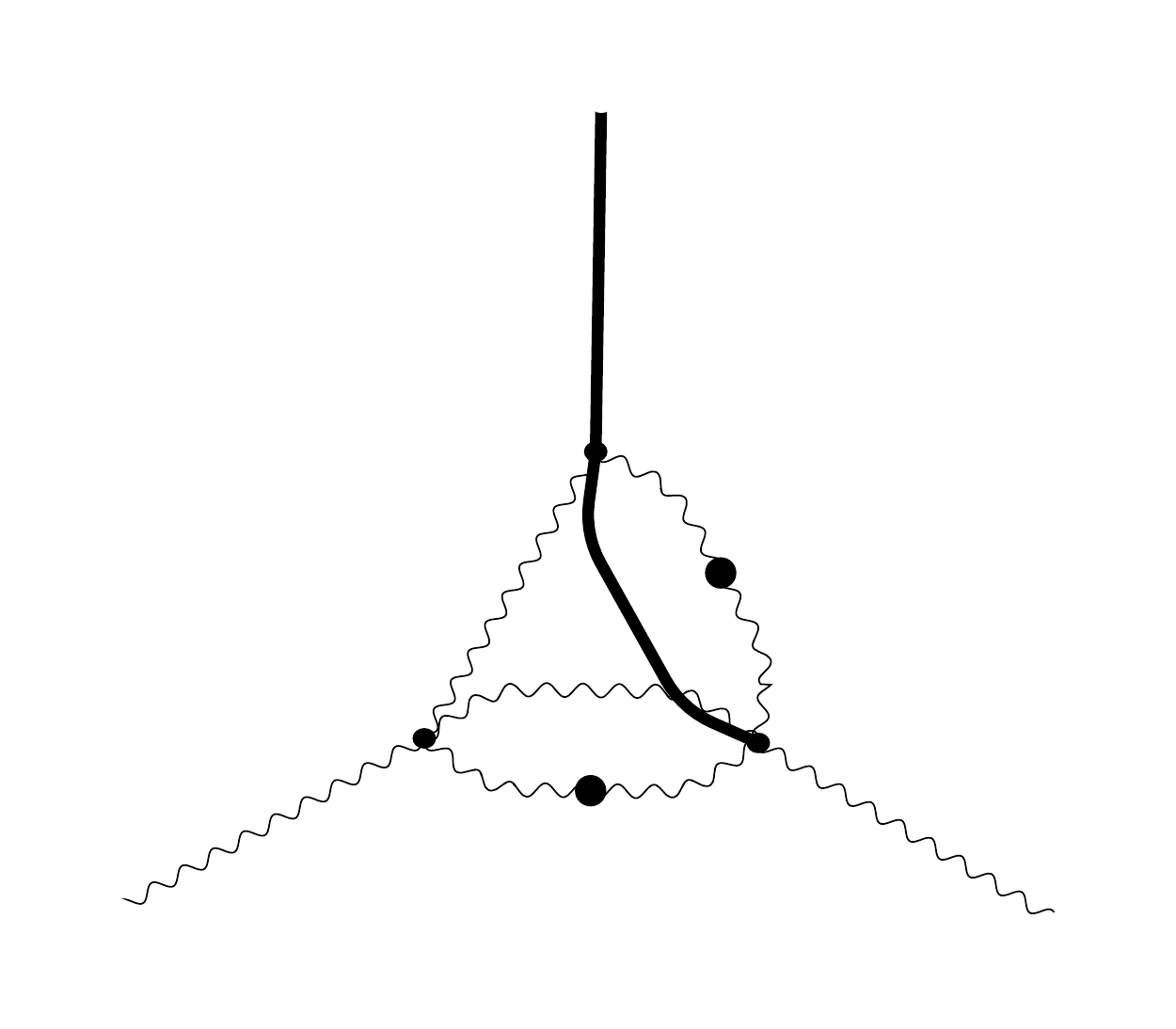}}
      \subfloat[][$I_{20}$ (PP)]
      {\includegraphics[width=0.16\textwidth]{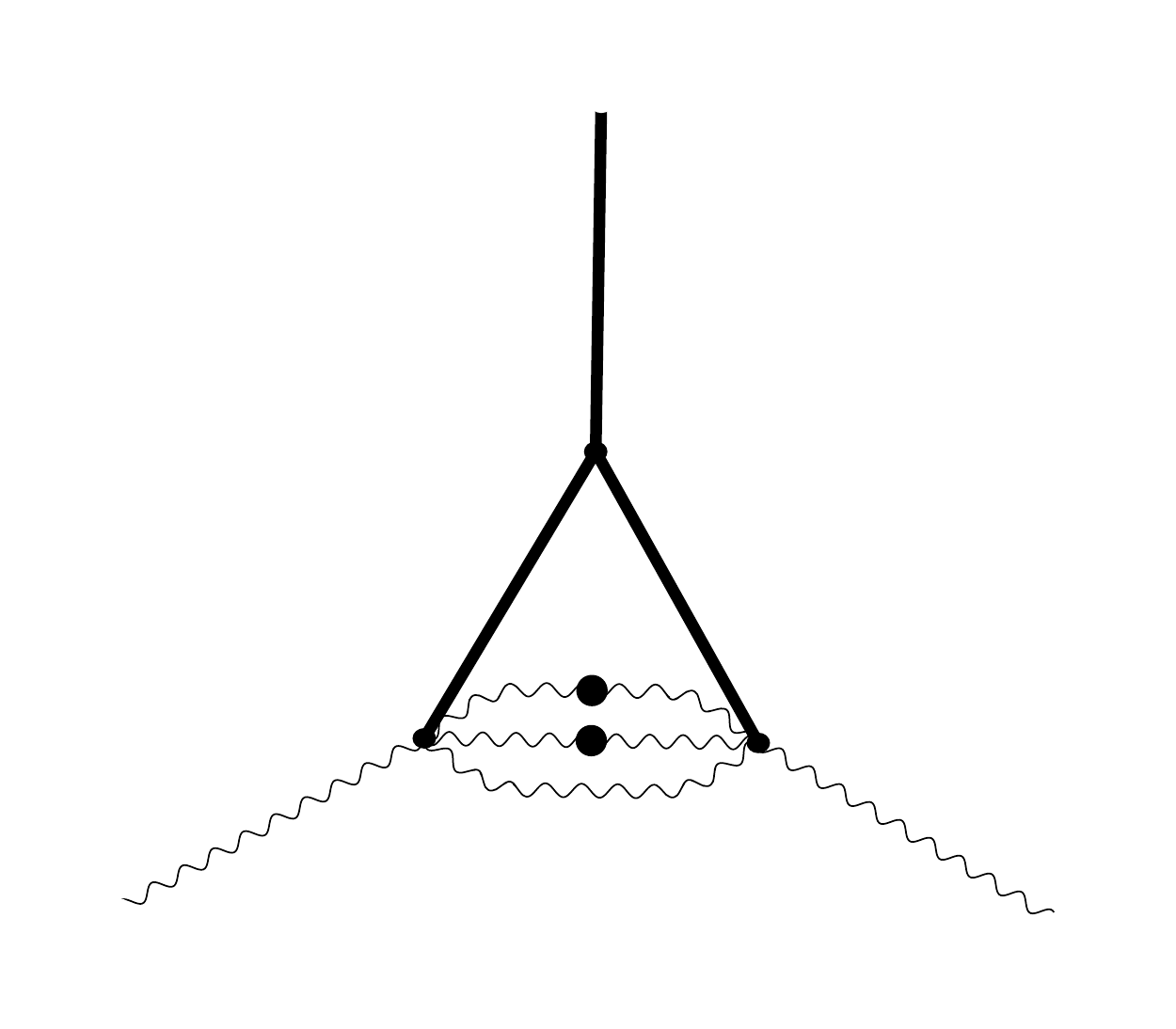}}
      \subfloat[][$I_{21}$ (PP)]
      {\includegraphics[width=0.16\textwidth]{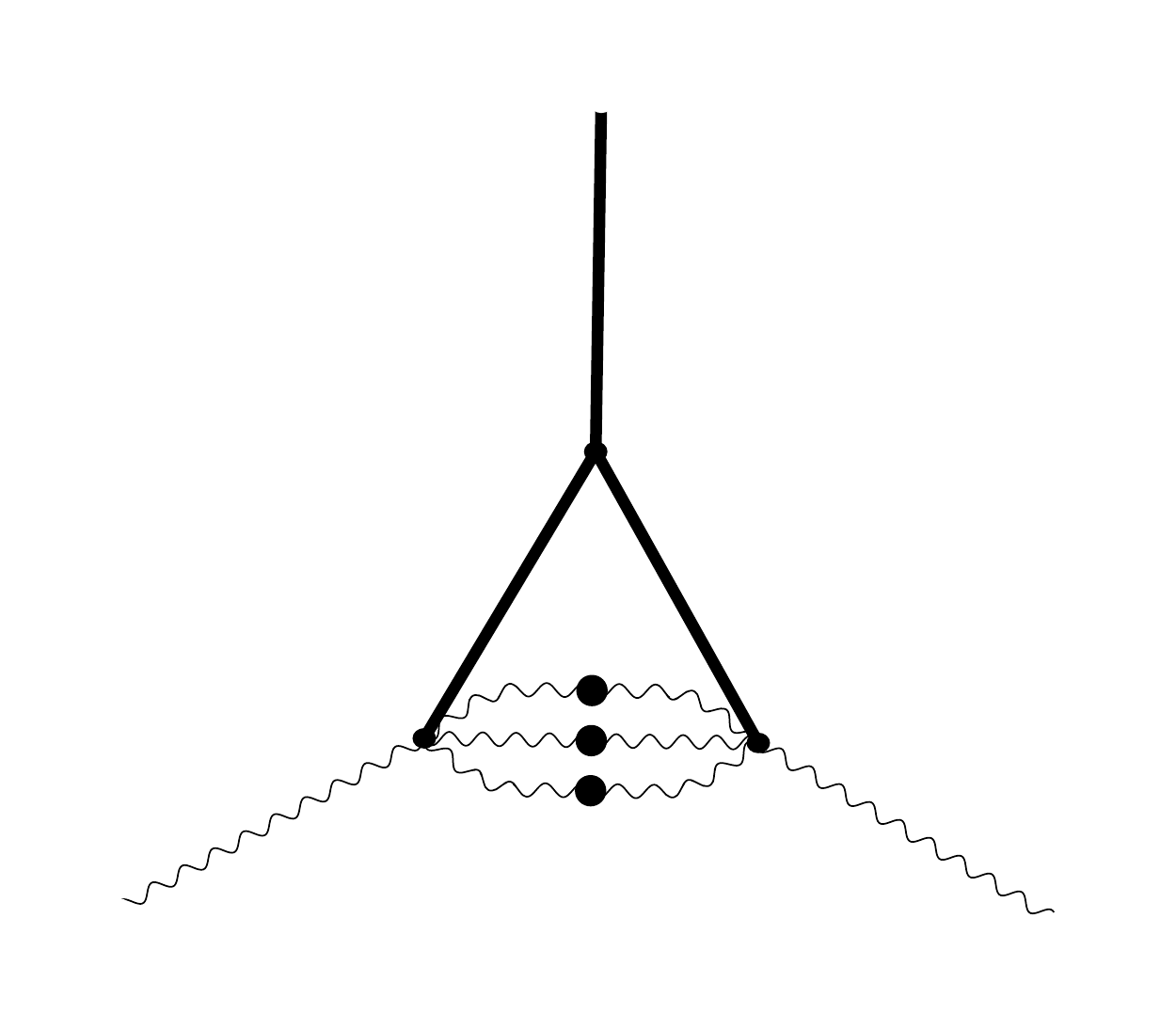}}
      \subfloat[][$I_{22}$ (PP)]
      {\includegraphics[width=0.16\textwidth]{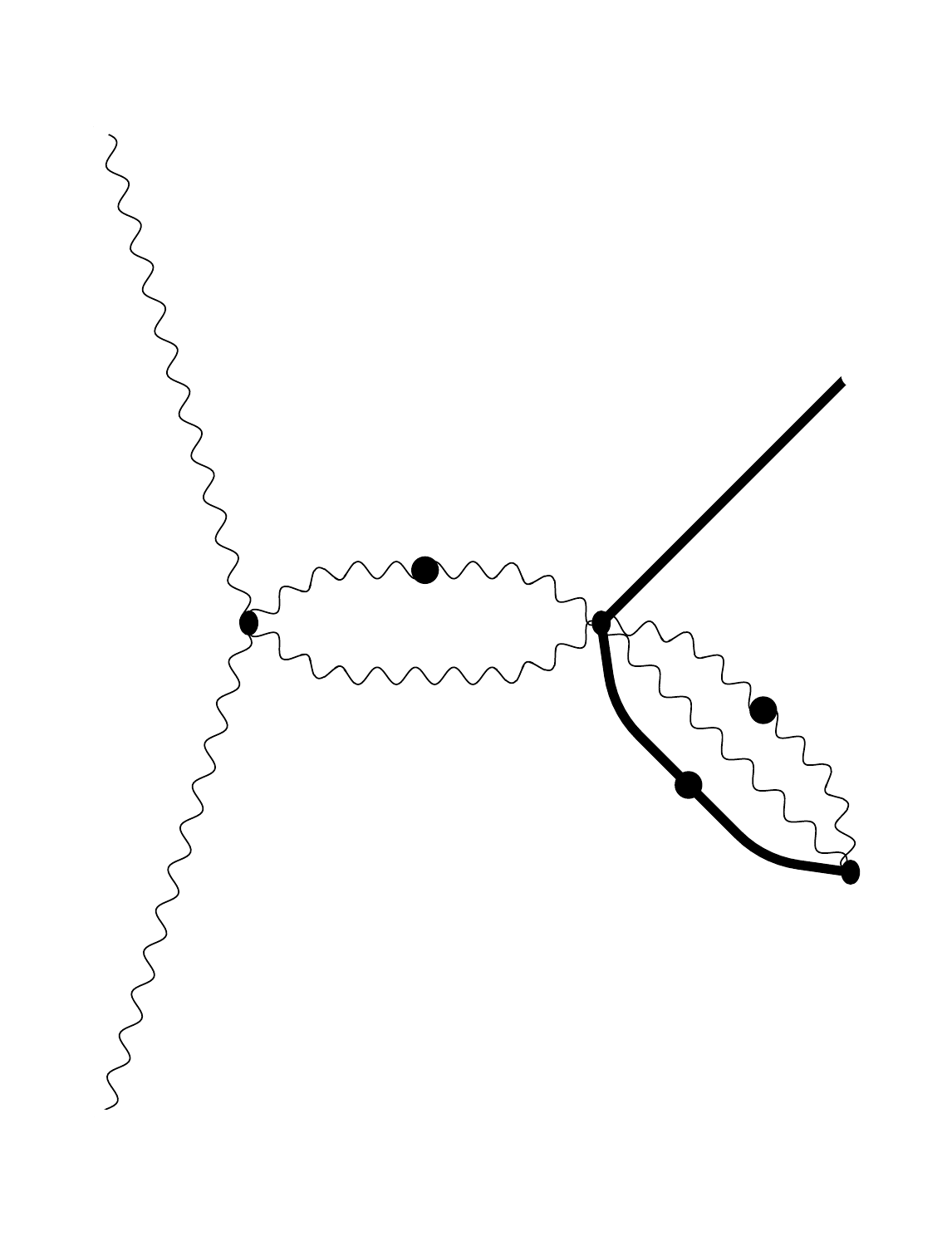}}
      \subfloat[][$I_{23}$ (PP)]
      {\includegraphics[width=0.16\textwidth]{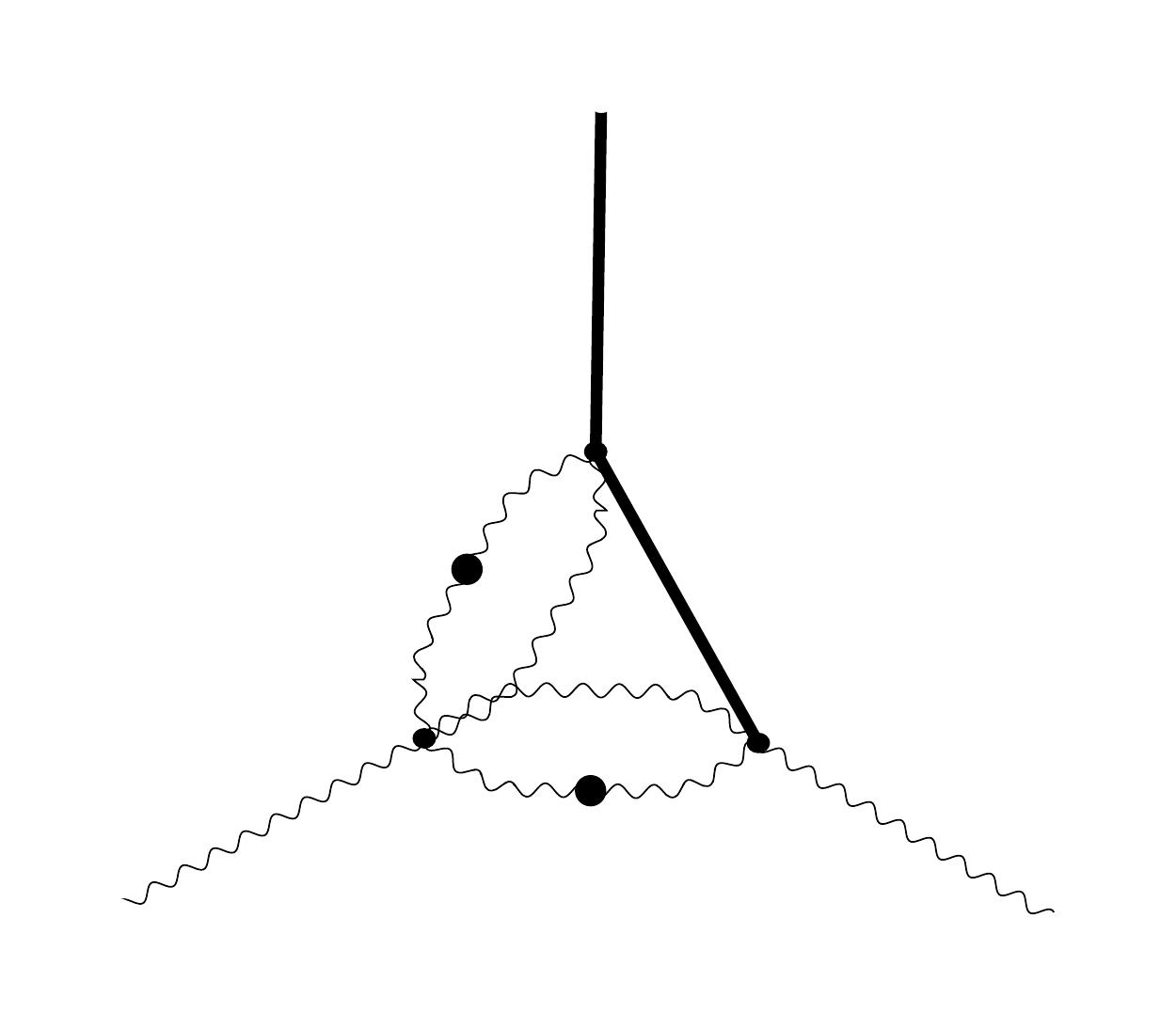}}
      \subfloat[][$I_{24}$ (PP)]
      {\includegraphics[width=0.16\textwidth]{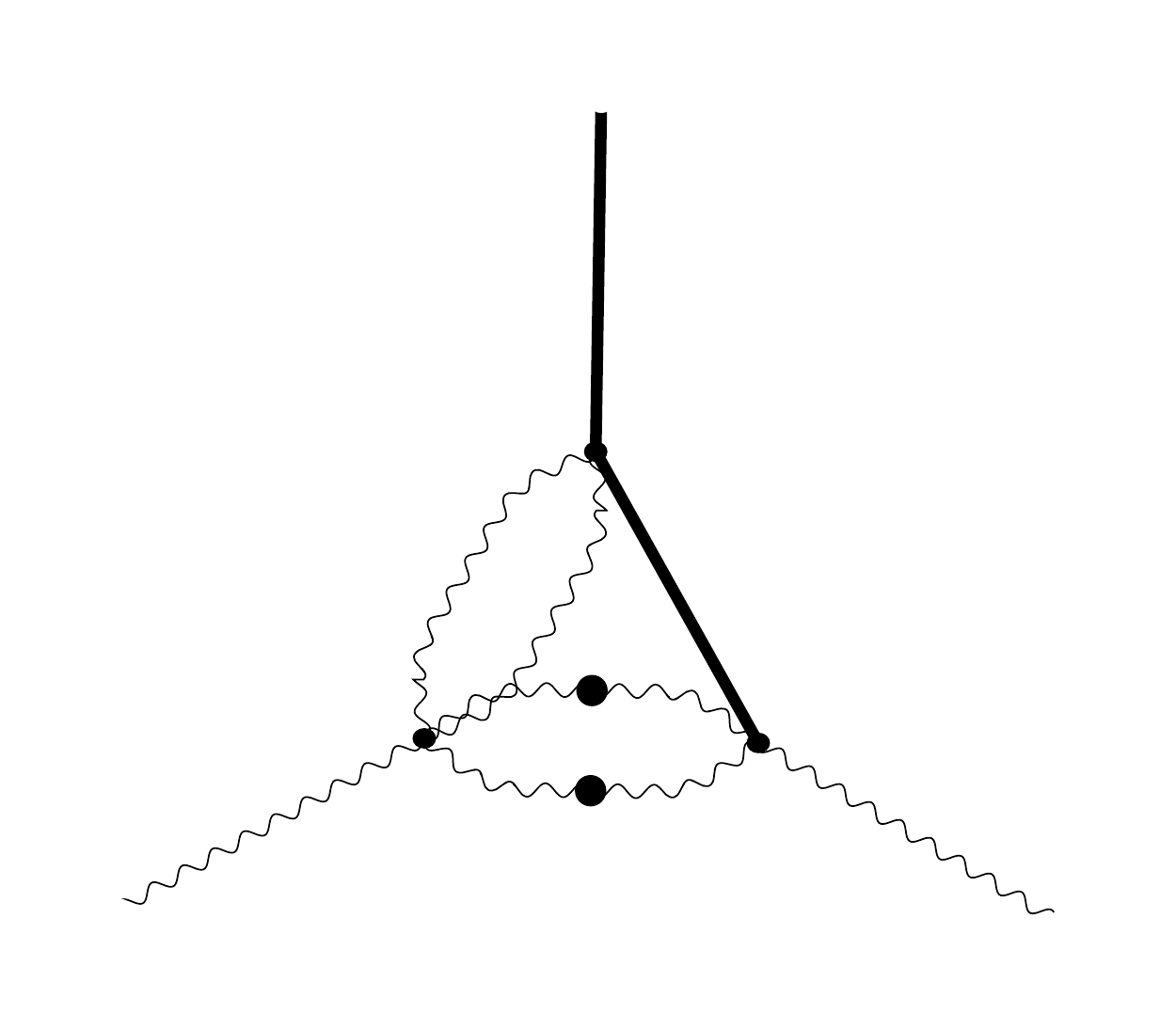}}
      \\

      \subfloat[][$I_{25}$ (PP)]
      {\includegraphics[width=0.16\textwidth]{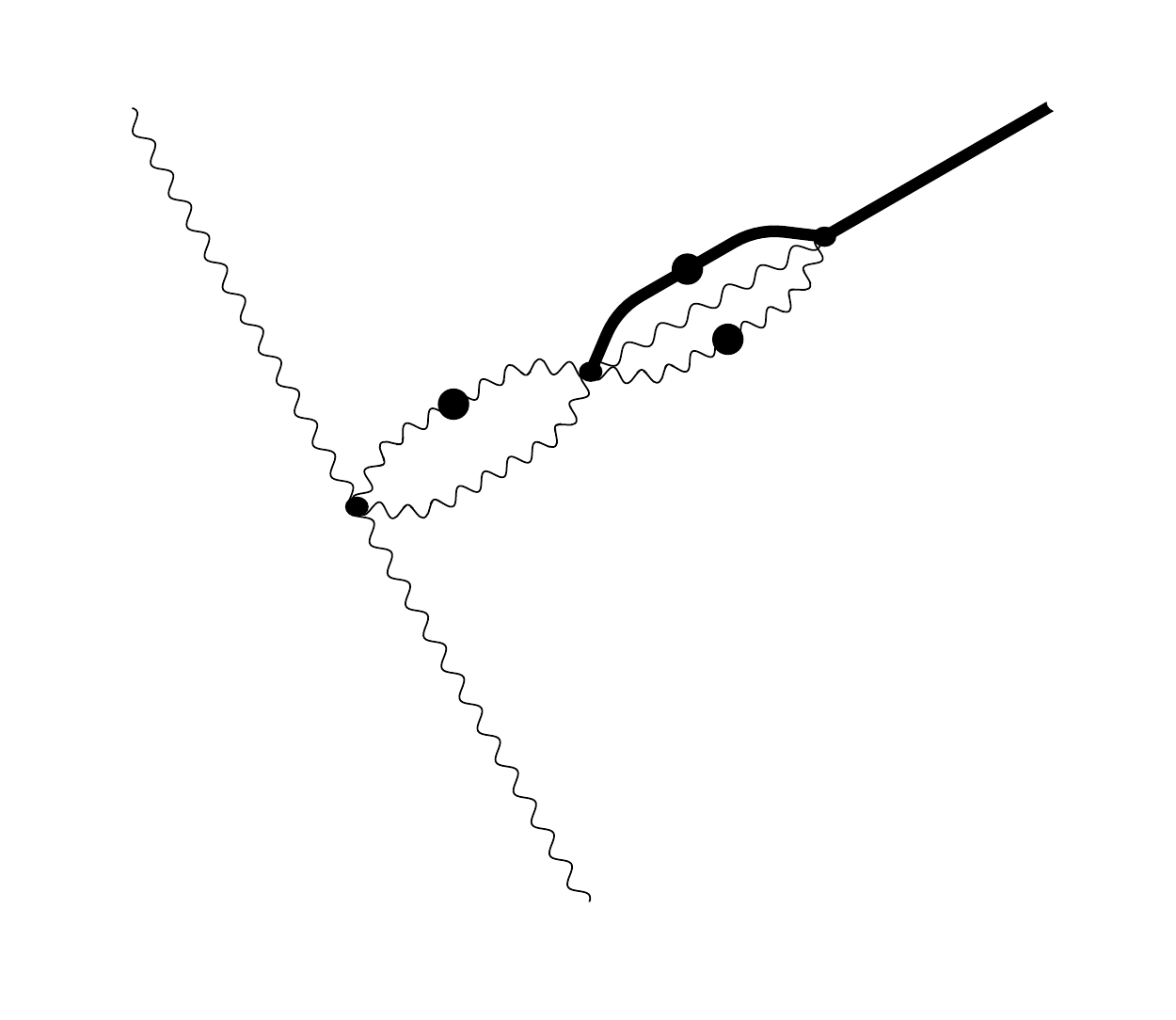}}
      \subfloat[][$I_{26}$ (PP)]
      {\includegraphics[width=0.16\textwidth]{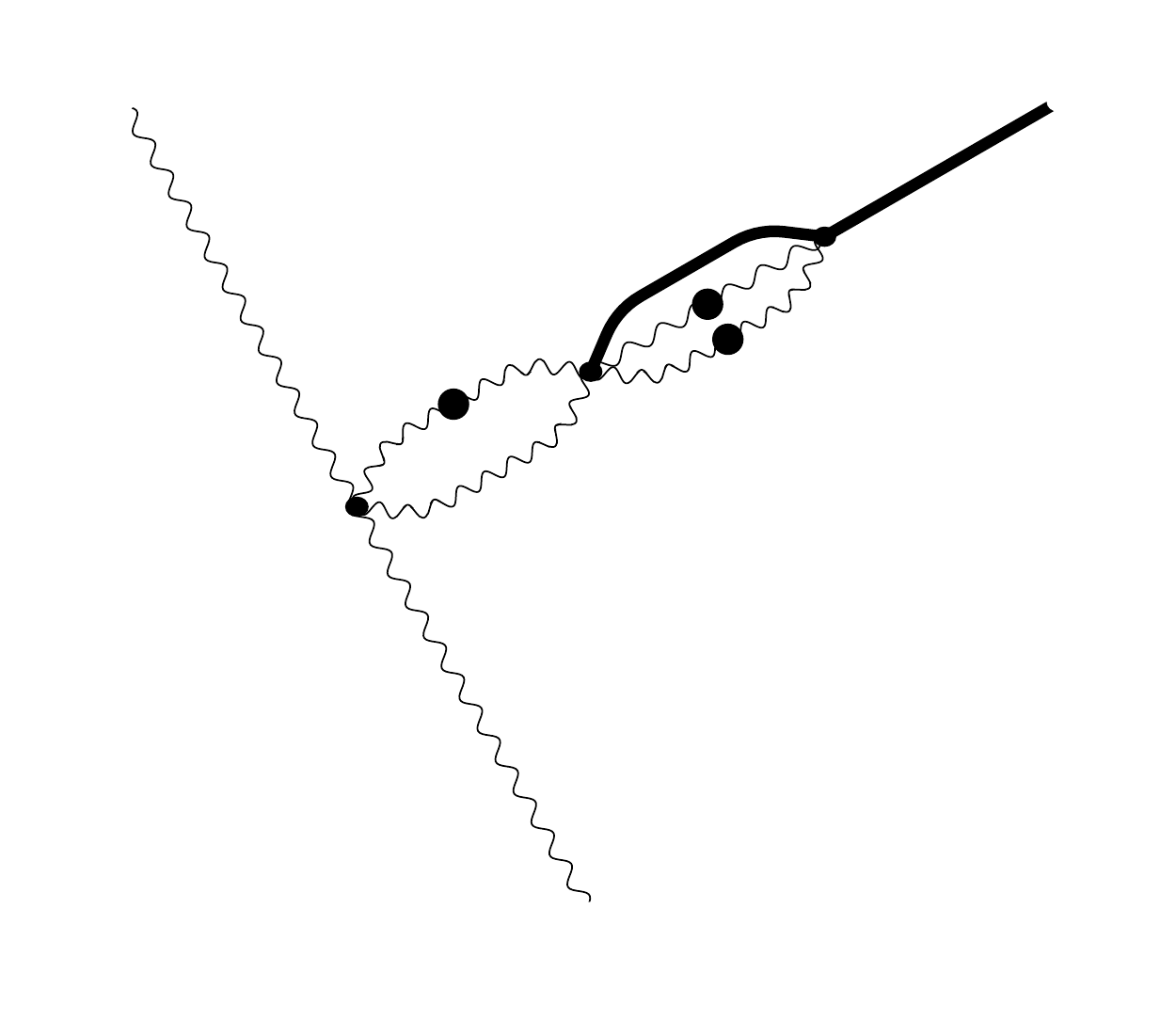}}
      \subfloat[][$I_{27}$ (PP)]
      {\includegraphics[width=0.16\textwidth]{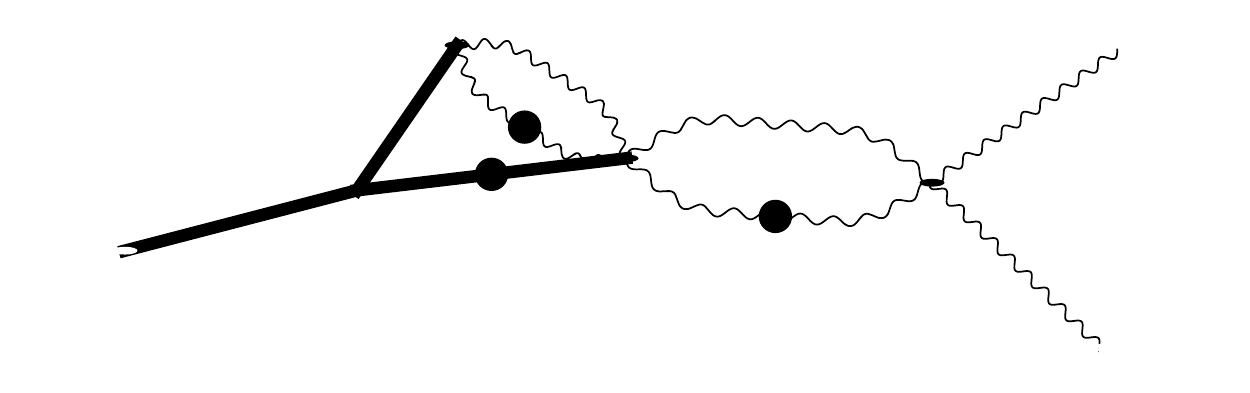}}
      \subfloat[][$I_{28}^*$ (PP)]
      {\includegraphics[width=0.16\textwidth]{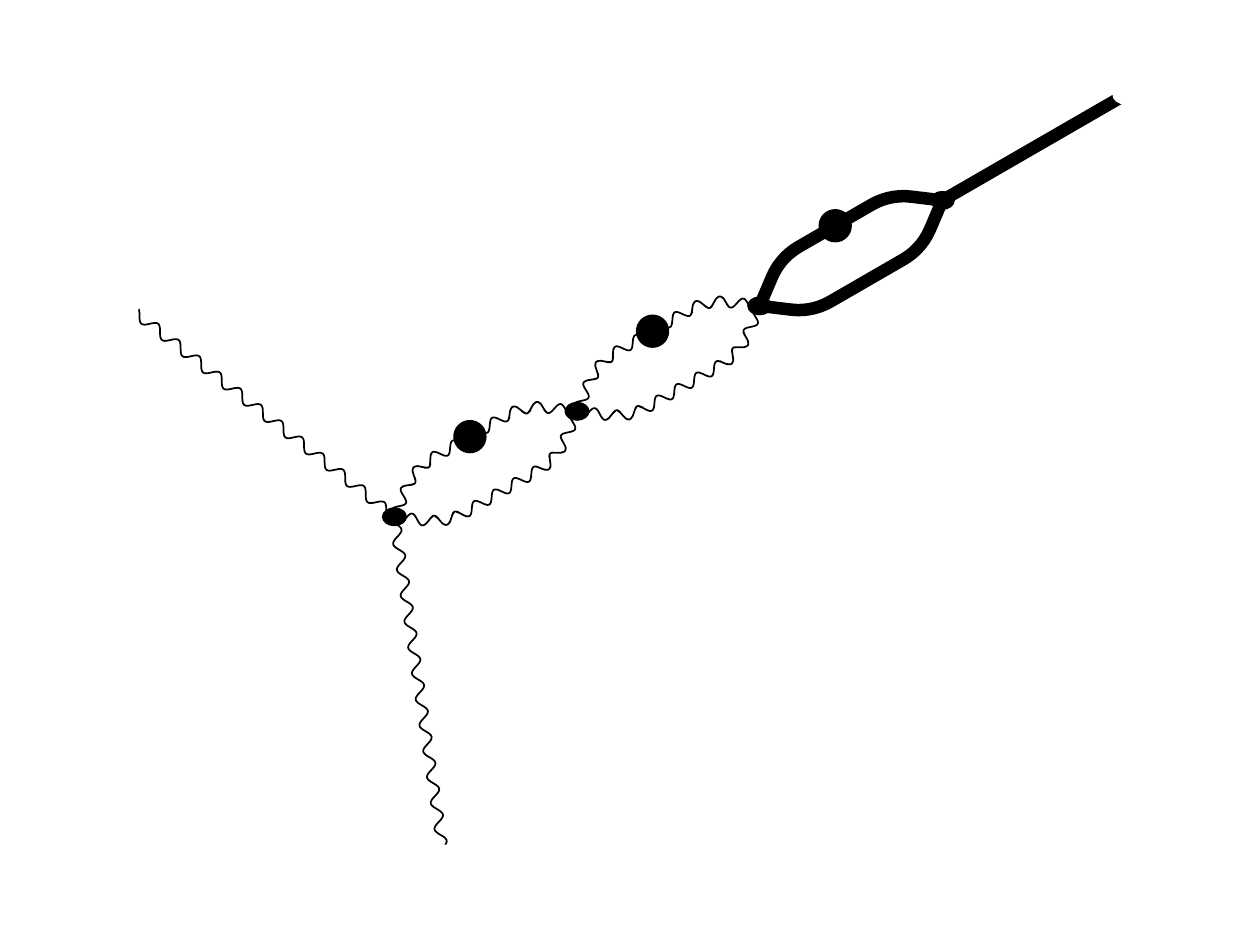}}
      \subfloat[][$I_{29}^*$ (PP)]
      {\includegraphics[width=0.16\textwidth]{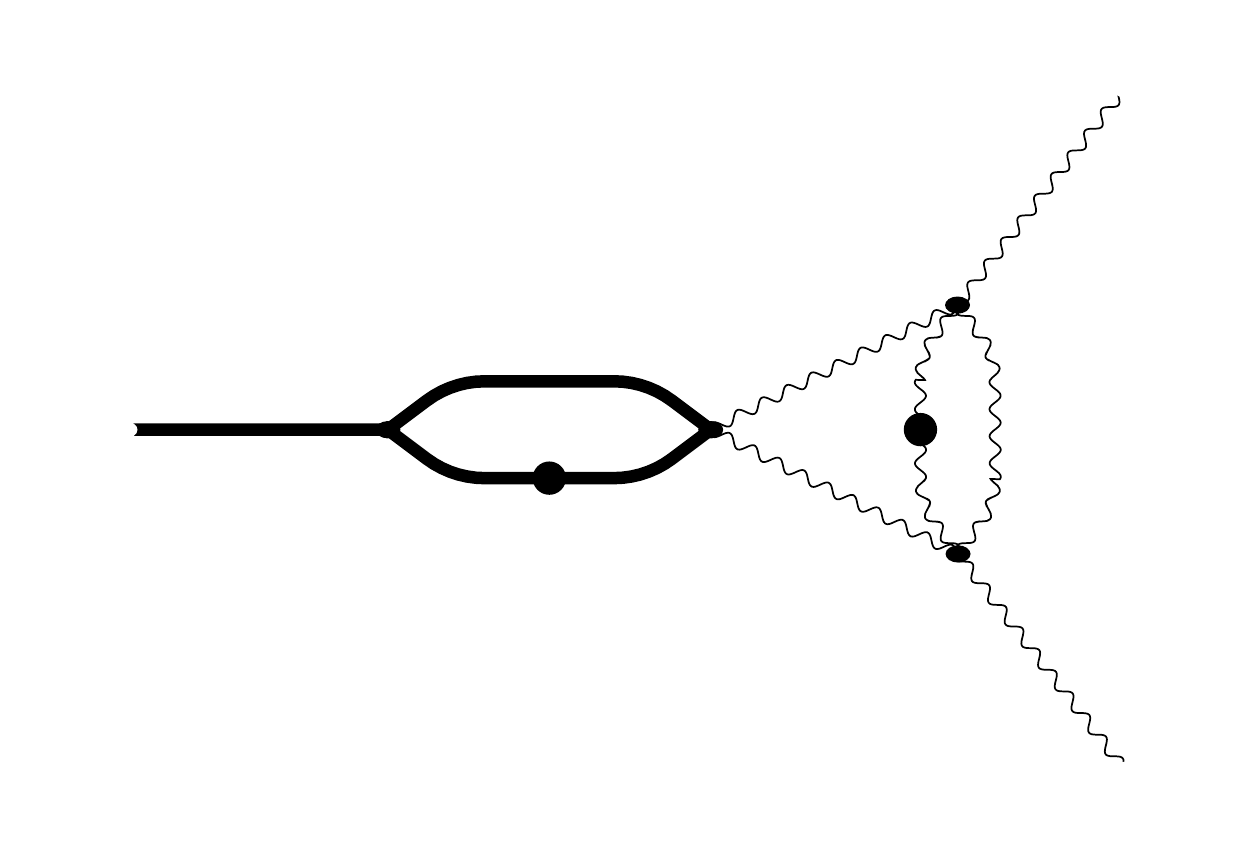}}
      \subfloat[][$I_{30}$ (PP)]
      {\includegraphics[width=0.16\textwidth]{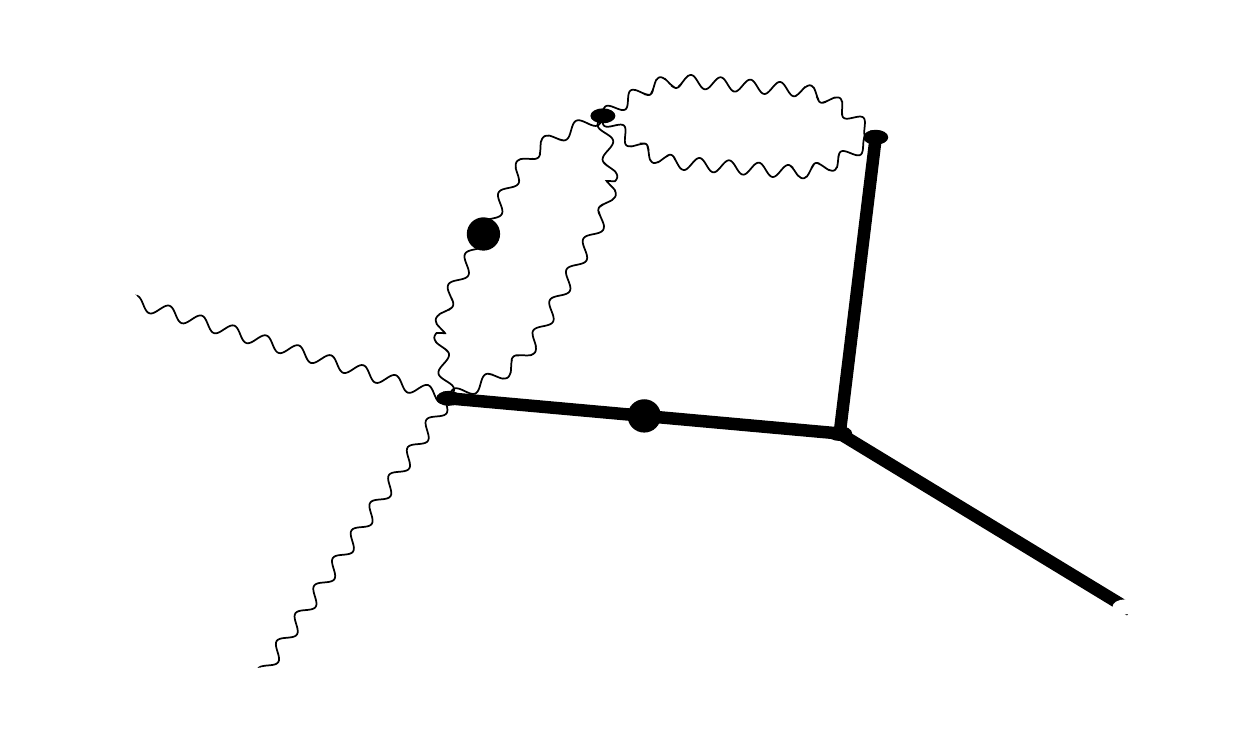}}
      \caption{Master Integrals (1/3).}
      \label{MI1_30}
\end{figure}

\begin{figure}[H]
      \captionsetup[subfigure]{labelformat=empty}
      \centering
      \subfloat[][$I_{31}$ (PP)]
      {\includegraphics[width=0.16\textwidth]{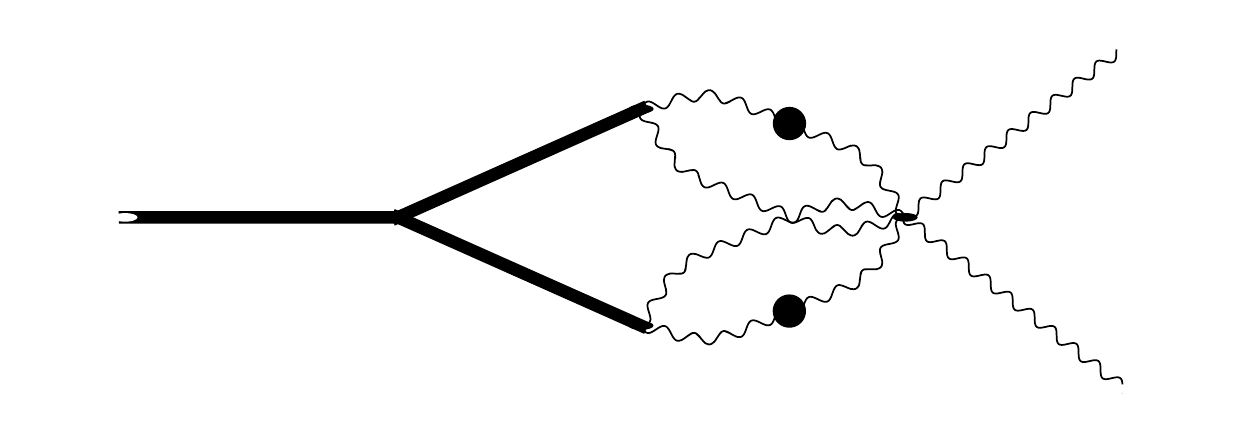}}
      \subfloat[][$I_{32}$ (NA)]
      {\includegraphics[width=0.16\textwidth]{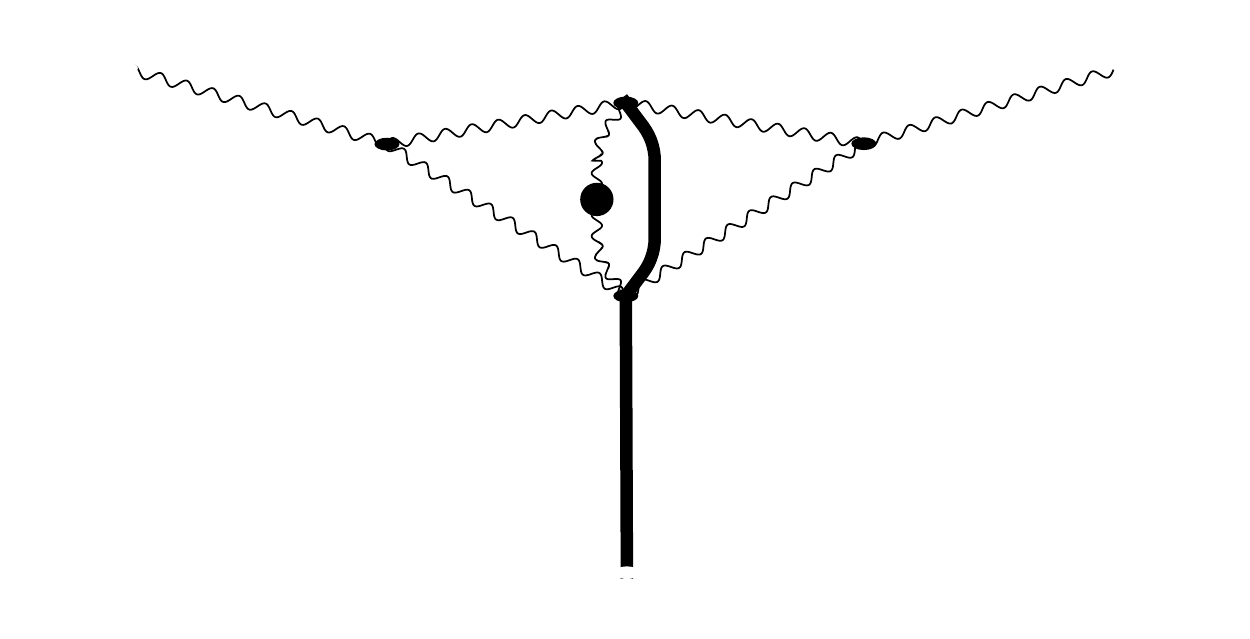}}
      \subfloat[][$I_{33}$ (NA)]
      {\includegraphics[width=0.16\textwidth]{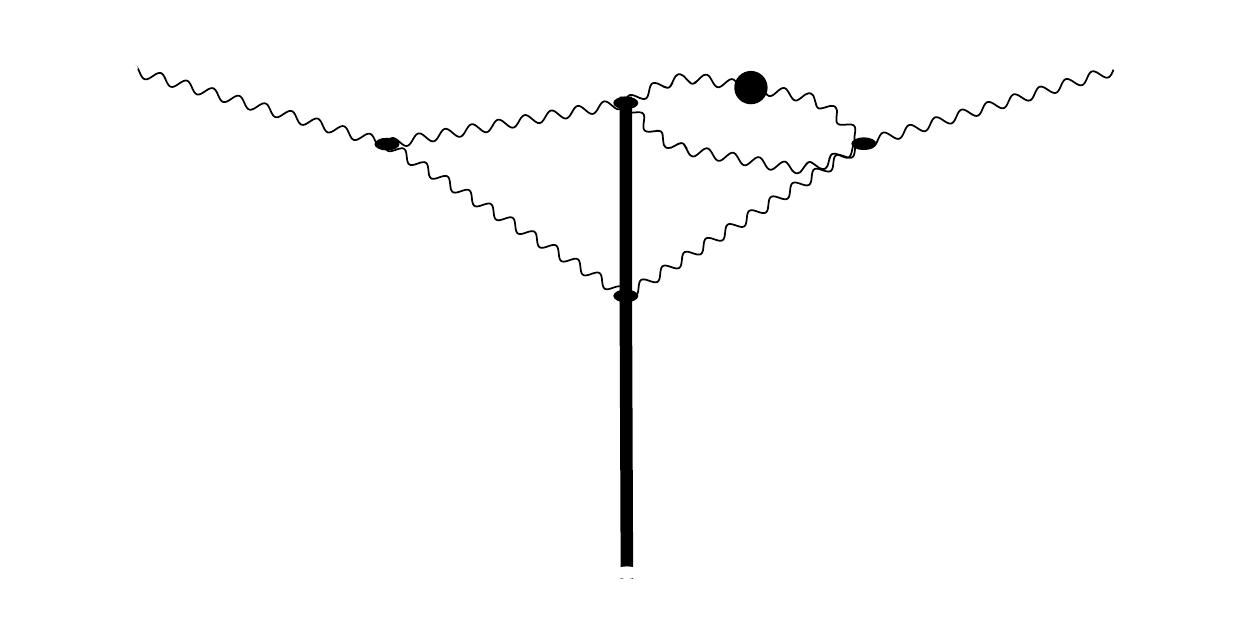}}
      \subfloat[][$I_{34}$ (PP)]
      {\includegraphics[width=0.16\textwidth]{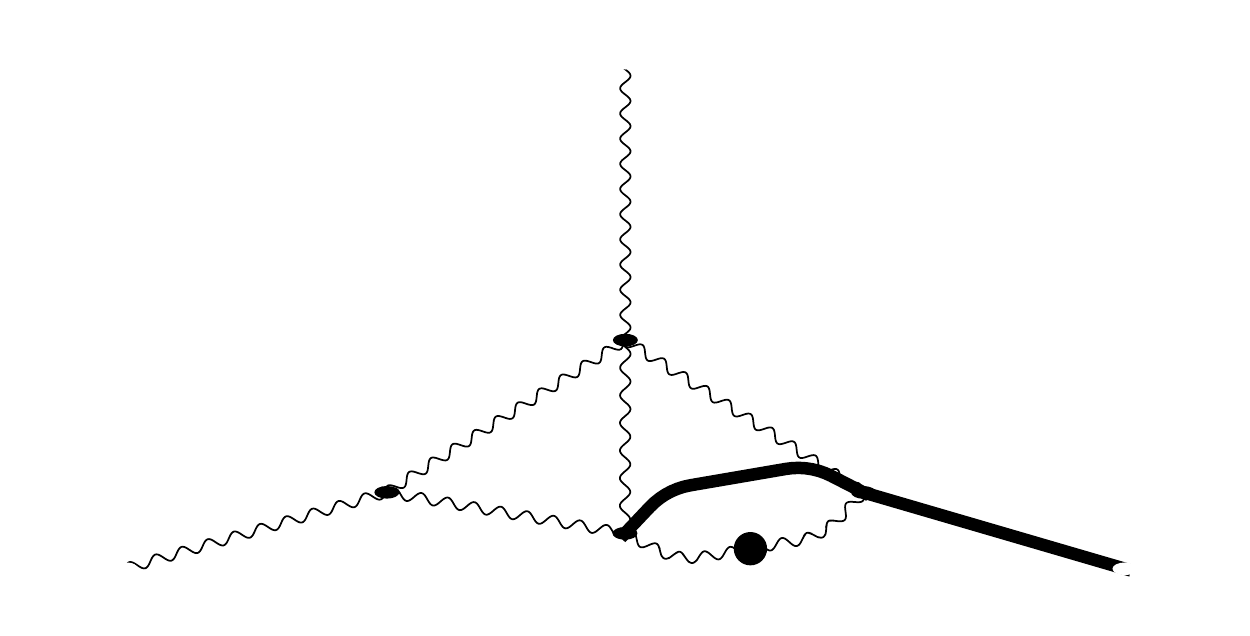}}
      \subfloat[][$I_{35}$ (PP)]
      {\includegraphics[width=0.16\textwidth]{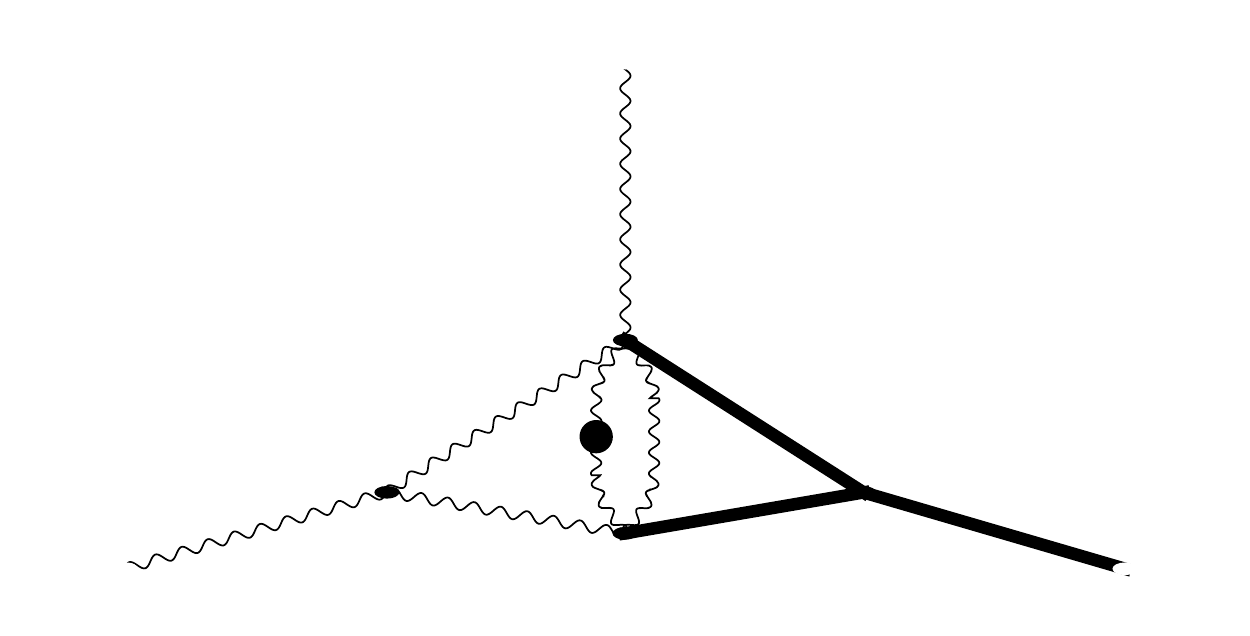}}
      \subfloat[][$I_{36}$ (PP)]
      {\includegraphics[width=0.16\textwidth]{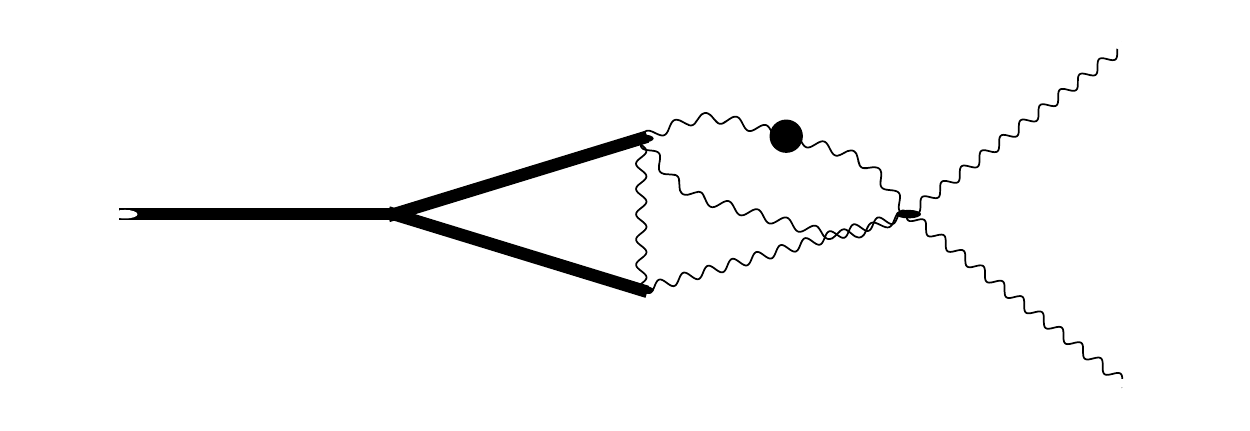}}
      \\

      \subfloat[][$I_{37}$ (PP)]
      {\includegraphics[width=0.16\textwidth]{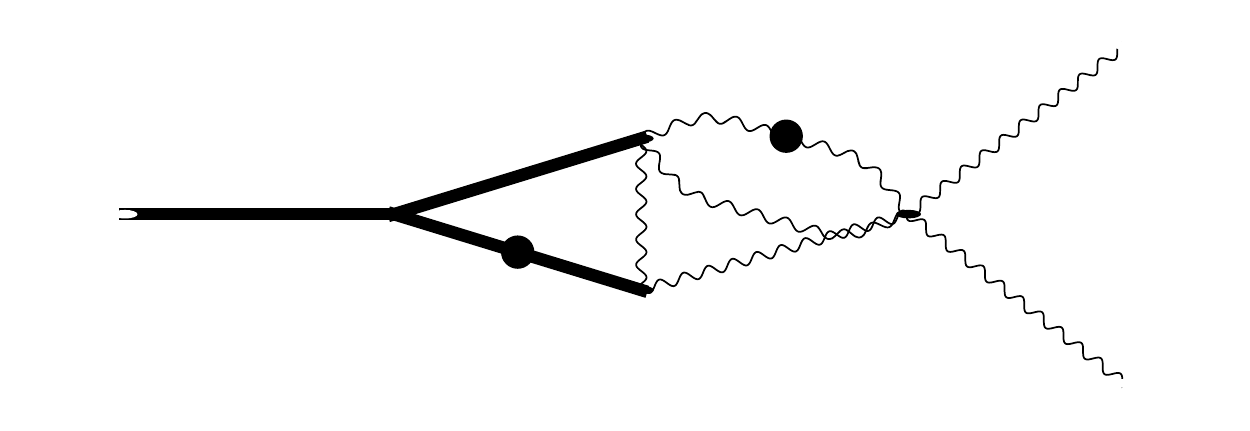}}
      \subfloat[][$I_{38}$ (PP)]
      {\includegraphics[width=0.16\textwidth]{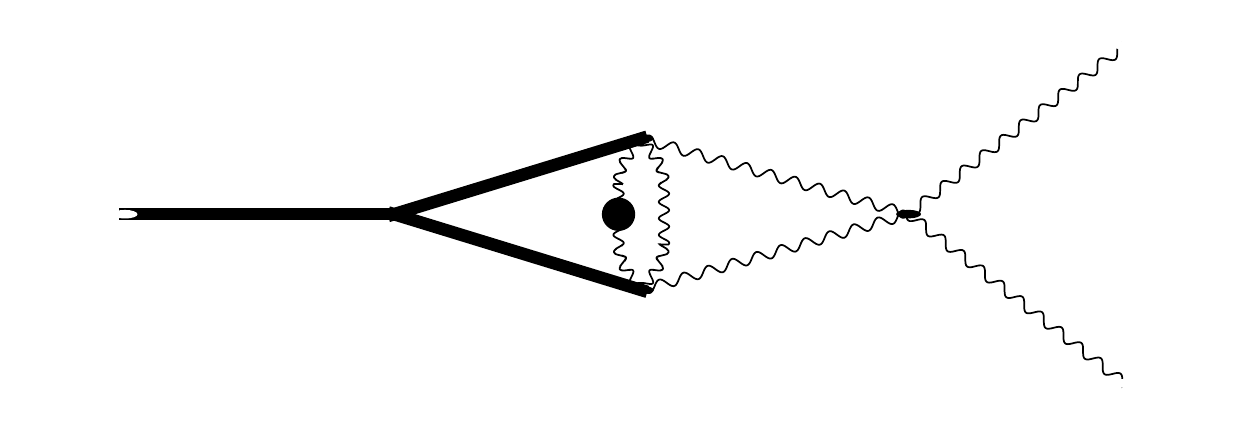}}
      \subfloat[][$I_{39}$ (PP)]
      {\includegraphics[width=0.16\textwidth]{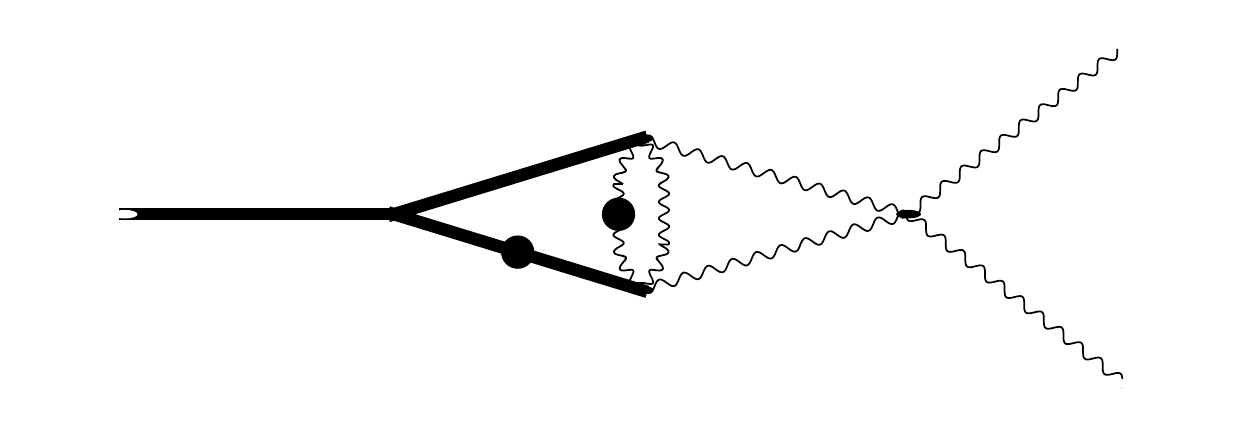}}
      \subfloat[][$I_{40}$ (PP)]
      {\includegraphics[width=0.16\textwidth]{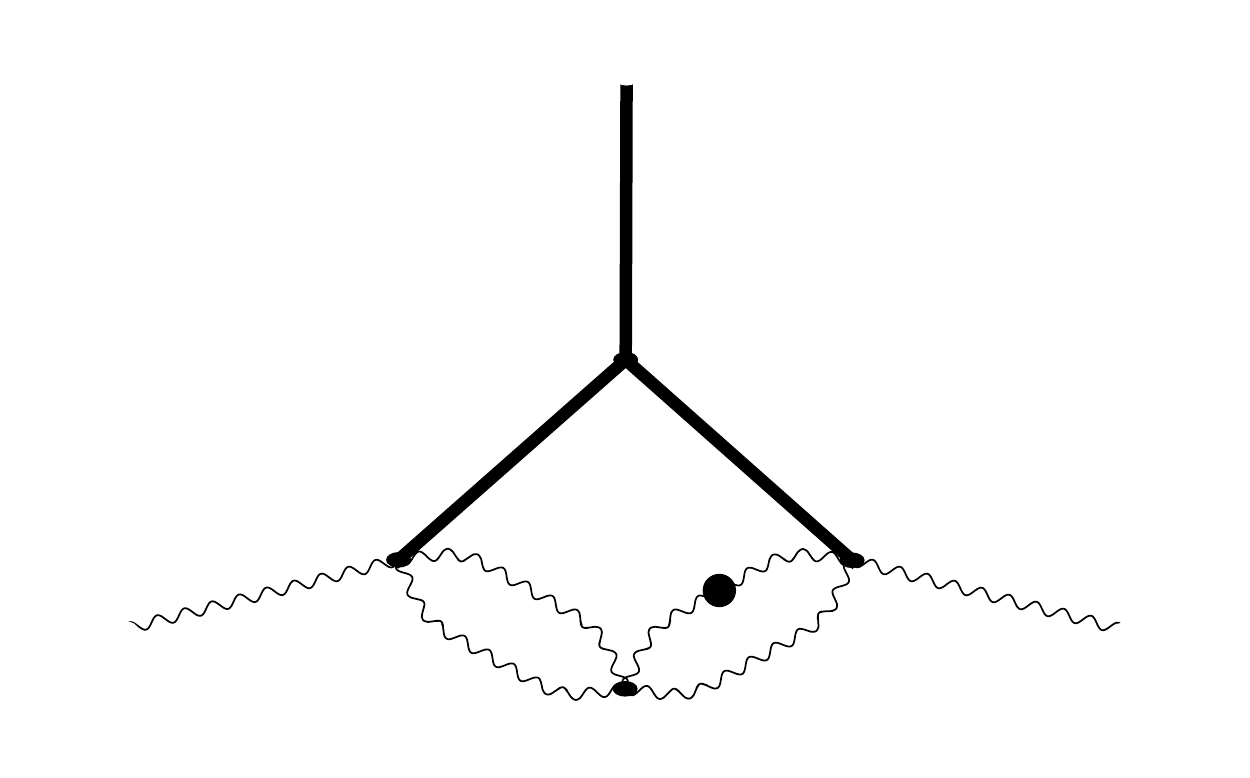}}
      \subfloat[][$I_{41}$ (PP)]
      {\includegraphics[width=0.16\textwidth]{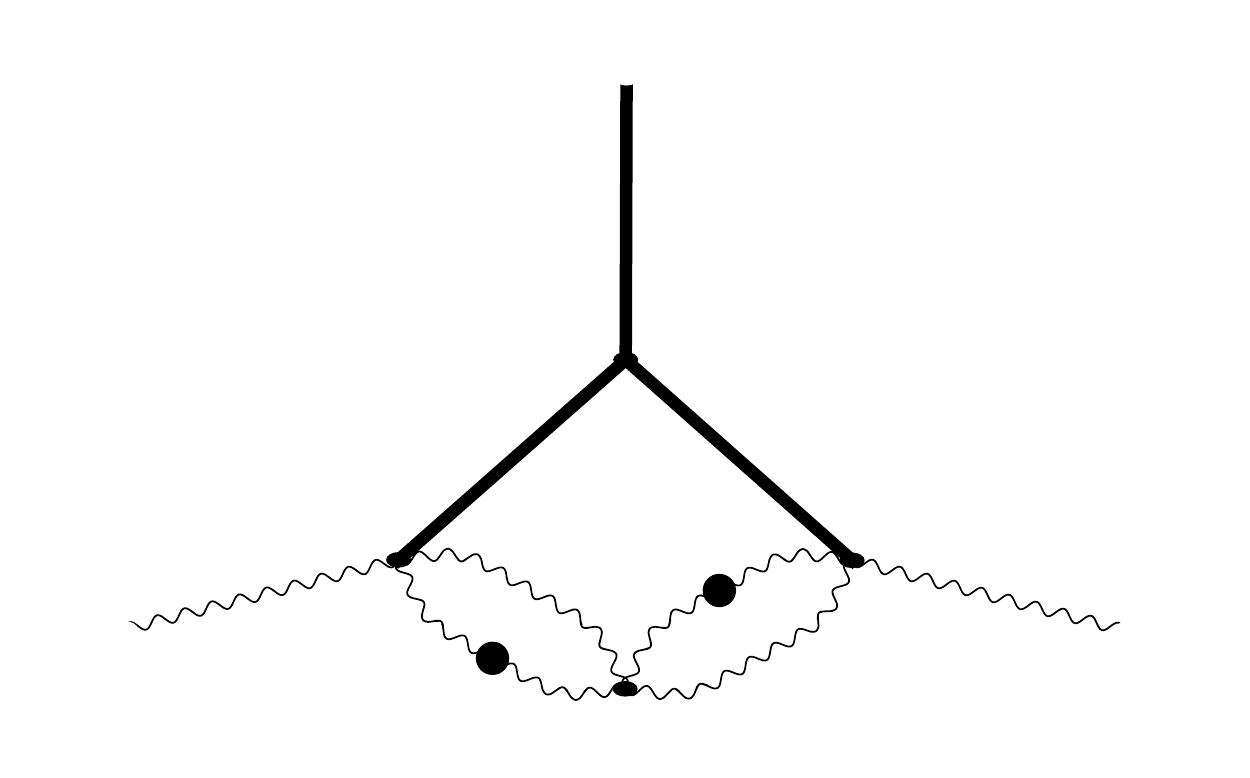}}
      \subfloat[][$I_{42}$ (PP)]
      {\includegraphics[width=0.16\textwidth]{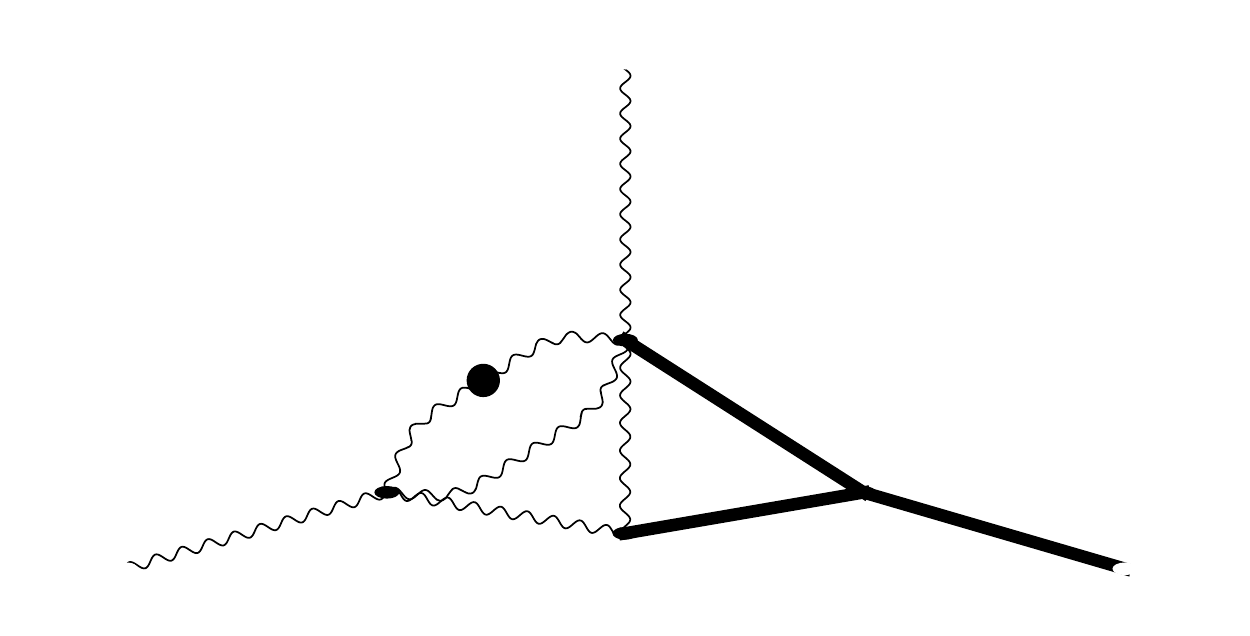}}
      \\

      \subfloat[][$I_{43}$ (PP)]
      {\includegraphics[width=0.16\textwidth]{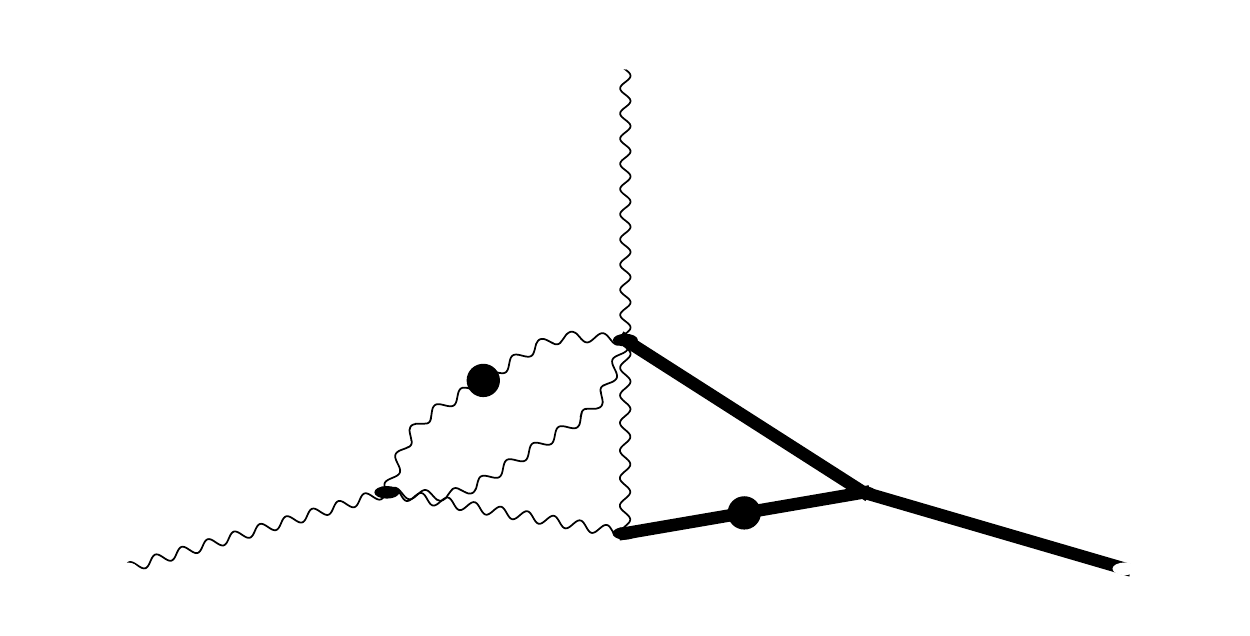}}
      \subfloat[][$I_{44}$ (PP)]
      {\includegraphics[width=0.16\textwidth]{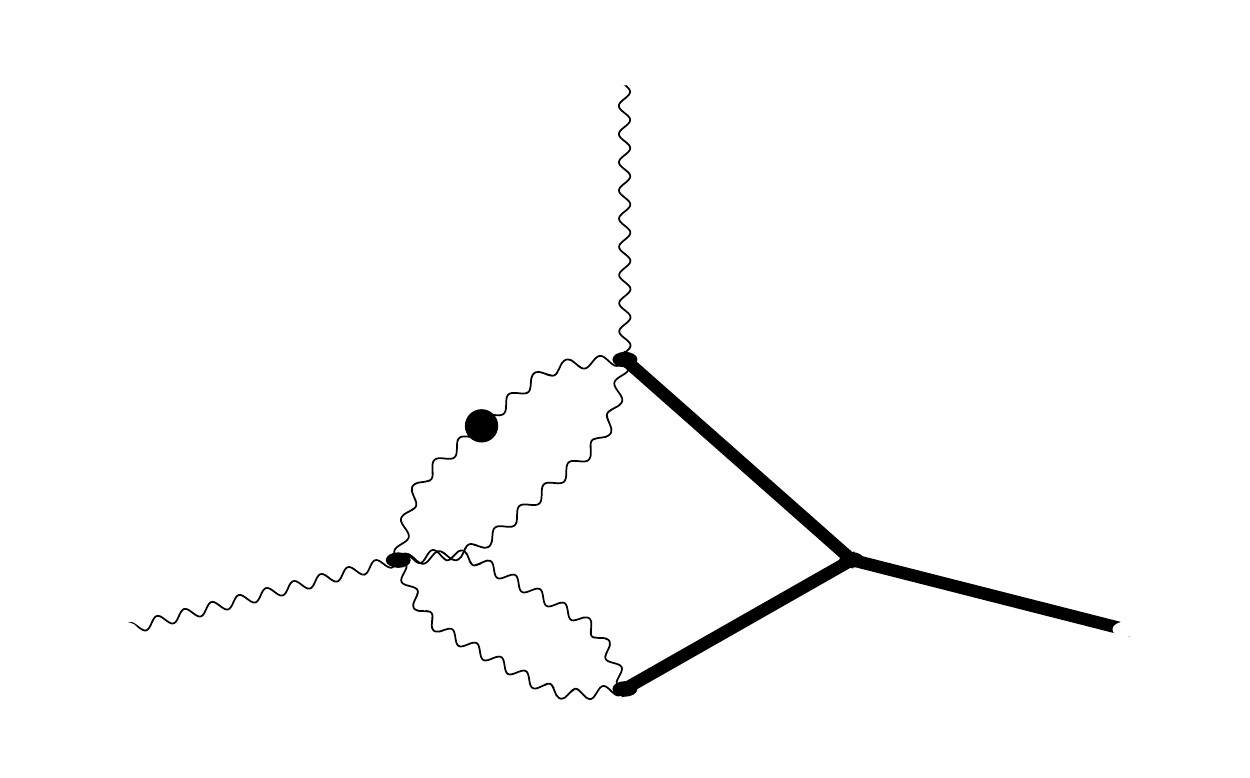}}
      \subfloat[][$I_{45}$ (PP)]
      {\includegraphics[width=0.16\textwidth]{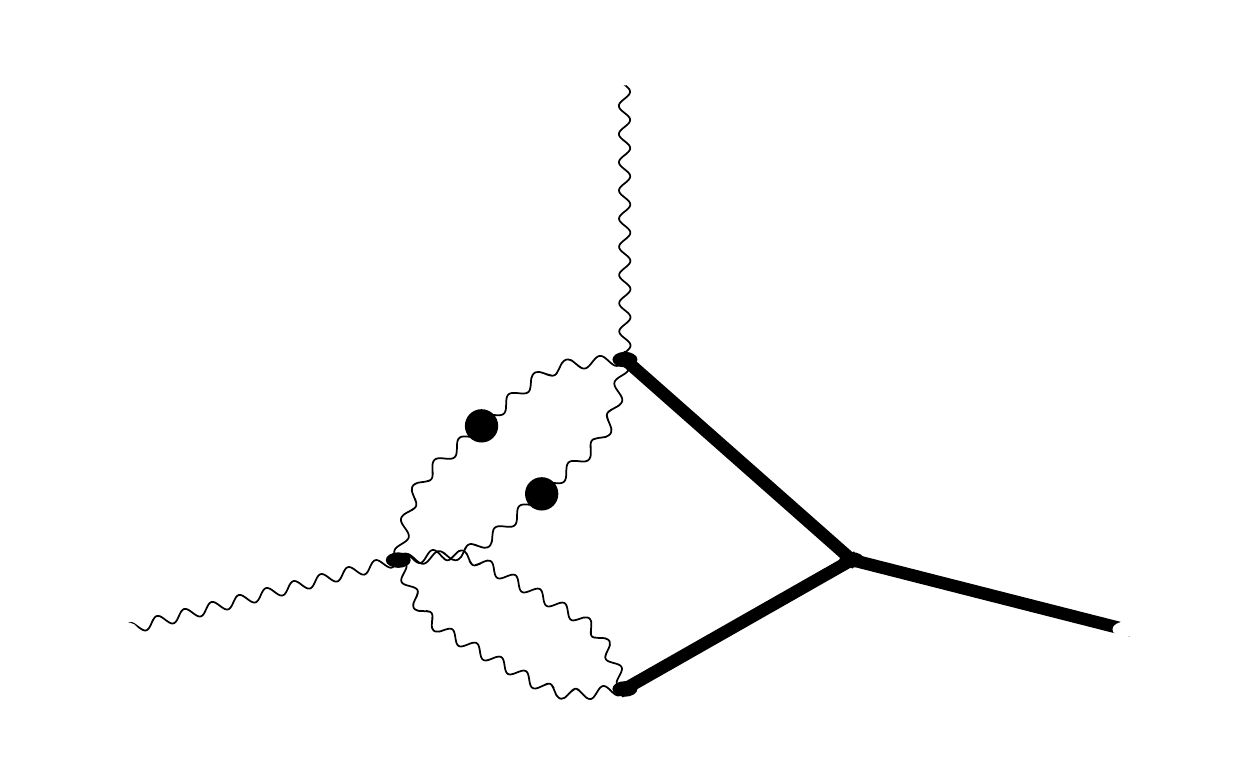}}
      \subfloat[][$I_{46}$ (PP)]
      {\includegraphics[width=0.16\textwidth]{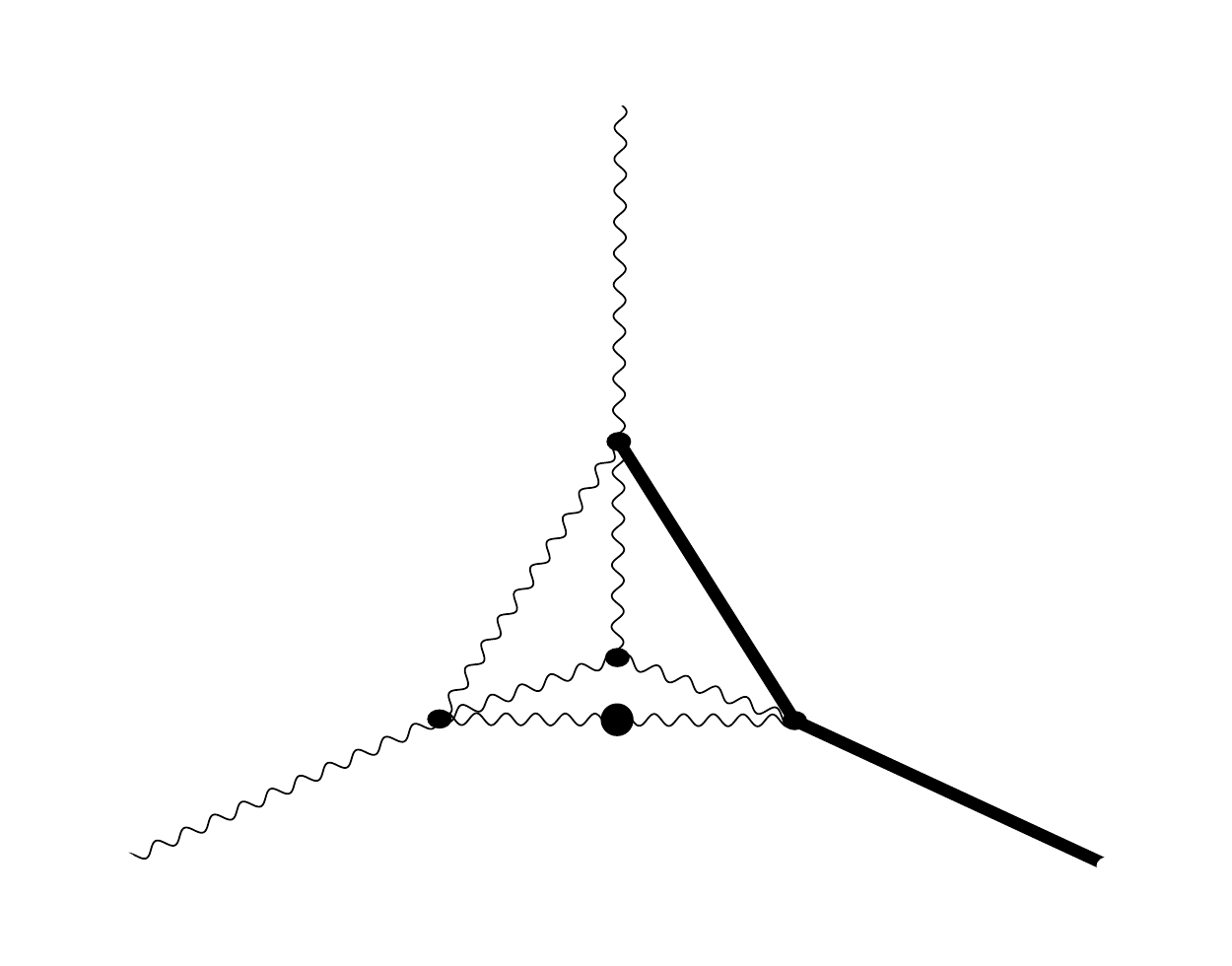}}
      \subfloat[][$I_{47}$ (PP)]
      {\includegraphics[width=0.16\textwidth]{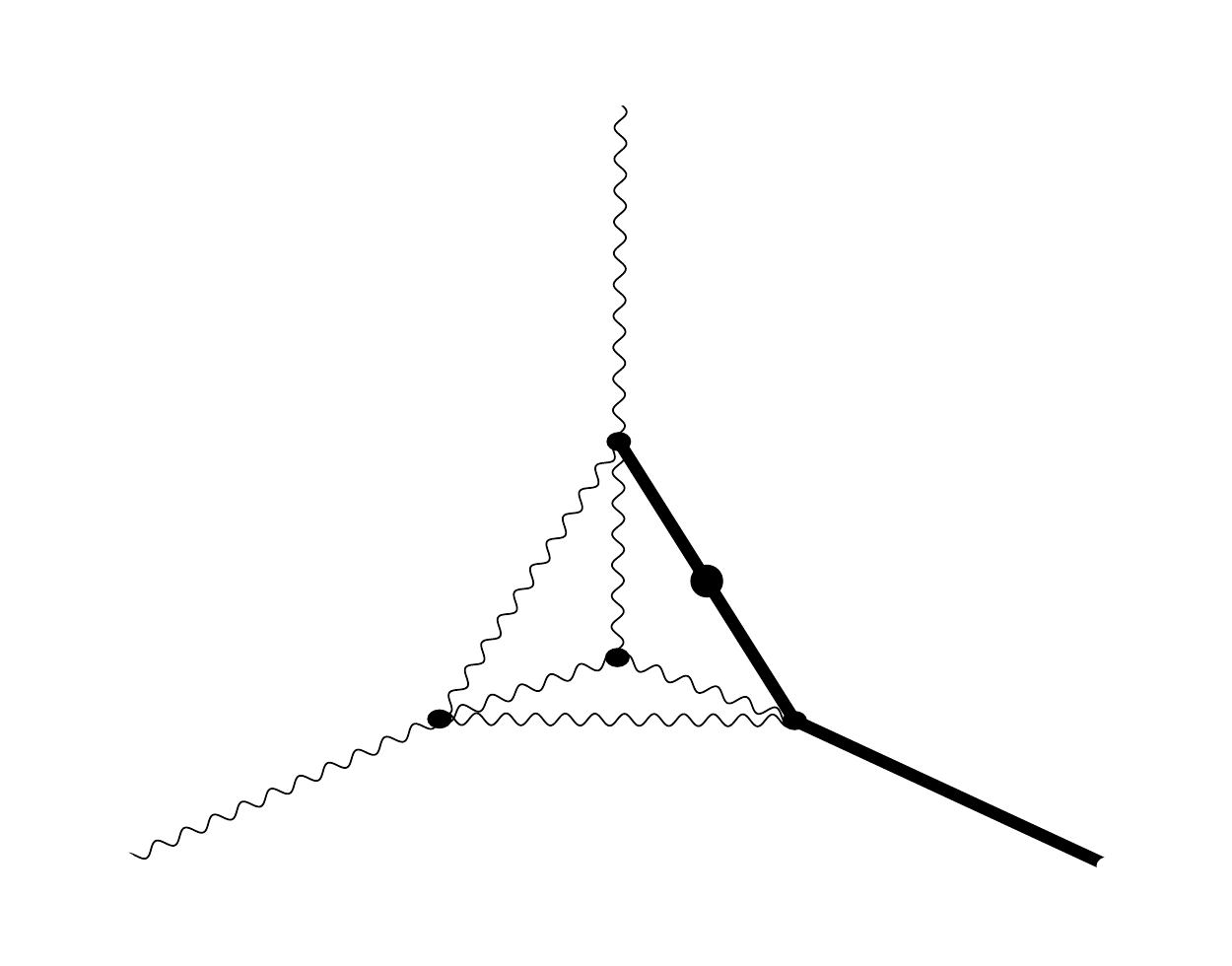}}
      \subfloat[][$I_{48}$ (PP)]
      {\includegraphics[width=0.16\textwidth]{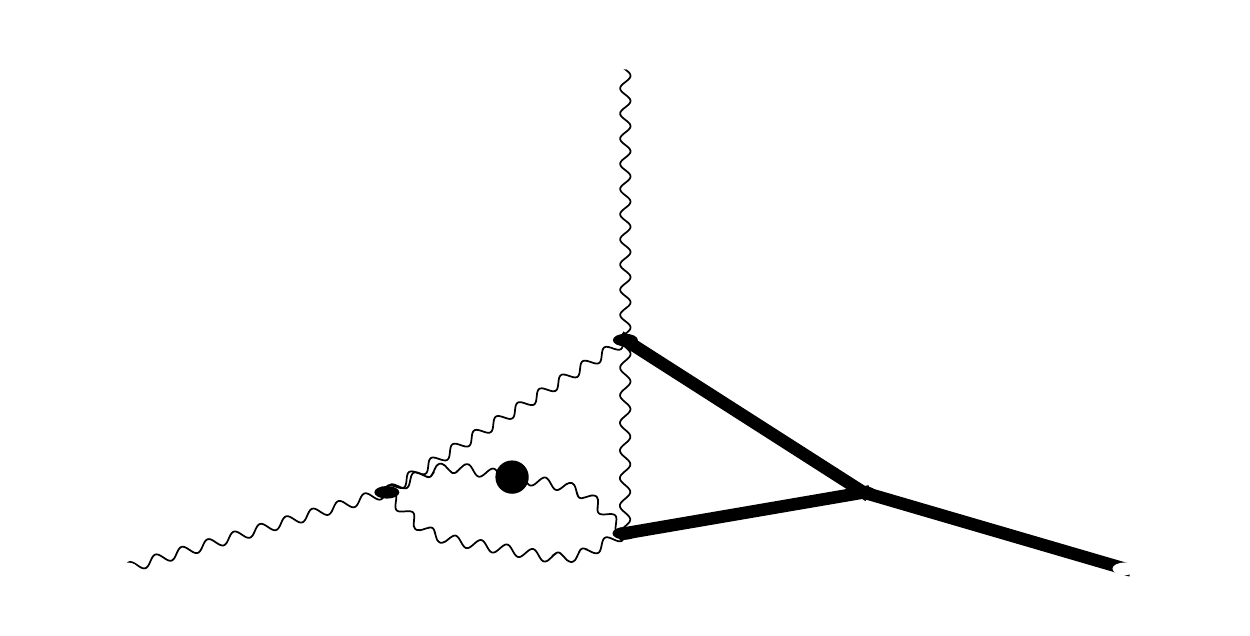}}
      \\

      \subfloat[][$I_{49}$ (PP)]
      {\includegraphics[width=0.16\textwidth]{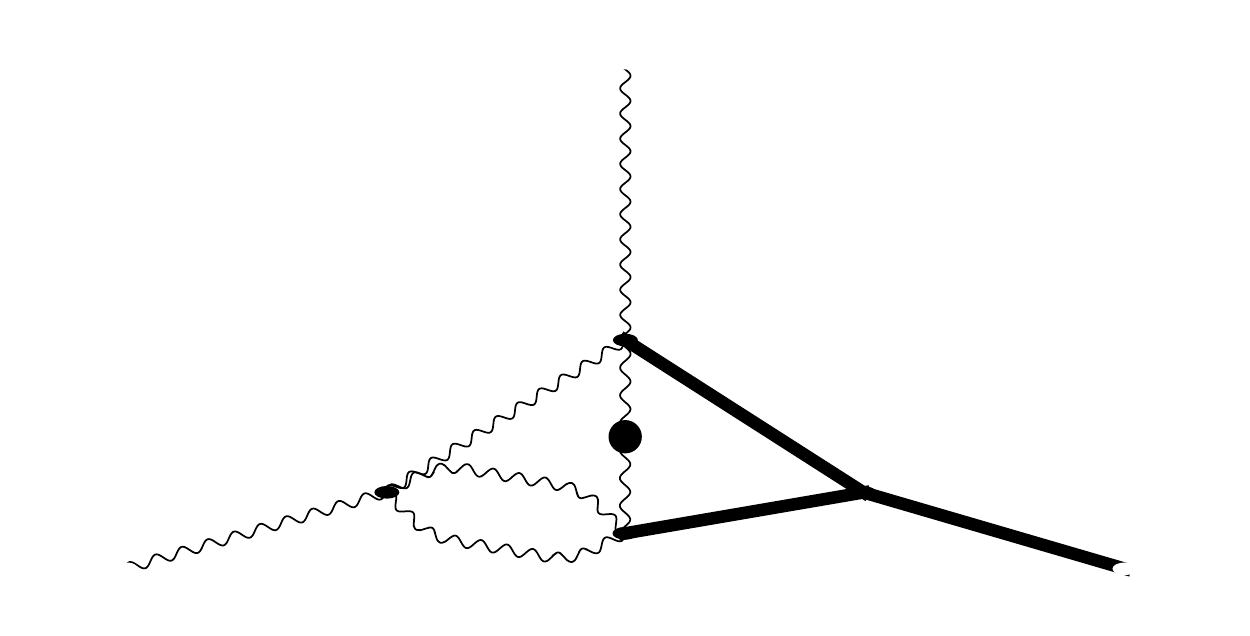}}
      \subfloat[][$I_{50}$ (PP)]
      {\includegraphics[width=0.16\textwidth]{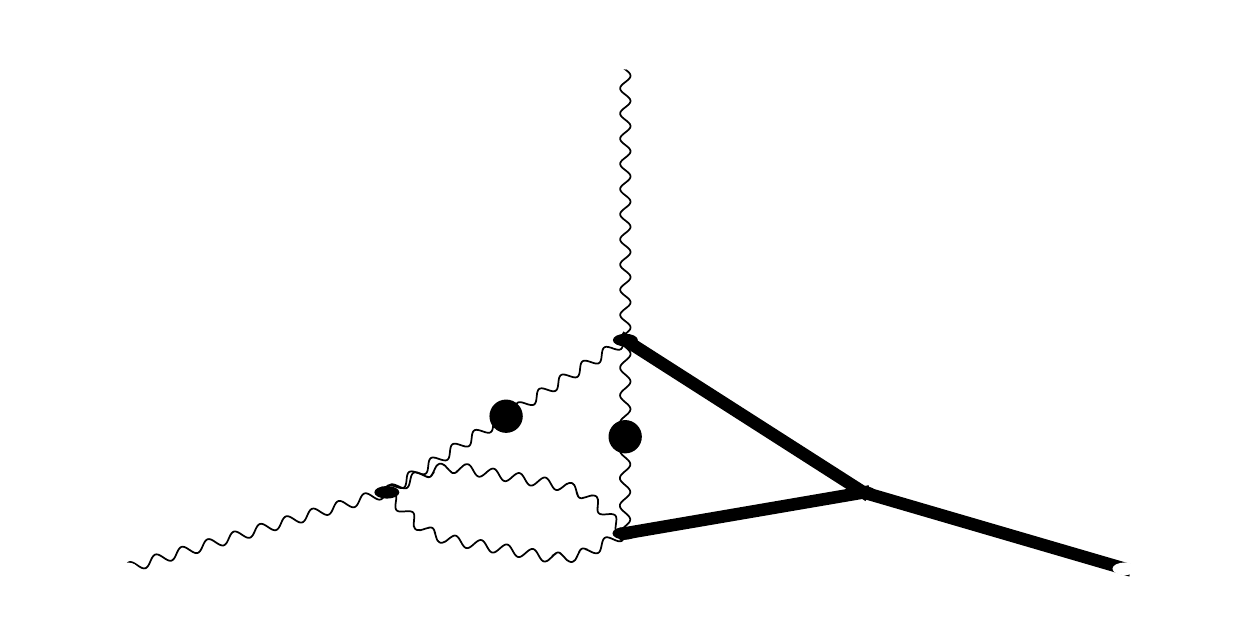}}
      \subfloat[][$I_{51}$ (NA)]
      {\includegraphics[width=0.16\textwidth]{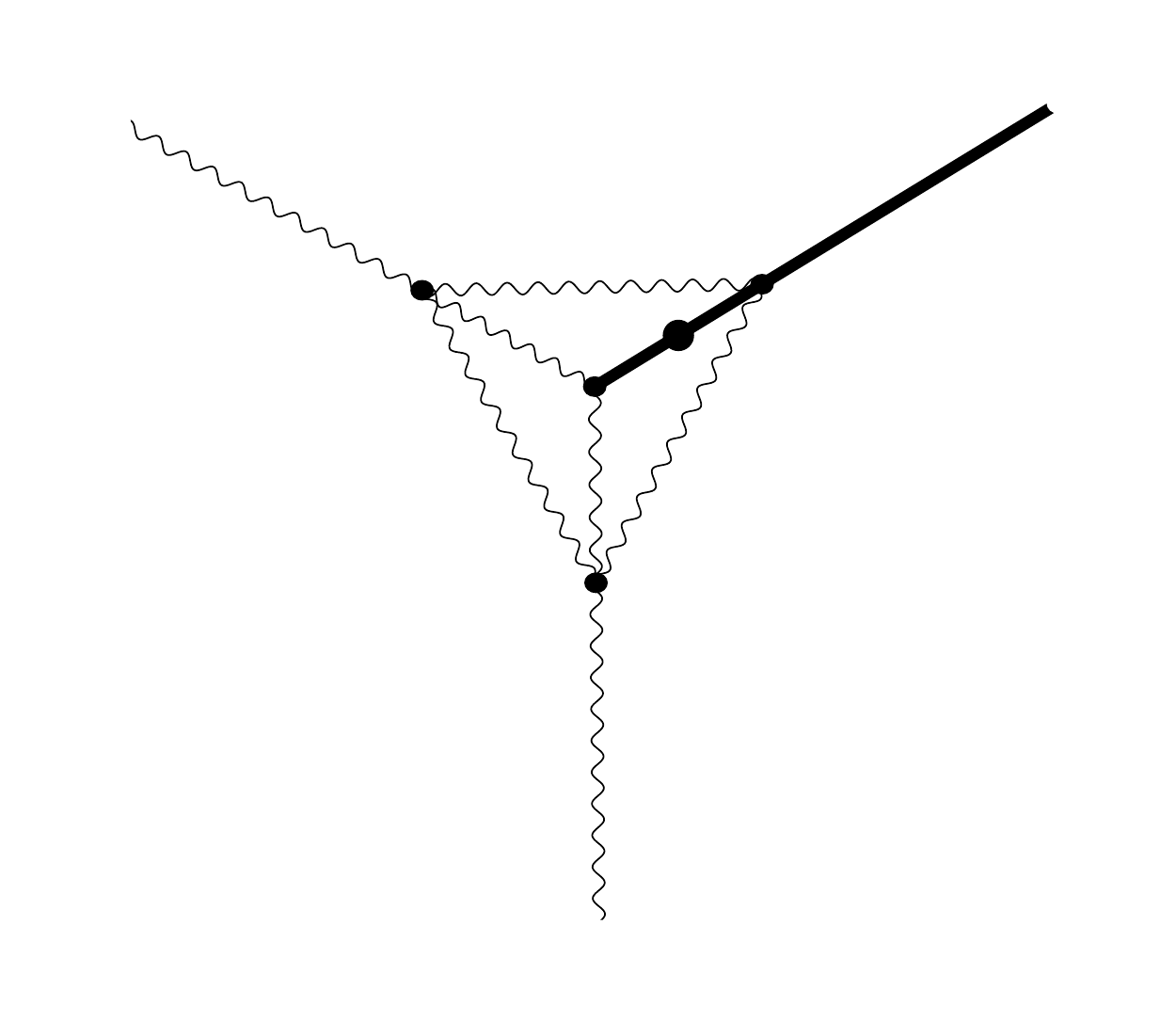}}
      \subfloat[][$I_{52}$ (NA)]
      {\includegraphics[width=0.16\textwidth]{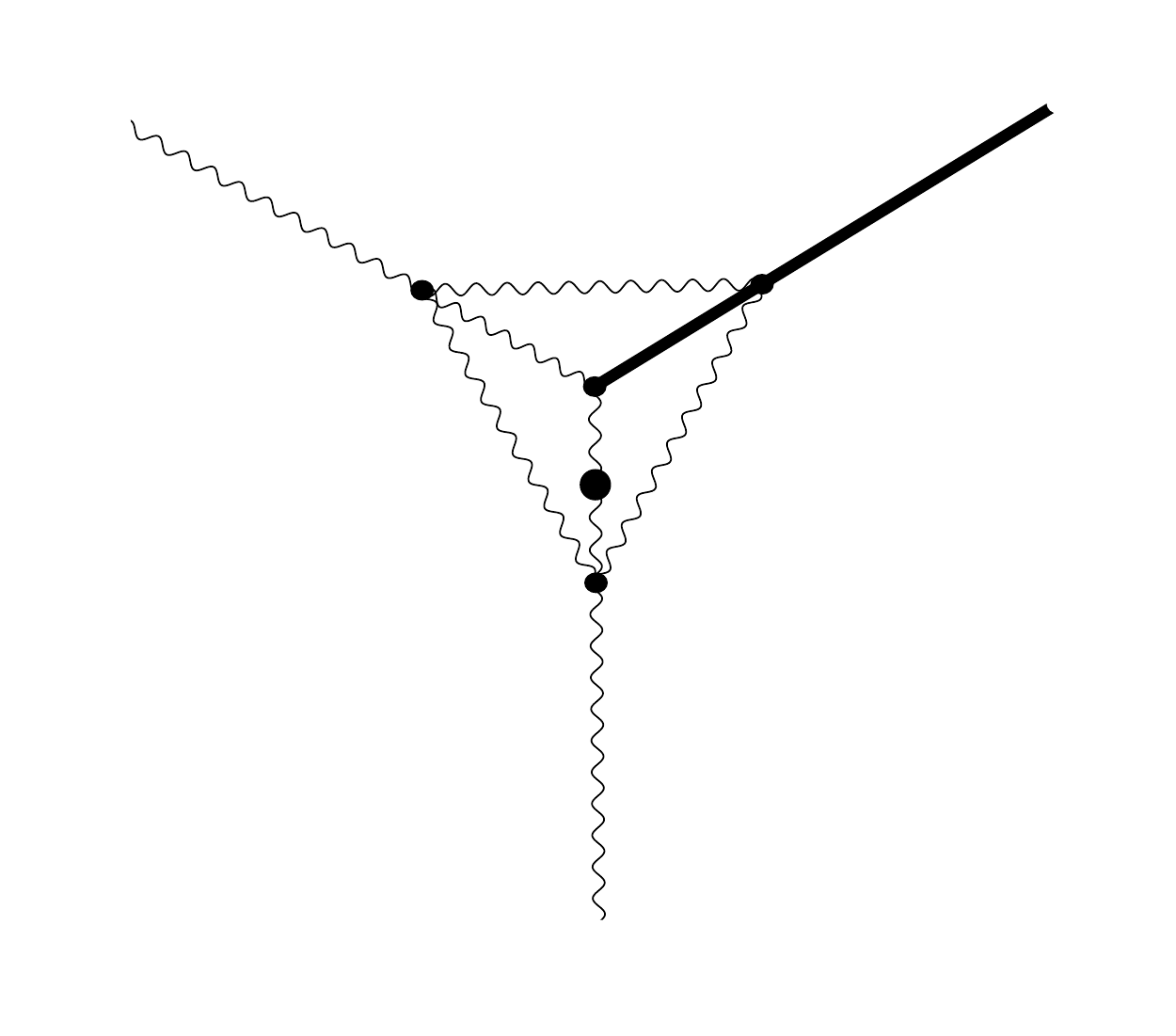}}
      \subfloat[][$I_{53}$ (NA)]
      {\includegraphics[width=0.16\textwidth]{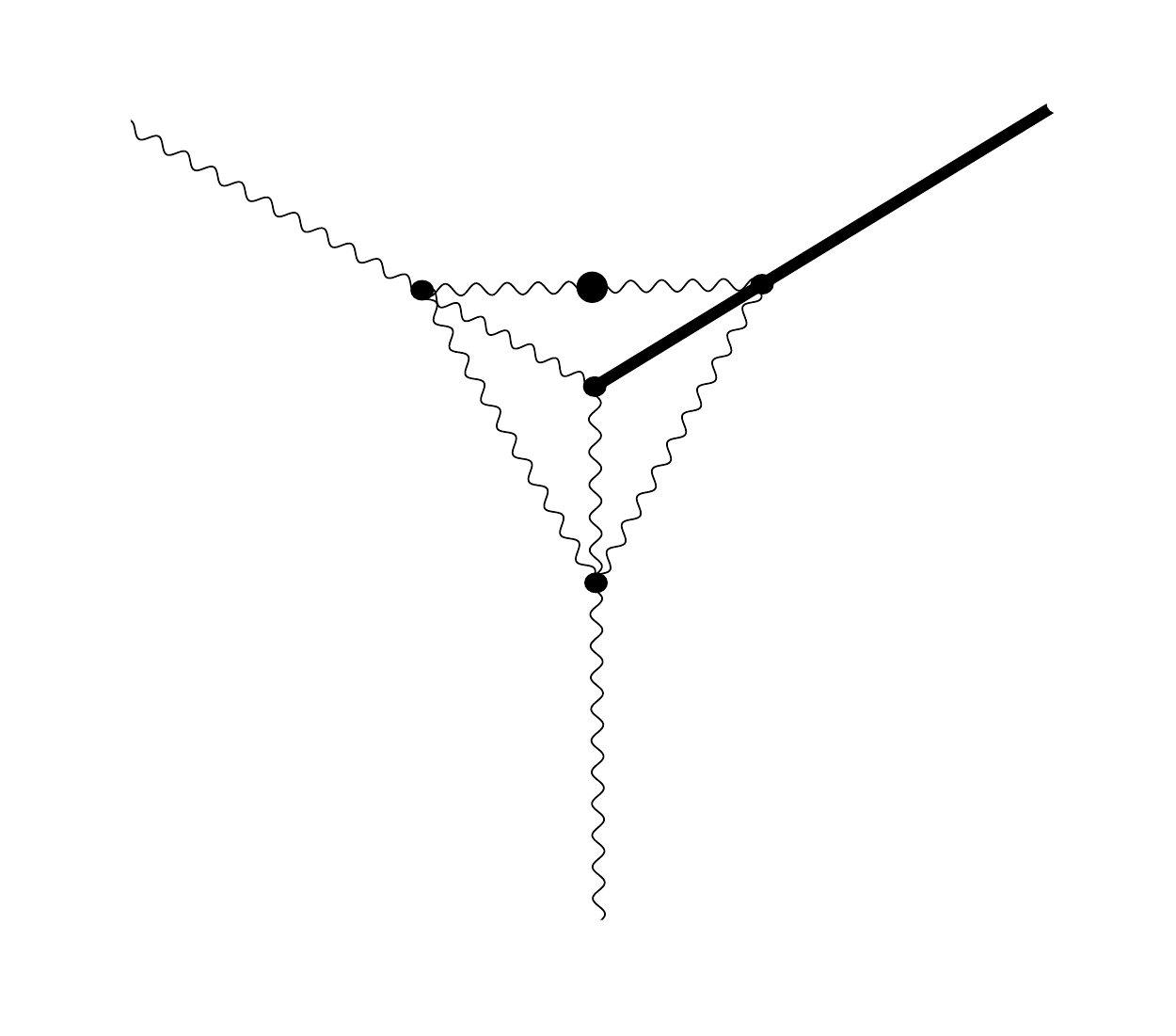}}
      \subfloat[][$I_{54}$ (NA)]
      {\includegraphics[width=0.16\textwidth]{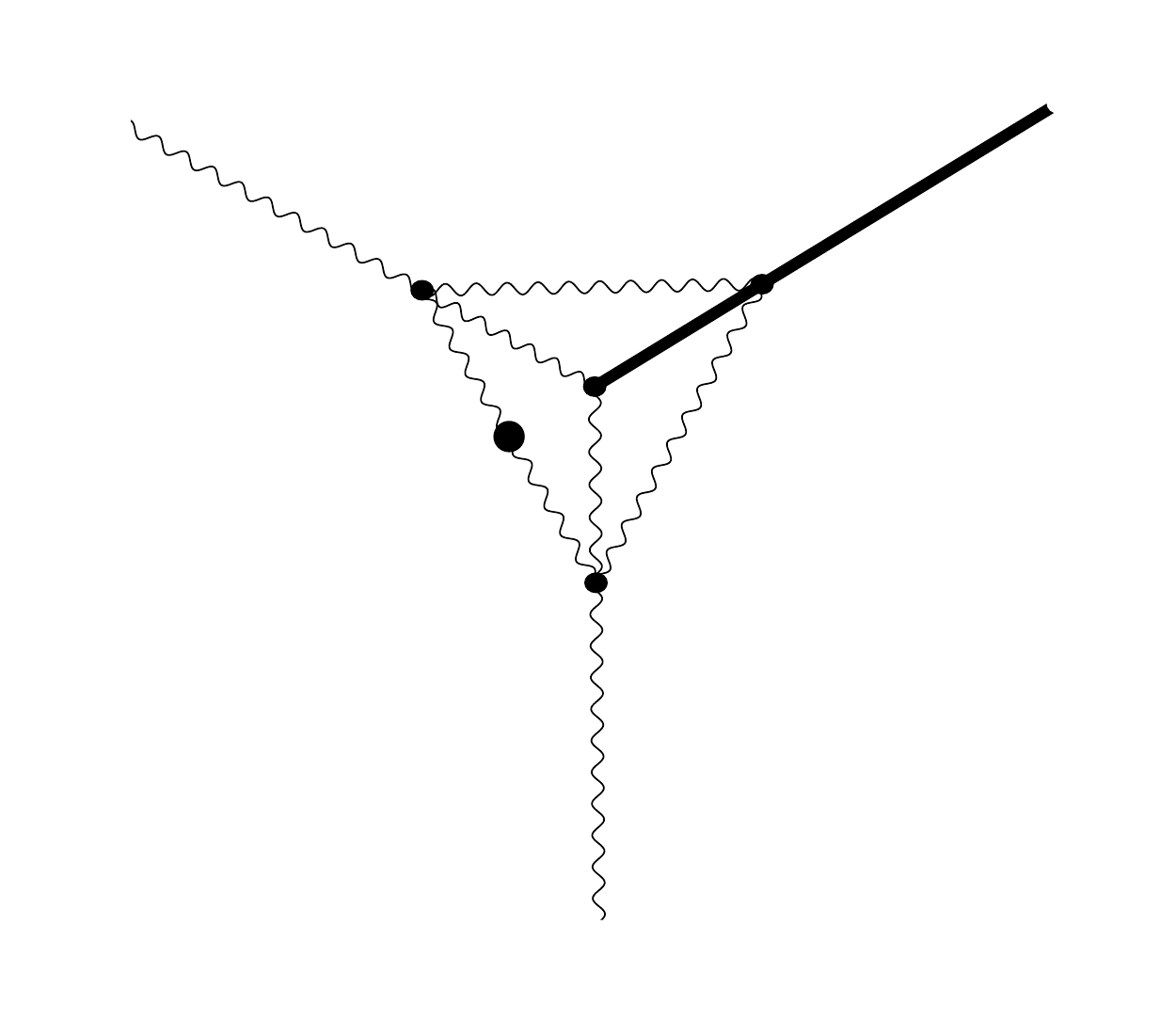}}
      \\

      \subfloat[][$I_{55}^*$ (PP)]
      {\includegraphics[width=0.16\textwidth]{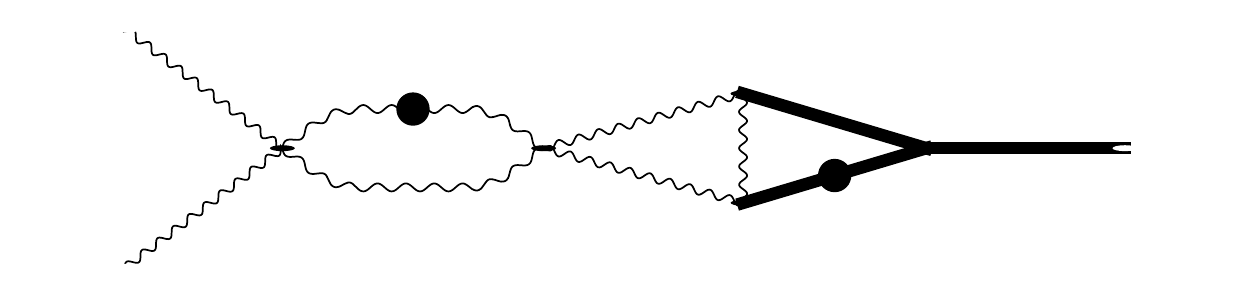}}
      \subfloat[][$I_{56}$ (NB)]
      {\includegraphics[width=0.16\textwidth]{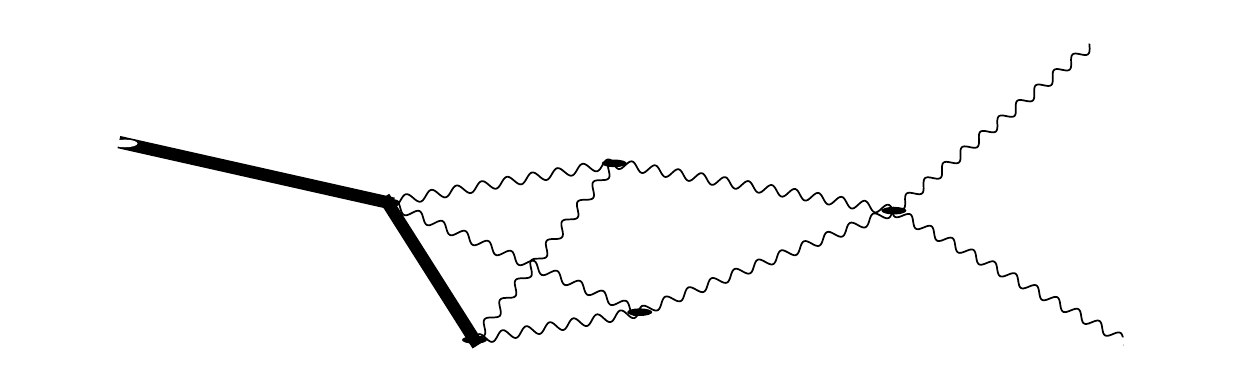}}
      \subfloat[][$I_{57}$ (PP)]
      {\includegraphics[width=0.16\textwidth]{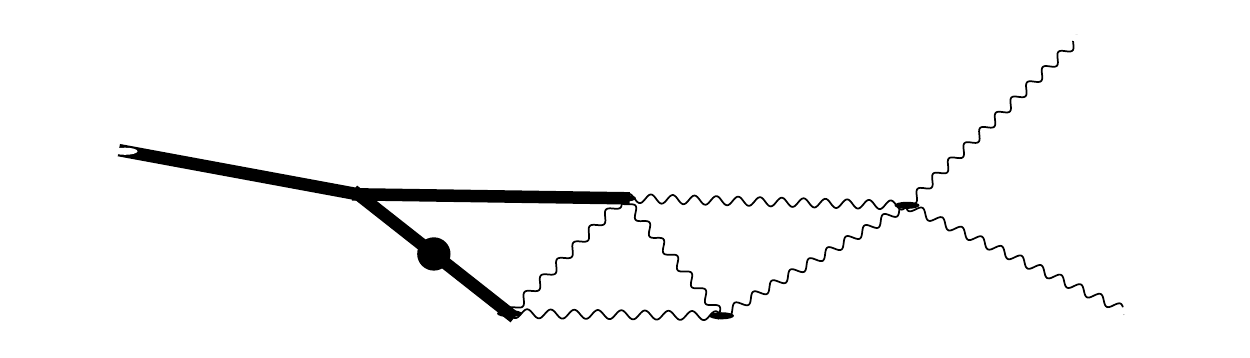}}
      \subfloat[][$I_{58}^*$ (PP)]
      {\includegraphics[width=0.16\textwidth]{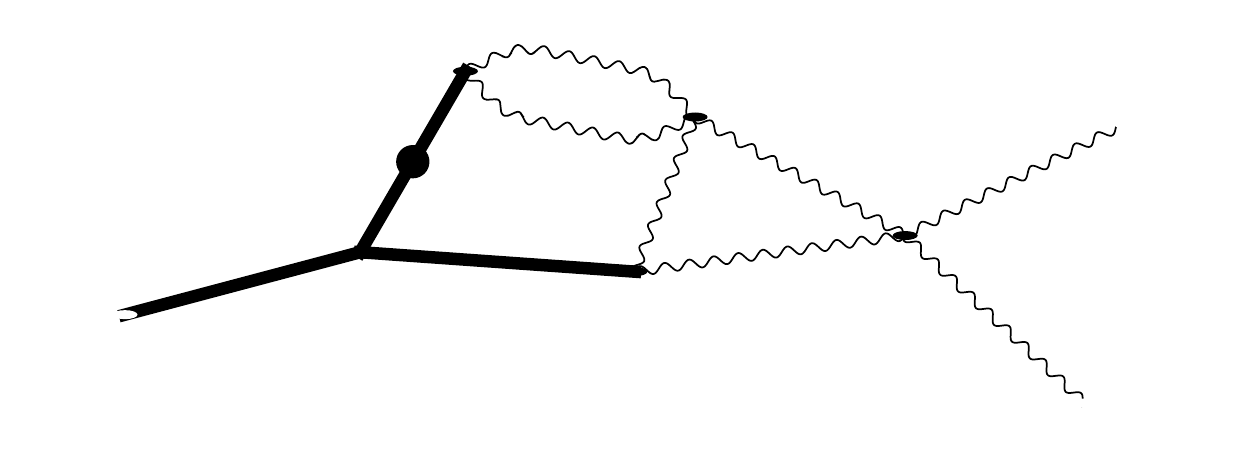}}
      \subfloat[][$I_{59}$ (PP)]
      {\includegraphics[width=0.16\textwidth]{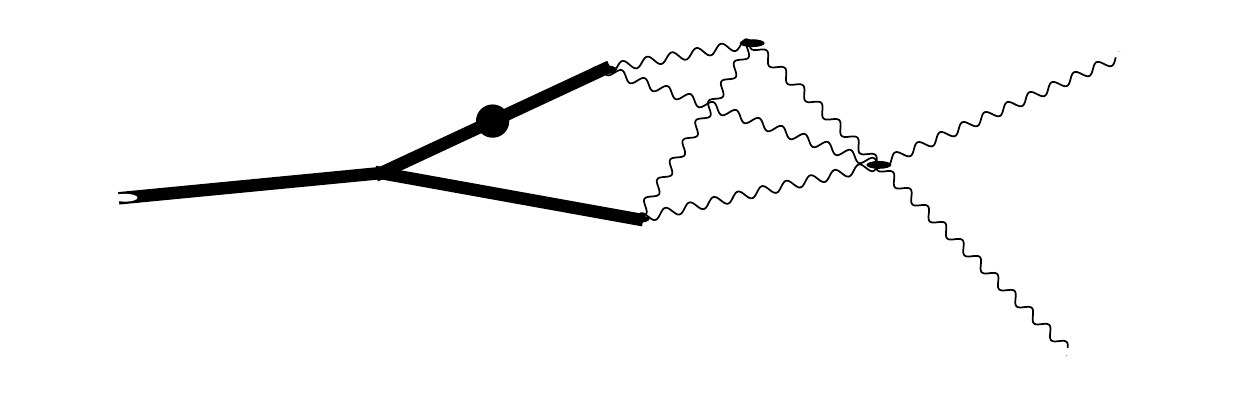}}
      \subfloat[][$I_{60}$ (NA)]
      {\includegraphics[width=0.16\textwidth]{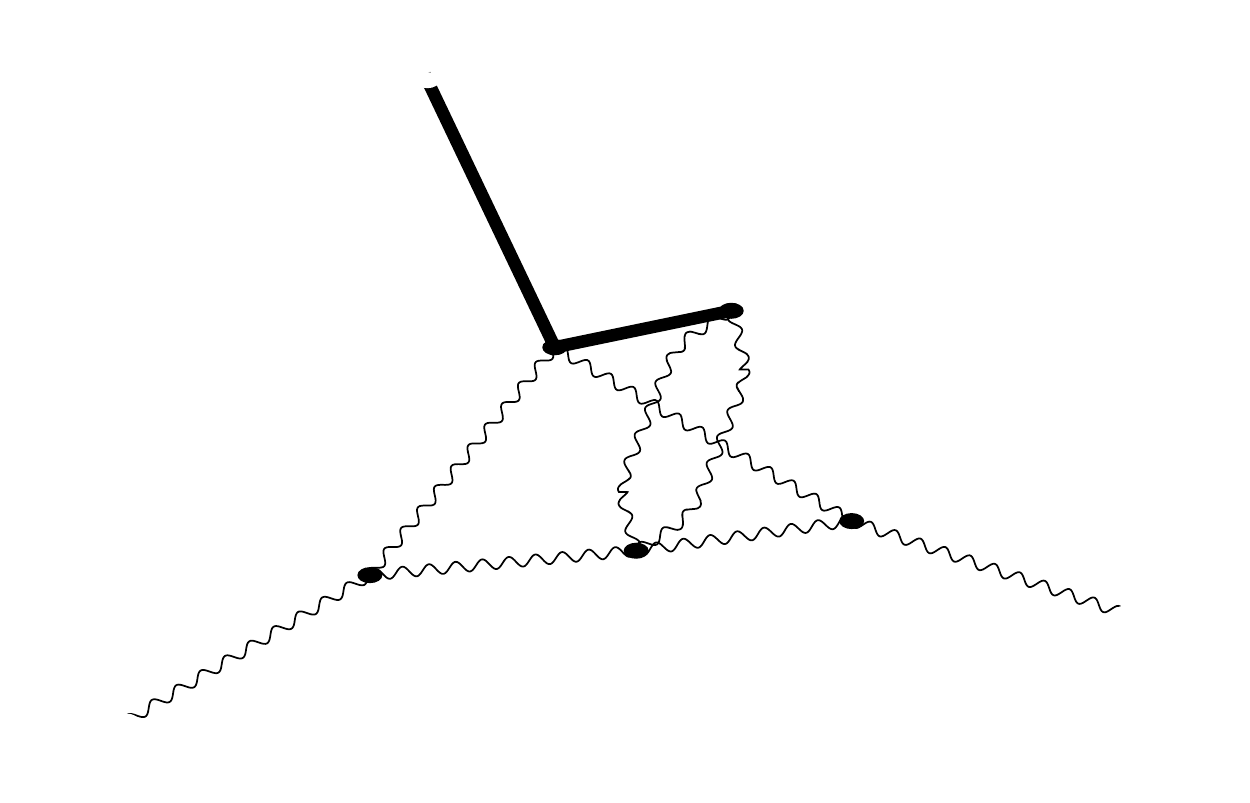}}
      \\

      \subfloat[][$I_{61}$ (PP)]
      {\includegraphics[width=0.16\textwidth]{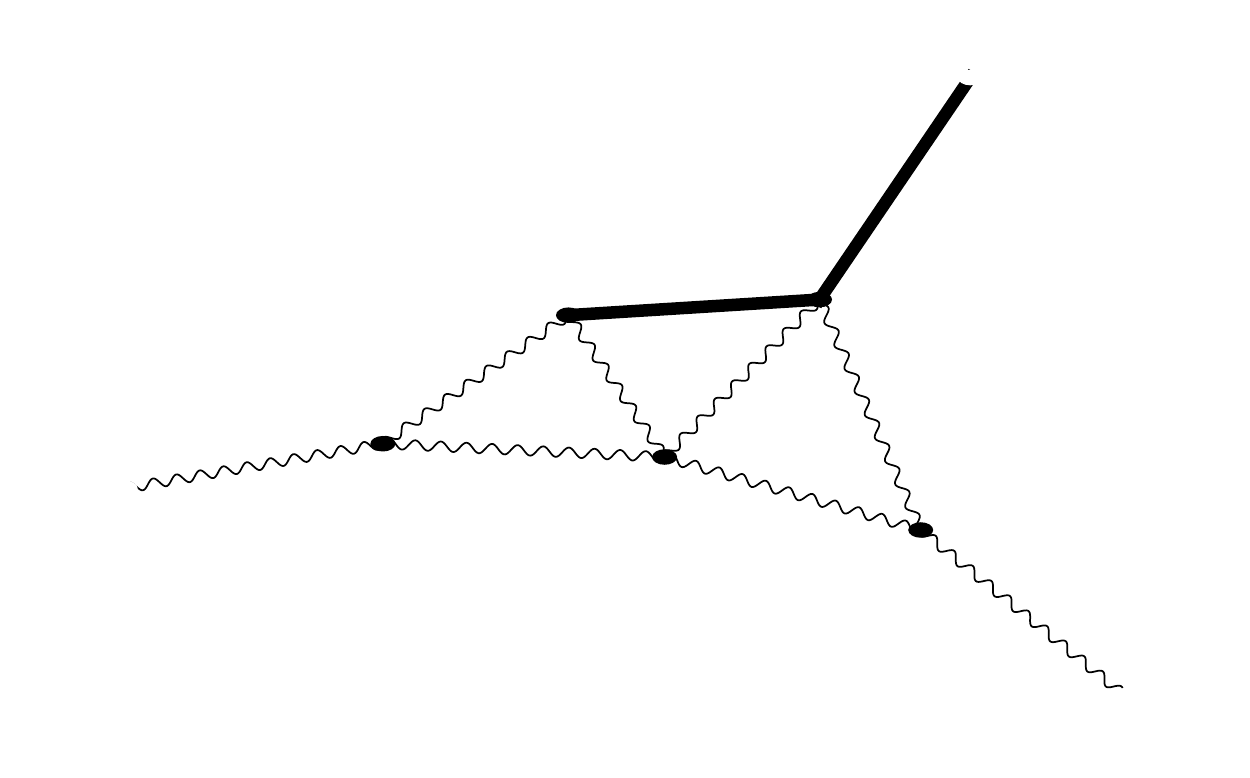}}
      \subfloat[][$I_{62}$ (PP)]
      {\includegraphics[width=0.16\textwidth]{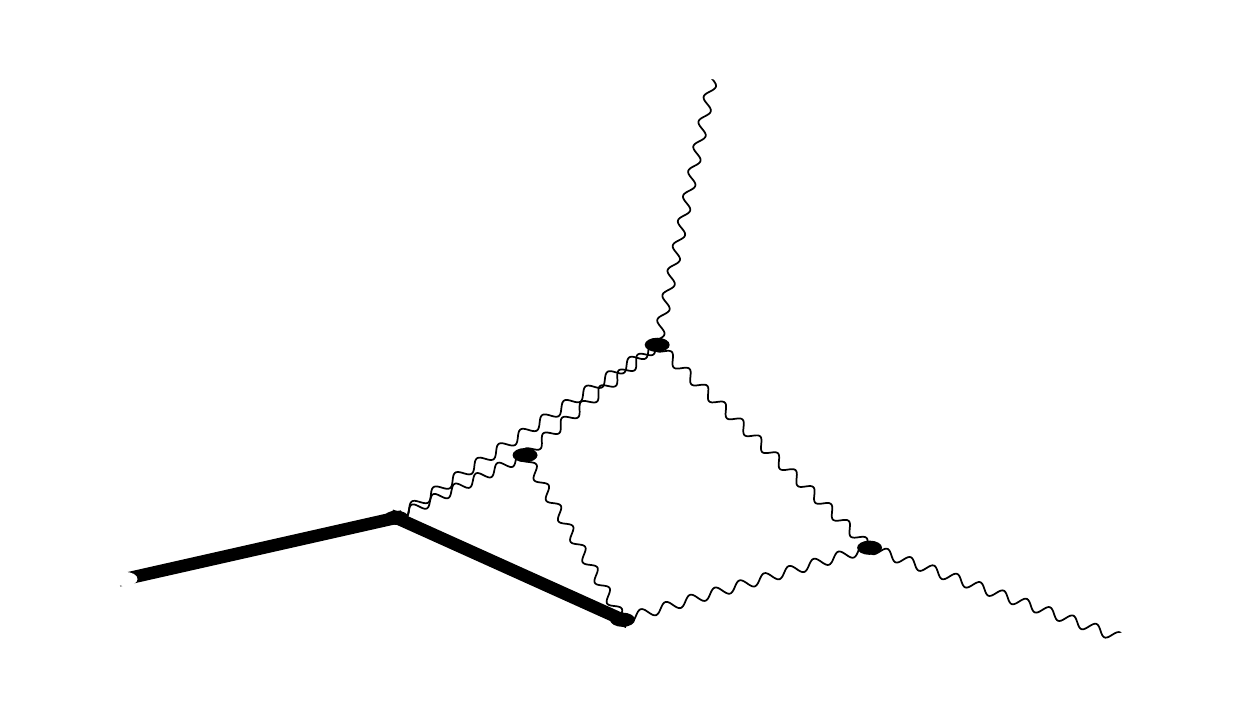}}
      \subfloat[][$I_{63}$ (PP)]
      {\includegraphics[width=0.16\textwidth]{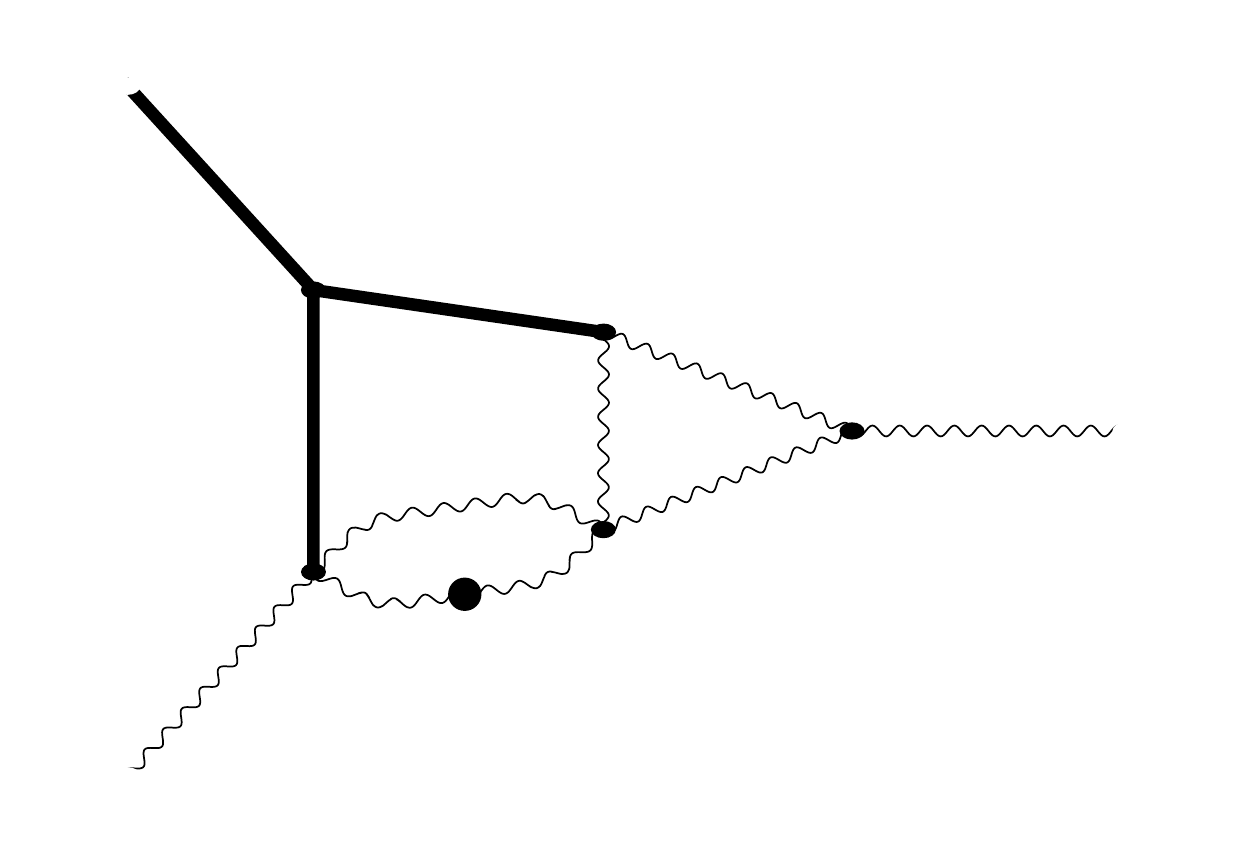}}
      \subfloat[][$I_{64}$ (PP)]
      {\includegraphics[width=0.16\textwidth]{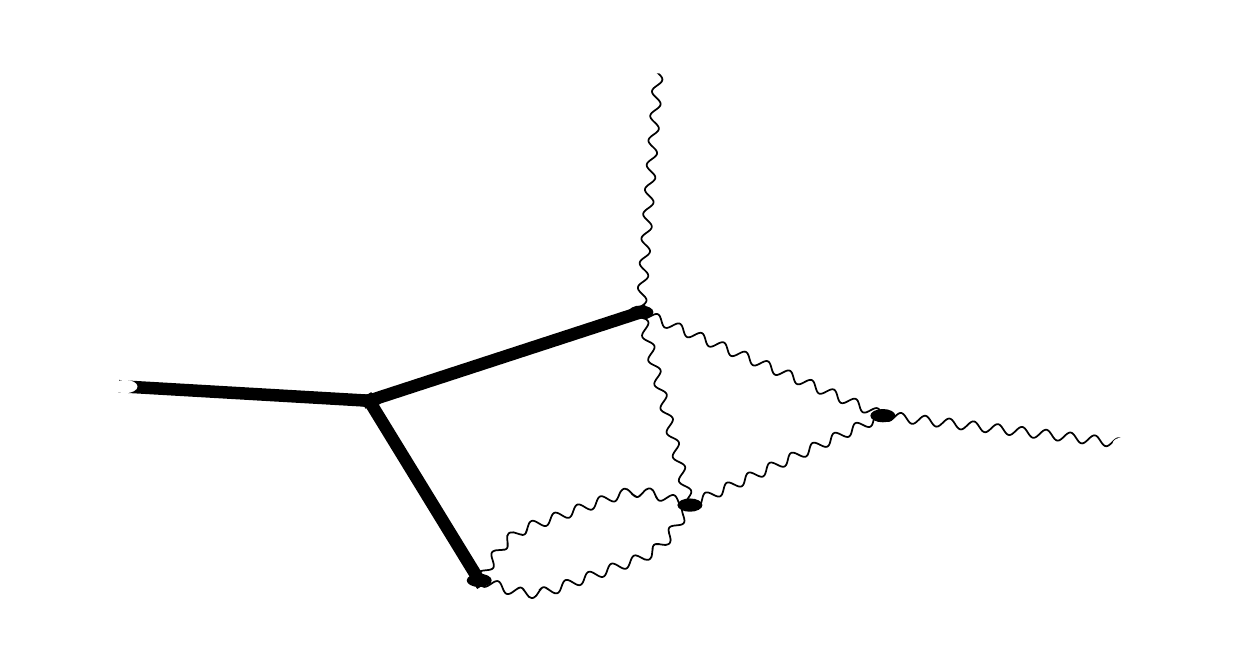}}
      \subfloat[][$I_{65}$ (PP)]
      {\includegraphics[width=0.16\textwidth]{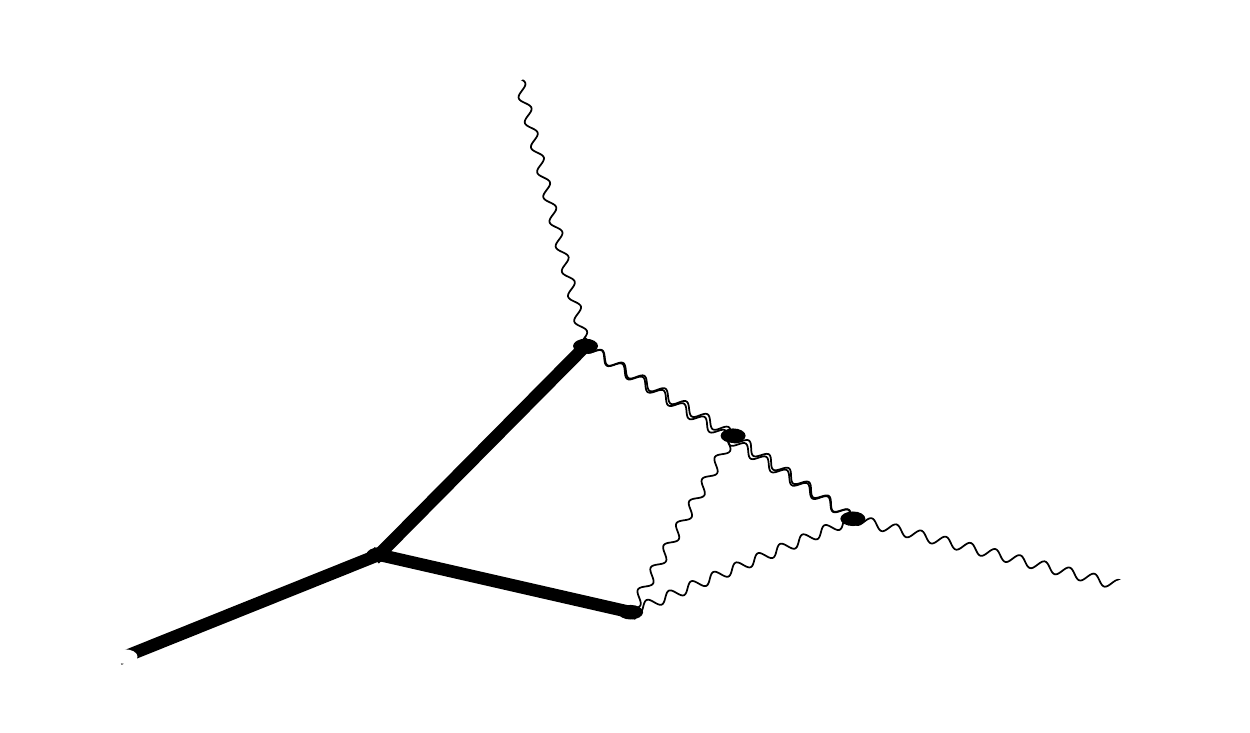}}
      \subfloat[][$I_{66}$ (PP)]
      {\includegraphics[width=0.16\textwidth]{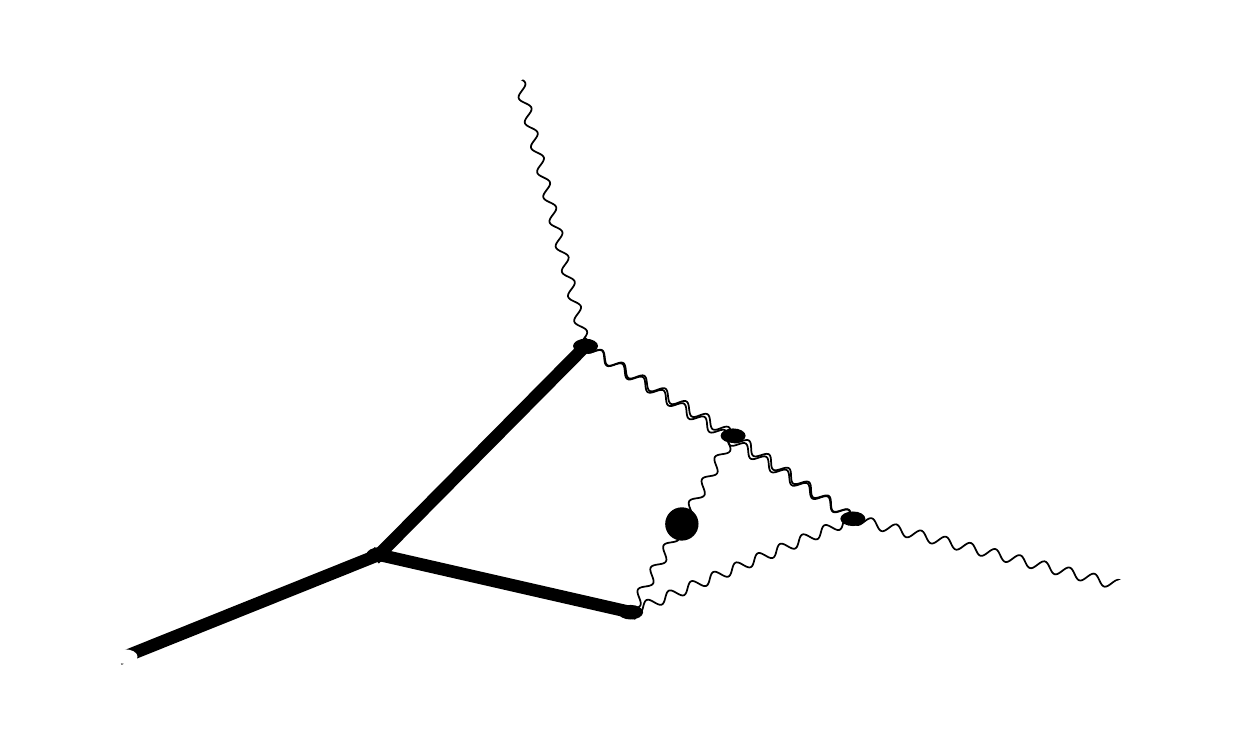}}
      \\

      \subfloat[][$I_{67}$ (PP)]
      {\includegraphics[width=0.16\textwidth]{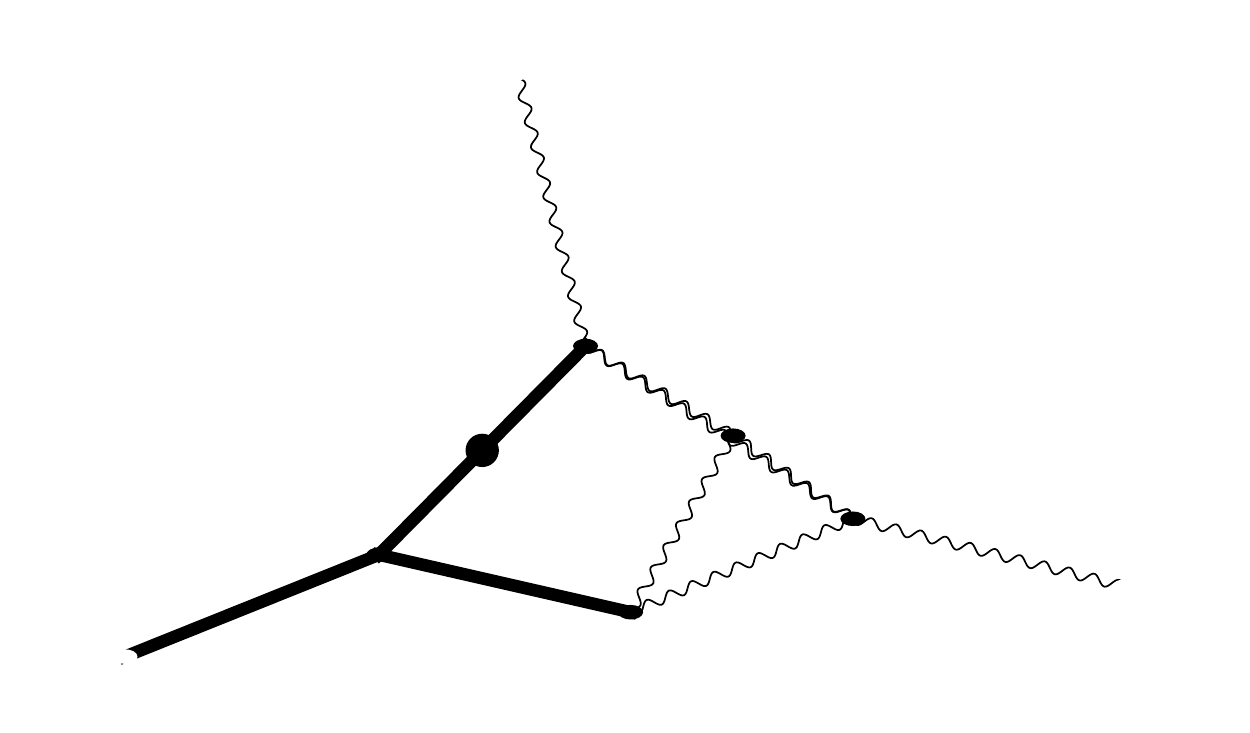}}
      \subfloat[][$I_{68}^*$ (PP)]
      {\includegraphics[width=0.16\textwidth]{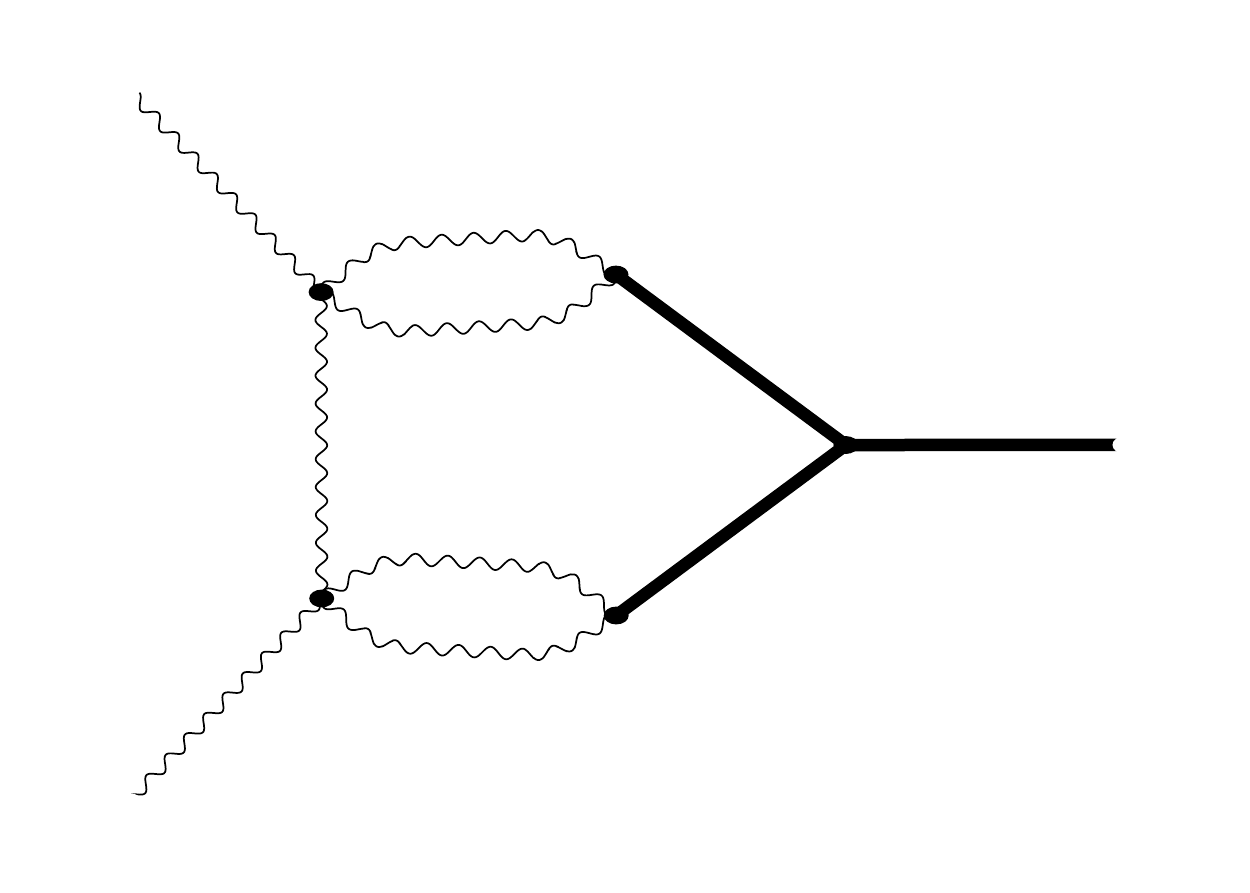}}
      \subfloat[][$I_{69}^*$ (PP)]
      {\includegraphics[width=0.16\textwidth]{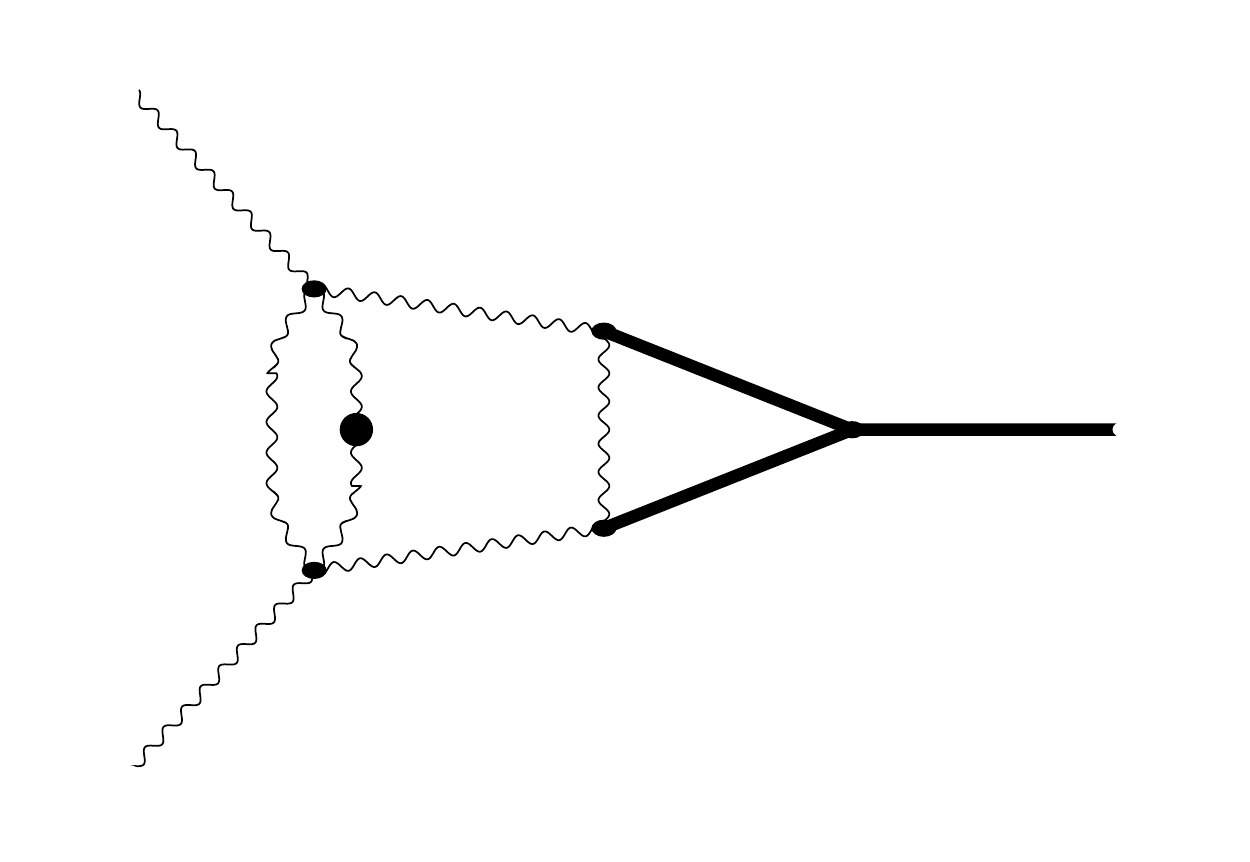}}
      \subfloat[][$I_{70}$ (PP)]
      {\includegraphics[width=0.16\textwidth]{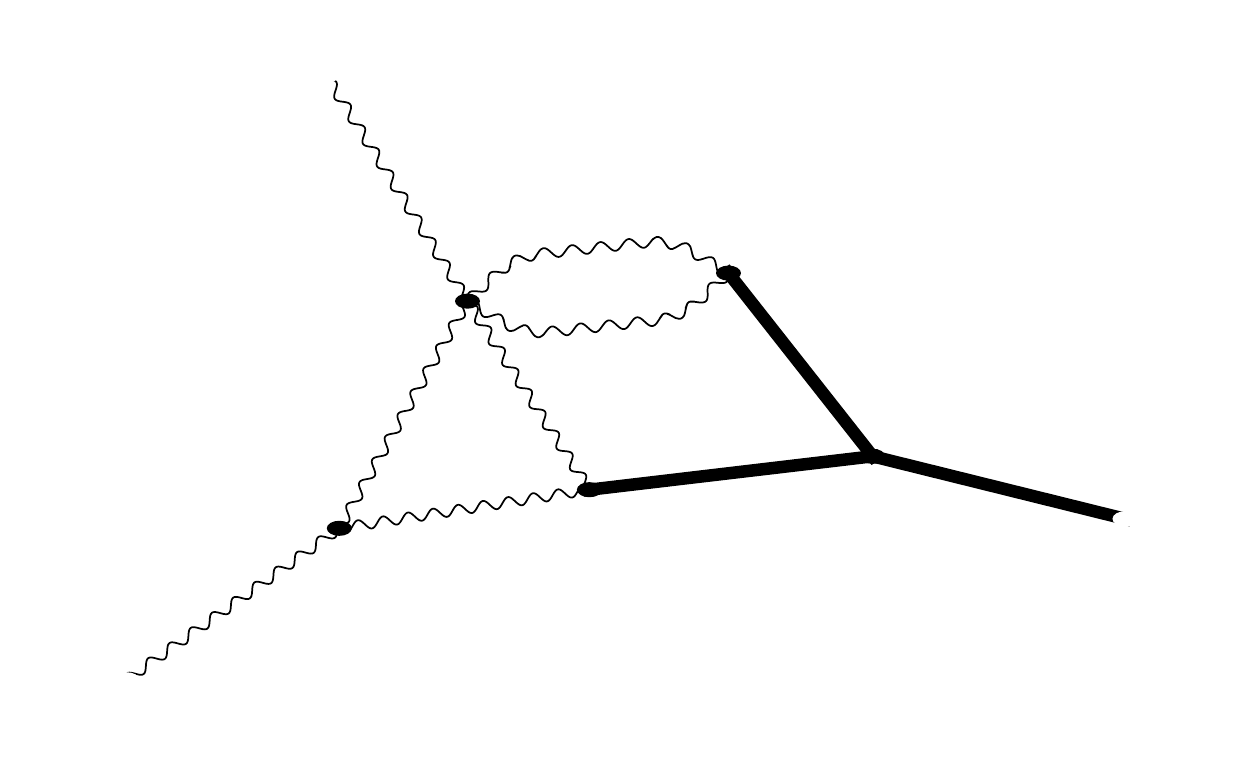}}
      \subfloat[][$I_{71}$ (PP)]
      {\includegraphics[width=0.16\textwidth]{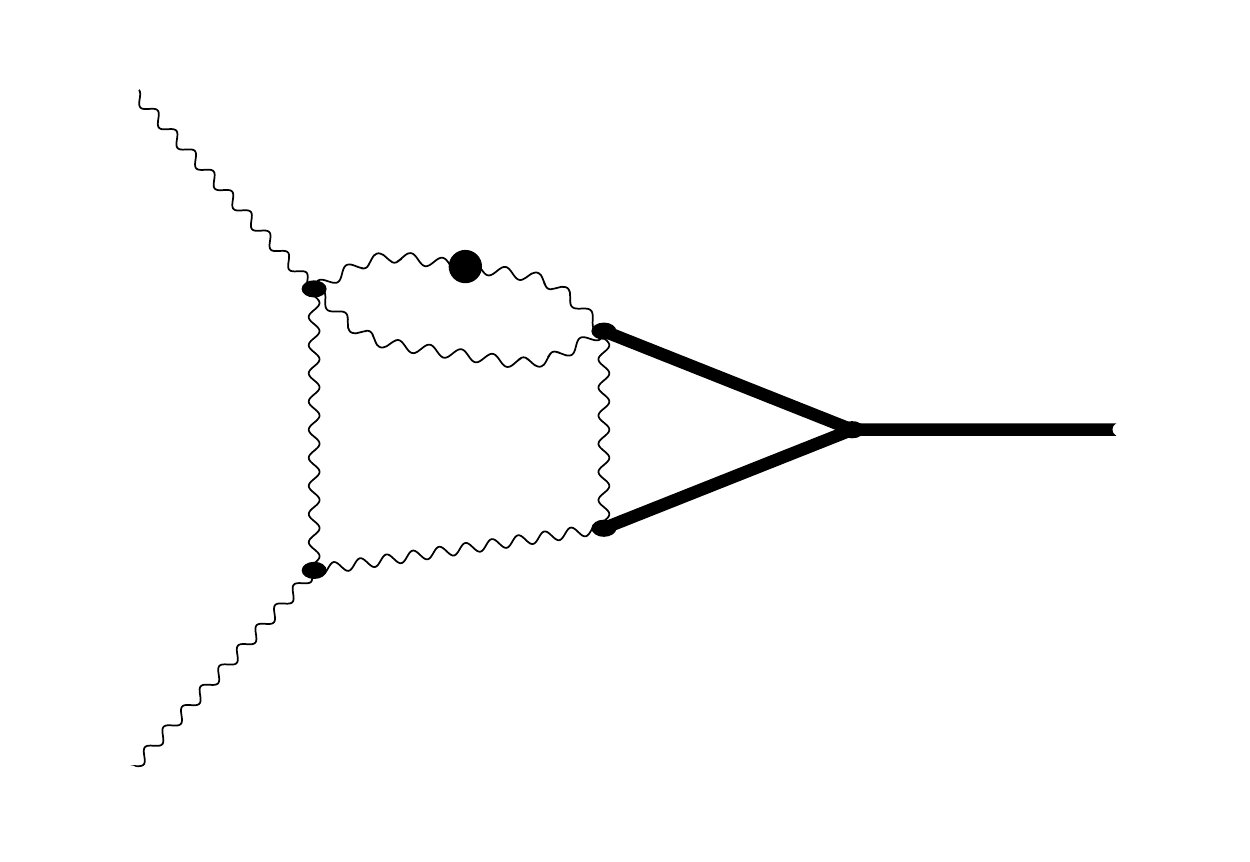}}
      \subfloat[][$I_{72}$ (PP)]
      {\includegraphics[width=0.16\textwidth]{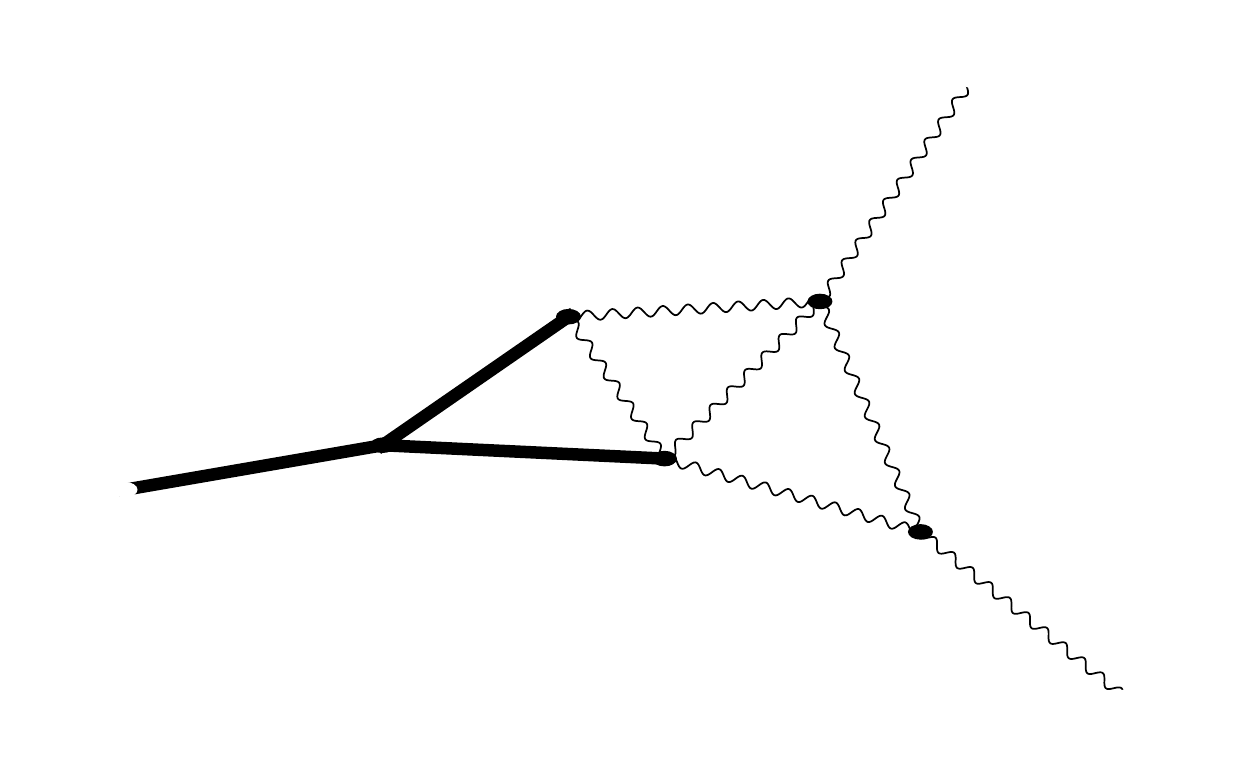}}
      \caption{Master Integrals (2/3).}
      \label{MI31_72}
\end{figure}

\begin{figure}[H]
      \captionsetup[subfigure]{labelformat=empty}
      \centering
      \subfloat[][$I_{73}$ (PP)]
      {\includegraphics[width=0.16\textwidth]{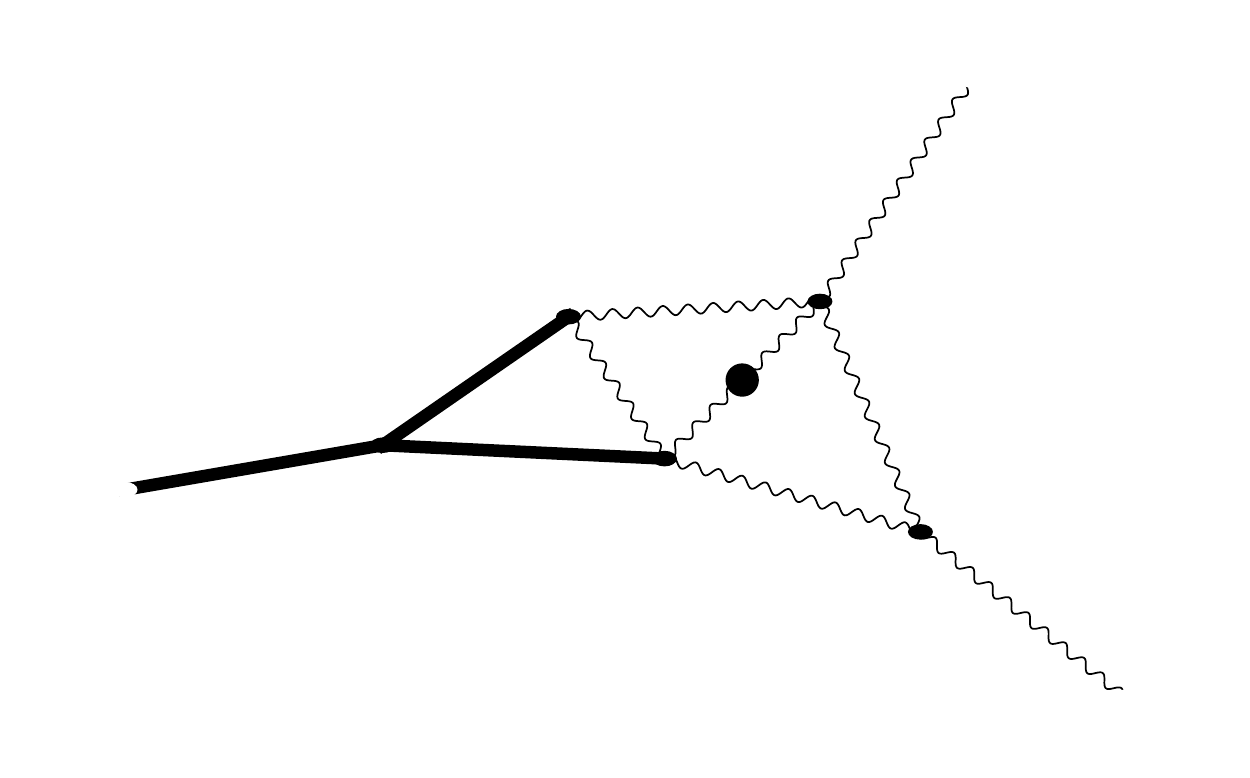}}
      \subfloat[][$I_{74}$ (PP)]
      {\includegraphics[width=0.16\textwidth]{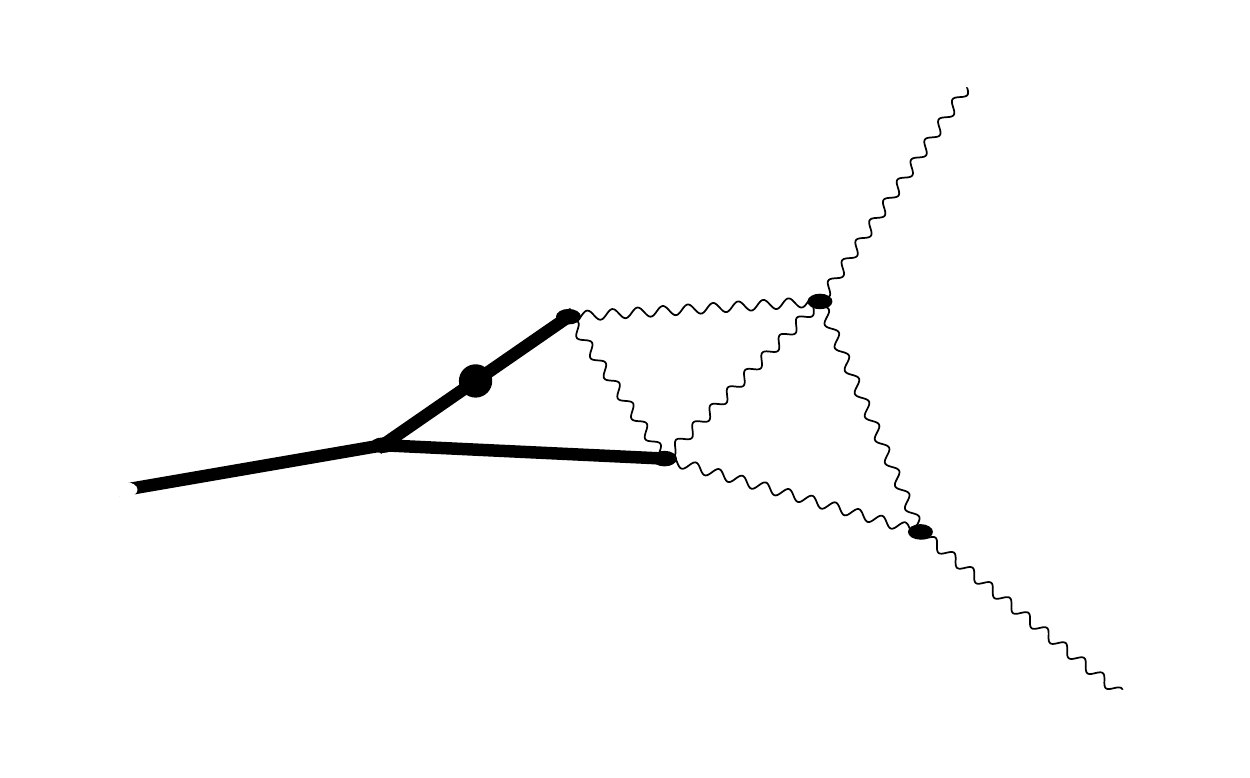}}
      \subfloat[][$I_{75}$ (NA)]
      {\includegraphics[width=0.16\textwidth]{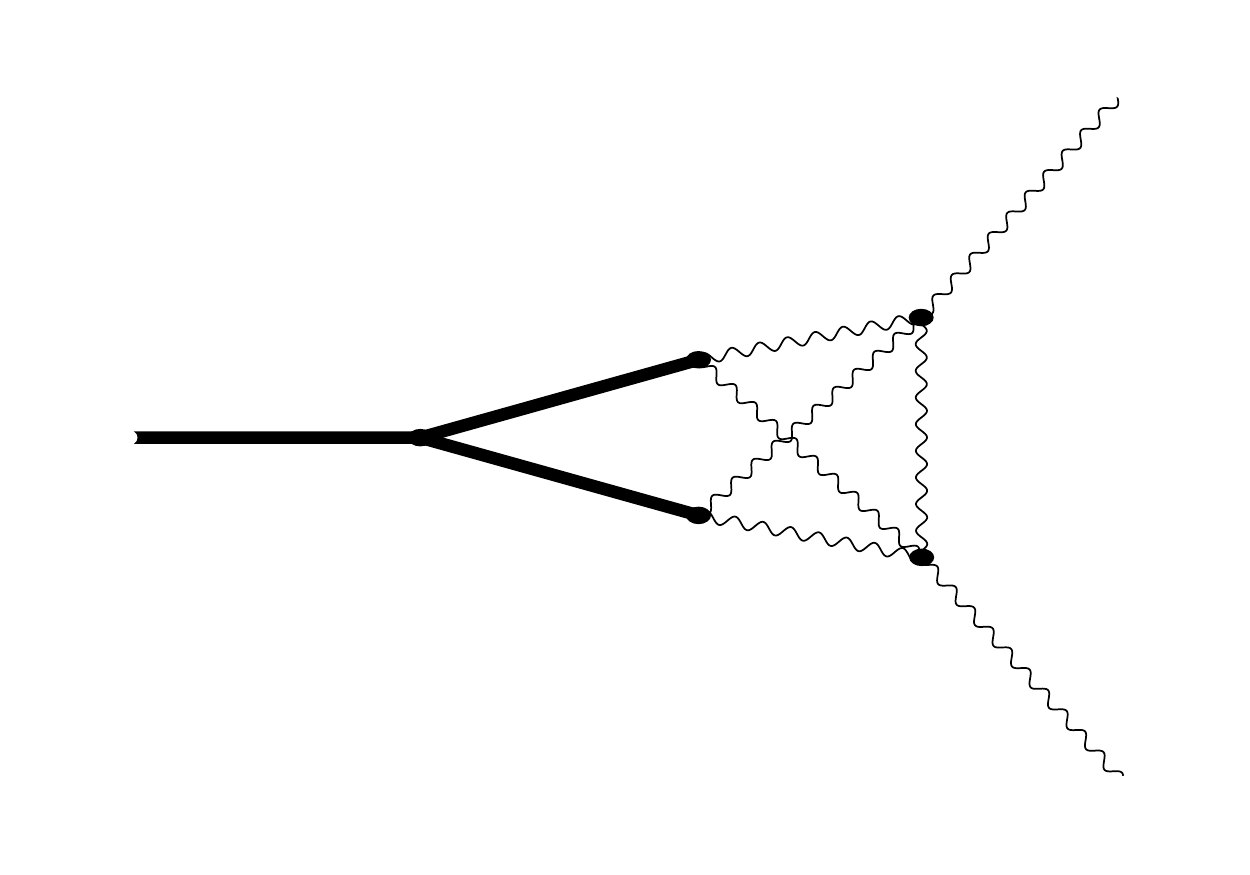}}
      \subfloat[][$I_{76}$ (NA)]
      {\includegraphics[width=0.16\textwidth]{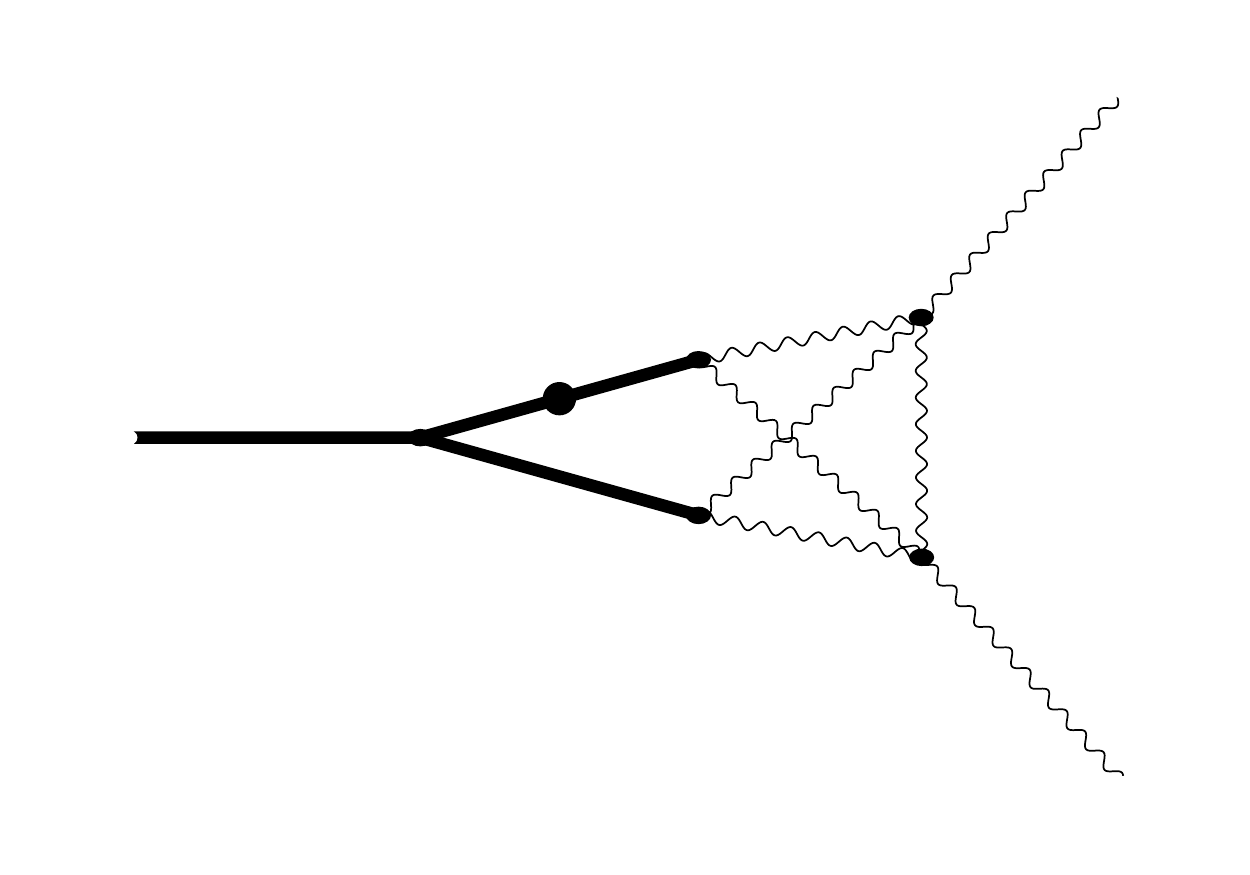}}
      \subfloat[][$I_{77}$ (NA)]
      {\includegraphics[width=0.16\textwidth]{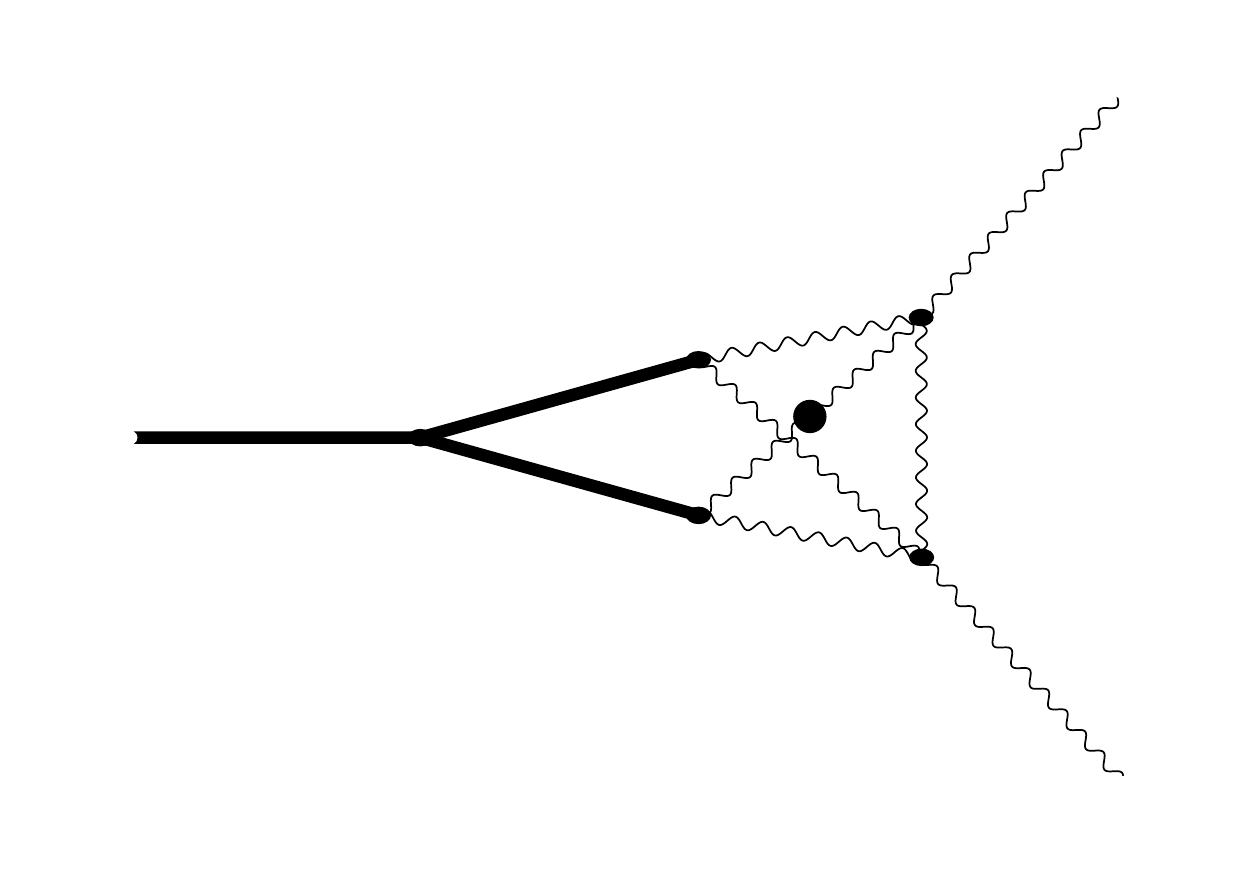}}
      \subfloat[][$I_{78}$ (NA)]
      {\includegraphics[width=0.16\textwidth]{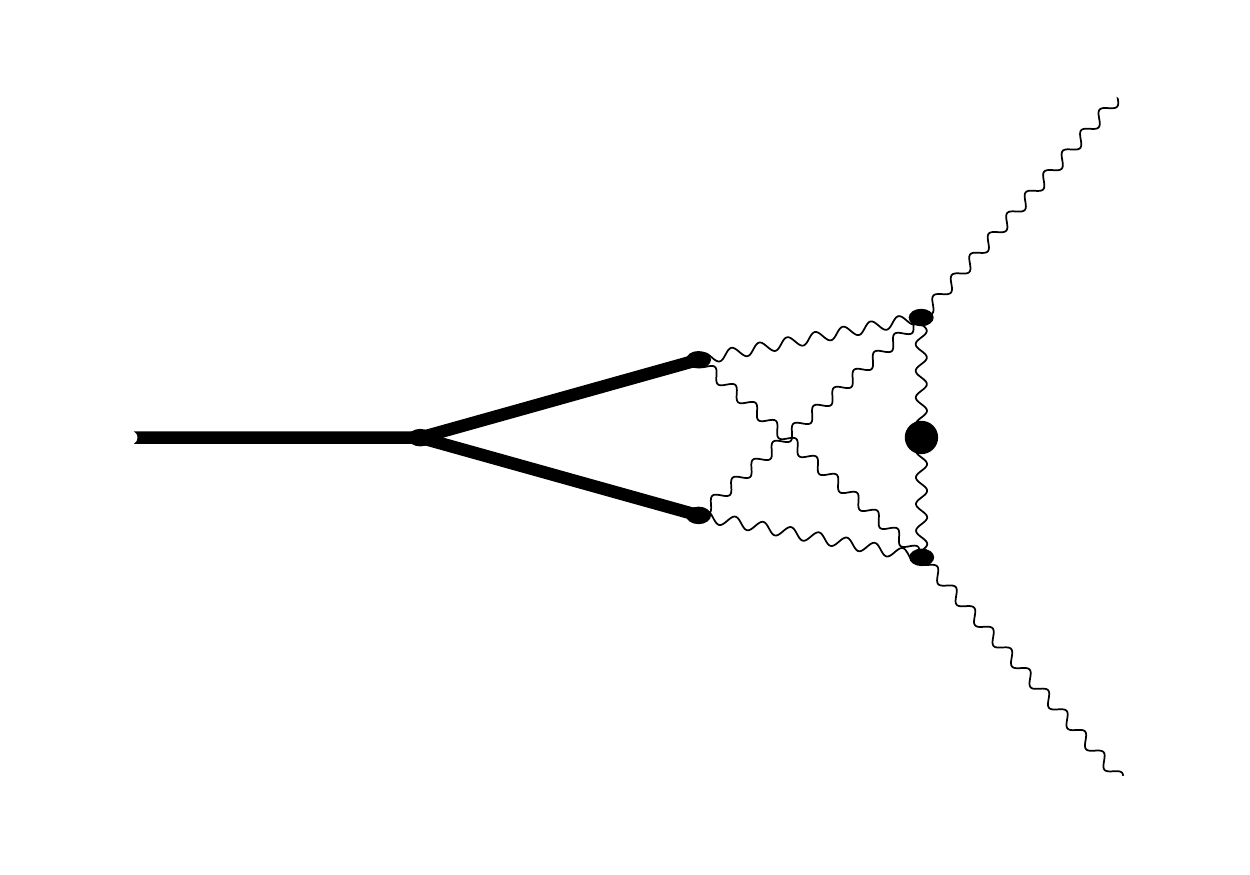}}
      \\

      \subfloat[][$I_{79}$ (NA)]
      {\includegraphics[width=0.16\textwidth]{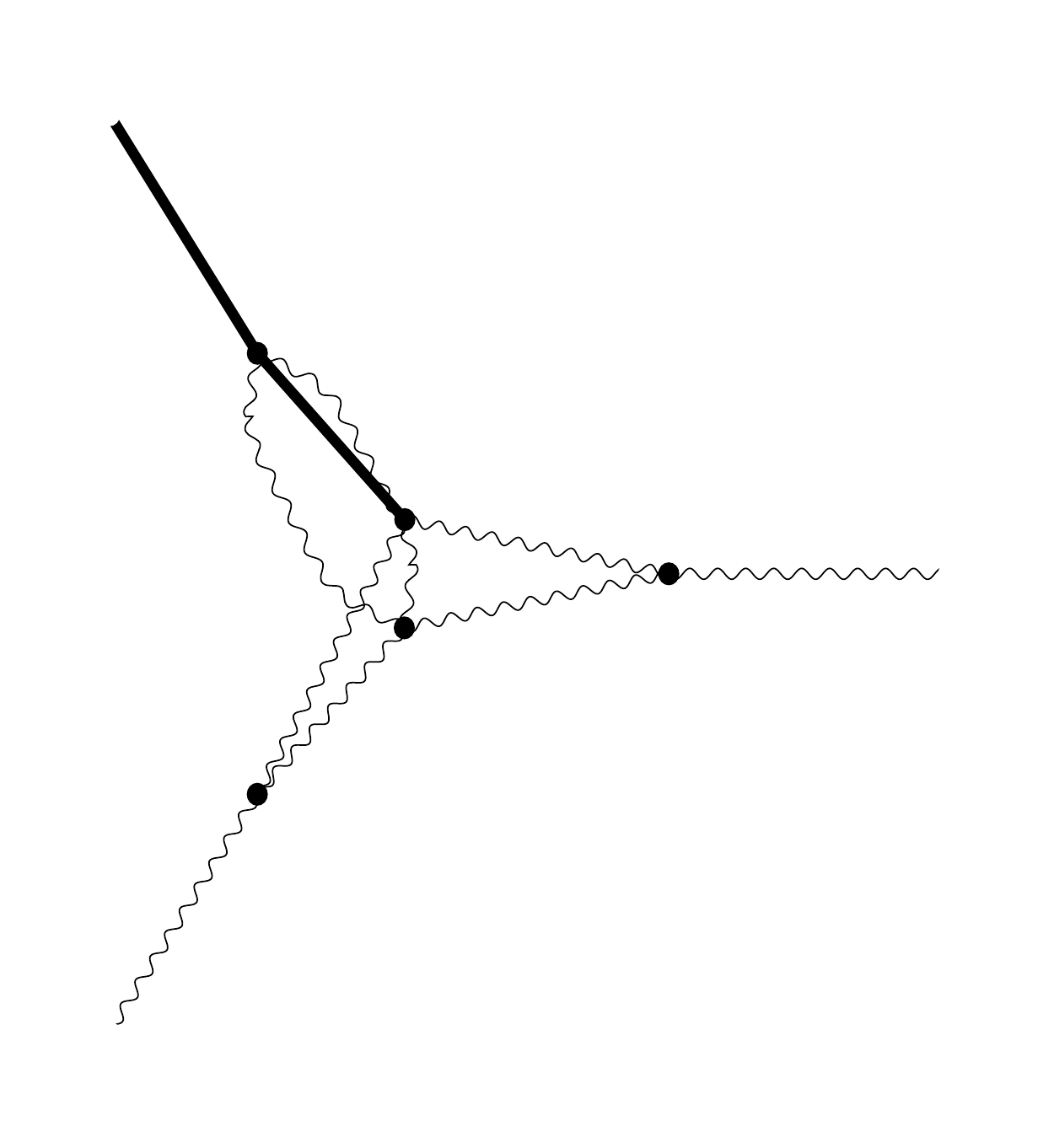}}
      \subfloat[][$I_{80}$ (NA)]
      {\includegraphics[width=0.16\textwidth]{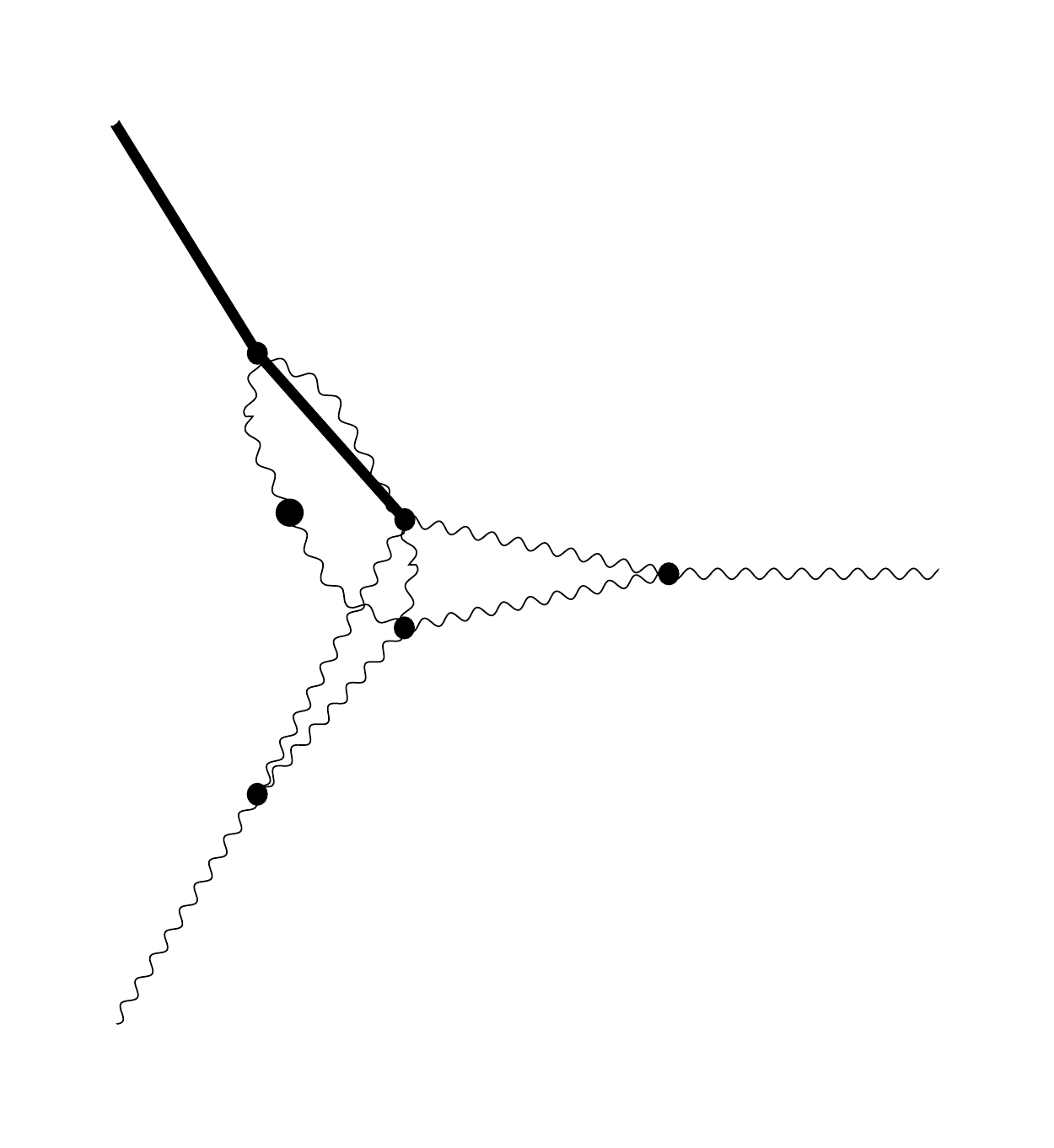}}
      \subfloat[][$I_{81}$ (NA)]
      {\includegraphics[width=0.16\textwidth]{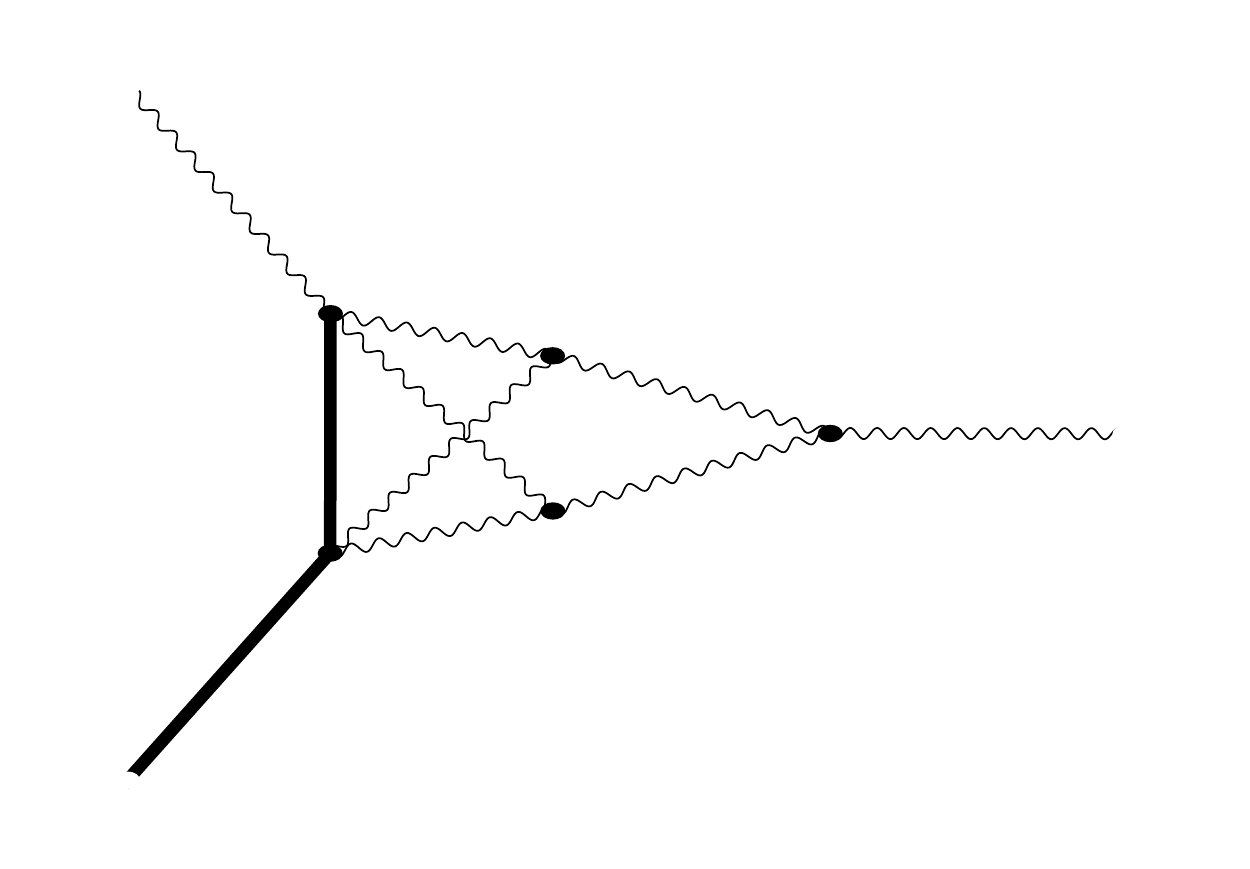}}
      \subfloat[][$I_{82}^*$ (NB)]
      {\includegraphics[width=0.16\textwidth]{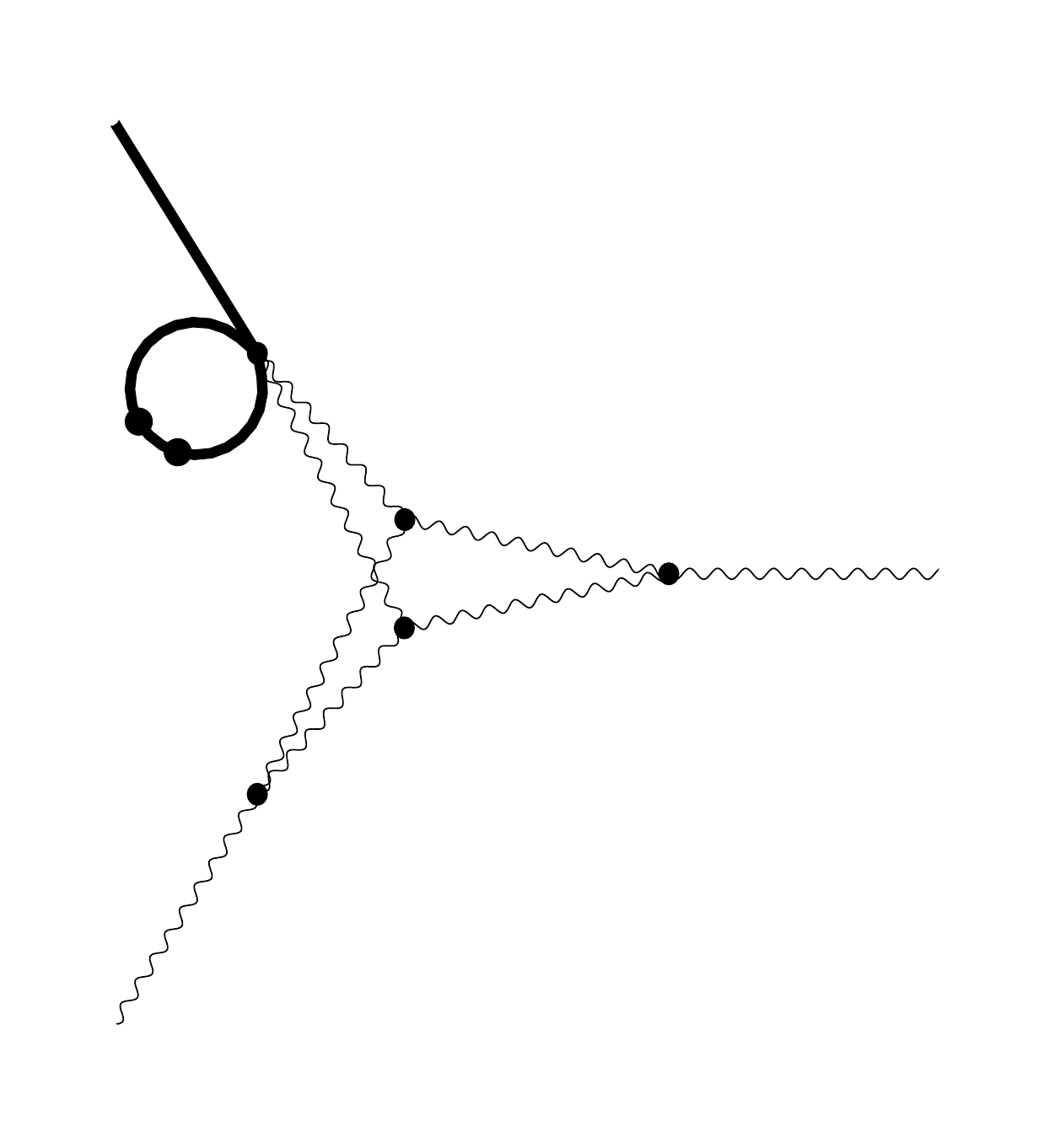}}
      \subfloat[][$I_{83}$ (NA)]
      {\includegraphics[width=0.16\textwidth]{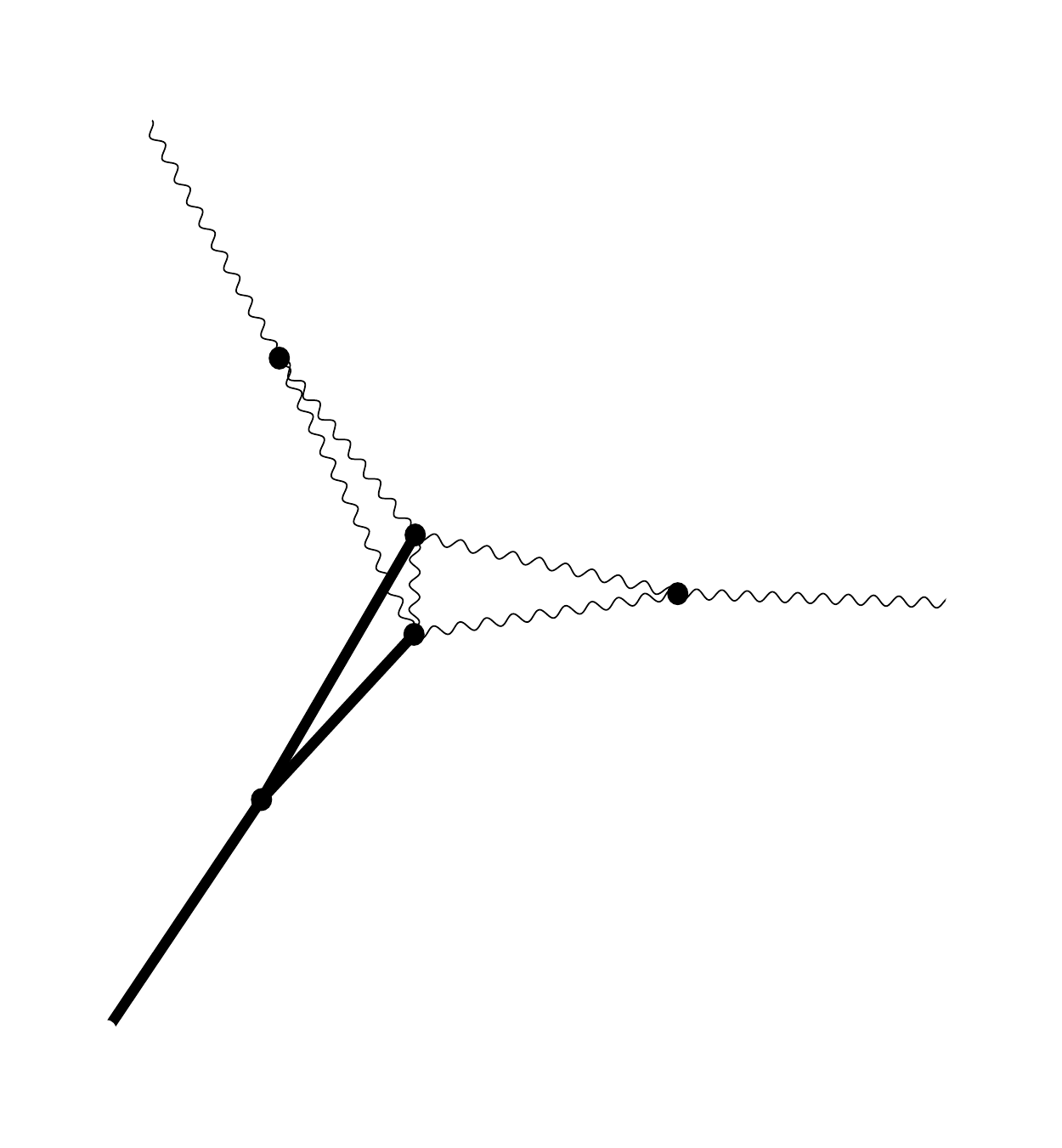}}
      \subfloat[][$I_{84}$ (NA)]
      {\includegraphics[width=0.16\textwidth]{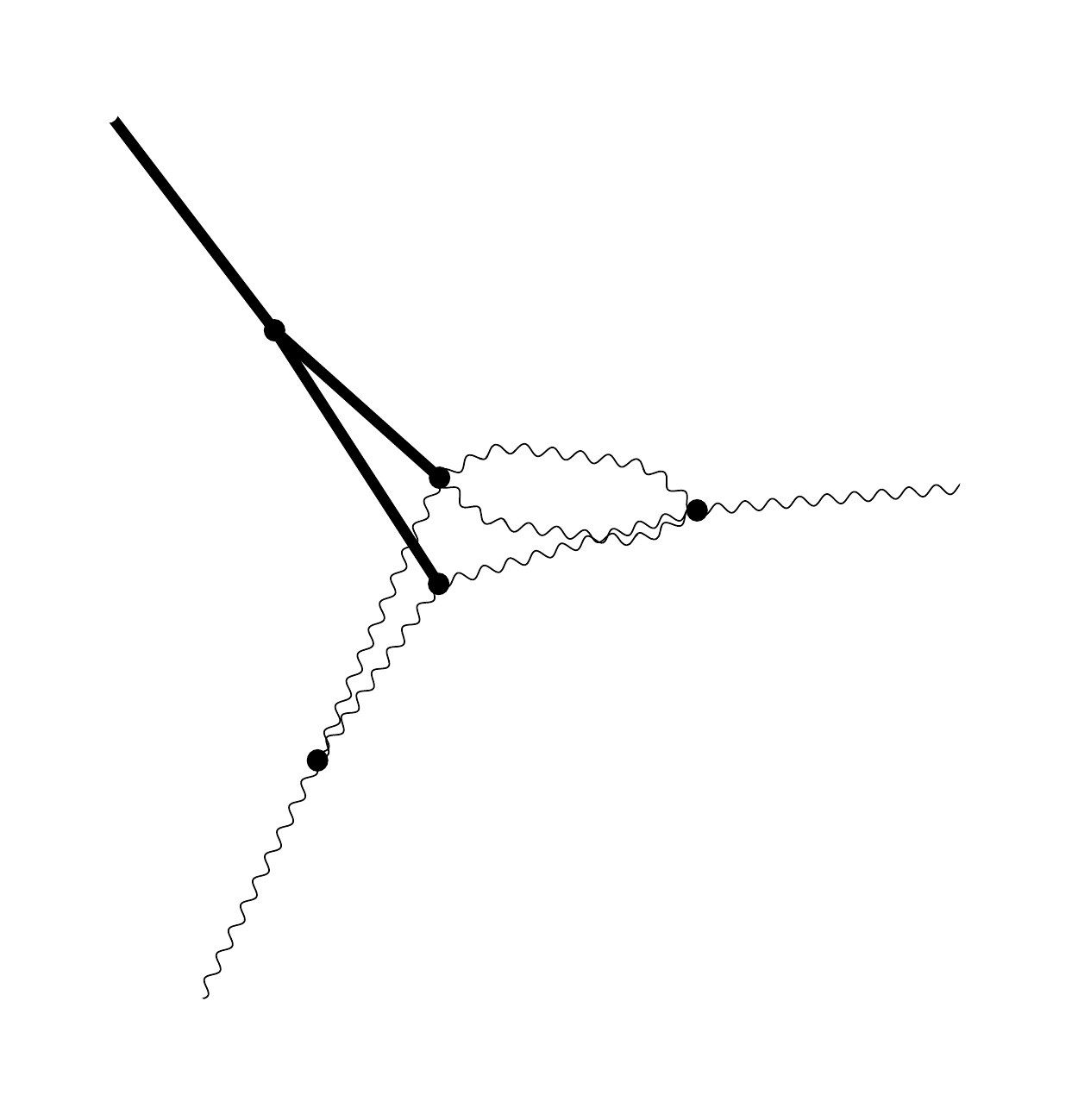}}
      \\

      \subfloat[][$I_{85}$ (NA)]
      {\includegraphics[width=0.16\textwidth]{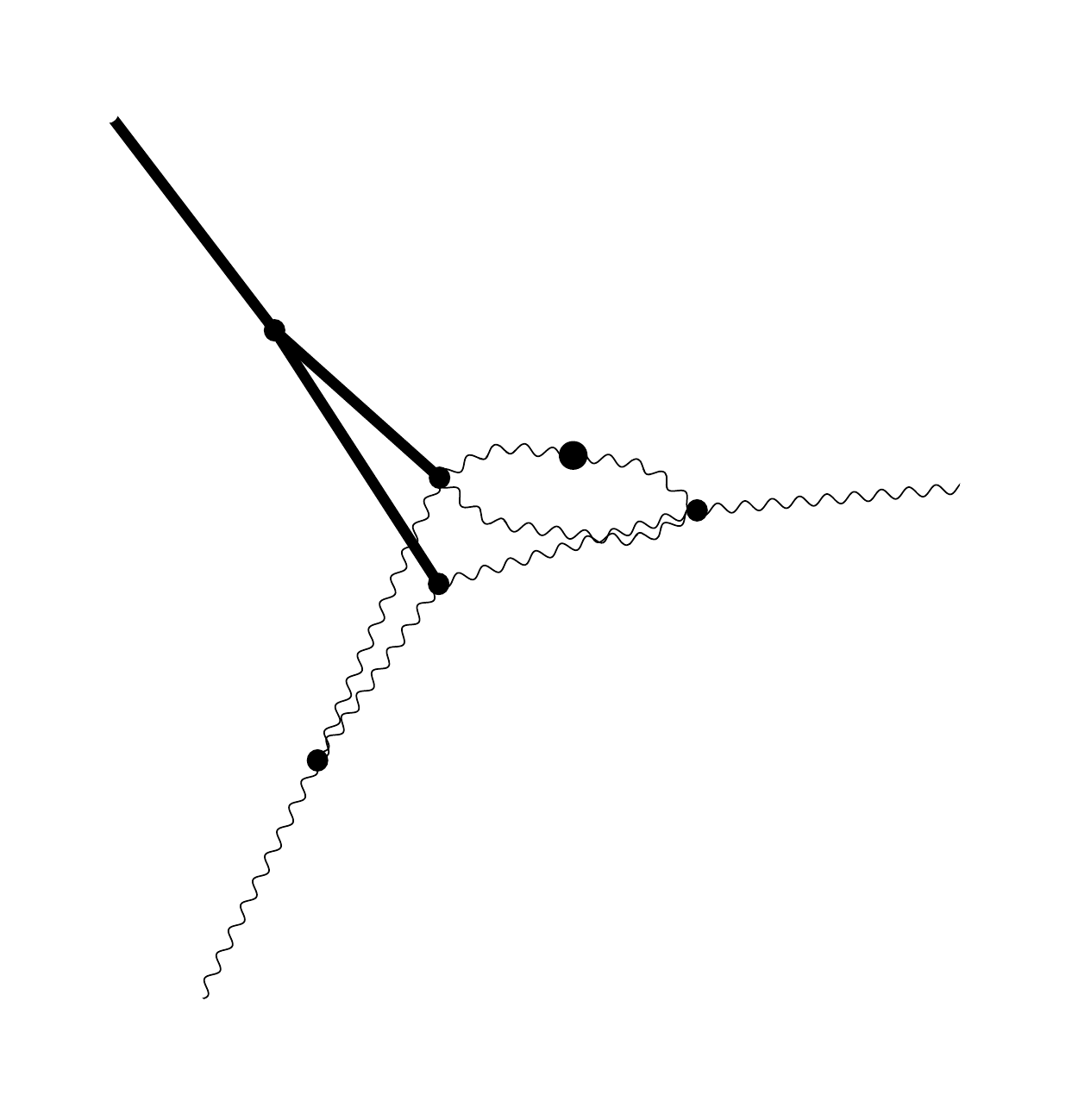}}
      \subfloat[][$I_{86}$ (NA)]
      {\includegraphics[width=0.16\textwidth]{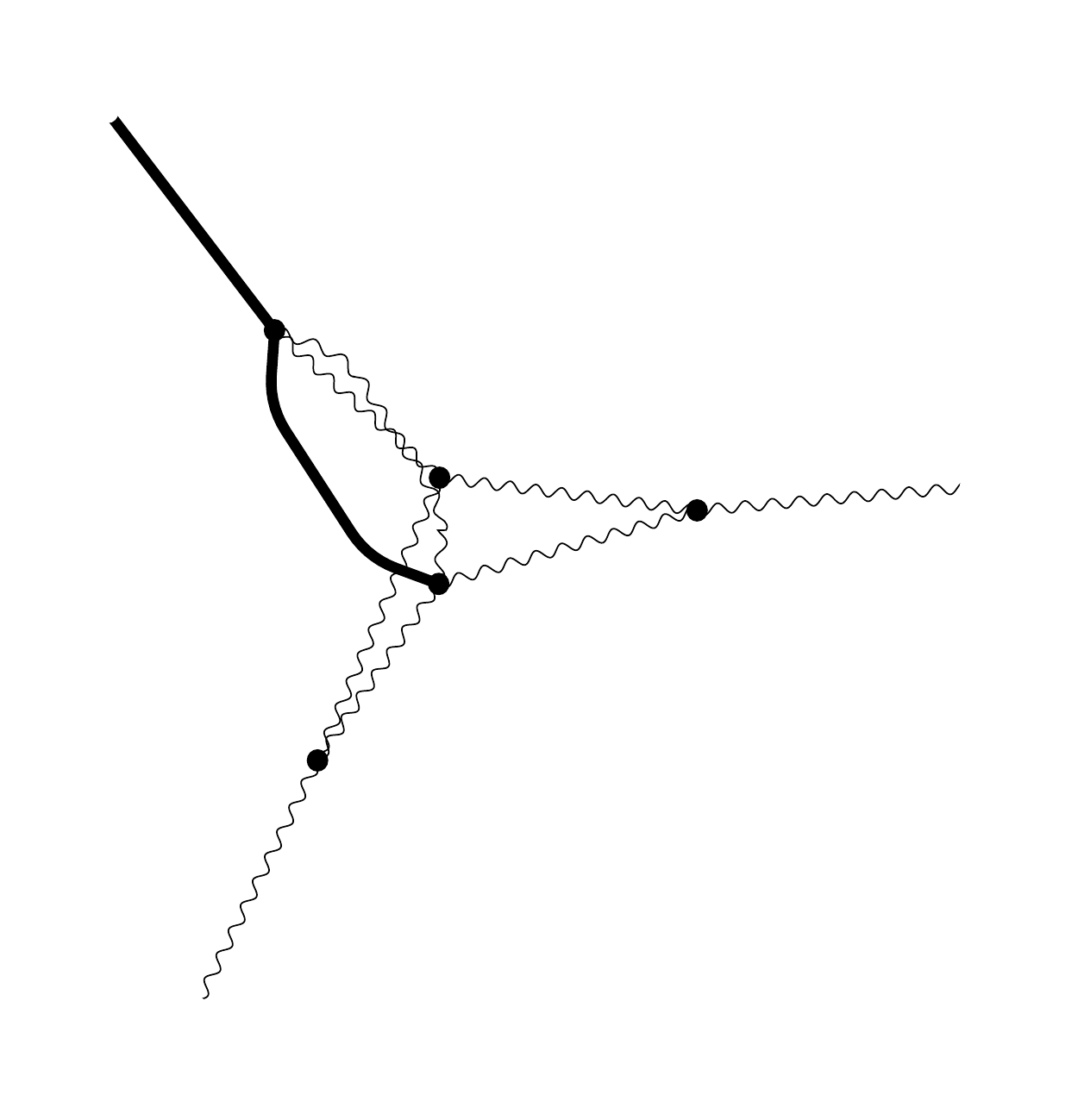}}
      \subfloat[][$I_{87}$ (NA)]
      {\includegraphics[width=0.16\textwidth]{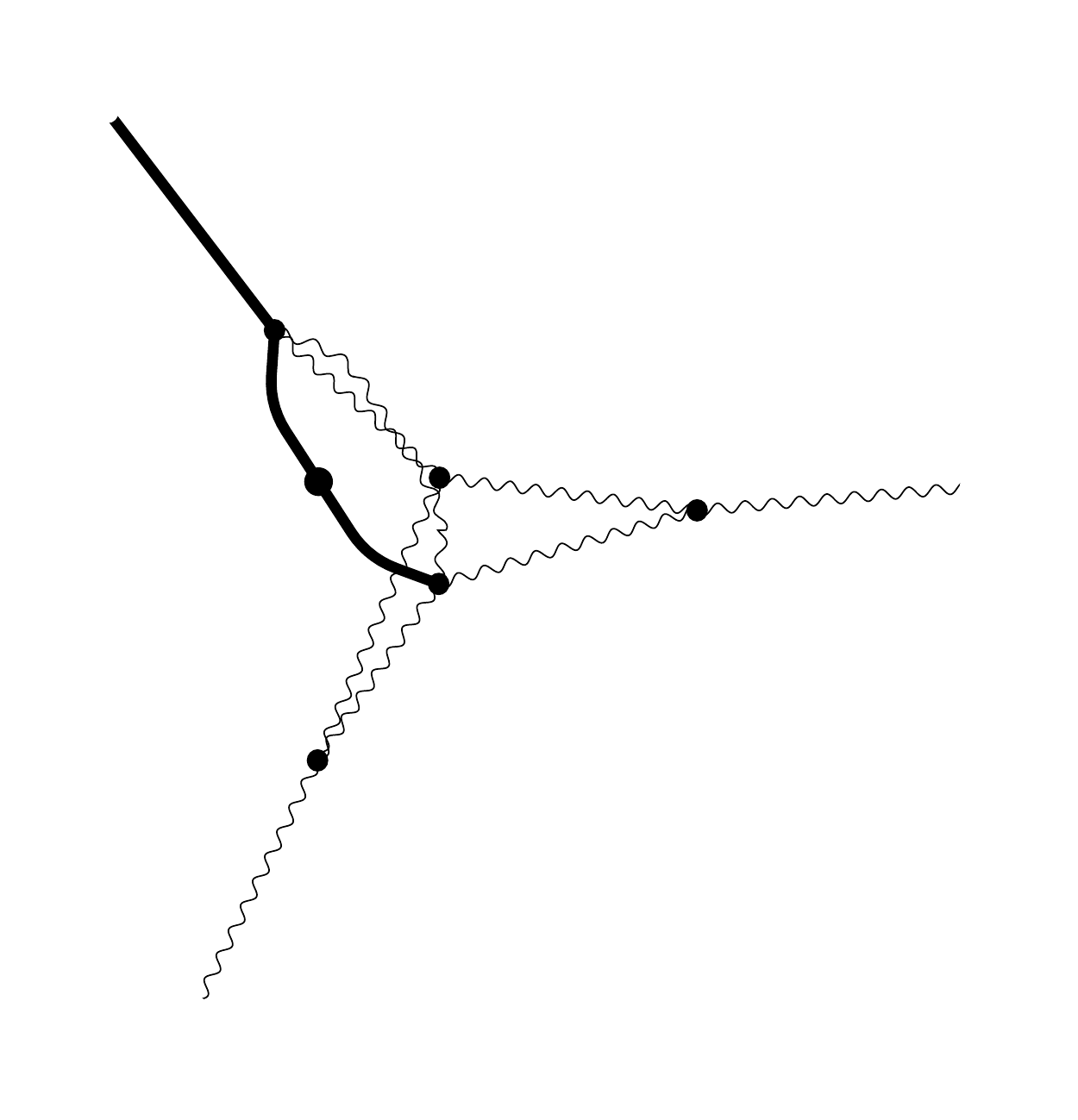}}
      \subfloat[][$I_{88}$ (NA)]
      {\includegraphics[width=0.16\textwidth]{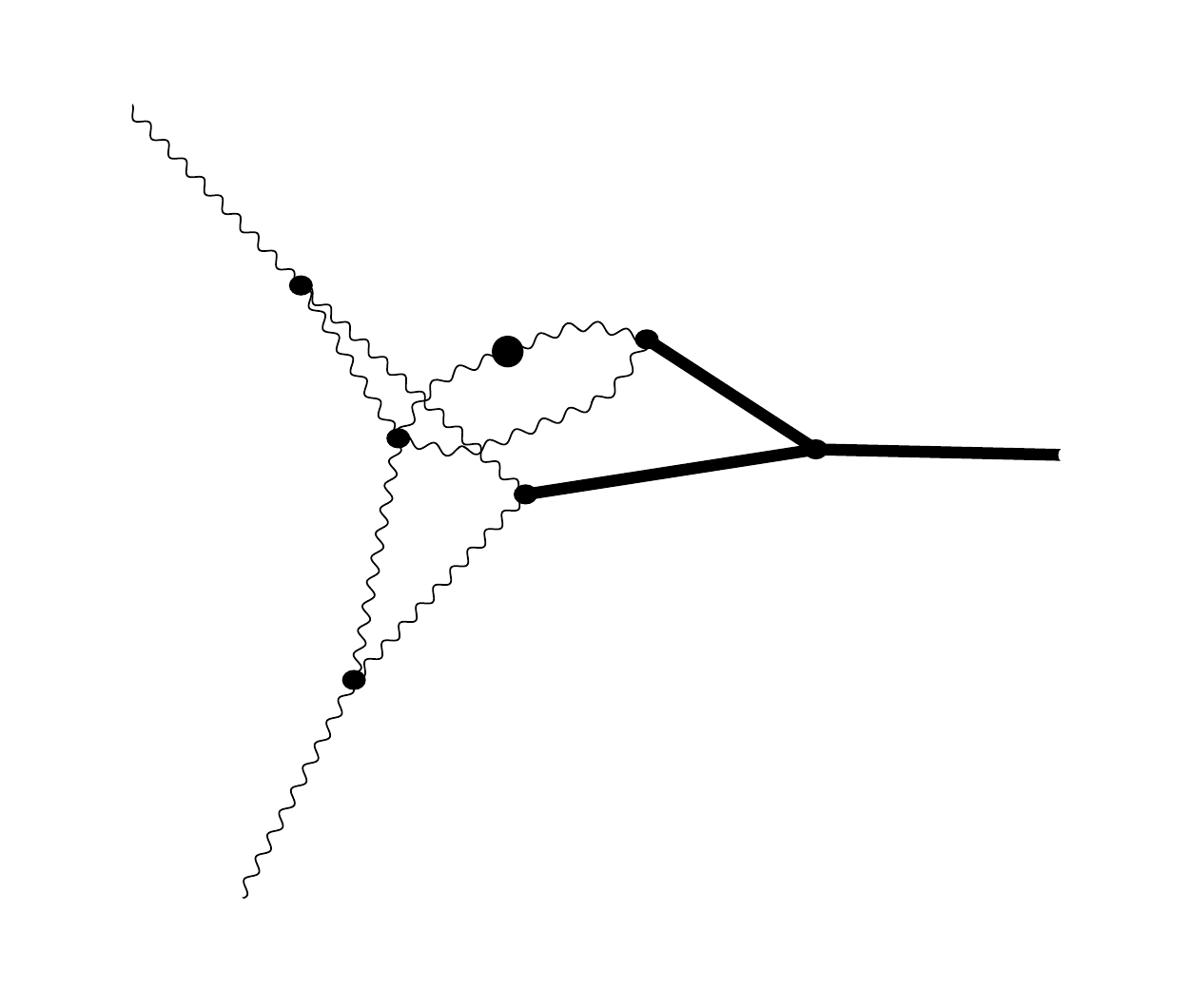}}
      \subfloat[][$I_{89}$ (NA)]
      {\includegraphics[width=0.16\textwidth]{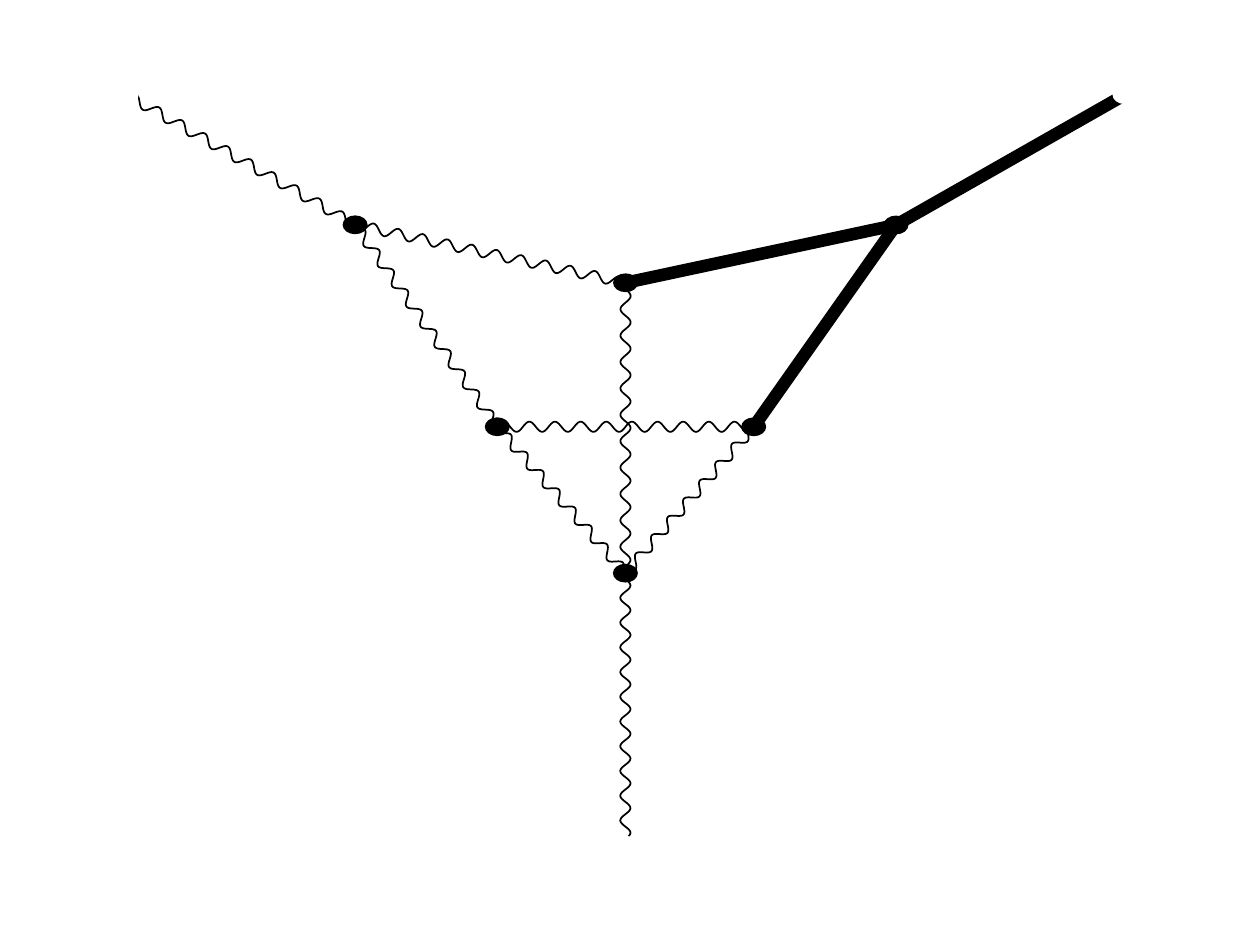}}
      \subfloat[][$I_{90}$ (NB)]
      {\includegraphics[width=0.16\textwidth]{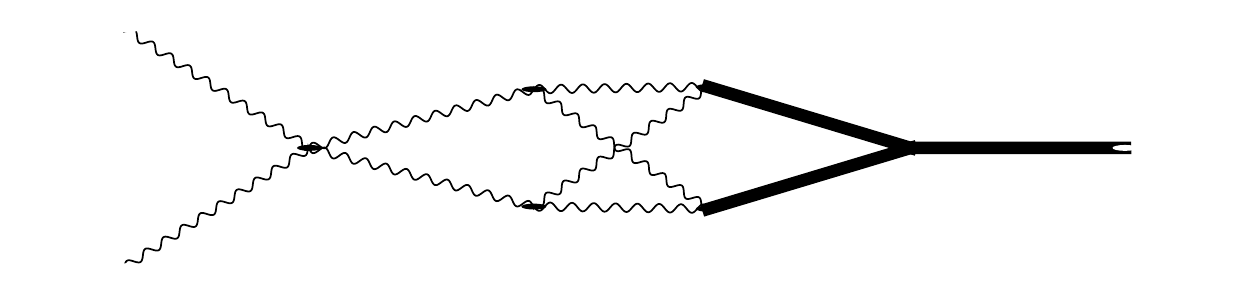}}
      \\

      \subfloat[][$I_{91}^*$ (NB)]
      {\includegraphics[width=0.16\textwidth]{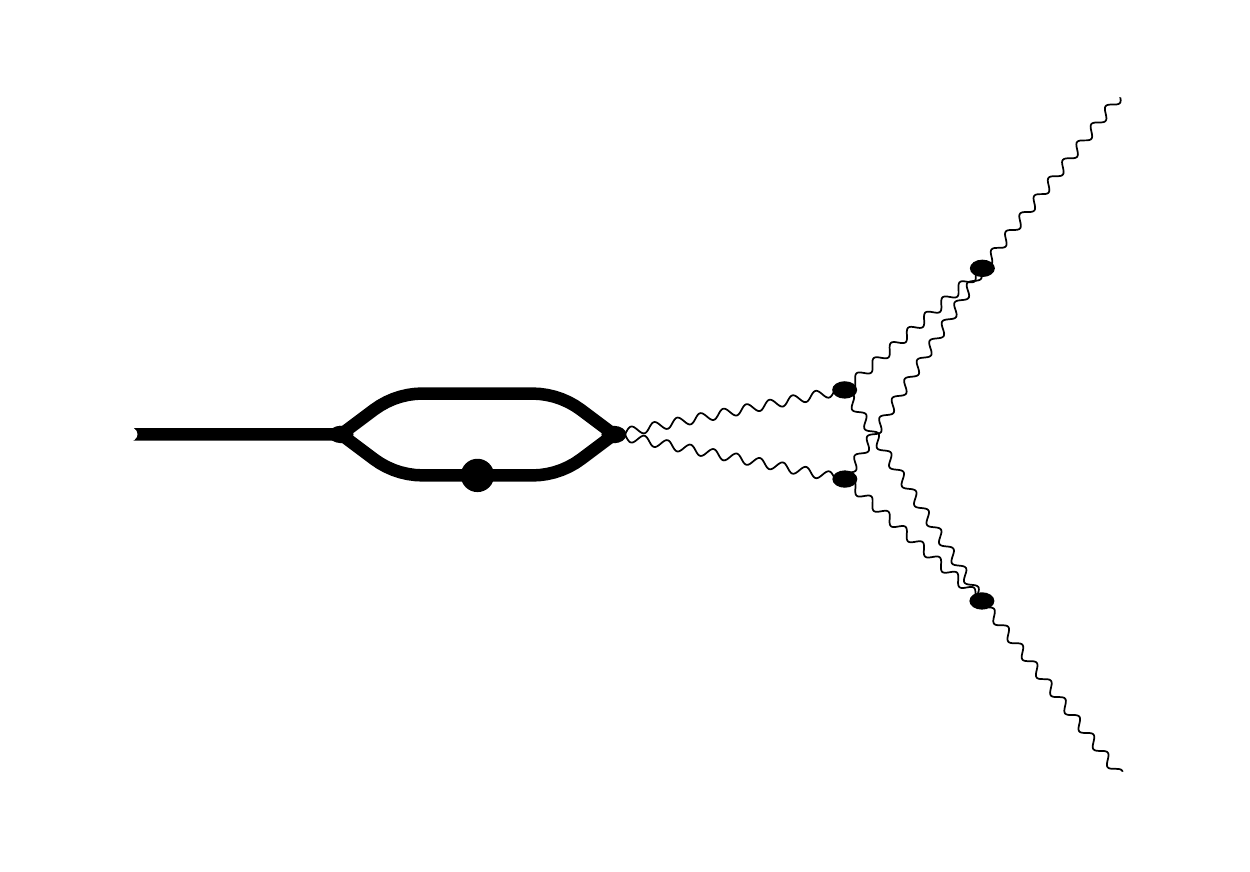}}
      \subfloat[][$I_{92}$ (NA)]
      {\includegraphics[width=0.16\textwidth]{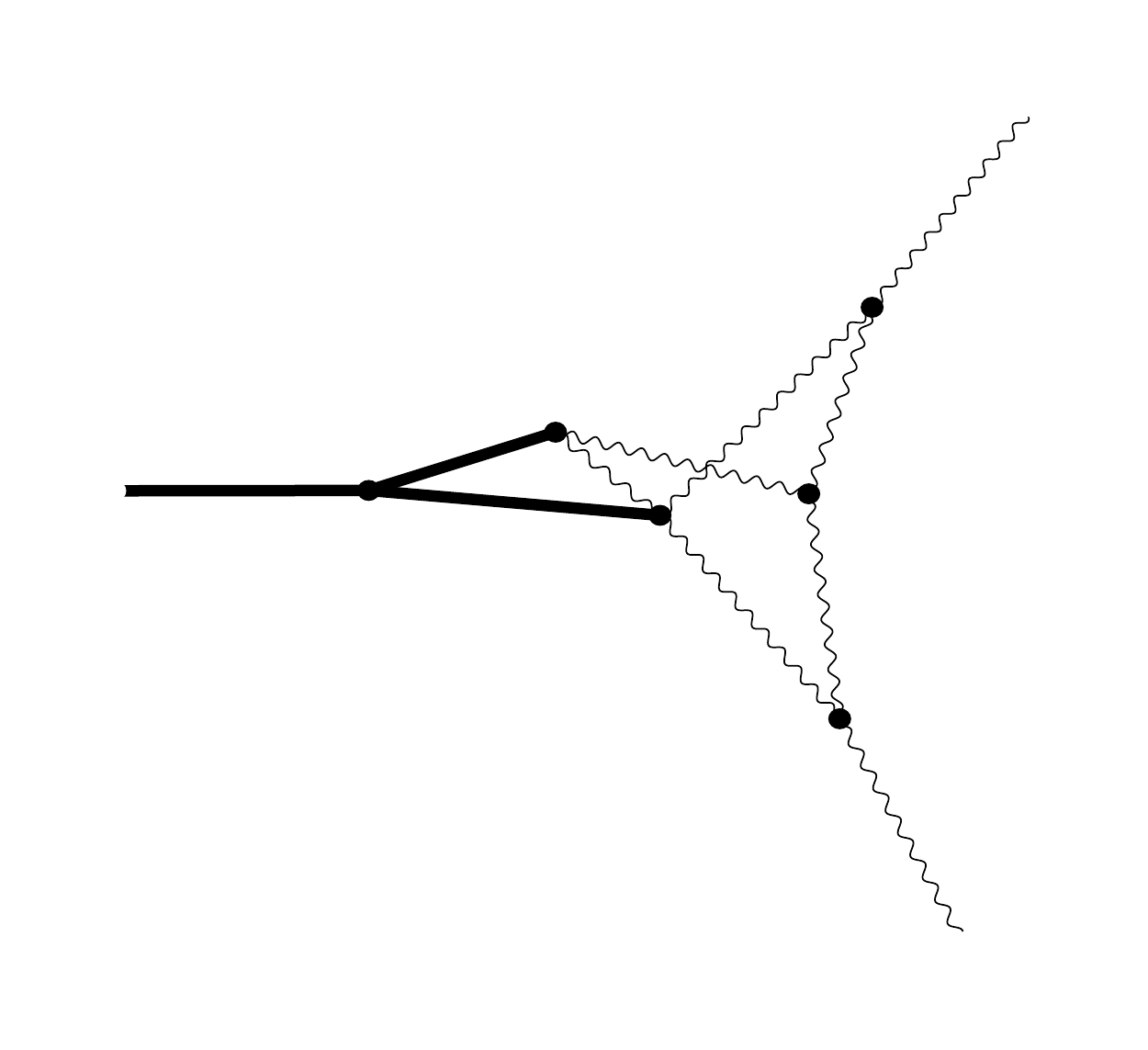}}
      \subfloat[][$I_{93}$ (NA)]
      {\includegraphics[width=0.16\textwidth]{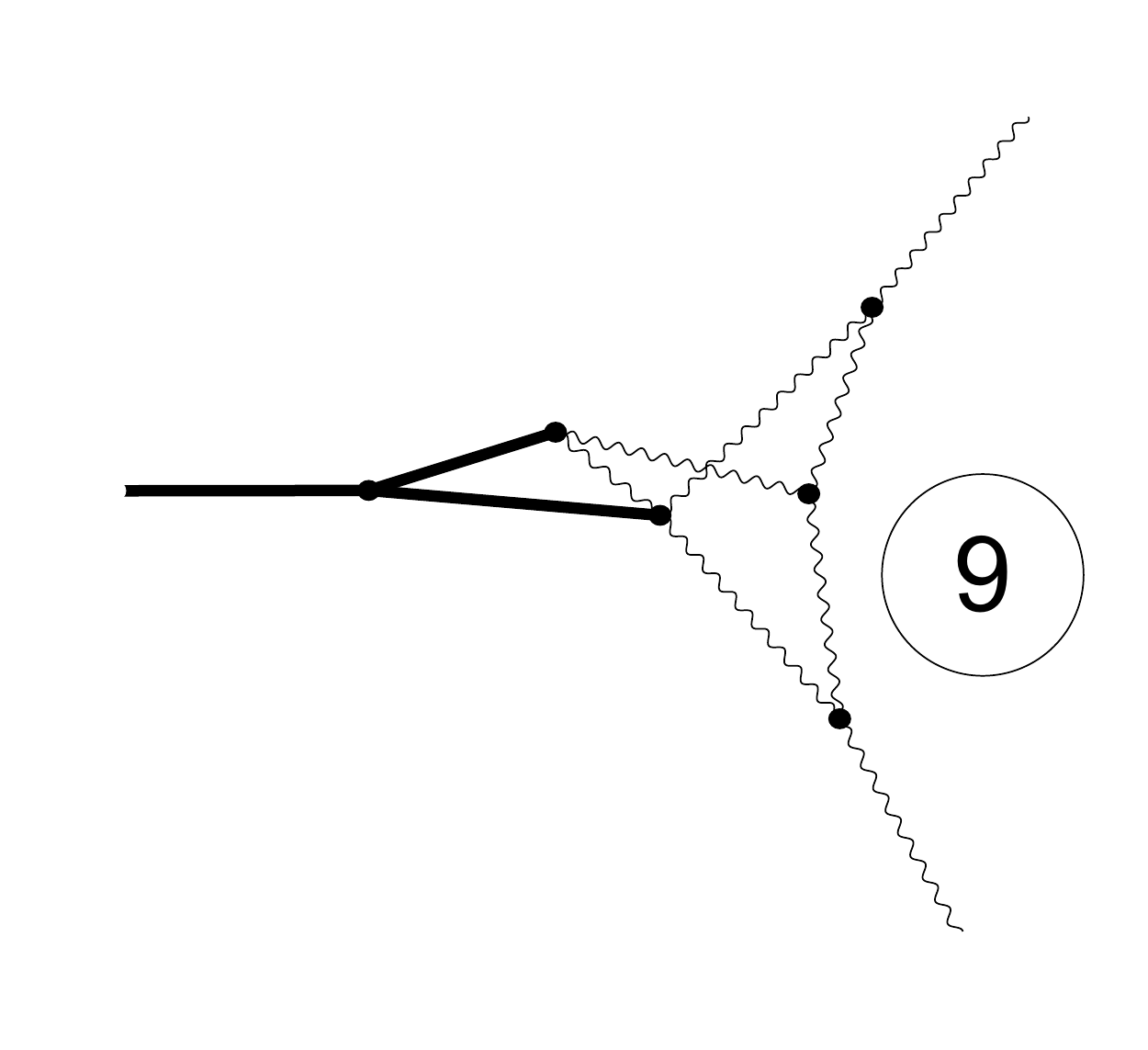}}
      \subfloat[][$I_{94}$ (NA)]
      {\includegraphics[width=0.16\textwidth]{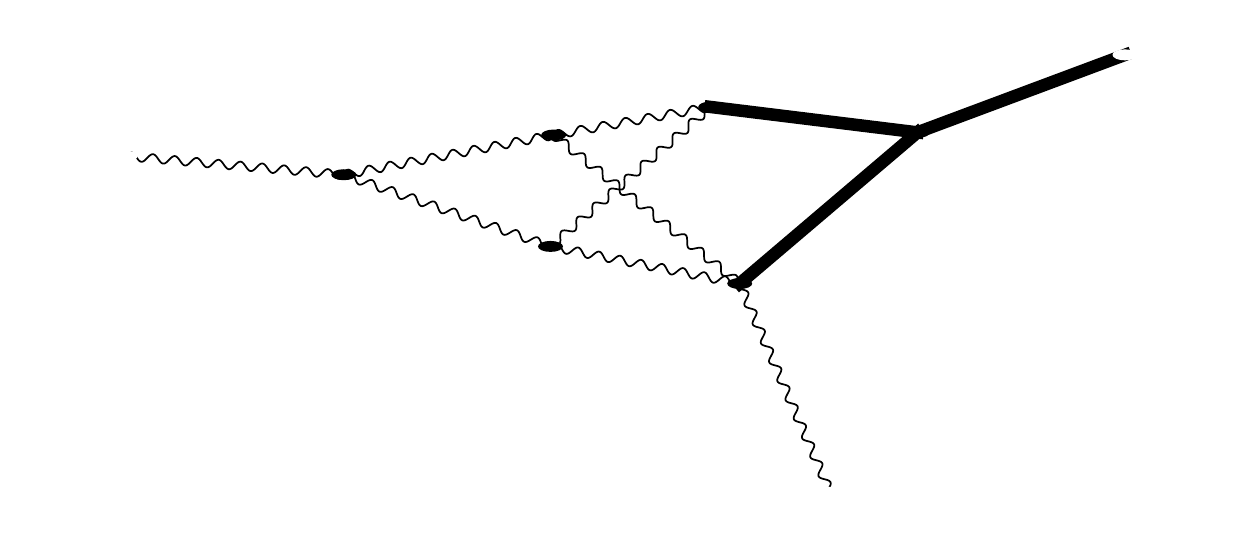}}
      \subfloat[][$I_{95}$ (NA)]
      {\includegraphics[width=0.16\textwidth]{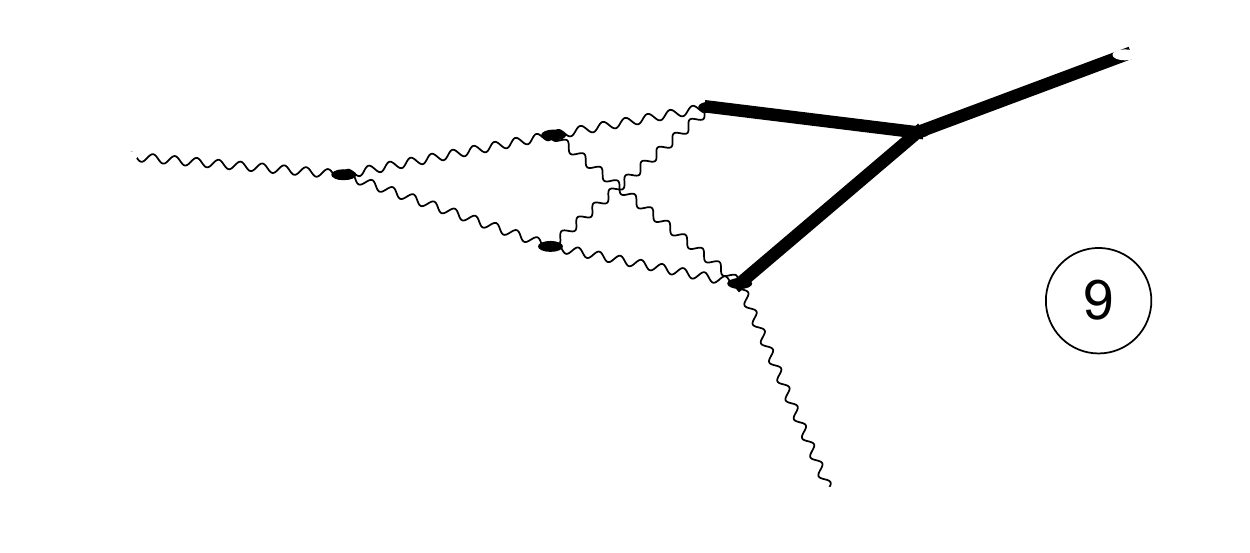}}
      \subfloat[][$I_{96}^*$ (NB)]
      {\includegraphics[width=0.16\textwidth]{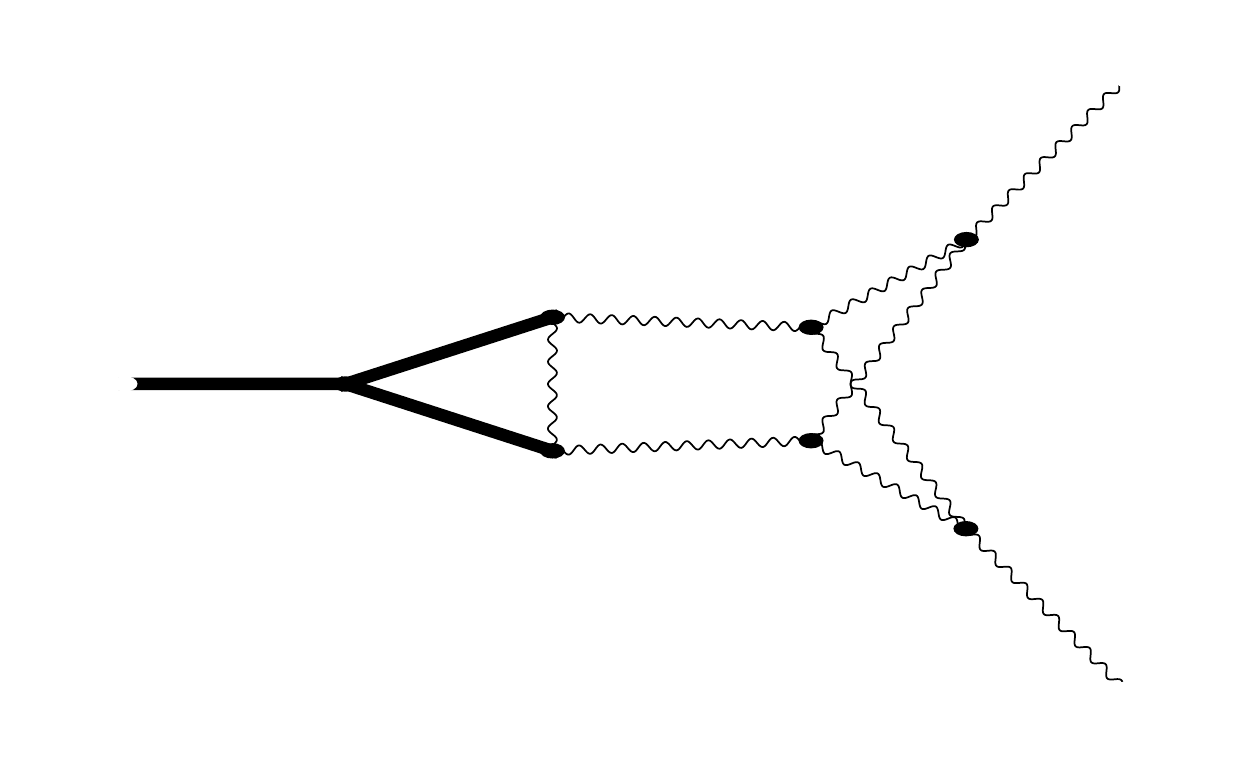}}
      \\

      \subfloat[][$I_{97}$ (PP)]
      {\includegraphics[width=0.16\textwidth]{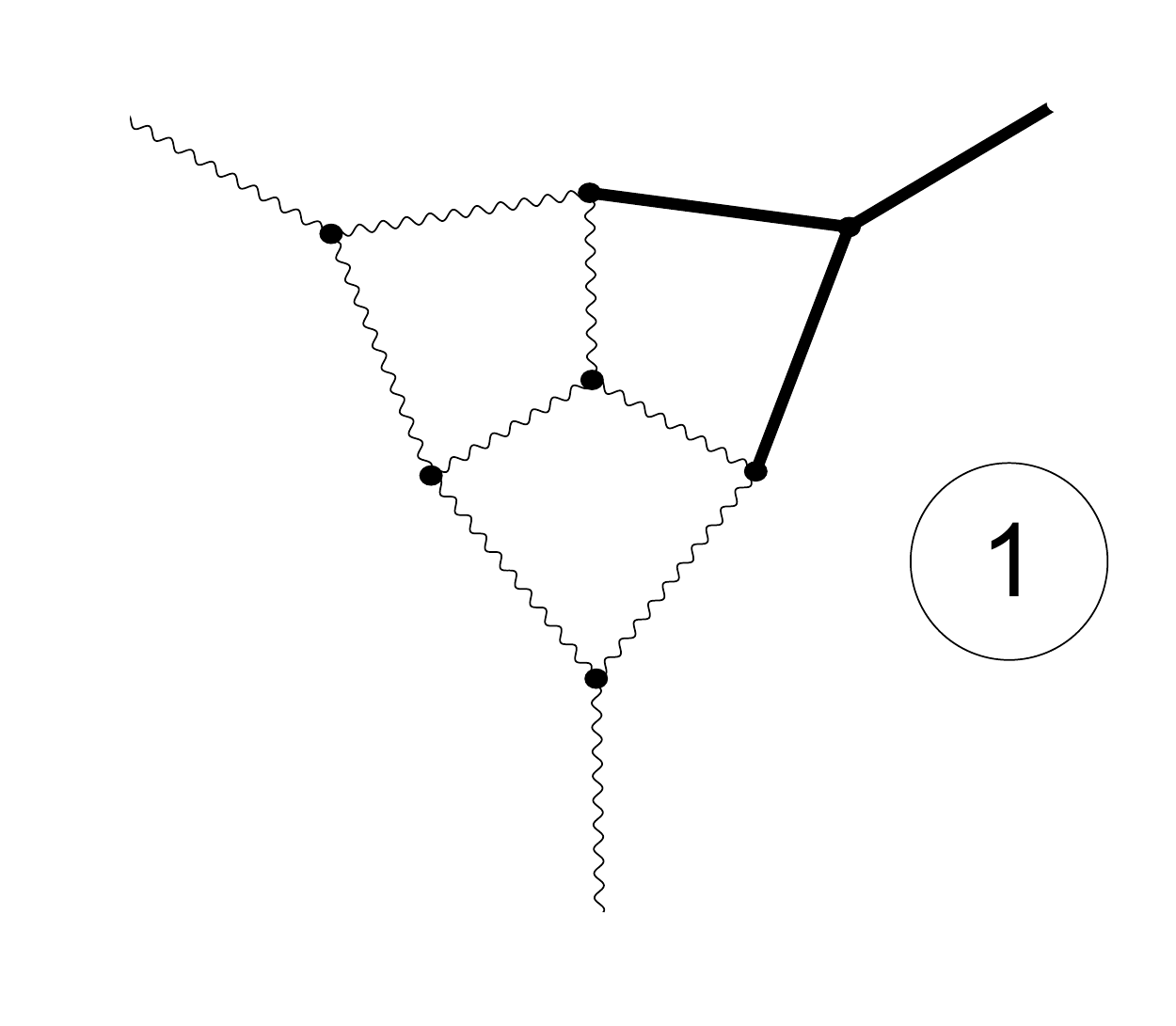}}
      \subfloat[][$I_{98}$ (PP)]
      {\includegraphics[width=0.16\textwidth]{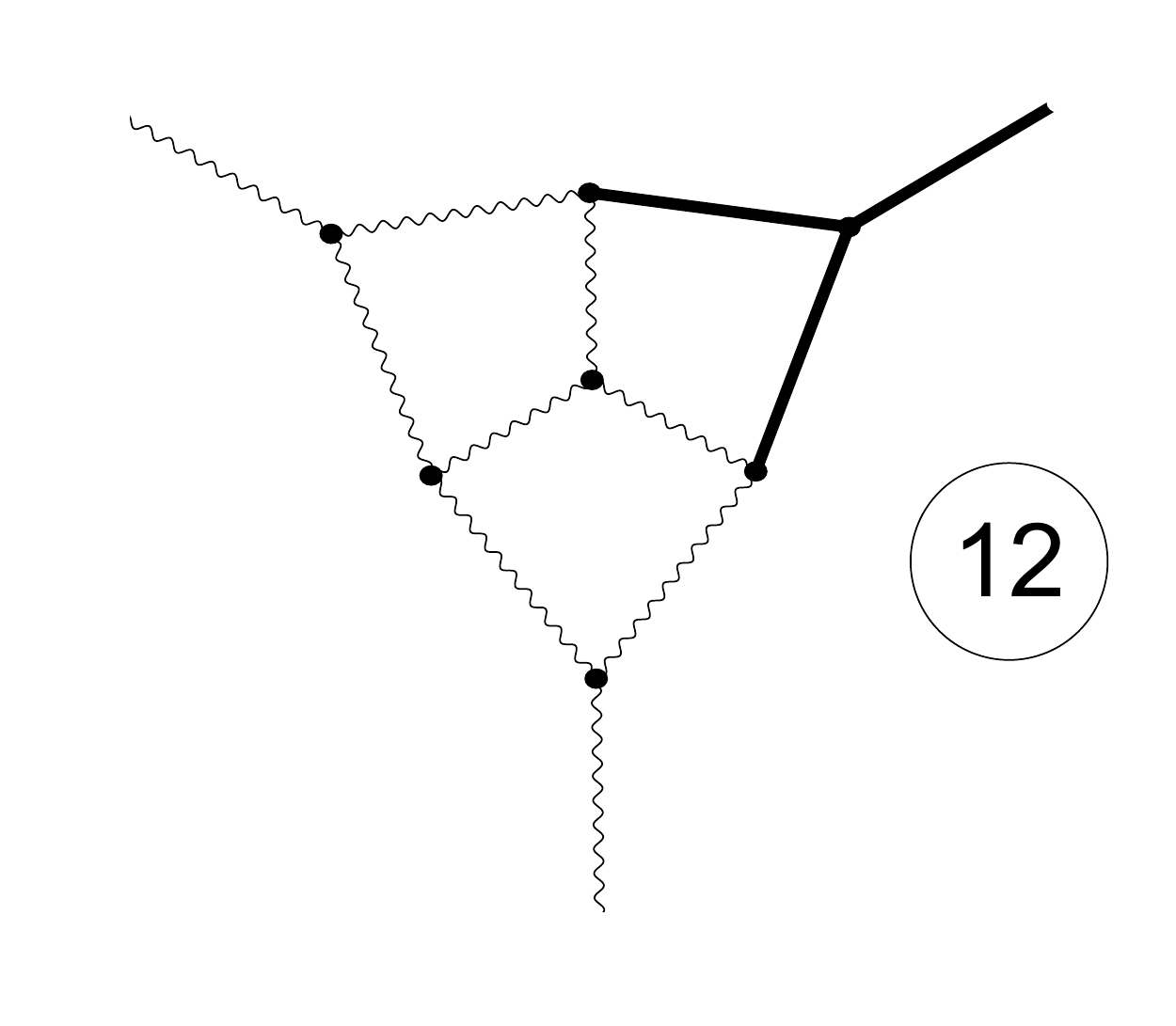}}
      \subfloat[][$I_{99}$ (PP)]
      {\includegraphics[width=0.16\textwidth]{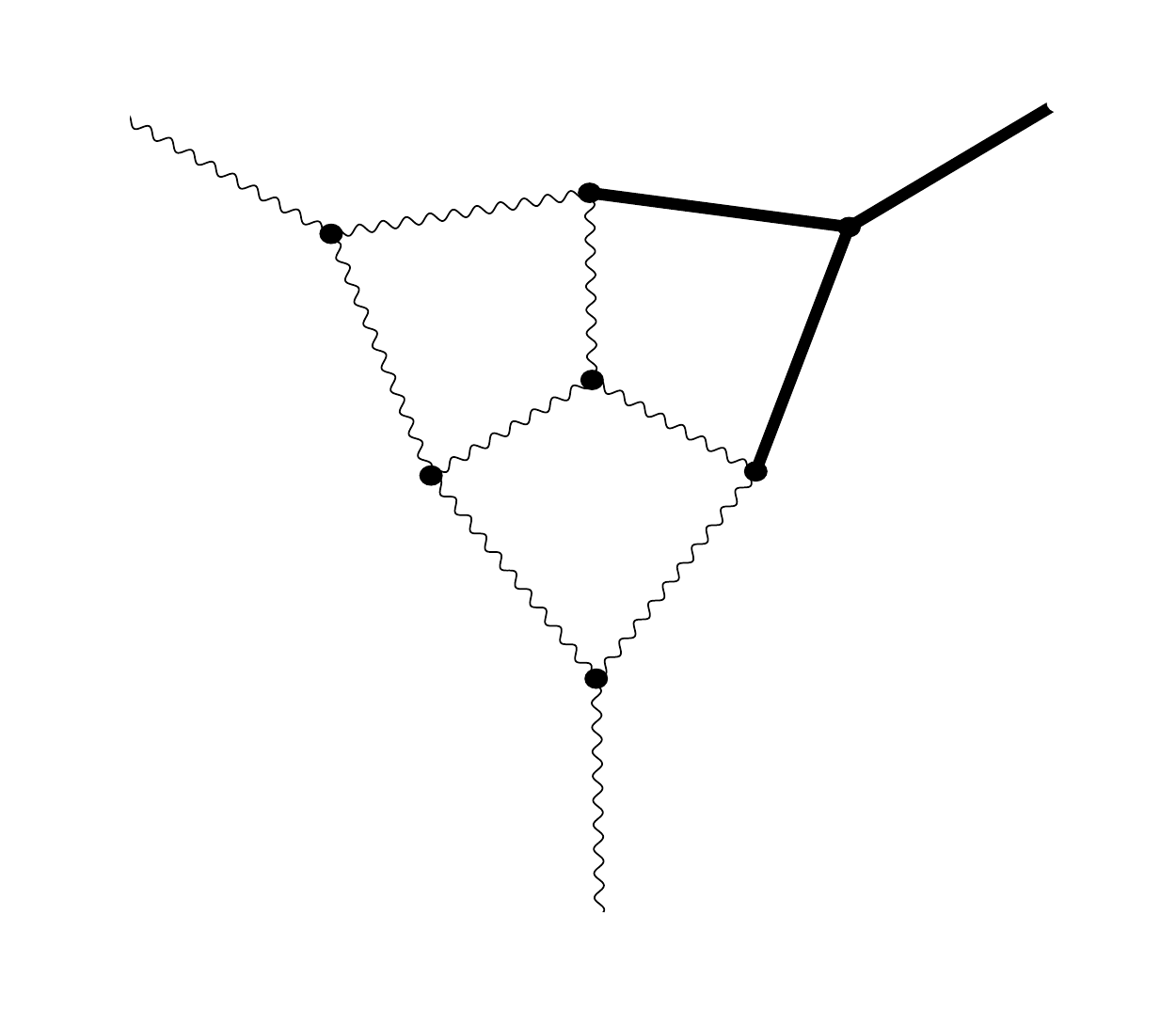}}
      \subfloat[][$I_{100}$ (NA)]
      {\includegraphics[width=0.16\textwidth]{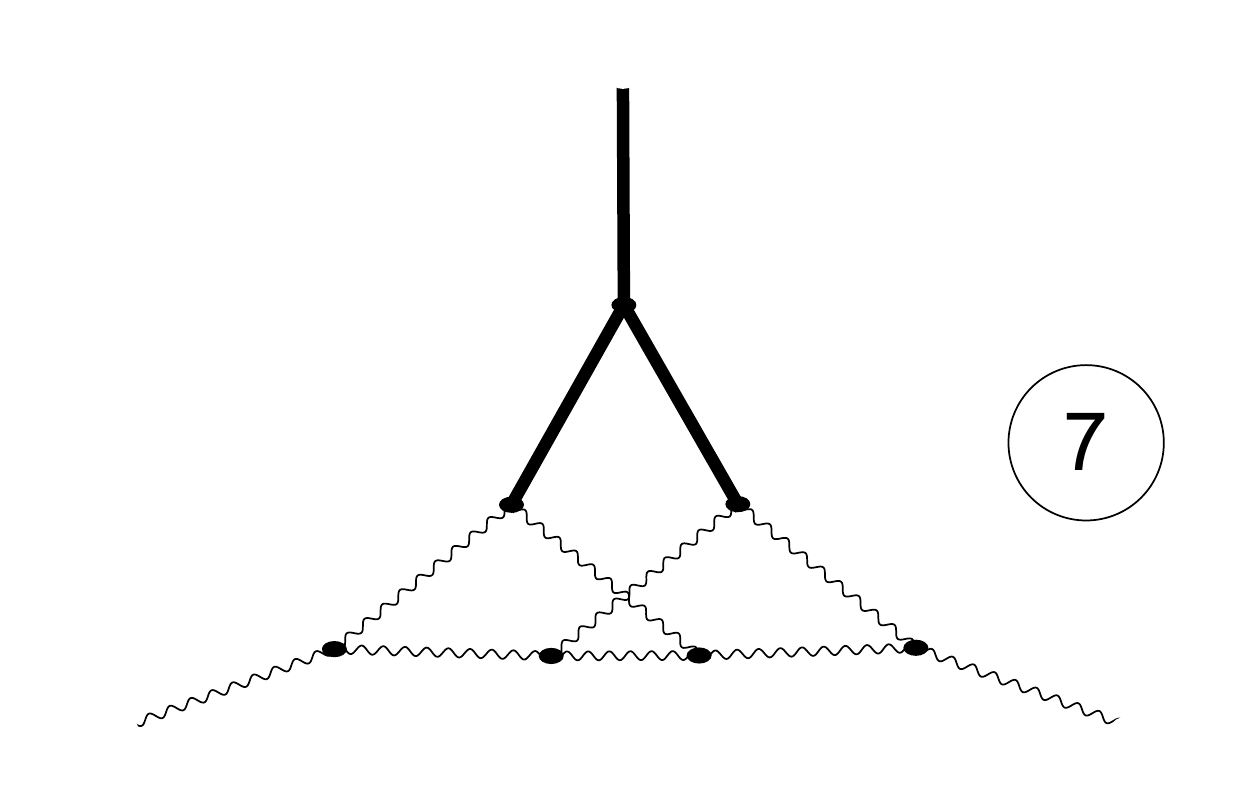}}
      \subfloat[][$I_{101}$ (NA)]
      {\includegraphics[width=0.16\textwidth]{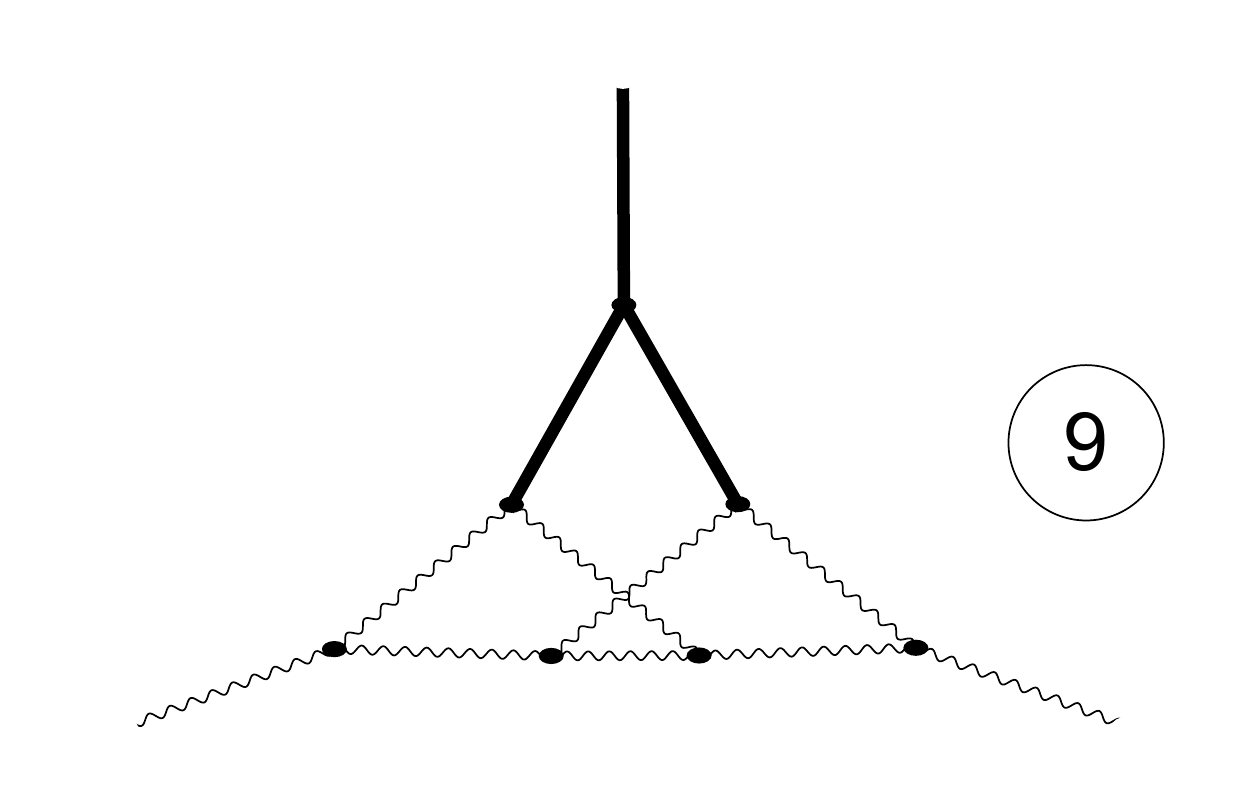}}
      \subfloat[][$I_{102}$ (NA)]
      {\includegraphics[width=0.16\textwidth]{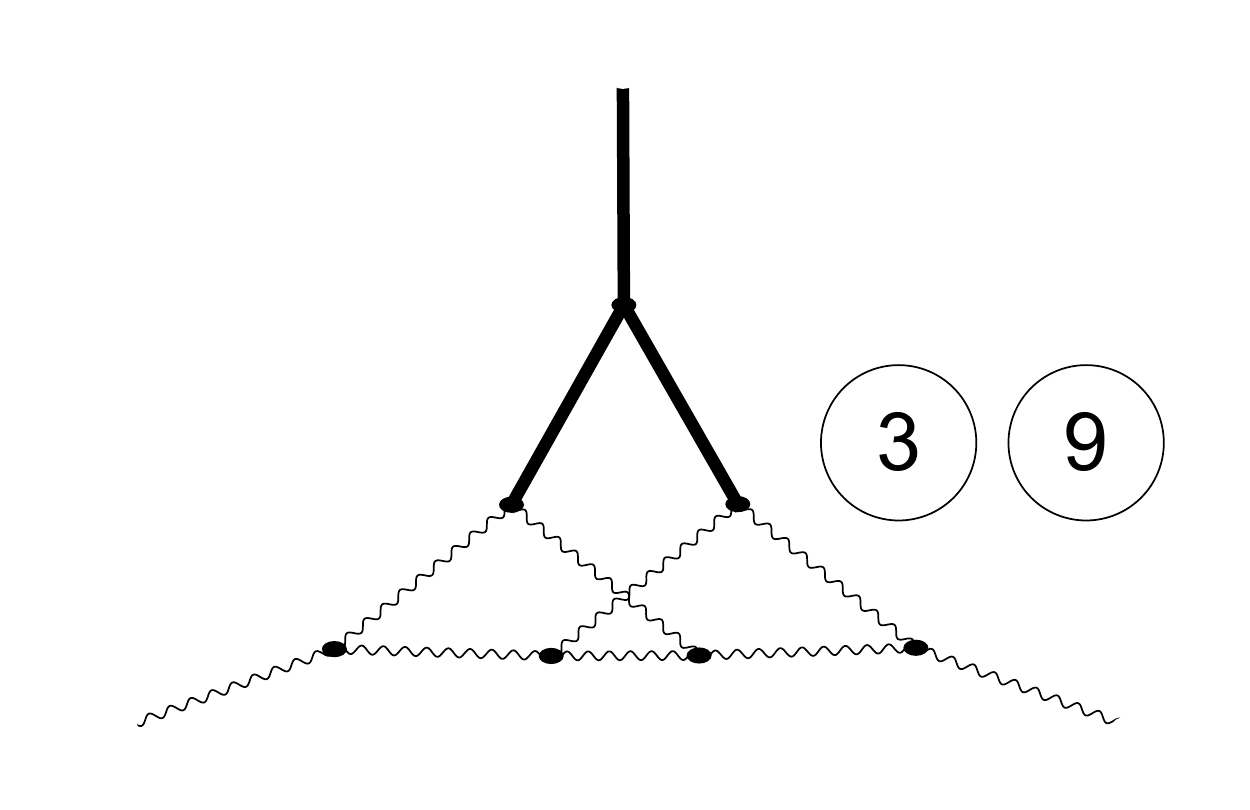}}
      \\

      \subfloat[][$I_{103}$ (NA)]
      {\includegraphics[width=0.16\textwidth]{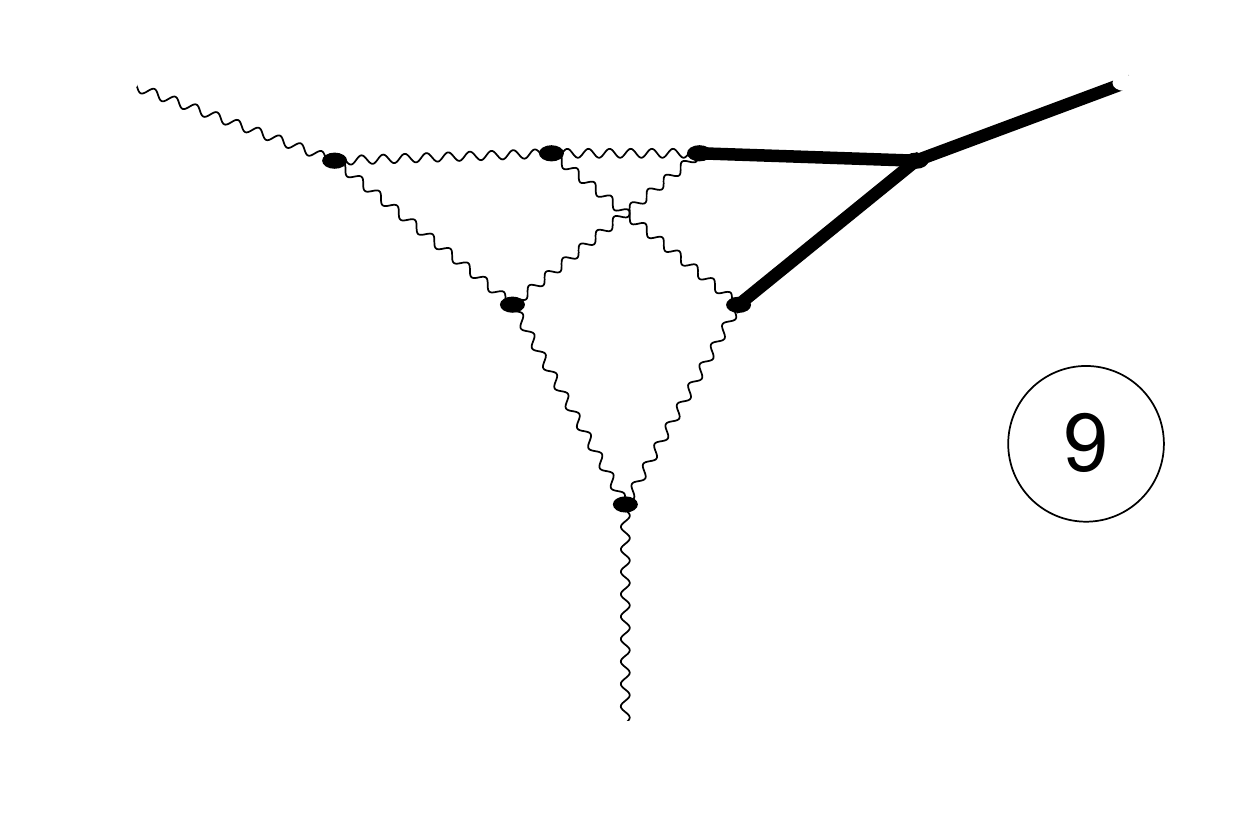}}
      \subfloat[][$I_{104}$ (NA)]
      {\includegraphics[width=0.16\textwidth]{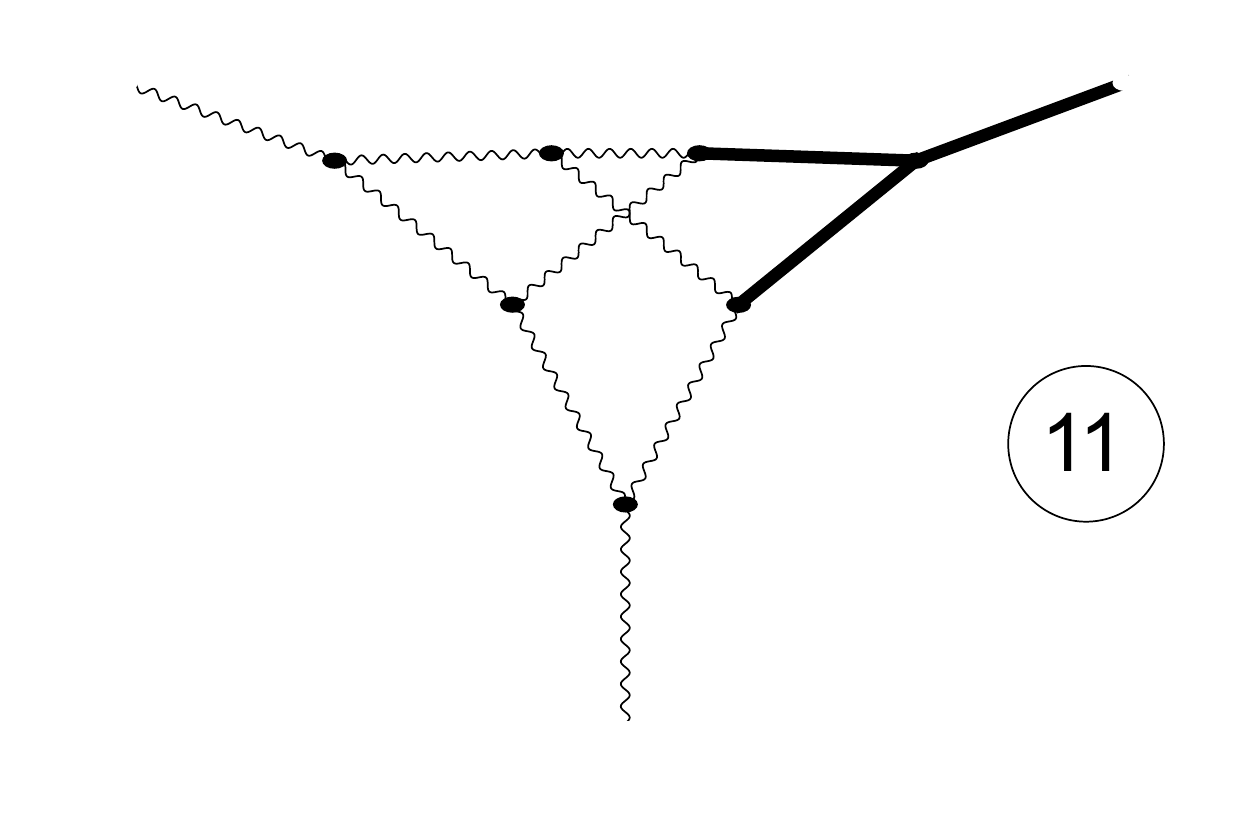}}
      \subfloat[][$I_{105}$ (NA)]
      {\includegraphics[width=0.16\textwidth]{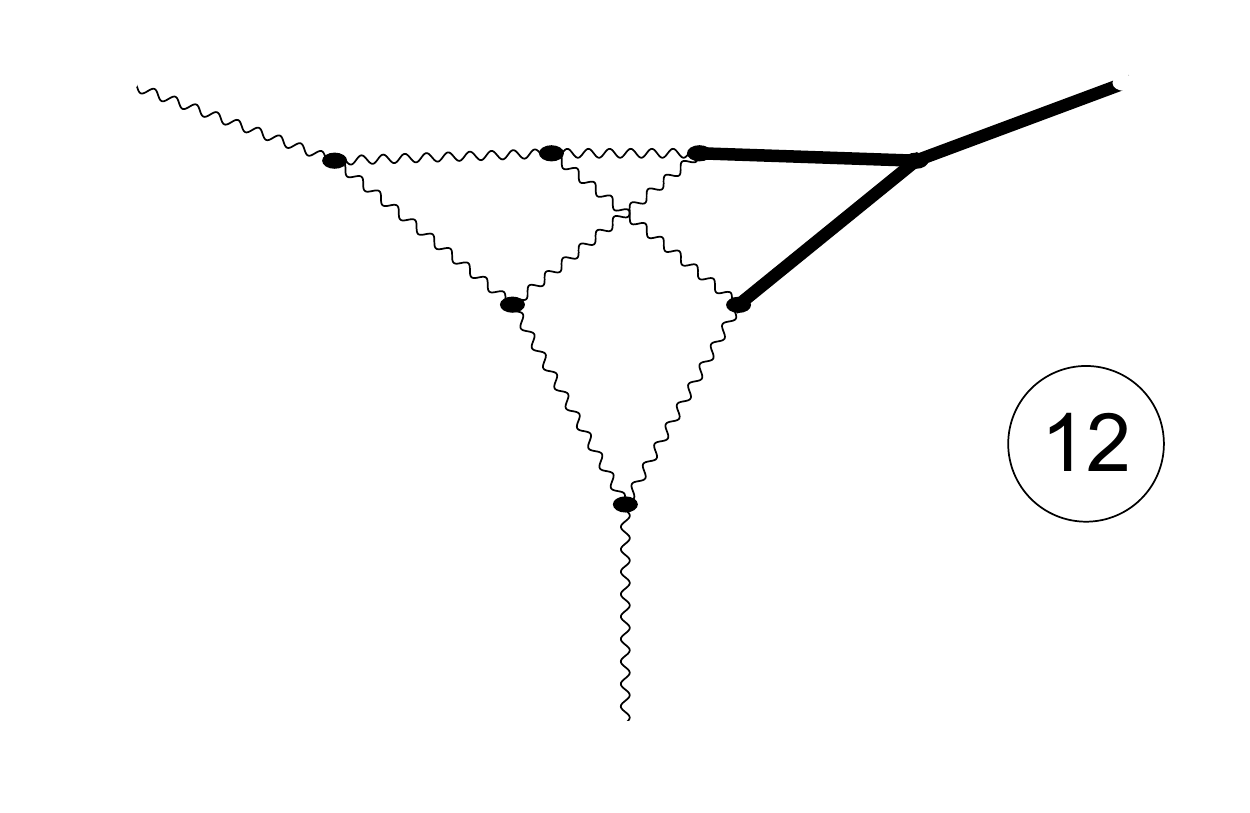}}
      \subfloat[][$I_{106}$ (NA)]
      {\includegraphics[width=0.16\textwidth]{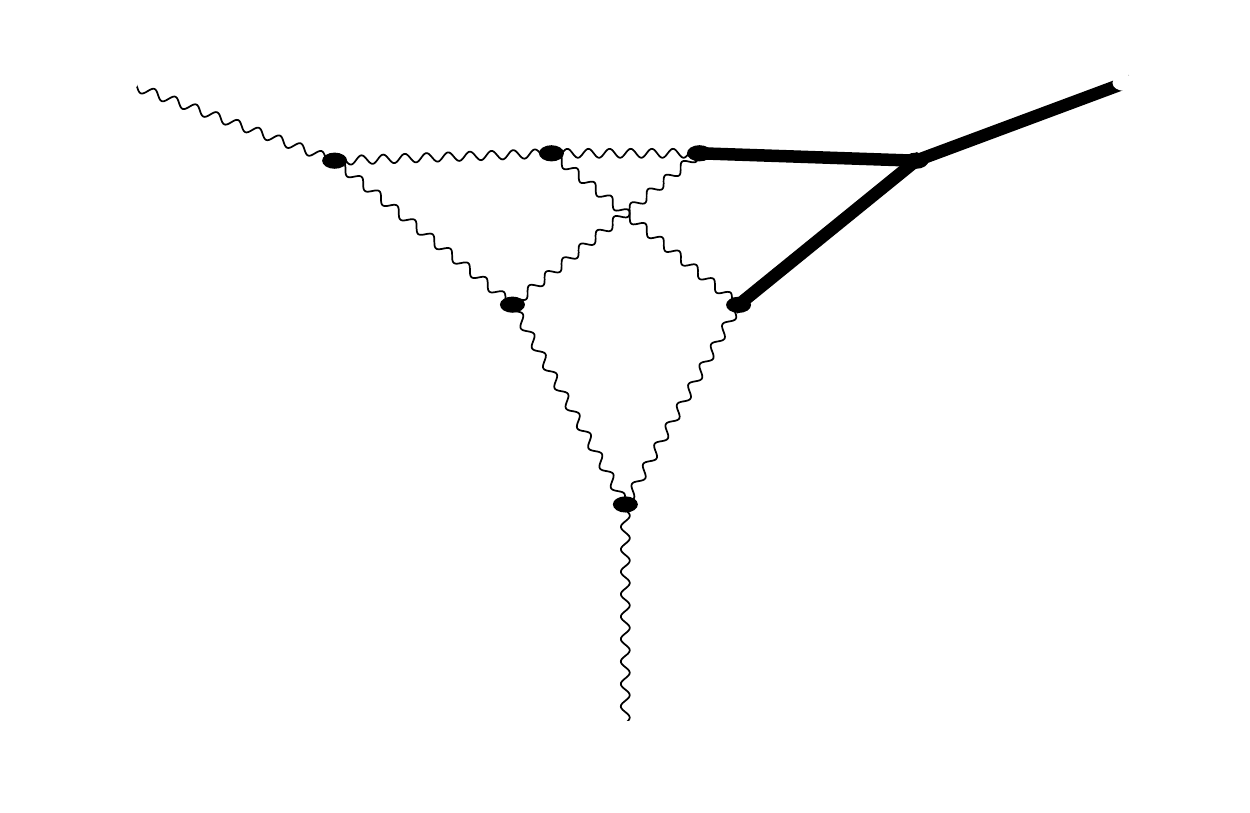}}
      \caption{Master Integrals (3/3).}
      \label{MI73_106}
\end{figure}

The form factor $\mathcal{F}$ can be expressed in terms of 95 MIs, depicted in Figs.~\ref{MI1_30}, \ref{MI31_72} and~\ref{MI73_106}.

A circled number besides the graph indicates that the corresponding denominator features in the numerator.

The MIs with an asterisk ($^*$) do not contribute to the NLO form factor, but appear in the differential equations.

\end{appendix}

\bibliographystyle{BIBLIOSTYLE}
\bibliography{biblio.bib}







\end{document}